\documentclass[preprint2,usenatbib]{aastex61}
\usepackage{natbib,longtable,amsmath}
\newcommand{\cosmo}{$H_0 = 67.8$~km~s$^{-1}$~Mpc$^{-1}$, $\Omega_M = 0.308$ and $\Omega_{\Lambda} = 0.692$ \citep{PlanckCollaboration16}}

\shorttitle{The Environments of the Most Energetic Gamma-Ray Bursts}
\shortauthors{Gompertz, Fruchter \& Pe'er}

\begin{document}

\title{The Environments of the Most Energetic Gamma-Ray Bursts}
\correspondingauthor{Benjamin Gompertz}
\email{b.gompertz@warwick.ac.uk}

\author{B. P. Gompertz}
\affil{Space Telescope Science Institute, Baltimore, MD 21218, USA}
\affil{Department of Physics, University of Warwick, Coventry, CV4 7AL, UK}

\author{A. S. Fruchter}
\affil{Space Telescope Science Institute, Baltimore, MD 21218, USA}

\author{A. Pe'er}
\affil{Physics Department, University College Cork, Cork, Ireland}

\begin{abstract}
We analyze the properties of a sample of long gamma-ray bursts (LGRBs) detected by the Fermi satellite that have a spectroscopic redshift and good follow-up coverage at both X-ray and optical/nIR wavelengths. The evolution of LGRB afterglows depends on the density profile of the external medium, enabling us to separate wind or ISM-like environments based on the observations. We do this by identifying the environment that provides the best agreement between estimates of $p$, the index of the underlying power-law distribution of electron energies, as determined by the behavior of the afterglow in different spectral/temporal regimes. At 11 rest-frame hours after trigger, we find a roughly even split between ISM-like and wind-like environments. We further find a 2$\sigma$ separation in the prompt emission energy distributions of wind-like and ISM-like bursts. We investigate the underlying physical parameters of the shock, and calculate the (degenerate) product of density and magnetic field energy ($\epsilon_B$). We show that $\epsilon_B$ must be $\ll 10^{-2}$ to avoid implied densities comparable to the intergalactic medium. Finally, we find that the most precisely constrained observations disagree on $p$ by more than would be expected based on observational errors alone. This suggests additional sources of error that are not incorporated in the standard afterglow theory. For the first time, we provide a measurement of this intrinsic error which can be represented as an error in the estimate of $p$ of magnitude $0.25 \pm 0.04$. When this error is included in the fits, the number of LGRBs with an identified environment drops substantially, but the equal division between the two types remains.
\end{abstract}

\section{Introduction}
Long gamma-ray bursts (LGRBs) are well established as core-collapse events following the deaths of massive stars, due to their close proximity to very young star forming regions \citep{Fruchter06,Levesque10a} and consistent association with type Ic supernovae \citep[e.g.][]{Hjorth03,Cano13}. Their emission is best modelled as a collimated jet of relativistic material that is launched from the central star by the initial collapse \citep[cf.][]{Woosley93}. The $\gamma$-rays are widely believed to be produced by dissipation of either kinetic \citep[e.g.][]{Paczynski86,Rees92} or magnetic \citep[e.g.][]{Usov94,Drenkhahn02,Zhang11} energy in the expanding jet.

The expanding jet collides with material in the circumstellar environment, causing it to decelerate. As it moves at relativistic speeds, a shock wave is formed \citep{Blandford76}. The shock is believed to accelerate particles to high energies as well as generate a strong magnetic field. The heated particles in turn emit a broad-band radiation (mainly via synchrotron and possibly Compton) known as the `afterglow', with emission ranging from X-rays to radio frequencies.

Broad-band fitting of LGRB afterglows suggests that they can occur in at least two different types of environment \citep{Chevalier99}: interstellar medium (ISM)-like environments, in which the density of the circumstellar material does not change with distance from the central object, or wind-like environments, in which the density falls as $r^{-2}$. The latter case should occur when the progenitor star possessed a strong stellar wind in its final stages of life before collapse \citep{Chevalier00}. The former may be the result of a progenitor star with a very weak stellar wind, so that the expanding fireball quickly crosses from the wind-like environment into a homogeneous density region at larger radii \citep[see e.g.][]{Peer06,vanMarle06}.

LGRB afterglows are generally well explained by synchrotron theory\citep{Rees92,Wijers97,Sari98,Wijers99,Granot02,Gao13}. It is assumed that electrons at the shock front are accelerated to a power-law distribution of the form $N(\gamma_e)d\gamma_e \propto \gamma_e^{-p}d\gamma_e$, above a minimum Lorentz factor $\gamma_m$. $\gamma_e$ is the electron's Lorentz factor. Here the power-law index is $p > 2$. The shape of the distribution means that the minimum Lorentz factor is also the most common, and represents the peak frequency ($\nu_m$) of the synchrotron spectrum. At sufficiently high frequencies, the radiative cooling time equals the dynamical time, which is the time available to cool. As a result, above the cooling break ($\nu_c$), the steady state distribution of the electrons, resulting from both rapid acceleration and cooling, steepens. Finally, at low frequencies the emitted photons are more easily re-absorbed by the emitting material, and this creates a third break in the synchrotron spectrum: the self-absorption frequency $\nu_a$.

The positions of these breaks are functions of the physical properties of the plasma flow, such as the energy, magnetic field, density and available time. While there is some knowledge about, e.g., the available energy, the values of parameters like the magnetic field strength and the density are highly uncertain. Nonetheless, these parameters combine to describe a spectrum comprised of four power-law segments smoothly connected by three spectral breaks: $\nu_a$, $\nu_m$ and $\nu_c$. At times of around half a day post-burst, $\nu_c$ is typically found close to X-ray frequencies in LGRBs, with $\nu_m$ at optical/near infra-red (nIR) frequencies and $\nu_a$ down at radio wavelengths.

Because the synchrotron spectrum is sensitive to density, its evolution can be used to diagnose the type of environment into which the GRB afterglow is expanding. The theoretical relation between $p$, the spectral index $\beta$, and the temporal index $\alpha$ is given by the synchrotron closure relations, which vary between environment types and in different regions of the synchrotron spectrum. We use the convention $F \propto t^{-\alpha}\nu^{-\beta}$ when discussing them.

As early as a few years after the detection of the first LGRB optical afterglow \citep{vanParadijs97} a split between ISM-like and wind-like environments was observed, with up to 50 per cent of bursts found to be consistent with a homogeneous medium \citep[e.g.][]{Chevalier00,Panaitescu01,Panaitescu02}. In later studies, \citep[e.g.][]{Starling08,Curran09} ISM-like environments continued to be found in LGRB afterglows. \citet{Schulze11} studied the environments of a sample of 26 \emph{Swift} LGRBs (and one SGRB), finding just 6 that showed a wind-like evolution at late times. 18 were classed as ISM-like. Measurements of the spectral and temporal indices for optical \citep{Oates12} and X-ray \citep{Oates15,Racusin16} afterglows of LGRBs all point to a split in environment types between wind and ISM. In the largest previous study to date, \citet{Li15} investigated the X-ray and optical slopes of a large sample of 87 GRBs (80 with redshift) discovered by the \emph{Swift} satellite up until 2013, finding that 61 per cent of them were consistent with an ISM-like medium, while just 39 per cent appeared wind-like.

Given the apparent prevalence of ISM-like environments, the signature of the afterglow emission site transitioning across a termination shock from a wind-like to an ISM-like medium \citep[e.g.][]{Dai02} should in theory be commonly observed within the LGRB afterglow population. However, despite several claims \citep{Dai03,Jin09,Feng11,Veres15}, such a signature has never been unambiguously identified. LGRBs therefore warrant further investigations into their environment types, their energies, and the underlying physical parameters of their afterglows. These properties may provide clues as to why a termination shock transition has not been observed. Obtaining a large sample with known redshift allows for well-defined energetics, which provides a previously unexplored attribute when making comparisons between the identified environment types.

In this paper, we investigate the energetics and environments of a sample of 56 \emph{Fermi}-discovered LGRBs, all of which have an identified redshift and detections in both X-rays and optical/nIR. We include bursts observed by both the Large Area Telescope \citep[LAT;][]{Atwood09} and the Gamma-ray Burst Monitor \citep[GBM;][]{Meegan09}, and those seen by the GBM only. Contrary to previous works, we discriminate between the different environment types based on several independent measures: the observed spectral and temporal indices, and the measured ratio of their optical-to-X-ray fluxes ($F_R/F_X$) where appropriate. The increased bandpass of GBM compared to \emph{Swift}-BAT also allows us to better constrain the isotropic equivalent $\gamma$-ray energy ($E_{\gamma,iso}$) of each burst and compare it across the sample. This enables us for the first time to make statistical comparisons between environment types based on burst energies.

In Section~\ref{sec:data} we introduce the data collected for the sample. Section~\ref{sec:environment} describes how we ascertain the environment type for each burst. Our results are presented in Section~\ref{sec:results}, and are discussed in Section~\ref{sec:discussion}. We outline our conclusions in Section~\ref{sec:conclusions}. We use a cosmology of \cosmo{} throughout this paper.

\section{Data}\label{sec:data}
We collected all LGRBs that were detected by the \emph{Fermi} satellite, had been observed by both \emph{Swift}-XRT and ground-based optical/nIR telescopes, and had an identified redshift. Our sample is the largest and most comprehensive collection of \emph{Fermi}-discovered GRBs with redshift and multi-wavelength observations, comprising $56$ events.

\subsection{Prompt Data}
To calculate the isotropic $\gamma$-ray energy release, $E_{\rm \gamma,iso}$, we take the GBM fluence measured in the $10$ - $1000$~keV bandpass from the \emph{Fermi}-GBM catalogue \citep{Gruber14,vonKienlin14,Bhat16}. Comparing energies between bursts of different redshifts requires a cosmological k-correction \citep{Bloom01} in order to account for the shift in the observer frame bandpass, and this requires knowledge of the spectral shape of the prompt emission. GRB prompt emission spectra are usually fitted with the Band function \citep{Band93}, comprised of a low-energy spectral index $a$, a peak energy $E_p$, and a high energy spectral index $b$. Taking the Band function spectral parameters from the \emph{Fermi}-GBM catalogue \citep{Gruber14,vonKienlin14,Bhat16}, we convert the observer frame $10$ - $1000$~keV fluence into $E_{\rm \gamma,iso}$ between $10$ - $1000$~keV in the \emph{rest} frame for each burst. This is one area in which our work has a significant advantage over studies done with \emph{Swift}-BAT, since the BAT bandpass of $15$ - $150$~keV often does not capture $E_p$, making the true prompt emission energy uncertain. Our prompt emission parameters, including k-corrected energies, are displayed in Table~\ref{tab:prompt}.

{\setlength{\tabcolsep}{1pt}
\begin{longtable*}{lccccccc}
\hline\hline
GRB & $z$ & Fluence & T$_{90}$ & E$_p$ & $a$ & $b$ & E$_{\rm \gamma,iso}$ \\
 & & (erg cm$^{-2}$) & (s) & (keV) & & & (erg) \\
\hline
\endhead
\bf{LAT} \\
080916C & $4.35^{(1)}$ & $(6.03 \pm 0.01) \times 10^{-5}$ & $63.0 \pm 0.81$ & $662 \pm 45$ & $1.08 \pm 0.01$ & $2.15 \pm 0.07$ & $(7.84 \pm 0.01) \times 10^{53}$ \\
090323 & $3.44^{(2)}$ & $(11.8 \pm 0.02) \times 10^{-5}$ & $135 \pm 1.45$ & $633 \pm 41$ & $1.28 \pm 0.01$ & $2.44 \pm 0.17$ & $(16.7 \pm 0.02) \times 10^{53}$ \\
090328A & $0.74^{(2)}$ & $(4.20 \pm 0.01) \times 10^{-5}$ & $62.0 \pm 1.81$ & $640 \pm 46$ & $1.09 \pm 0.02$ & $2.37 \pm 0.18$ & $(4.90 \pm 0.01) \times 10^{52}$ \\
090902B & $1.24^{(2)}$ & $(22.2 \pm 0.03) \times 10^{-5}$ & $19.3 \pm 0.29$ & $1055 \pm 17$ & $1.01 \pm 0.004$ & $11.8 \pm 4018$ & $(8.98 \pm 0.01) \times 10^{53}$ \\
090926A & $2.11^{(3)}$ & $(14.7 \pm 0.03) \times 10^{-5}$ & $13.8 \pm 0.29$ & $340 \pm 6$ & $0.86 \pm 0.01$ & $2.40 \pm 0.04$ & $(9.50 \pm 0.02) \times 10^{53}$ \\
091003 & $0.90^{(4)}$ & $(2.33 \pm 0.01) \times 10^{-5}$ & $20.2 \pm 0.36$ & $367 \pm 27$ & $1.07 \pm 0.02$ & $2.23 \pm 0.11$ & $(4.20 \pm 0.01) \times 10^{52}$ \\
091208B & $1.06^{(5)}$ & $(6.19 \pm 0.19) \times 10^{-6}$ & $12.5 \pm 5.02$ & $38 \pm 6$ & $0.15 \pm 0.39$ & $1.90 \pm 0.04$ & $(1.52 \pm 0.05) \times 10^{52}$ \\
100414A  & $1.37^{(6)}$ & $(8.85 \pm 0.02) \times 10^{-5}$ & $26.5 \pm 2.07$ & $664 \pm 16$ & $0.62 \pm 0.01$ & $3.54 \pm 0.49$ & $(21.0 \pm 0.04) \times 10^{52}$ \\
100728A & $1.57^{(7)}$ & $(1.28 \pm 0.01) \times 10^{-4}$ & $165 \pm 2.90$ & $290 \pm 8$ & $0.64 \pm 0.02$ & $2.70 \pm 0.15$ & $(5.34 \pm 0.02) \times 10^{53}$ \\
110731A & $2.83^{(8)}$ & $(2.29 \pm 0.01) \times 10^{-8}$ & $7.49 \pm 0.57$ & $319 \pm 20$ & $0.87 \pm 0.03$ & $2.44 \pm 0.16$ & $(2.19 \pm 0.01) \times 10^{53}$ \\
120711A & $1.41^{(9)}$ & $(19.4 \pm 0.02) \times 10^{-5}$ & $44.0 \pm 0.72$ & $1319 \pm 46$ & $0.98 \pm 0.01$ & $2.80 \pm 0.09$ & $(5.17 \pm 0.01) \times 10^{53}$ \\
130427A & $0.34^{(10)}$ & $(24.6 \pm 0.01) \times 10^{-4}$ & $138 \pm 3.24$ & $830 \pm 5^a$ & $0.79 \pm 0.003^a$ & $3.06 \pm 0.02^a$ & $(60.8 \pm 0.03) \times 10^{52}$ \\
130518A & $2.49^{(11)}$ & $(9.46 \pm 0.02) \times 10^{-5}$ & $48.6 \pm 0.92$ & $398 \pm 16$ & $0.91 \pm 0.02$ & $2.25 \pm 0.07$ & $(7.21 \pm 0.01) \times 10^{53}$ \\
130907A & $1.24^{(12)}$ & $(7.9 \pm 0.5) \times 10^{-4 \, \dagger}$ & - & $394 \pm 11^b$ & $0.65 \pm 0.03^b$ & $2.22 \pm 0.05^b$ & $(2.32 \pm 0.15) \times 10^{54}$ \\
131108A & $2.4^{(13)}$ & $(3.57 \pm 0.01) \times 10^{-5}$ & $18.2 \pm 0.57$ & $367 \pm 18$ & $0.91 \pm 0.02$ & $2.46 \pm 0.15$ & $(2.73 \pm 0.01) \times 10^{53}$ \\
131231A & $0.64^{(14)}$ & $(15.2 \pm 0.01) \times 10^{-5}$ & $31.2 \pm 0.57$ & $178 \pm 4$ & $1.21 \pm 0.01$ & $2.30 \pm 0.03$ & $(16.1 \pm 0.01) \times 10^{52}$ \\
141028A & $2.33^{(15)}$ & $(3.48 \pm 0.01) \times 10^{-5}$ & $31.5 \pm 2.43$ & $294 \pm 16$ & $0.84 \pm 0.03$ & $1.97 \pm 0.04$ & $(2.51 \pm 0.01) \times 10^{53}$ \\
150314A & $1.76^{(16)}$ & $(8.16 \pm 0.01) \times 10^{-5}$ & $10.7 \pm 0.14$ & $347 \pm 7$ & $0.68 \pm 0.01$ & $2.60 \pm 0.08$ & $(3.73 \pm 0.01) \times 10^{53}$ \\
150403A & $2.06^{(17)}$ & $(5.47 \pm 0.01) \times 10^{-5}$ & $22.3 \pm 0.81$ & $429 \pm 17$ & $0.87 \pm 0.02$ & $2.11 \pm 0.04$ & $(30.9 \pm 0.03) \times 10^{52}$ \\
150514A & $0.81^{(18)}$ & $(4.74 \pm 0.05) \times 10^{-6}$ & $10.8 \pm 1.07$ & $65 \pm 6$ & $1.21 \pm 0.10$ & $2.43 \pm 0.13$ & $(8.67 \pm 0.09) \times 10^{51}$ \\
160509A & $1.17^{(19)}$ & $(17.9 \pm 0.02) \times 10^{-5}$ & $370 \pm 0.81$ & $355 \pm 10$ & $1.02 \pm 0.01$ & $2.23 \pm 0.04$ & $(50.7 \pm 0.04) \times 10^{52}$ \\
160623A & $0.37^{(20)}$ & $(3.96 \pm 0.07) \times 10^{-6}$ & $108 \pm 8.69$ & $999 \pm 4630$ & $1.49 \pm 0.17$ & $1.61 \pm 0.15$ & $(1.25 \pm 0.02) \times 10^{51}$ \\
160625B & $1.41^{(21)}$ & $(5.69 \pm 0.01) \times 10^{-4}$ & $455 \pm 11.0$ & $649 \pm 9$ & $0.95 \pm 0.003$ & $2.37 \pm 0.03$ & $(17.7 \pm 0.03) \times 10^{53}$ \\
\hline
\bf{GBM} \\
080916A & $0.69^{(22)}$ & $(7.81 \pm 0.08) \times 10^{-6}$ & $46.3 \pm 7.17$ & $107 \pm 19$ & $0.82 \pm 0.13$ & $1.78 \pm 0.05$ & $(8.38 \pm 0.09) \times 10^{51}$ \\
081007 & $0.53^{(23)}$ & $(1.20 \pm 0.10) \times 10^{-6 \, c}$ & $\sim 12^c$ & $40 \pm 10^c$ & $1.4 \pm 0.4^c$ & $8^{c \, +}$ & $(1.06 \pm 0.09) \times 10^{51}$ \\
081121 & $2.51^{(24)}$ & $(1.53 \pm 0.02) \times 10^{-5}$ & $42.0 \pm 8.51$ & $161 \pm 17$ & $0.43 \pm 0.13$ & $2.09 \pm 0.09$ & $(1.26 \pm 0.02) \times 10^{53}$ \\
090423 & $8.23^{(25)}$ & $(8.16 \pm 0.72) \times 10^{-7}$ & $7.17 \pm 2.42$ & $66 \pm 16$ & $0.59 \pm 0.50$ & $2.67 \pm 0.74$ & $(4.45 \pm 0.39) \times 10^{52}$ \\
090424 & $0.54^{(26)}$ & $(46.3 \pm 0.04) \times 10^{-6}$ & $14.1 \pm 0.26$ & $154 \pm 4$ & $1.04 \pm 0.02$ & $2.76 \pm 0.12$ & $(36.2 \pm 0.03) \times 10^{51}$ \\
090618 & $0.54^{(27)}$ & $(26.8 \pm 0.04) \times 10^{-5}$ & $112 \pm 1.09$ & $147 \pm 4$ & $1.13 \pm 0.01$ & $2.22 \pm 0.02$ & $(20.0 \pm 0.03) \times 10^{52}$ \\
091020 & $1.71^{(28)}$ & $(1.07 \pm 0.01) \times 10^{-5}$ & $37.5 \pm 0.91$ & $1193 \pm 239$ & $0.99 \pm 0.05$ & $2.71 \pm 0.61$ & $(5.78 \pm 0.10) \times 10^{52}$ \\
091127 & $0.49^{(29)}$ & $(20.7 \pm 3.70) \times 10^{-6}$ & $8.70 \pm 0.57$ & $35 \pm 2$ & $1.26 \pm 0.07$ & $2.22 \pm 0.02$ & $(13.8 \pm 0.02) \times 10^{51}$ \\
100906A & $1.73^{(30)}$ & $(2.33 \pm 0.01) \times 10^{-5}$ & $111 \pm 2.83$ & $70 \pm 10$ & $0.90 \pm 0.13$ & $1.86 \pm 0.03$ & $(13.3 \pm 0.03) \times 10^{52}$ \\
101219B & $0.55^{(31)}$ & $(3.99 \pm 0.05) \times 10^{-6}$ & $51.0 \pm 1.78$ & $56 \pm 7$ & $-1.37 \pm 0.72$ & $2.26 \pm 0.14$ & $(2.85 \pm 0.03) \times 10^{51}$ \\
110213A & $1.46^{(32)}$ & $(9.37 \pm 0.05) \times 10^{-6}$ & $34.3 \pm 1.64$ & $75 \pm 13$ & $1.42 \pm 0.09$ & $2.13 \pm 0.09$ & $(5.13 \pm 0.03) \times 10^{52}$ \\
111228A & $0.71^{(33)}$ & $(1.81 \pm 0.01) \times 10^{-5}$ & $99.8 \pm 2.11$ & $27 \pm 1$ & $1.58 \pm 0.08$ & $2.44 \pm 0.06$ & $(3.00 \pm 0.01) \times 10^{52}$ \\
120119A & $1.73^{(34)}$ & $(3.87 \pm 0.01) \times 10^{-5}$ & $55.3 \pm 6.23$ & $183 \pm 8$ & $0.96 \pm 0.03$ & $2.37 \pm 0.09$ & $(2.36 \pm 0.01) \times 10^{53}$ \\
120729A & $0.80^{(35)}$ & $(5.08 \pm 0.05) \times 10^{-6}$ & $25.5 \pm 2.61$ & $26 \pm 168$ & $0.06 \pm 30.5$ & $1.62 \pm 0.04$ & $(6.83 \pm 0.07) \times 10^{51}$ \\
120811C & $2.67^{(36)}$ & $(3.45 \pm 0.21) \times 10^{-6}$ & $14.3 \pm 6.56$ & $56 \pm 5$ & $0.71 \pm 0.27$ & $2.85 \pm 0.32$ & $(5.40 \pm 0.33) \times 10^{52}$ \\
121211A & $1.02^{(37)}$ & $(6.41 \pm 0.40) \times 10^{-7}$ & $5.62 \pm 1.72$ & $100 \pm 16$ & $0.27 \pm 0.37$ & $4.80 \pm 15.9$ & $(1.79 \pm 0.11) \times 10^{51}$ \\
130420A & $1.30^{(38)}$ & $(1.16 \pm 0.02) \times 10^{-5}$ & $105 \pm 8.81$ & $57 \pm 3$ & $1.13 \pm 0.12$ & $10.6 \pm 1405$ & $(5.99 \pm 0.13) \times 10^{52}$ \\
140213A & $1.21^{(39)}$ & $(2.12 \pm 0.01) \times 10^{-5}$ & $18.6 \pm 0.72$ & $87 \pm 4$ & $1.13 \pm 0.03$ & $2.26 \pm 0.05$ & $(7.82 \pm 0.02) \times 10^{52}$ \\
140423A & $3.26^{(40)}$ & $(1.81 \pm 0.01) \times 10^{-5}$ & $95.2 \pm 11.6$ & $121 \pm 15$ & $0.60 \pm 0.12$ & $1.83 \pm 0.05$ & $(2.09 \pm 0.01) \times 10^{53}$ \\
140506A & $0.89^{(41)}$ & $(6.59 \pm 0.12) \times 10^{-6}$ & $64.1 \pm 2.01$ & $198 \pm 33$ & $1.18 \pm 0.11$ & $9.80 \pm 2845$ & $(1.43 \pm 0.03) \times 10^{52}$ \\
140512A & $0.73^{(42)}$ & $(2.93 \pm 0.01) \times 10^{-5}$ & $148 \pm 2.36$ & $683 \pm 70$ & $1.22 \pm 0.02$ & $3.24 \pm 1.64$ & $(3.46 \pm 0.01) \times 10^{52}$ \\
140606B & $0.38^{(43)}$ & $(7.59 \pm 0.04) \times 10^{-6}$ & $22.8 \pm 2.06$ & $555 \pm 165$ & $1.24 \pm 0.05$ & $2.20 \pm 0.52$ & $(2.71 \pm 0.02) \times 10^{51}$ \\
140620A & $2.04^{(44)}$ & $(6.15 \pm 0.06) \times 10^{-6}$ & $45.8 \pm 12.1$ & $76 \pm 11$ & $0.93 \pm 0.16$ & $2.15 \pm 0.09$ & $(5.24 \pm 0.05) \times 10^{52}$ \\
140703A & $3.14^{(45)}$ & $(7.57 \pm 0.05) \times 10^{-6}$ & $84.0 \pm 3.00$ & $208 \pm 36$ & $1.27 \pm 0.06$ & $2.61 \pm 0.79$ & $(1.32 \pm 0.01) \times 10^{53}$ \\
140801A & $1.32^{(46)}$ & $(12.4 \pm 0.04) \times 10^{-6}$ & $7.17 \pm 0.57$ & $119 \pm 3$ & $0.38 \pm 0.04$ & $3.93 \pm 0.66$ & $(5.32 \pm 0.02) \times 10^{52}$ \\
140907A & $1.21^{(47)}$ & $(6.45 \pm 0.06) \times 10^{-6}$ & $35.8 \pm 5.47$ & $141 \pm 10$ & $1.03 \pm 0.06$ & $14.7 \pm 10^6$ & $(2.58 \pm 0.03) \times 10^{52}$ \\
141004A & $0.57^{(48)}$ & $(1.18 \pm 0.03) \times 10^{-6}$ & $2.56 \pm 0.61$ & $28 \pm 11$ & $-0.10 \pm 1.68$ & $1.91 \pm 0.08$ & $(9.27 \pm 0.26) \times 10^{50}$ \\
150301B & $1.52^{(49)}$ & $(3.09 \pm 0.03) \times 10^{-6}$ & $13.3 \pm 1.56$ & $183 \pm 36$ & $1.05 \pm 0.10$ & $2.22 \pm 0.28$ & $(1.52 \pm 0.01) \times 10^{52}$ \\
150821A & $0.76^{(50)}$ & $(5.21 \pm 0.03) \times 10^{-5}$ & $103 \pm 5.75$ & $281 \pm 18$ & $1.24 \pm 0.02$ & $2.13 \pm 0.07$ & $(7.21 \pm 0.04) \times 10^{52}$ \\
151027A & $0.81^{(51)}$ & $(1.41 \pm 0.01) \times 10^{-5}$ & $123 \pm 1.15$ & $203 \pm 34$ & $1.25 \pm 0.05$ & $1.96 \pm 0.08$ & $(2.21 \pm 0.01) \times 10^{52}$ \\
160804A & $0.74^{(52)}$ & $(1.62 \pm 0.02) \times 10^{-5}$ & $132 \pm 21.7$ & $71 \pm 4$ & $1.03 \pm 0.09$ & $2.82 \pm 0.90$ & $(2.47 \pm 0.04) \times 10^{52}$ \\
161017A & $2.01^{(53)}$ & $(5.42 \pm 0.13) \times 10^{-6}$ & $32.3 \pm 8.08$ & $239 \pm 41$ & $1.03 \pm 0.10$ & $2.37 \pm 0.78$ & $(4.05 \pm 0.09) \times 10^{52}$ \\
170113A & $1.97^{(54)}$ & $(2.04 \pm 0.08) \times 10^{-6}$ & $49.2 \pm 4.14$ & $113 \pm 59$ & $1.70 \pm 0.22$ & $10^{11}$ & $(2.59 \pm 0.10) \times 10^{52}$ \\
\hline\hline
\caption{The prompt emission properties of the sample. Fluence is $10$ -- $1000$~keV in the observer frame unless marked. $E_{\rm \gamma,iso}$ values have been k-corrected \citep{Bloom01} to $10$ -- $1000$~keV in the rest frame. Fluence, T$_{90}$, E$_p$, $\alpha$ and $\beta$ come from the \emph{Fermi}-GBM catalogue \citep{Gruber14,vonKienlin14,Bhat16} unless otherwise marked.\\$^a$\citet{vonKienlin13}; $^b$\citet{Golenetskii13}; $^c$\citet{Bissaldi08}; $^{\dagger}$Seen by LAT but not by GBM, so the tabulated fluence is from \emph{Konus-Wind} in the $20$~keV -- $10$~MeV range; $^+$The best fit is an exponential cutoff, which we model with an index of 8.\\Redshift references: (1) - \citet{Greiner09}; (2) - \citet{McBreen10}; (3) - \citet{D'Elia10}; (4) - \citet{Cucchiara09}; (5) - \citet{Wiersema09b}; (6) - \citet{Cucchiara10}; (7) - \citet{Kruhler15a}; (8) - \citet{Tanvir11}; (9) - \citet{Tanvir12}; (10) - \citet{Levan14b}; (11) - \citet{Sanchez-Ramirez13}; (12) - \citet{deUgartePostigo13}; (13) - \citet{deUgartePostigo13b}; (14) - \citet{Xu14}; (15) - \citet{Xu14b}; (16) - \citet{deUgartePostigo15}; (17) - \citet{Pugliese15}; (18) - \citet{deUgartePostigo15b}; (19) - \citet{Tanvir16}; (20) - \citet{Malesani16}; (21) - \citet{Xu16}; (22) - \citet{Fynbo08}; (23) - \citet{Berger08}; (24) - \citet{Berger08b}; (25) - \citet{Tanvir09}; (26) - \citet{Chornock09}; (27) - \citet{Cenko09}; (28) - \citet{Xu09b}; (29) - \citet{Cucchiara09b}; (30) - \citet{Tanvir10}; (31) - \citet{deUgartePostigo11}; (32) - \citet{Milne11}; (33) - \citet{Dittmann11}; (34) - \citet{Cucchiara12}; (35) - \citet{Tanvir12b}; (36) - \citet{Thoene12}; (37) - \citet{Perley12}; (38) - \citet{deUgartePostigo13c}; (39) - \citet{Schulze14}; (40) - \citet{Tanvir14}; (41) - \citet{Fynbo14a}; (42) - \citet{deUgartePostigo14b}; (43) - \citet{Singer15}; (44) - \citet{Kasliwal14}; (45) - \citet{Castro-Tirado14}; (46) - \citet{deUgartePostigo14d}; (47) - \citet{Castro-Tirado14b}; (48) - \citet{deUgartePostigo14c}; (49) - \citet{deUgartePostigo15c}; (50) - \citet{D'Elia15}; (51) - \citet{Perley15}; (52) - \citet{Xu16b}; (53) - \citet{deUgartePostigo16}; (54) - \citet{Xu17} \label{tab:prompt}}
\end{longtable*}
}

\subsection{X-ray Data}
The X-ray data (Table~\ref{tab:x-ray}) come almost exclusively from the UK \emph{Swift} Science Data Centre \citep[UKSSDC;][]{Evans07,Evans09}\footnote{www.swift.ac.uk}. For each GRB, the data point closest to $11$~h in the rest frame is identified. The \emph{Swift}-XRT GRB catalog automatically fits the X-ray light curves with a broken power law model, and we take the temporal and spectral\footnote{The UKSSDC actually gives the photon index, $\Gamma_x$, where the spectral index $\beta_x = \Gamma_x - 1$} indices for the power law segment local to the selected data point. Absorption is accounted for by taking the ratio of the unabsorbed counts-to-flux over the observed counts-to-flux from the spectrum of the local power law, and applying it as a multiplication factor to the flux. We then convert to flux density via \citep[cf.][]{Gehrels08}:
\begin{equation}
F_{\rm \nu,x} = 4.13 \times 10^{11} \frac{F_{\rm x}(2 - \Gamma_x)E_0^{1-\Gamma_x}}{(E_2^{2-\Gamma_x} - E_1^{2-\Gamma_x})} \mbox{ $\mu$Jy,}
\end{equation}
where $E_1$ and $E_2$ are the lower and upper bounds of the bandpass in keV, $E_0$ is the target energy for the flux density in keV, $F_{\rm x}$ is the measured flux in erg~cm$^{-2}$~s$^{-1}$, and $\Gamma_x$ is the estimated unabsorbed X-ray photon index. A k-correction \citep[cf.][]{Bloom01} is applied to the flux density to account for the disparate redshifts of the rest frame bandpass between bursts, which keeps the results comparable between GRBs at different cosmological distances. Note that these may not be the same values for k as in the $E_{\rm iso}$ calculations in Table~\ref{tab:prompt} due to the spectral break in the Band function. The flux densities are then extrapolated to $11$ hours rest-frame using the identified temporal index $\alpha_x$, and to $5$~keV in the rest frame ($5 \times (1 + z)$~keV) using $\beta_x$. We use 5~keV because at this energy the error in the flux density, caused by the uncertainty in the estimate of intrinsic absorption, is relatively small.

Due to the mis-identification of flares, for some bursts the UKSSDC automatic fitting routine gives obviously erroneous results. For these cases, we manually fit the X-ray light curves to obtain the local temporal index, then run the UKSSDC time slice spectrum routine for the identified power law segment to ascertain the local spectral index and counts-to-flux ratio. The affected bursts are GRB 100728A and GRB 120119A. For four bursts in the sample, we have additional X-ray observations from \emph{Chandra} (Fruchter et al. in prep). In these cases, we again fit the light curves manually to find the temporal indices, and use the UKSSDC time slice spectrum tool to find the spectral index for the local power law segment. The four bursts are GRBs 110731A, 120711A, 130427A and 150314A.

\begin{longtable*}{lcccccccc}
\hline\hline
GRB & $t_{\rm x,rest}$ & F$_{\rm x\, [0.3-10keV]}$ & $\alpha_{\rm x}$ & $\Gamma_x$ & Absorption & k & F$_{\rm \nu,x\, [5keV]}$ \\
& (h) & (erg cm$^{-2}$ s$^{-1}$) & & & Correction & & ($\mu$Jy) \\
\hline
\endhead
\bf{LAT} \\
080916C & $9.66$ & $(1.41 \pm 0.39) \times 10^{-13}$ & $1.31^{+0.09}_{-0.08}$ & $1.80^{+0.30}_{-0.24}$ & $1.44$ & $0.13$ & $(2.43 \pm 0.66) \times 10^{-3}$ \\
090323 & $11.20$ & $(2.88 \pm 0.71) \times 10^{-13}$ & $1.61^{+0.09}_{-0.09}$ & $1.88^{+0.12}_{-0.13}$ & $1.04$ & $0.25$ & $(5.64 \pm 1.38) \times 10^{-3}$ \\
090328A & $10.11$ & $(2.80 \pm 0.58) \times 10^{-12}$ & $1.69^{+0.08}_{-0.07}$ & $1.55^{+0.19}_{-0.18}$ & $1.19$ & $0.61$ & $(6.02 \pm 1.24) \times 10^{-2}$ \\
090902B & $10.90$ & $(1.58 \pm 0.30) \times 10^{-12}$ & $1.40^{+0.04}_{-0.03}$ & $1.74^{+0.09}_{-0.08}$ & $1.27$ & $0.27$ & $(3.45 \pm 0.64) \times 10^{-2}$ \\
090926A & $12.91$ & $(7.90 \pm 2.08) \times 10^{-13}$ & $1.41^{+0.03}_{-0.03}$ & $1.98^{+0.09}_{-0.09}$ & $1.13$ & $0.37$ & $(2.54 \pm 0.67) \times 10^{-2}$ \\
091003 & $9.36$ & $(2.28 \pm 0.53) \times 10^{-12}$ & $1.36^{+0.04}_{-0.04}$ & $1.69^{+0.10}_{-0.10}$ & $1.14$ & $0.43$ & $(4.35 \pm 1.00) \times 10^{-2}$ \\
091208B & $10.63$ & $(6.77 \pm 1.08) \times 10^{-13}$ & $1.09^{+0.03}_{-0.04}$ & $1.89^{+0.09}_{-0.09}$ & $1.37$ & $0.44$ & $(2.02 \pm 0.32) \times 10^{-2}$ \\
100414A  & $20.56$ & $(9.28 \pm 2,32) \times 10^{-13}$ & $2.53^{+0.12}_{-0.27}$ & $1.50^{+0.24}_{-0.24}$ & $1.11$ & $0.27$ & $(7.34 \pm 1.84) \times 10^{-2}$ \\
100728A & $9.40$ & $(1.09 \pm 0.26) \times 10^{-12}$ & $1.56^{+0.03\, \dagger}_{-0.03}$ & $1.79^{+0.05}_{-0.05}$ & $1.41$ & $0.30$ & $(2.33 \pm 0.56) \times 10^{-2}$ \\
110731A & $10.47$ & $(3.10 \pm 0.86) \times 10^{-13}$ & $1.21^{+0.01\, \dagger}_{-0.01}$ & $1.83^{+0.07}_{-0.06}$ & $1.35$ & $0.20$ & $(7.01 \pm 1.93) \times 10^{-3}$ \\
120711A & $11.62$ & $(6.25 \pm 1.41) \times 10^{-12}$ & $1.64^{+0.01\, \dagger}_{-0.01}$ & $1.77^{+0.04}_{-0.04}$ & $1.34$ & $0.34$ & $(1.80 \pm 0.41) \times 10^{-1}$ \\
130427A & $11.00$ & $(5.34 \pm 1.21) \times 10^{-11}$ & $1.32^{+0.01\, \dagger}_{-0.01}$ & $1.69^{+0.02}_{-0.02}$ & $1.12$ & $0.68$ & $(1.55 \pm 0.35) \times 10^0$ \\
130518A & $5.94$ & $(7.96 \pm 1.52) \times 10^{-13}$ & $1.26^{+0.13}_{-0.10}$ & $2.06^{+0.22}_{-0.15}$ & $1.50$ & $0.31$ & $(1.41 \pm 0.27) \times 10^{-2}$ \\
130907A & $10.93$ & $(7.16 \pm 1.36) \times 10^{-12}$ & $1.69^{+0.01}_{-0.01}$ & $1.87^{+0.02}_{-0.02}$ & $1.27$ & $0.46$ & $(1.94 \pm 0.37) \times 10^{-1}$ \\
131108A & $12.04$ & $(7.88 \pm 1.82) \times 10^{-14}$ & $1.33^{+0.04}_{-0.04}$ & $1.97^{+0.12}_{-0.11}$ & $1.22$ & $0.32$ & $(2.41 \pm 0.56) \times 10^{-3}$ \\
131231A & $24.30$ & $(4.50 \pm 0.97) \times 10^{-12}$ & $1.45^{+0.04}_{-0.03}$ & $1.79^{+0.08}_{-0.08}$ & $1.17$ & $0.55$ & $(3.90 \pm 0.84) \times 10^{-1}$ \\
141028A & $10.04$ & $(3.44 \pm 0.88) \times 10^{-13}$ & $0.70^{+0.36}_{-0.49}$ & $2.40^{+0.55}_{-0.43}$ & $1.67$ & $0.49$ & $(2.03 \pm 0.52) \times 10^{-2}$ \\
150314A & $10.69$ & $(9.96 \pm 1.92) \times 10^{-13}$ & $1.50^{+0.06\, \dagger}_{-0.06}$ & $1.84^{+0.05}_{-0.05}$ & $1.27$ & $0.31$ & $(2.43 \pm 0.47) \times 10^{-2}$ \\
150403A & $11.04$ & $(4.33 \pm 0.98) \times 10^{-12}$ & $1.51^{+0.02}_{-0.02}$ & $1.80^{+0.04}_{-0.04}$ & $1.17$ & $0.23$ & $(9.23 \pm 2.08) \times 10^{-2}$ \\
150514A & $7.23$ & $(1.27 \pm 0.33) \times 10^{-12}$ & $1.26^{+0.18}_{-0.14}$ & $1.80^{+0.33}_{-0.12}$ & $1.04$ & $0.49$ & $(1.75 \pm 0.46) \times 10^{-2}$ \\
160509A & $11.20$ & $(9.11 \pm 2.06) \times 10^{-12}$ & $1.26^{+0.04}_{-0.04}$ & $2.02^{+0.06}_{-0.06}$ & $1.71$ & $0.48$ & $(3.80 \pm 0.86) \times 10^{-1}$ \\
160623A & $10.49$ & $(3.44 \pm 0.61) \times 10^{-11}$ & $1.67^{+0.06}_{-0.06}$ & $1.85^{+0.09}_{-0.09}$ & $1.90$ & $0.70$ & $(1.51 \pm 0.27) \times 10^0$ \\
160625B & $11.16$ & $(9.37 \pm 2.11) \times 10^{-12}$ & $1.21^{+0.02}_{-0.02}$ & $1.85^{+0.06}_{-0.06}$ & $1.22$ & $0.37$ & $(2.41 \pm 0.54) \times 10^{-1}$ \\
\hline
\bf{GBM} \\
080916A & $11.44$ & $(5.68 \pm 1.23) \times 10^{-13}$ & $1.24^{+0.05}_{-0.05}$ & $2.13^{+0.10}_{-0.10}$ & $1.61$ & $0.63$ & $(2.22 \pm 0.48) \times 10^{-2}$ \\
081007 & $10.86$ & $(1.05 \pm 0.23) \times 10^{-12}$ & $1.18^{+0.07}_{-0.04}$ & $2.02^{+0.12}_{-0.15}$ & $1.55$ & $0.65$ & $(3.76 \pm 0.82) \times 10^{-2}$ \\
081121 & $10.84$ & $(9.52 \pm 2.49) \times 10^{-13}$ & $1.42^{+0.01}_{-0.01}$ & $1.80^{+0.05}_{-0.04}$ & $1.11$ & $0.24$ & $(1.78 \pm 0.47) \times 10^{-2}$ \\
090423 & $9.44$ & $(1.91 \pm 0.59) \times 10^{-14}$ & $1.41^{+0.05}_{-0.04}$ & $1.86^{+0.12}_{-0.12}$ & $1.18$ & $0.08$ & $(2.67 \pm 0.82) \times 10^{-4}$ \\
090424 & $6.57$ & $(7.19 \pm 1.42) \times 10^{-12}$ & $1.09^{+0.02}_{-0.02}$ & $1.84^{+0.05}_{-0.05}$ & $1.31$ & $0.60$ & $(1.28 \pm 0.25) \times 10^{-1}$ \\
090618 & $10.76$ & $(7.02 \pm 1.36) \times 10^{-12}$ & $1.74^{+0.04\, (1)}_{-0.04}$ & $1.80^{+0.03}_{-0.02}$ & $1.23$ & $0.60$ & $(2.03 \pm 0.39) \times 10^{-1}$ \\
091020 & $11.70$ & $(2.74 \pm 0.53) \times 10^{-13}$ & $1.37^{+0.02}_{-0.02}$ & $2.06^{+0.06}_{-0.06}$ & $1.26$ & $0.39$ & $(9.30 \pm 1.81) \times 10^{-3}$ \\
091127 & $9.69$ & $(9.95 \pm 1.65) \times 10^{-12}$ & $1.53^{+0.02}_{-0.02}$ & $1.71^{+0.05}_{-0.05}$ & $1.14$ & $0.60$ & $(2.28 \pm 0.38) \times 10^{-1}$ \\
100906A & $9.68$ & $(2.52 \pm 0.55) \times 10^{-13}$ & $1.99^{+0.04}_{-0.04}$ & $1.97^{+0.07}_{-0.07}$ & $1.53$ & $0.36$ & $(6.92 \pm 1.50) \times 10^{-3}$ \\
101219B & $10.45$ & $(4.58 \pm 0.91) \times 10^{-13}$ & $0.65^{+0.03}_{-0.03}$ & $2.16^{+0.15}_{-0.13}$ & $1.30$ & $0.69$ & $(1.30 \pm 0.26) \times 10^{-2}$ \\
110213A & $11.19$ & $(6.97 \pm 1.83) \times 10^{-13}$ & $1.93^{+0.04}_{-0.04}$ & $1.98^{+0.05}_{-0.05}$ & $1.51$ & $0.40$ & $(2.53 \pm 0.67) \times 10^{-2}$ \\
111228A & $11.03$ & $(2.56 \pm 0.67) \times 10^{-12}$ & $1.14^{+0.02}_{-0.02}$ & $1.94^{+0.05}_{-0.05}$ & $1.29$ & $0.56$ & $(7.73 \pm 2.01) \times 10^{-2}$ \\
120119A & $11.58$ & $(4.75 \pm 0.94) \times 10^{-13}$ & $1.89^{+0.11\, \dagger}_{-0.11}$ & $1.73^{+0.07}_{-0.09}$ & $1.33$ & $0.28$ & $(1.23 \pm 0.24) \times 10^{-2}$ \\
120729A & $6.57$ & $(3.49 \pm 0.96) \times 10^{-14}$ & $2.96^{+0.17}_{-0.15}$ & $2.08^{+0.22}_{-0.16}$ & $1.55$ & $0.58$ & $(2.77 \pm 0.76) \times 10^{-4}$ \\
120811C & $5.87$ & $(4.36 \pm 1.07) \times 10^{-13}$ & $1.19^{+0.09}_{-0.08}$ & $2.12^{+0.13}_{-0.12}$ & $1.27$ & $0.32$ & $(7.43 \pm 1.82) \times 10^{-3}$ \\
121211A & $7.90$ & $(9.59 \pm 1.82) \times 10^{-13}$ & $1.30^{+0.18}_{-0.07}$ & $1.83^{+0.15}_{-0.15}$ & $1.24$ & $0.48$ & $(1.70 \pm 0.32) \times 10^{-2}$ \\
130420A & $11.39$ & $(6.38 \pm 1.68) \times 10^{-13}$ & $1.14^{+0.04}_{-0.04}$ & $2.07^{+0.12}_{-0.11}$ & $1.24$ & $0.46$ & $(2.00 \pm 0.53) \times 10^{-2}$ \\
140213A & $10.76$ & $(4.09 \pm 0.81) \times 10^{-12}$ & $0.99^{+0.02}_{-0.02}$ & $1.86^{+0.04}_{-0.02}$ & $1.27$ & $0.41$ & $(1.11 \pm 0.22) \times 10^{-1}$ \\
140423A & $11.91$ & $(2.04 \pm 0.53) \times 10^{-13}$ & $1.49^{+0.10}_{-0.09}$ & $1.94^{+0.15}_{-0.13}$ & $1.11$ & $0.22$ & $(5.47 \pm 1.41) \times 10^{-3}$ \\
140506A & $11.77$ & $(3.12 \pm 0.83) \times 10^{-12}$ & $0.98^{+0.02}_{-0.02}$ & $1.87^{+0.06}_{-0.05}$ & $1.41$ & $0.49$ & $(1.08 \pm 0.29) \times 10^{-1}$ \\
140512A & $10.38$ & $(6.48 \pm 1.17) \times 10^{-12}$ & $1.67^{+0.05}_{-0.05}$ & $1.88^{+0.05}_{-0.05}$ & $1.33$ & $0.54$ & $(1.81 \pm 0.33) \times 10^{-1}$ \\
140606B & $37.22$ & $(6.16 \pm 1.25) \times 10^{-13}$ & $0.90^{+0.43}_{-0.36}$ & $1.90^{+0.55}_{-0.43}$ & $1.56$ & $0.70$ & $(7.02 \pm 1.42) \times 10^{-2}$ \\
140620A & $11.39$ & $(7.68 \pm 2.02) \times 10^{-13}$ & $1.53^{+0.10}_{-0.09}$ & $2.00^{+0.15}_{-0.15}$ & $1.29$ & $0.33$ & $(2.47 \pm 0.65) \times 10^{-2}$ \\
140703A & $5.29$ & $(3.01 \pm 0.49) \times 10^{-13}$ & $1.74^{+0.10}_{-0.09}$ & $1.90^{+0.09}_{-0.06}$ & $1.24$ & $0.21$ & $(2.04 \pm 0.33) \times 10^{-3}$ \\
140801A & $15.09$ & $(1.41 \pm 0.36) \times 10^{-13}$ & $0.92^{+0.15}_{-0.13}$ & $1.80^{+0.36}_{-0.24}$ & $1.24$ & $0.36$ & $(4.71 \pm 1.22) \times 10^{-3}$ \\
140907A & $10.44$ & $(9.94 \pm 1.99) \times 10^{-13}$ & $1.05^{+0.11}_{-0.06}$ & $1.98^{+0.10}_{-0.07}$ & $1.47$ & $0.45$ & $(3.27 \pm 0.65) \times 10^{-2}$ \\
141004A & $4.14$ & $(1.87 \pm 0.33) \times 10^{-13}$ & $1.14^{+0.03}_{-0.03}$ & $1.79^{+0.13}_{-0.09}$ & $1.35$ & $0.58$ & $(1.98 \pm 0.35) \times 10^{-3}$ \\
150301B & $7.76$ & $(1.49 \pm 0.57) \times 10^{-13}$ & $1.11^{+0.03}_{-0.02}$ & $1.81^{+0.09}_{-0.07}$ & $1.12$ & $0.33$ & $(2.23 \pm 0.86) \times 10^{-3}$ \\
150821A & $6.67$ & $(8.36 \pm 1.75) \times 10^{-13}$ & $1.33^{+0.05}_{-0.04}$ & $2.44^{+0.14}_{-0.13}$ & $2.39$ & $0.73$ & $(2.26 \pm 0.47) \times 10^{-2}$ \\
151027A & $9.76$ & $(2.65 \pm 0.62) \times 10^{-12}$ & $1.85^{+0.07}_{-0.05}$ & $2.08^{+0.13}_{-0.12}$ & $1.30$ & $0.58$ & $(6.54 \pm 1.52) \times 10^{-2}$ \\
160804A & $11.94$ & $(7.63 \pm 1.26) \times 10^{-13}$ & $0.92^{+0.06}_{-0.05}$ & $1.97^{+0.12}_{-0.12}$ & $1.24$ & $0.57$ & $(2.41 \pm 0.40) \times 10^{-2}$ \\
161017A & $10.57$ & $(3.70 \pm 0.84) \times 10^{-13}$ & $1.80^{+0.10}_{-0.10}$ & $1.87^{+0.12}_{-0.09}$ & $1.06$ & $0.29$ & $(7.83 \pm 1.68) \times 10^{-3}$ \\
170113A & $10.86$ & $(7.74 \pm 1.86) \times 10^{-13}$ & $1.20^{+0.02}_{-0.02}$ & $1.78^{+0.05}_{-0.05}$ & $1.23$ & $0.27$ & $(1.71 \pm 0.41) \times 10^{-2}$ \\
\hline\hline
\caption{X-ray properties of the sample. $F_{\rm x}$ is in the range $0.3$ -- $10$~keV in the observer frame. The absorption correction is the ratio of the counts-to-flux unabsorbed over the counts-to-flux observed from the spectrum on the UKSSDC. $F_{\rm \nu,x}$ has been extrapolated to $11$ hours and $5$~keV in the rest frame, and has had a k-factor applied to account for the redshift of the observed bandpass. k is the ratio of the observer frame fluence to rest frame fluence, as in \citet{Bloom01}. All tabulated values are from the UKSSDC unless marked otherwise. $^{\dagger}$Manual fit. (1) - \citet{Cano11} \label{tab:x-ray}}
\end{longtable*}

\subsection{Optical Data}
Our optical data (Table~\ref{tab:optical}) are collected from the literature, as well as GCN circulars. We select data that is as close to the rest-frame R-band as possible, which usually means the J, H or K nIR bands in the observer frame where available. We also aim to collect data as close to $11$~hours in the rest frame as possible. Because optical/nIR coverage is never as comprehensive as it is for the X-ray, it is rarer to find values for the temporal, and in particular the spectral, indices in the literature. Where no published value exists, we fit light curves and SEDs from GCN circulars to provide our own values where possible. Light curves are fitted with both a single power law and a broken power law, which is assessed for an improvement in the fit using an f-test, and a break is accepted at the 3$\sigma$ level. It is the index local to the observed data point that is reported in Table~\ref{tab:optical}. SEDs are fitted with a power law multiplied by the parameterised extinction curves of \citet{Cardelli89}. If the model does not converge when both the spectral index $\beta_o$ and rest frame V-band extinction $A_V$ are free parameters (usually due to a lack of nIR detections in the fitted SED), we fit the data with $\beta_o = \beta_x$ to account for the case where the synchrotron cooling break ($\nu_c$) does not lie between the two bands, then fix $\beta_o = \beta_x - 0.5$ to account for the case where it does. We then report the extinction for the best fit, but do not report the spectral index, since this would effectively double-count $\beta_x$ in our analysis.

We correct the magnitudes reported in the literature for galactic extinction using the maps from \citet{Schlafly11}, and for intrinsic extinction where it can be identified. Magnitudes are then converted to flux densities and extrapolated from their observed wavelength to the rest frame R-band wavelength ($\lambda_{\rm R,rest} = 6400 \times (1 + z)$\AA) using $\beta_o$, and to $11$~h in the rest frame using $\alpha_o$. In the absence of an SED or sufficient data for a light curve, we take the mean spectral and temporal indices from the sample of known values when extrapolating to the desired time/frequency, and introduce the associated standard deviation into the uncertainty of the flux density. This is indicated by the bracketed values in Table~\ref{tab:optical}.

{\setlength{\tabcolsep}{2pt}
\begin{longtable*}{lcccccccc}
\hline\hline
GRB & $t_{\rm o,rest}$ & m & Filter & $A_{\rm \lambda,obs}$ & $A_{\rm \lambda,rest}$ & $\beta_o$ & $\alpha_o$ & F$_{\rm \nu,o}$ \\
 & (h) & & & (Gal) & (Int) & & & ($\mu$Jy) \\
\hline
\endhead
\bf{LAT} \\
080916C & $6.10$ & $21.10 \pm 0.15^{(1)}$ & K$_s$ (AB) & $0.10$ & $0.00^{+0.00\, (1)}_{-0.00}$ & $0.38 \pm 0.20^{(1)}$ & $1.40 \pm 0.05^{(1)}$ & $7.55^{+1.04}_{-1.04}$ \\
090323 & $10.82$ & $21.01 \pm 0.04^{(2)}$ & R & $0.05$ & $0.46^{+0.13\, (3)}_{-0.10}$ & $0.65 \pm 0.13^{(3)}$ & $1.74 \pm 0.05^{(2)}$ & $50.43^{+4.94}_{-6.38}$ \\
090328A & $22.12$ & $19.54 \pm 0.06^{(4)}$ & J (AB) & $0.04$ & $0.15^{+0.04\, (3)}_{-0.12}$ & $1.19 \pm 0.20^{(3)}$ & $1.85 \pm 0.13^{(2)}$ & $207.14^{+26.06}_{-13.85}$ \\
090902B & $12.16$ & $19.99 \pm 0.15^{(5)}$ & J & $0.03$ & $0.00^{+0.00\, (3)}_{-0.00}$ & $0.82 \pm 0.10^{(2)}$ & $0.89 \pm 0.03^{(2)}$ & $24.39^{+3.37}_{-3.37}$ \\
090926A & $9.35$ & $18.29 \pm 0.10^{(2)}$ & I & $0.04$ & $0.22^{+0.15}_{-0.15}$ & $0.94 \pm 0.12$ & $1.63 \pm 0.01^{(6)}$ & $244.43^{+25.17}_{-36.04}$ \\
091003 & $21.38$ & $21.33 \pm 0.11^{(7)}$ & r & $0.05$ & - & ($0.76 \pm 0.29$) & $1.04 \pm 0.05$ & $34.99^{+7.04}_{-7.04}$ \\
091208B & $7.44$ & $20.61 \pm 0.10^{(8)}$ & J (AB) & $0.04$ & $0.36^{+0.07}_{-0.07}$ & ($0.76 \pm 0.29$) & $0.75 \pm 0.02^{(9)}$ & $23.14^{+2.58}_{-2.58}$ \\
100414A  & $30.81$ & $20.9 \pm 0.1^{(10)}$ & r' & $0.05$ & - & $1.2 \pm 0.2^{(11)}$ & $2.6 \pm 0.1^{(11)}$ & $692.63^{+63.79}_{-63.79}$ \\
100728A & $2.90$ & $21.2 \pm 0.3^{(12)}$ & H (AB) & $0.09$ & - & ($0.76 \pm 0.29$) & ($1.22 \pm 0.44$) & $2.62^{+1.69}_{-1.69}$ \\
110731A & $17.17$ & $22.31 \pm 0.28^{(13)}$ & H (AB) & $0.09$ & $0.35^{+0.09\, (13)}_{-0.09}$ & $0.66 \pm 0.03^{(13)}$ & $1.08 \pm 0.01$ & $13.90^{+3.76}_{-3.76}$ \\
120711A & $12.47$ & $19.74 \pm 0.27^{(14)}$ & H (AB) & $0.05$ & $0.68^{+0.05\, (14),a}_{-0.05}$ & $0.53 \pm 0.02^{(14)}$ & $1.25 \pm 0.03$ & $103.60^{+26.17}_{-26.17}$ \\
130427A & $11.82$ & $16.99 \pm 0.03^{(15)}$ & z & $0.03$ & $0.10^{+0.05\, (15),b}_{-0.05}$ & $0.60 \pm 0.02^{(15)}$ & $1.36 \pm 0.02^{(16)}$ & $696.80^{+35.33}_{-35.33}$ \\
130518A & $12.86$ & $20.19 \pm 0.04^{(17)}$ & i' & $0.23$ & $0.04^{+0.08}_{-0.04}$ & $0.90 \pm 0.07$ & $1.33 \pm 0.05$ & $122.60^{+6.39}_{-10.10}$ \\
130907A & $23.81$ & $21.38 \pm 0.09^{(18)}$ & i & $0.02$ & $2.18^{+0.17\, (19)}_{-0.17}$ & ($0.76 \pm 0.29$) & $1.37 \pm 0.38^{(19)}$ & $349.37^{+86.32}_{-86.32}$ \\
131108A & $4.31$ & $19.09 \pm 0.08^{(20)}$ & H (AB) & $0.02$ & $0.00^{+0.00}_{-0.00}$ & $0.86 \pm 0.05$ & $1.60 \pm 0.07$ & $24.84^{+1.83}_{-1.83}$ \\
131231A & $12.72$ & $18.48 \pm 0.02^{(21)}$ & R & $0.05$ & - & ($0.76 \pm 0.29$) & $1.30 \pm 0.04$ & $228.77^{+33.47}_{-33.47}$ \\
141028A & $11.81$ & $21.00 \pm 0.05^{(22)}$ & r' & $0.11$ & $0.00^{+0.00\, (23)}_{-0.00}$ & $0.92 \pm 0.04$ & $0.97 \pm 0.03^{(22)}$ & $52.63^{+2.42}_{-2.42}$ \\
150314A & $5.74$ & $22.7 \pm 0.3^{(24)}$ & R & $0.08$ & - & ($0.76 \pm 0.29$) & $0.89 \pm 0.13$ & $3.32^{+1.35}_{-1.35}$ \\
150403A & $3.53$ & $19.1 \pm 0.1^{(25)}$ & r & $0.13$ & - & ($0.76 \pm 0.29$) & ($1.22 \pm 0.44$) & $54.63^{+32.94}_{-32.94}$ \\
150514A & $16.07$ & $19.2 \pm 0.2^{(26)}$ & J & $0.02$ & $0.00^{+0.00}_{-0.00}$ & ($0.76 \pm 0.29$) & ($1.22 \pm 0.44$) & $48.27^{+11.97}_{-11.97}$ \\
160509A & $11.39$ & $24.05 \pm 0.14^{(27)}$ & r' & $0.67$ & $6.85^{+0.16\, (28)}_{-0.14}$ & ($0.76 \pm 0.29$) & $1.09 \pm 0.45^{(28)}$ & $1678.07^{+496.44}_{-511.30}$ \\
160623A & $10.36$ & $20.32 \pm 0.13^{(29)}$ & R & $3.27$ & $1.32^{+0.03}_{-0.03}$ & ($0.76 \pm 0.29$) & $1.29 \pm 0.06$ & $1834.57^{+280.24}_{-280.24}$ \\
160625B & $14.66$ & $18.98 \pm 0.03^{(30)}$ & H (AB) & $0.06$ & $0.06^{+0.03}_{-0.03}$ & $0.47 \pm 0.07$ & $0.92 \pm 0.01$ & $132.76^{+5.57}_{-5.57}$ \\
\hline
\bf{GBM} \\
080916A & $12.22$ & $22.1 \pm 0.07^{(31)}$ & r' & $0.05$ & - & ($0.76 \pm 0.29$) & $0.81 \pm 0.13^{(31)}$ & $9.01^{+1.54}_{-1.54}$ \\
081007 & $9.89$ & $20.65 \pm 0.15^{(32)}$ & r' & $0.03$ & $0.46^{+0.37}_{-0.34}$ & $0.43 \pm 0.36^{(33)}$ & $1.25 \pm 0.13^{(32)}$ & $33.14^{+11.38}_{-12.21}$ \\
081121 & $1.54$ & $16.75 \pm 0.15^{(34)}$ & K & $0.01$ & $0.00^{+0.00}_{-0.00}$ & $0.42 \pm 0.24$ & $1^{(34)} \, (\pm 0.44$) & $19.12^{+2.64}_{-2.64}$ \\
090423 & $9.28$ & $20.58 \pm 0.06^{(35)}$ & K & $0.01$ & - & $0.30 \pm 0.06^{(35)}$ & $1.36 \pm 0.33^{(35)}$ & $4.24^{+0.23}_{-0.23}$ \\
090424 & $9.79$ & $19.70 \pm 0.04^{(36)}$ & I & $0.04$ & $0.72^{+0.10}_{-0.10}$ & $0.89 \pm 0.10$ & $1.13 \pm 0.03$ & $72.44^{+6.94}_{-6.94}$ \\
090618 & $11.69$ & $19.36 \pm 0.01^{(37)}$ & i & $0.13$ & $0.29^{+0.11\, (37)}_{-0.11}$ & $0.55 \pm 0.07^{(37)}$ & $1.57 \pm 0.07^{(37)}$ & $119.74^{+12.04}_{-12.04}$ \\
091020 & $14.52$ & $21.81 \pm 0.23^{(38)}$ & R & $0.05$ & $2.84^{+0.09\, (33)}_{-0.27}$ & ($0.76 \pm 0.29$) & $1.12 \pm 0.06^{(39)}$ & $240.80^{+105.35}_{-89.07}$ \\
091127 & $10.95$ & $18.76 \pm 0.05^{(40)}$ & i' & $0.06$ & $0.12^{+0.05\, (40),c}_{-0.05}$ & $0.30 \pm 0.05^{(40)}$ & $1.64 \pm 0.06^{(40)}$ & $141.05^{+9.19}_{-9.19}$ \\
100906A & $13.11$ & $22.15 \pm 0.05^{(41)}$ & R & $0.81$ & $0.99^{+0.25\, (41),d}_{-0.25}$ & $1.34 \pm 0.04^{(41)}$ & $2.03 \pm 0.02$ & $121.63^{+29.06}_{-29.06}$ \\
101219B & $6.96$ & $19.81 \pm 0.04^{(42\,*)}$ & z' & $0.02$ & - & $0.92 \pm 0.09^{(43)}$ & $1.01 \pm 0.01^{(42)}$ & $30.11^{+1.11}_{-1.11}$ \\
110213A & $10.05$ & $19.61 \pm 0.05^{(44)}$ & i' & $0.53$ & - & $1.22 \pm 0.18^{(44)}$ & $1.80 \pm 0.15^{(44)}$ & $166.99^{+7.69}_{-7.69}$ \\
111228A & $11.06$ & $20.3 \pm 0.1^{(45)}$ & r' & $0.08$ & $0.00^{+0.00\, (46)}_{-0.00}$ & $0.90 \pm 0.05^{(46)}$ & $1.04 \pm 0.09$ & $49.13^{+4.52}_{-4.52}$ \\
120119A & $2.29$ & $17.14 \pm 0.24^{(47)}$ & H & $0.05$ & $0.81^{+0.01\, (46)}_{-0.01}$ & $0.89 \pm 0.01^{(47)}$ & $1.30 \pm 0.01^{(47)}$ & $44.02^{+9.74}_{-9.74}$ \\
120729A & $10.27$ & $23.27 \pm 0.27^{(48)}$ & z' & $0.20$ & $0.17^{+0.06\, (48)}_{-0.06}$ & $1.0 \pm 0.1^{(48)}$ & $2.7 \pm 0.18^{(48)}$ & $2.64^{+0.67}_{-0.67}$ \\
120811C & $7.39$ & $22.3 \pm 0.2^{(49)}$ & R & $0.08$ & - & ($0.76 \pm 0.29$) & $1.04 \pm 0.05^{(49)}$ & $7.02^{+2.97}_{-2.97}$ \\
121211A & $11.03$ & $22.1 \pm 0.4^{(50)}$ & z & $0.01$ & - & ($0.76 \pm 0.29$) & $1.18 \pm 0.47$ & $6.94^{+2.66}_{-2.66}$ \\
130420A & $11.18$ & $20.38 \pm 0.23^{(51)}$ & H (AB) & $0.01$ & $0.07^{+0.02\, (52)}_{-0.02}$ & ($0.76 \pm 0.29$) & $0.95 \pm 0.03$ & $26.23^{+5.61}_{-5.61}$ \\
140213A & $10.44$ & $18.6 \pm 0.1^{(53)}$ & R & $0.34$ & $0.00^{+0.00\, (54)}_{-0.00}$ & $0.8 \pm 0.1^{(54)}$ & ($1.22 \pm 0.44$) & $268.87^{+25.53}_{-25.53}$ \\
140423A & $10.99$ & $21.87 \pm 0.35^{(55)}$ & H (AB) & $0.01$ & - &$0.73 \pm 0.19$ & $1.05 \pm 0.02$ & $9.65^{+3.11}_{-3.11}$ \\
140506A & $30.75$ & $21.09 \pm 0.13^{(56)}$ & J (AB) & $0.21$ & $0.72^{+0.08\, (56),e}_{-0.08}$ & $0.40 \pm 0.38$ & $0.60 \pm 0.03$ & $57.11^{+8.03}_{-8.03}$ \\
140512A & $4.59$ & $19.0 \pm 0.1^{(57)}$ & J (AB) & $0.12$ & $0.00^{+0.00}_{-0.00}$ & $0.89 \pm 0.09$ & ($1.22 \pm 0.44$) & $31.16^{+12.27}_{-12.27}$ \\
140606B & $20.98$ & $21.30 \pm 0.11^{(58)}$ & i & $0.19$ & $0.46^{+0.08}_{-0.08}$ & ($0.76 \pm 0.29$) & $1.49 \pm 0.43^{(58)}$ & $56.96^{+7.50}_{-7.50}$ \\
140620A & $13.82$ & $21.9 \pm 0.15^{(59)}$ & R & $0.13$ & - & ($0.76 \pm 0.29$) & $1.51 \pm 0.17$ & $19.67^{+6.95}_{-6.95}$ \\
140703A & $2.49$ & $18.98 \pm 0.09^{(60)}$ & H (AB) & $0.02$ & $0.03^{+0.30\, (52)}_{-0.03}$ & ($0.76 \pm 0.29$) & $1.30 \pm 0.16$ & $20.70^{+3.55}_{-6.71}$ \\
140801A & $6.52$ & $20.46 \pm 0.36^{(61)}$ & H (AB) & $0.12$ & $0.00^{+0.00\, (61)}_{-0.00}$ & $0.81 \pm 0.02^{(61)}$ & $0.82 \pm 0.01^{(61)}$ & $16.27^{+5.39}_{-5.39}$ \\
140907A & $10.97$ & $21.26 \pm 0.08^{(62)}$ & R & $0.65$ & - & ($0.76 \pm 0.29$) & $1.2 \pm 0.1^{(63)}$ & $31.76^{+7.73}_{-7.73}$ \\
141004A & $21.82$ & $23.4 \pm 0.2^{(64)}$ & z & $0.39$ & - & ($0.74 \pm 0.29$) & $1.08 \pm 0.16$ & $5.13^{+0.96}_{-0.96}$ \\
150301B & $2.00$ & $20.6 \pm 0.4^{(65)}$ & H (AB) & $0.04$ & - & ($0.76 \pm 0.29$) & ($1.22 \pm 0.44$) & $2.72^{+2.27}_{-2.27}$ \\
150821A & $9.71$ & $22.0 \pm 0.2^{(66)}$ & z & $0.01$ & - & ($0.76 \pm 0.29$) & ($1.22 \pm 0.44$) & $5.85^{+1.18}_{-1.18}$ \\
151027A & $9.41$ & $16.4 \pm 0.1^{(67)}$ & J & $0.03$ & $0.00^{+0.00}_{-0.00}$ & $0.62 \pm 0.07$ & $1.89 \pm 0.02$ & $320.42^{+29.51}_{-29.51}$ \\
160804A & $13.13$ & $20.20 \pm 0.07^{(68)}$ & J (AB) & $0.02$ & $0.06^{+0.11}_{-0.06}$ & $1.05 \pm 0.23$ & $1.54 \pm 1.16$ & $37.43^{+3.18}_{-4.35}$ \\
161017A & $11.31$ & $19.80 \pm 0.03^{(69)}$ & i & $0.04$ & - & ($0.76 \pm 0.29$) & $1.46 \pm 0.05$ & $92.78^{+24.37}_{-24.37}$ \\
170113A & $5.26$ & $20.8 \pm 0.3^{(70)}$ & H (AB) & $ 0.06$ & $0.00^{+0.00}_{-0.00}$ & $0.86 \pm 0.17$ & $0.94 \pm 0.04$ & $10.63^{+2.94}_{-2.94}$ \\
\hline\hline
\caption{Optical/nIR properties of our sample. Lower case filters give magnitudes in the AB system. Upper case filters give Vega magnitudes unless otherwise indicated. Galactic extinction comes from \citet{Schlafly11}. Intrinsic extinction uses an SMC-like extinction curve ($R_v = 2.74$) unless otherwise indicated. F$_{\rm \nu,o}$ has been extrapolated to the rest frame R-band using the spectral index $\beta_o$, and to $11$~h in the rest frame using the temporal index $\alpha_o$. Bracketed values of $\alpha_o$ and $\beta_o$ are the mean and standard deviation of the sample of known values. The standard deviation is introduced into the errors of F$_{\rm \nu,o}$ when they are used. Results without references were calculated as part of this work.\\(1) - \citet{Greiner09}; (2) - \citet{Cenko11a}; (3) - \citet{McBreen10}; (4) - \citet{Updike09b}; (5) - \citet{Pandey10}; (6) - \citet{Rau10}; (7) - \citet{Wiersema09a}; (8) - \citet{Updike09a}; (9) - \citet{Uehara12}; (10) - \citet{Filgas10}; (11) - \citet{Urata12} (12) - \citet{Olivares10}; (13) - \citet{Ackermann13}; (14) - \citet{Martin-Carrillo14}; (15) - \citet{Perley14}; (16) - \citet{Maselli14}; (17) - \citet{Troja13a};  (18) - \citet{Butler13a}; (19) - \citet{Veres15}; (20) - \citet{Troja13b}; (21) - \citet{Halpern14}; (22) - \citet{Cenko14a}; (23) - \citet{Graham14b}; (24) - \citet{Xu15}; (25) - \citet{Pugliese15}; (26) - \citet{Yates15}; (27) - \citet{Cenko16}; (28) - \citet{Laskar16}; (29) - \citet{Pozanenko16}; (30) - \citet{Watson16}; (31) - \citet{Rossi08}; (32) - \citet{Jin13}; (33) - \citet{Covino13}; (34) - \citet{Cobb08}; (35) - \citet{Tanvir09}; (36) - \citet{Cobb09}; (37) - \citet{Cano11}; (38) - \citet{Perley09b}; (39) - \citet{Kann09}; (40) - \citet{Troja12}; (41) - \citet{Gorbovskoy12}; (42) - \citet{Olivares15}; (43) - \citet{Sparre11}; (44) - \citet{Cucchiara11b}; (45) - \citet{Cenko11b}; (46) - \citet{NicuesaGuelbenzu11}: (47) - \citet{Morgan14}; (48) - \citet{Cano14}; (49) - \citet{Galeev12}; (50) - \citet{Butler12}; (51) - \citet{Butler13b}; (52) - \citet{Littlejohns15}; (53) - \citet{Trotter14}; (54) - \citet{Elliott14}; (55) - \citet{Butler14a}; (56) - \citet{Fynbo14b}; (57) - \citet{Graham14a}; (58) - \citet{Cano15a}; (59) - \citet{Keleman14}; (60) - \citet{Butler14b}; (61) - \citet{Lipunov16}; (62) - \citet{Volnova14}; (63) - \citet{Cenko14b}; (64) - \citet{Schmidl14}; (65) - \citet{Kann15}; (66) - \citet{Kruhler15b}; (67) - \citet{Cano15b}; (68) - \citet{Bolmer16}; (69) - \citet{Guidorzi16}; (70) - \citet{Kruhler17}; a - $R_v = 3.41$; b - $R_v = 3$;  c - $R_v = 3.3$; d - $R_v = 2.93$; e - $R_v = 3.1$; *we subtract out the extinction reported in \citet{Olivares15}. \label{tab:optical}}
\end{longtable*}
}

\subsection{Data Trends}
The weighted mean of the X-ray temporal indices is $1.37 \pm 0.24$ and the weighted mean spectral index is $0.83 \pm 0.11$. In the optical, the weighted mean temporal index is $1.14 \pm 0.31$ and the weighted mean spectral index is $0.79 \pm 0.22$. The errors here are the weighted standard deviation, and the weights themselves are one over the errors squared. Bursts with temporal indices $> 2$ are excluded, as they indicate a jet break has occurred. Interpretation of these index distributions is complicated due to the different environment types and spectral regimes, but the spread in the temporal indices suggests that we are seeing both ISM-like and wind-like GRBs, since the steeper end of the distribution is difficult to interpret as ISM-like due to the high value of $p$ ($p\gtrsim 3$)  implied by the relevant closure relations. Similarly, the shallower indices do not favor a wind-like environment, since the closure relations would require $p \ll 2$.

\begin{figure}
\begin{center}
\includegraphics[width=8.9cm]{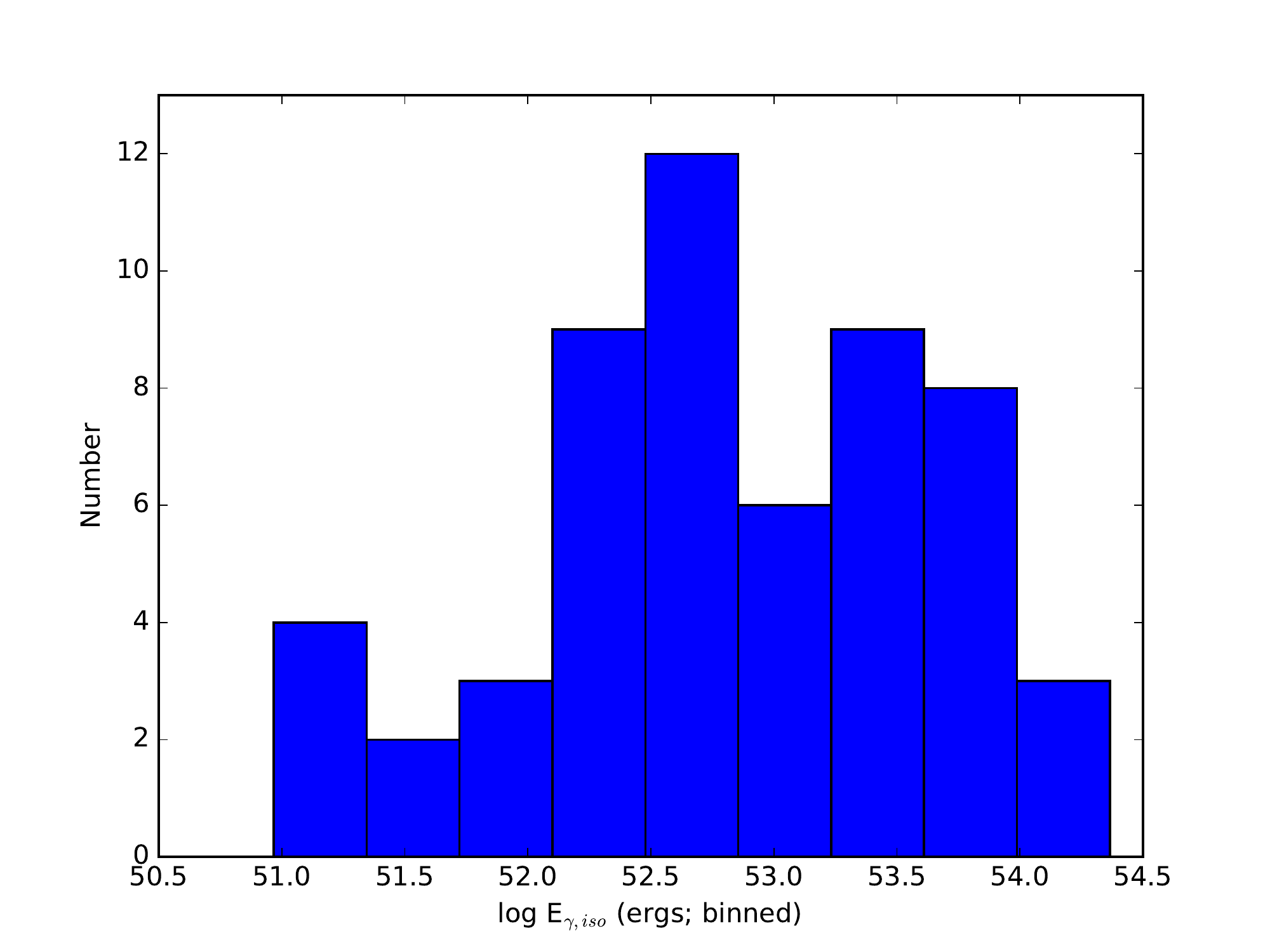}
\caption{Histogram of the isotropic $\gamma$-ray energy release $E_{\rm \gamma,iso}$ for all bursts in the sample. We find a mean $\gamma$-ray energy of $E_{\rm \gamma,iso} = 10^{53.46 \pm 0.54}$~ergs. \label{fig:distributions}}
\end{center}
\end{figure}

The distribution of $E_{\rm \gamma,iso}$ for all GRBs in our sample is shown in Figure~\ref{fig:distributions}. We find a mean $\gamma$-ray energy of $E_{\rm \gamma,iso} = 10^{53.46 \pm 0.54}$~ergs.

\section{Determining the Environment Type}\label{sec:environment}

Our aim is to ascertain whether the X-ray and optical afterglow data are best described by deceleration in a wind-like medium or an ISM-like medium. To do this, we take our 4 power law measurements ($\alpha_x, \beta_x, \alpha_o, \beta_o$) and use the synchrotron closure relations to calculate the implied value of the electron distribution power law index $p$ for each of the 5 possible scenarios, which are:\\ \\
I $\nu_c < \nu_R < \nu_x$, where:
\begin{center}
$\alpha_o = \alpha_x = \frac{3p-2}{4}$; $\beta_o = \beta_x = \frac{p}{2}$
\end{center}
(In this case, the synchrotron spectrum is identical in the observed regime for both wind and ISM, and the environment type cannot be determined)\\ \\
II $\nu_R < \nu_c < \nu_x$ in an ISM-like environment, where:
\begin{center}
$\alpha_o = \frac{3(p-1)}{4}$; $\alpha_x = \frac{3p-2}{4}$; $\beta_o = \frac{p-1}{2}$; $\beta_x = \frac{p}{2}$
\end{center}
III $\nu_R < \nu_c < \nu_x$ in a wind-like environment, where:
\begin{center}
$\alpha_o = \frac{3p-1}{4}$; $\alpha_x = \frac{3p-2}{4}$; $\beta_o = \frac{p-1}{2}$; $\beta_x = \frac{p}{2}$
\end{center}
IV $\nu_R < \nu_x < \nu_c$ in an ISM-like environment, where:
\begin{center}
$\alpha_o = \alpha_x = \frac{3(p-1)}{4}$; $\beta_o = \beta_x = \frac{p-1}{2}$
\end{center}
V $\nu_R < \nu_x < \nu_c$ in a wind-like environment, where:
\begin{center}
$\alpha_o = \alpha_x = \frac{3p-1}{4}$; $\beta_o = \beta_x = \frac{p-1}{2}$
\end{center}
For scenarios I, IV and V, where $\nu_x$ and $\nu_R$ are on the same power law segment of the synchrotron spectrum, we can also use the calculated value of $F_R/F_X$ as an indicator of $p$. Since $F_R = k\nu_R^{-\beta}$ and $F_X = k\nu_x^{-\beta}$, we find
\begin{equation}\label{eq:FrFx}
F_R/F_X = (\nu_R/\nu_X)^{-\beta} = 2580^{\beta} \mbox{,\\}
\end{equation}

where $k$ is the normalisation constant and $\beta = (p-1)/2$ if $\nu_c > \nu_x$, or $\beta = p/2$ if $\nu_R > \nu_c$.

The electron distribution index $p$ is assumed to be a single value for a given burst \citep{Sari98}. We can therefore assess how well the values of $p$ derived from each of our four or five different metrics converge in each environment type and spectral regime. Based on their agreement, we can assign probabilities to how likely each environment is for a given burst. To demonstrate the concept, we take an assumed value of $p = 2.5$ and tabulate the expected theoretical values of the spectral and temporal indices in the GRB afterglow for each environment type and spectral regime listed above (see Table~\ref{tab:pexample}).

When fitting to the data, we calculate the individual values of $p$ from each measured index, and take the weighted mean of these as the best fit value for $p$. Our weights are $1/$errors$^2$. The error presented for $p$ is the uncertainty of the weighted mean, rather than the standard deviation of the sample. The uncertainty of the weighted mean measures the dispersion of the mean of the sampling distribution around the mean of the population distribution. This provides an estimate of how well our value of $p$, derived from a limited sample, represents the true value of $p$, the mean of the overall population (i.e. the mean when $N = \infty$). We calculate $\chi^2_{\nu}$ (reduced $\chi^2$), and assign probabilities that each environment is the correct one by comparing the calculated $\chi^2_{\nu}$ to the $\chi^2$ distribution for the appropriate degrees of freedom.

\begin{table}
\begin{center}
\begin{tabular}{llcccc}
\hline \hline
$p = 2.5$ & \vline & $\alpha_o$ & $\alpha_x$ & $\beta_o$ & $\beta_x$ \\
\hline
$\nu_c < \nu_R < \nu_x$ & \vline & $1.375$ & $1.375$ & $1.25$ & $1.25$ \\
$\nu_R < \nu_c < \nu_x$ (ISM) & \vline & $1.125$ & $1.375$ & $0.75$ & $1.25$ \\
$\nu_R < \nu_c < \nu_x$ (wind) & \vline & $1.625$ & $1.375$ & $0.75$ & $1.25$ \\
$\nu_R < \nu_x < \nu_c$ (ISM) & \vline & $1.125$ & $1.125$ & $0.75$ & $0.75$ \\
$\nu_R < \nu_x < \nu_c$ (wind) & \vline & $1.625$ & $1.625$ & $0.75$ & $0.75$ \\
\hline \hline
\end{tabular}
\caption{The expected values of the afterglow spectral and temporal indices at X-ray and optical frequencies for an assumed $p = 2.5$ for each of the 5 cases outlined in Section~\ref{sec:environment}. \label{tab:pexample}}
\end{center}
\end{table}

\begin{figure}
\begin{center}
\includegraphics[width=8.9cm]{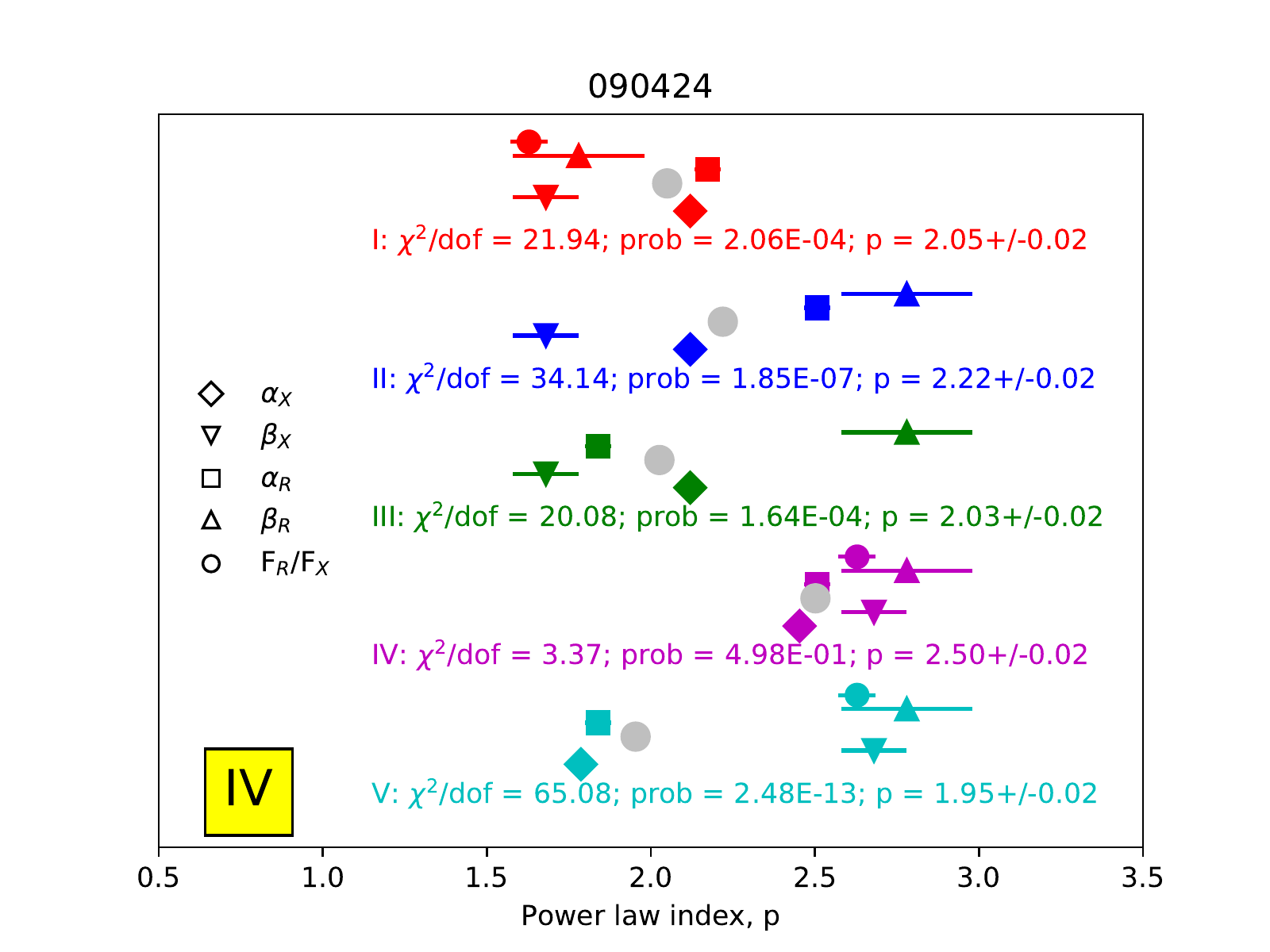}
\caption{An example of our fitting procedure, using GRB 090424. Scenarios I-V are listed from top to bottom in different colors (see Table~\ref{tab:pexample}). Different symbols represent individual measures of p (i.e. temporal and spectral indices at X-ray and optical frequencies and $F_R/F_X$) with their 1$\sigma$ errors. Grey circles represent the weighted mean of the data, taken as the best-fitting single value of p, and the errors are the uncertainty of the weighted mean (as opposed to the standard deviation of the data). Fit probabilities are derived by comparing reduced $\chi^2$ to the $\chi^2$ distribution for the appropriate degrees of freedom. Here, scenario IV ($\nu_R < \nu_x < \nu_c$; ISM) is the best fit because its probability is more than 3x better than the best fitting scenario of an alternate type. This is apparent in the clustering of the 5 data points when compared to the other scenarios. \label{fig:fitexample}}
\end{center}
\end{figure}

Using this method, each burst is assigned a best-fit scenario and a best-fit value of $p$ (with errors). An example using GRB 090424 is shown in Figure~\ref{fig:fitexample}. We require that the best-fit $p$ is consistent within 1 standard deviation with the range $1.8 \leq p \leq 3.0$. The best-fit scenario is accepted as the true environment type if the fit probability is at least $3$ times better than the best fitting scenario of an alternative type that produces a value of $p$ consistent with our accepted range. We make no minimum requirements on the {\it absolute} value of the probability an individual fit has, though the effects of using a minimum best-fit probability are discussed at the end of Section~\ref{sec:erreffects}.

\subsection{Constraining the Physical Parameters}\label{sec:params}
Once we have identified the environment type and found a best-fit value of $p$ for a burst, we are able to investigate the physical parameters underlying the emission.

For cases II and III, in which $\nu_c$ lies between the optical and X-ray bands, we can write \citep[cf.][]{Blandford76,Sari98,Peer05}:
\begin{equation}\label{eqn:FrFx}
F_R/F_X = \bigg(\frac{\nu_R}{\nu_X}\bigg)^{-p/2}\nu_R^{1/2}\nu_c^{-1/2} \mbox{ .}
\end{equation}
We can gain information on the physical parameters of the shock from $\nu_c$, which is
\begin{multline}\label{eqn:case2}
\nu_c = 5.6\times10^{13}\frac{1}{(1+Y)^2}\bigg(\frac{1+z}{2}\bigg)^{-1/2} \\ \times E_{53}^{-1/2}t_{obs,day}^{-1/2}\epsilon_{B,-2}^{-3/2}n_0^{-1} \mbox{ Hz}
\end{multline}
in the ISM case (case II), or
\begin{multline}\label{eqn:case3}
\nu_c = 3.7\times10^{13}\frac{1}{(1+Y)^2}\bigg(\frac{1+z}{2}\bigg)^{-3/2} \\ \times E_{53}^{1/2}t_{obs,day}^{1/2}\epsilon_{B,-2}^{-3/2}A_*^{-2} \mbox{ Hz}
\end{multline}
for the wind-like environment (case III). Here and below, for any parameter, $Q$, we define $Q_x$ such that $Q=10^x Q_x$.  $(1+Y)^2$ is the comptonisation fraction, which for simplicity we set as 1. $\epsilon_B$ is the fraction of the available energy contained in the magnetic fields. $n$ is the number density of particles in the ISM environment in cm$^{-3}$. The wind density $A_* = 5\times 10^{11}A$, where $A = \dot{M}_w/4\pi v_w$~g~cm$^{-1}$, is expressed as a function of the (unknown) mass loss rate and wind velocity from the star that created the wind.

Substituting Equation~\ref{eqn:case2} or Equation~\ref{eqn:case3} into Equation~\ref{eqn:FrFx} provides a value for $\epsilon_B \times$density, our only unknowns. These properties are degenerate, and we must assume one to get the other.

In cases IV and V, the synchrotron cooling break $\nu_c$ lies above the X-ray frequency $\nu_x$, so we can't use the ratio $F_R/F_X$ to gain information about the physical parameters. In these cases, we take the flux at the optical band \citep[cf.][]{Blandford76,Sari98,Peer05}:
\begin{multline}\label{eqn:case4}
F_{\nu,obs} (\nu_m < \nu < \nu_c) = 3.68\times10^5 \\ \times \bigg(\frac{1.1\times10^{14}}{\nu_{obs}}\bigg)^{\frac{(p-1)}{2}} \bigg(\frac{1+z}{2}\bigg)^{\frac{(p+3)}{4}} d_{L,28.3}^{-2} \\ \times E_{54}^{\frac{(p+3)}{4}} n_0^{1/2} \epsilon_{e,-1}^{p-1} \epsilon_{B,-2}^{(p+1)/4} t_{obs,day}^{(3-3p)/4} \mbox{ $\mu$Jy}
\end{multline}
for the ISM (case IV), and
\begin{multline}\label{eqn:case5}
F_{\nu,obs} (\nu_m < \nu < \nu_c) = 3.12\times10^5 \\ \times \bigg(\frac{2.3\times10^{14}}{\nu_{obs}}\bigg)^{\frac{(p-1)}{2}} \bigg(\frac{1+z}{2}\bigg)^{\frac{(p+5)}{4}} d_{L,28.3}^{-2} \\ \times E_{54}^{\frac{(p+1)}{4}} A_* \epsilon_{e,-1}^{p-1} \epsilon_{B,-2}^{(p+1)/4} t_{obs,day}^{(1-3p)/4} \mbox{ $\mu$Jy}
\end{multline}
for the wind (case V). Unlike for Equations~\ref{eqn:case2} and \ref{eqn:case3}, solving these equations now gives a value for a combination of three unknown degenerate parameters: density, $\epsilon_B$ and $\epsilon_e$. We must therefore assume a value of $\epsilon_e$, the fraction of the available energy contained in the emitting electrons. We choose $\epsilon_e = 0.1$ for all cases, as \citeauthor{Beniamini17} show that  $\epsilon_e$ appears to be restricted to within about a factor of three of this value in their relatively large sample of bursts.

\subsection{Intrinsic Error}\label{sec:interr}

It is noticeable in our analysis that a number of bursts with very precise measurements for their spectral and temporal indices have very large values of $\chi^2_{\nu}$ and low probabilities because they do not agree on a single value of $p$. This is likely symptomatic of not taking all sources of error into account.

The biggest source of error not accounted for is the fact that the underlying synchrotron theory is a simplified version of a complex system, making it unreasonable to expect all measured indices to perfectly converge to a single value of $p$. Assuming this is the source of error, we attempt to quantify this intrinsic uncertainty by adding in a standard error to represent complexity in the theory. We perform our fitting experiment multiple times, each time adding a constant error in quadrature with the measured errors. This is done individually to the error for each index (excluding $F_R/F_X$) where the functional dependence of the parameter's error on the error in $p$ is determined using the relations at the beginning of Section~\ref{sec:environment}. We explore a standard error ranging from $0$ to $0.5$.

\begin{longtable*}{lcccccccc}
\hline\hline
GRB & $F_R/F_X$ & Environment & Spectral & $p$ & $\sigma_p$ & SE$_p$ & $\chi^2_{\nu}$ & Probability \\
 & & type & regime & & & \\
\hline
\endhead
\bf{LAT} \\
090323 & $8944.11^{+2361.19}_{-2467.40}$ & Wind & $\nu_R < \nu_c < \nu_x$ & $2.63$ & $0.32$ & $0.06$ & $5.39$ & $0.15$ \\
090328A & $3411.50^{+830.94}_{-745.62}$ & ISM & $\nu_R < \nu_x < \nu_c$ & $3.15$ & $0.25$ & $0.05$ & $3.43$ & $0.49$ \\
090902B & $707.87 \pm 164.41$ & ISM & $\nu_R < \nu_c < \nu_x$ & $2.31$ & $0.31$ & $0.03$ & $19.65$ & $2.01\times10^{-4}$ \\
090926A & $9640.05^{+2725.36}_{-2909.00}$ & Wind & $\nu_R < \nu_c < \nu_x$ & $2.51$ & $0.10$ & $0.01$ & $4.20$ & $0.24$ \\
091003 & $804.33 \pm 246.27$ & ISM & $\nu_R < \nu_x < \nu_c$ & $2.65$ & $0.23$ & $0.04$ & $9.19$ & $0.03$ \\
091208B & $1145.25 \pm 222.81$ & ISM & $\nu_R < \nu_c < \nu_x$ & $2.03$ & $0.10$ & $0.062$ & $3.44$ & $0.18$ \\
100728A & $112.12 \pm 77.24$ & Wind & $\nu_R < \nu_x < \nu_c$ & $2.43$ & $0.13$ & $0.04$ & $1.96$ & $0.38$ \\
110731A & $1982.10 \pm 765.28$ & ISM & $\nu_R < \nu_x < \nu_c$ & $2.53$ & $0.14$ & $0.01$ & $28.92$ & $8.10\times10^{-6}$ \\
120711A & $576.80 \pm 195.34$ & Wind & $\nu_R < \nu_x < \nu_c$ & $2.44$ & $0.31$ & $0.01$ & $63.43$ & $5.52\times10^{-13}$ \\
130427A & $448.46 \pm 104.13$ & Wind & $\nu_R < \nu_x < \nu_c$ & $2.15$ & $0.15$ & $0.01$ & $25.48$ & $4.04\times10^{-5}$ \\
130907A & $1798.34 \pm 560.45$ & Wind & $\nu_R < \nu_x < \nu_c$ & $2.61$ & $0.15$ & $0.01$ & $9.99$ & $0.02$ \\
131108A & $10294.02 \pm 2495.61$ & Wind & $\nu_R < \nu_c < \nu_x$ & $2.48$ & $0.19$ & $0.04$ & $3.95$ & $0.27$ \\
131231A & $586.77 \pm 153.18$ & ISM & $\nu_R < \nu_x < \nu_c$ & $2.79$ & $0.16$ & $0.03$ & $5.80$ & $0.12$ \\
141028A & $2591.39 \pm 673.57$ & Unknown & $\nu_c < \nu_R < \nu_x$ & $1.95$ & $0.09$ & $0.03$ & $1.04$ & $0.90$ \\
150403A & $591.93 \pm 380.98$ & Wind & $\nu_R < \nu_x < \nu_c$ & $2.38$ & $0.18$ & $0.03$ & $5.80$ & $0.06$ \\
160623A & $1216.66 \pm 285.83$ & ISM & $\nu_R < \nu_x < \nu_c$ & $2.89$ & $0.24$ & $0.04$ & $8.16$ & $0.04$ \\
160625B & $549.82 \pm 125.94$ & ISM & $\nu_R < \nu_c < \nu_x$ & $2.23$ & $0.11$ & $0.01$ & $10.36$ & $0.03$ \\
\hline
\bf{GBM} \\
080916A & $406.08 \pm 111.96$ & ISM & $\nu_R < \nu_c < \nu_x$ & $2.29$ & $0.13$ & $0.06$ & $0.84$ & $0.66$ \\
081121 & $1076.56 \pm 318.45$ & ISM & $\nu_R < \nu_x < \nu_c$ & $2.88$ & $0.24$ & $0.01$ & $8.79$ & $0.07$ \\
090424 & $565.49 \pm 124.12$ & ISM & $\nu_R < \nu_x < \nu_c$ & $2.50$ & $0.10$ & $0.02$ & $3.37$ & $0.50$ \\
090618 & $593.57 \pm 129.57$ & Wind & $\nu_R < \nu_x < \nu_c$ & $2.59$ & $0.14$ & $0.03$ & $4.40$ & $0.35$ \\
091127 & $619.13 \pm 110.30$ & Wind & $\nu_R < \nu_x < \nu_c$ & $2.41$ & $0.28$ & $0.02$ & $24.33$ & $6.87\times10^{-5}$ \\
100906A & $17588.58 \pm 5668.57$ & Unknown & $\nu_c < \nu_R < \nu_x$ & $2.50$ & $0.31$ & $0.05$ & $10.54$ & $0.01$ \\
101219B & $2273.75 \pm 460.45$ & Wind & $\nu_R < \nu_c < \nu_x$ & $1.67$ & $0.22$ & $0.01$ & $19.95$ & $8.31\times10^{-4}$ \\
110213A & $6587.84 \pm 1756.07$ & Wind & $\nu_R < \nu_x < \nu_c$ & $3.02$ & $0.22$ & $0.04$ & $5.13$ & $0.27$ \\
111228A & $635.51 \pm 175.62$ & ISM & $\nu_R < \nu_x < \nu_c$ & $2.56$ & $0.17$ & $0.02$ & $5.56$ & $0.23$ \\
120119A & $3584.06 \pm 1063.29$ & ISM & $\nu_R < \nu_x < \nu_c$ & $2.76$ & $0.12$ & $0.01$ & $13.90$ & $0.01$ \\
120729A & $9530.69 \pm 3564.97$ & Unknown & $\nu_c < \nu_R < \nu_x$ & $2.28$ & $0.23$ & $0.08$ & $1.33$ & $0.51$ \\
130420A & $1309.41 \pm 444.11$ & ISM & $\nu_R < \nu_c < \nu_x$ & $2.24$ & $0.06$ & $0.03$ & $0.81$ & $0.67$ \\
140213A & $2419.08 \pm 529.16$ & Unknown & $\nu_c < \nu_R < \nu_x$ & $1.95$ & $0.15$ & $0.02$ & $6.83$ & $0.08$ \\
140506A & $530.34 \pm 158.82$ & ISM & $\nu_R < \nu_c < \nu_x$ & $1.91$ & $0.13$ & $0.02$ & $5.20$ & $0.16$ \\
140512A & $172.39 \pm 74.66$ & Wind & $\nu_R < \nu_x < \nu_c$ & $2.58$ & $0.19$ & $0.05$ & $3.29$ & $0.35$ \\
140703A & $10128.06^{+2388.19}_{-3668.34}$ & ISM & $\nu_R < \nu_x < \nu_c$ & $3.23$ & $0.31$ & $0.06$ & $5.82$ & $0.12$ \\
140801A & $3455.25 \pm 1451.96$ & Unknown & $\nu_c < \nu_R < \nu_x$ & $1.75$ & $0.13$ & $0.01$ & $5.47$ & $0.24$ \\
150301B & $1223.06 \pm 1121.23$ & ISM & $\nu_R < \nu_x < \nu_c$ & $2.49$ & $0.16$ & $0.03$ & $1.39$ & $0.50$ \\
151027A & $4900.49 \pm 1227.01$ & Wind & $\nu_R < \nu_x < \nu_c$ & $2.88$ & $0.25$ & $0.02$ & $11.60$ & $0.02$ \\
170113A & $621.81 \pm 227.70$ & ISM & $\nu_R < \nu_x < \nu_c$ & $2.54$ & $0.20$ & $0.02$ & $8.93$ & $0.06$ \\
\hline\hline
\caption{The best fit results from trying to assign environment types to each GRB. Results tabulated here are at least a factor of three times better in probability than the next-best fit of an alternative environment type. $\sigma_p$ is the weighted, unbiased sample standard deviation, representing the 1$\sigma$ confidence interval of the sampling distribution. This error measures the spread of data points in the sample. SE$_p$ is the uncertainty of the weighted mean (the standard error of the mean), representing the dispersion of the sample mean about the population mean. This error describes the uncertainty in our measured $p$ (the sample mean) as an estimate of the true value of $p$ (the population mean) due to the limited sample size.
\label{tab:bestfits}}
\end{longtable*}

\begin{figure*}
\begin{center}
\includegraphics[width=8.9cm]{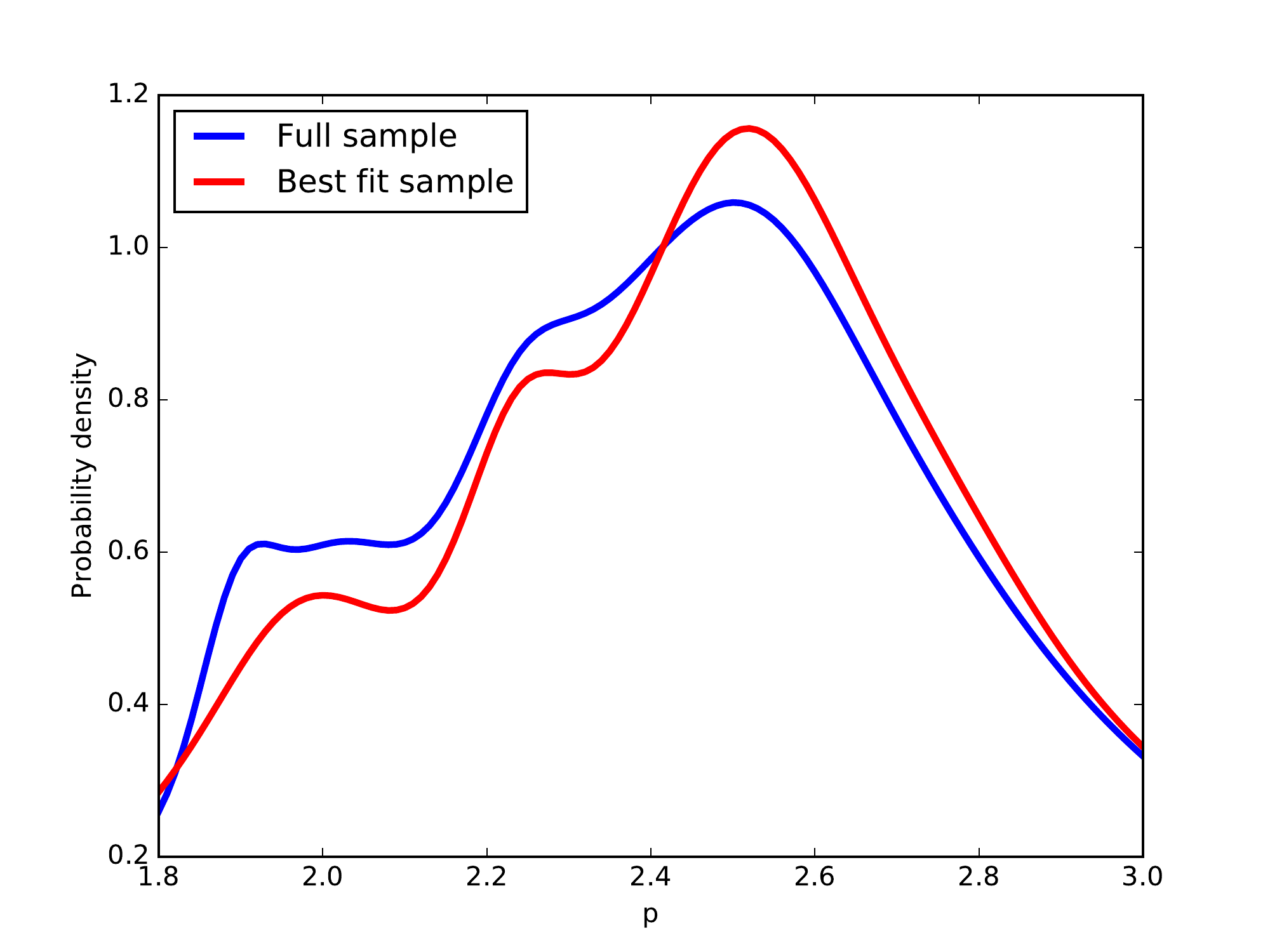}
\includegraphics[width=8.9cm]{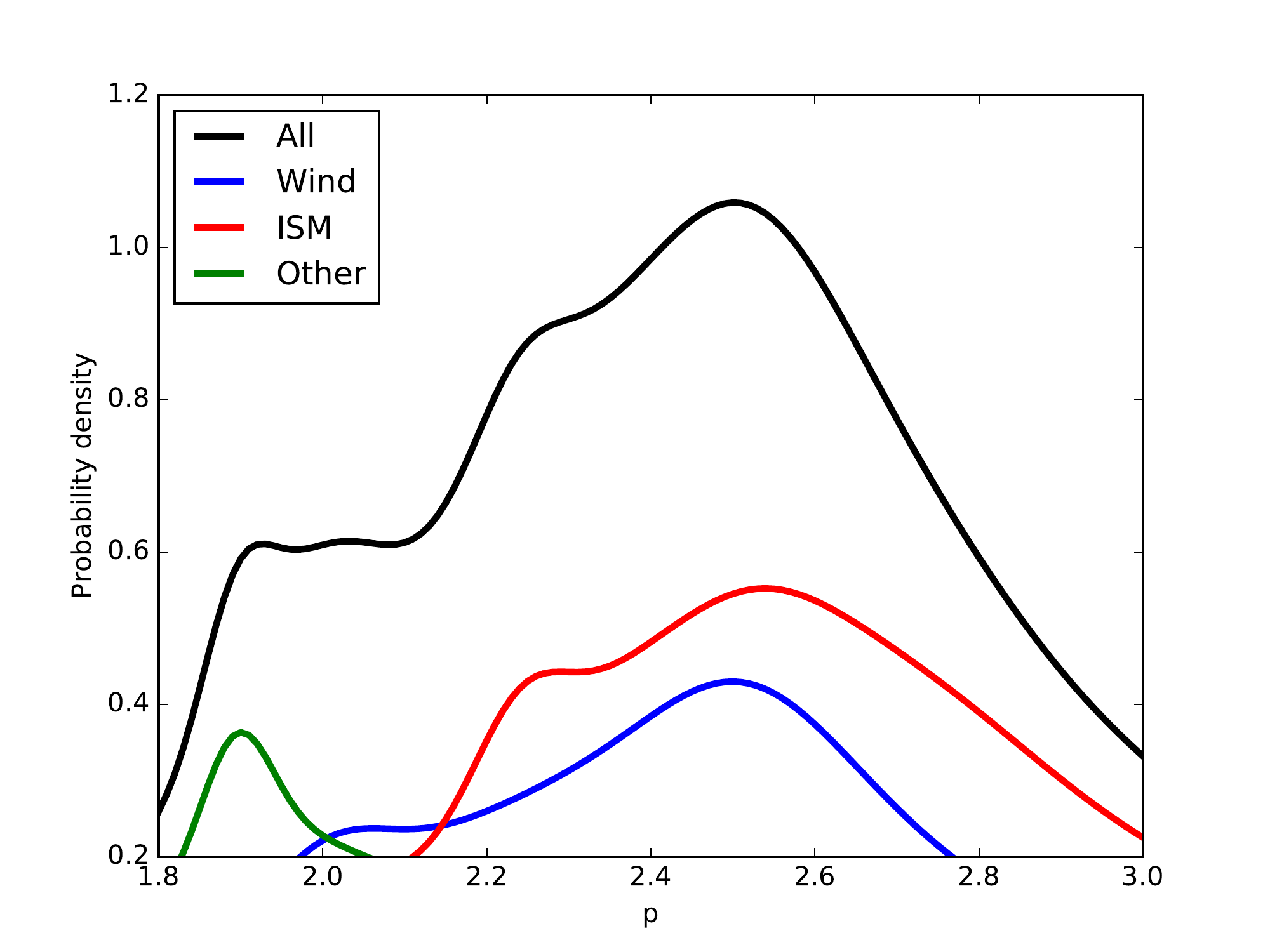}
\caption{The maximum likelihood curve of the electron energy distribution power law index, $p$. This was found by summing the probability density functions for each best fit and normalising by the number of bursts. \\ {\it Left:} The distribution of the best fit sample (red), in which the best fitting model has to be at least three times higher in probability than the best fitting model of an alternate environment type. The distribution for all $56$ GRBs is shown in blue. Both curves are normalised by the number of bursts they contain ($37$ and $56$, respectively). The two distributions are highly similar.\\ {\it Right:} All $56$ GRBs broken down into individual wind (blue), ISM (red) and $\nu_c < \nu_R$ (green) probability density functions. These curves are normalised by the total number of bursts ($56$). ISM and wind-like bursts share very similar probability density functions, with a most likely value of $p \sim 2.5$. \label{fig:pdist}}
\end{center}
\end{figure*}

The true standard error is estimated by investigating its effect on $\chi^2_{\nu}$ and obtaining the maximum likelihood value. To do this, we take the $\chi^2_{\nu}$ of each best fit and calculate its probability from the $\chi^2$ probability density function with the relevant degrees of freedom. We then take the product of the probabilities for all $56$ GRBs. This is done for all values of standard error in the range $0$ to $0.5$. We normalise the resulting function with its own integral via trapezium rule integration. Note that the probabilities used in this process are not the same as those used for identifying best fits, which are calculated using the $\chi^2$ cumulative distribution function, rather than the $\chi^2$ probability density function used here.

An important point to clarify is that this need for an intrinsic error is driven primarily by bursts with very good datasets \citep[see e.g.][]{Ackermann13,Martin-Carrillo14,Perley14}. These bursts have spectral and temporal indices that are well sampled over long baselines, and hence have small associated errors. Conversely, those bursts with poor data availability (e.g. those for which we rely on GCN circulars) have larger measured errors in their indices. While these bursts may have comparable levels of unseen intrinsic error and be more susceptible to intrinsic scatter, these effects are not felt in the fitting process because the larger measurement errors mean they dominate the statistics far less than in the well measured cases. We discuss the influence of intrinsic scatter in the data in Section~\ref{sec:intscat}.

\section{Results}\label{sec:results}
\subsection{Identification of Environments}

We are able to find a best fit with a probability at least three times greater than the best fit of an alternative environment type for $37$ out of $56$ GRBs ($66$ per cent). Of these, $14$ ($38$ per cent) are best fitted with a wind-like environment, and $18$  ($49$ per cent) with an ISM. We also find $5$ bursts ($14$ per cent) that are best fitted with $\nu_c < \nu_R$, and for these the environment type cannot be identified. The sum to $101$ per cent is the result of rounding. We therefore find a nearly even split between wind-like and ISM-like bursts. Our results are displayed in Table~\ref{tab:bestfits}, and individual plots can be seen in Appendix~\ref{app:fits}. Our results generally agree with \citet{Schulze11} where they overlap. The one exception is GRB 090926A, which we identify as wind-like.

For each best fit, we take the Gaussian probability density function with a mean of the measured value and standard deviation of the $1\sigma$ errors. We then take the sum of these distributions and normalise by the number of bursts to construct a maximum likelihood curve of $p$ for our results, which is shown in Figure~\ref{fig:pdist}. Both wind and ISM-like bursts have a maximum likelihood $p \sim 2.5$, consistent with the findings of \citet{Curran10}, who showed that their sample of \emph{Swift} GRBs exhibited values of $p$ consistent with a Gaussian distribution centred at $p = 2.36$ with a width of $0.59$. Bursts with $\nu_c < \nu_R$ have a maximum likelihood $p \sim 1.9$. The distribution with best fits three times better than the next best fit is highly similar to the overall distribution, indicating that this criterion is not biasing our results.

\begin{figure*}
\begin{center}
\includegraphics[width=8.9cm]{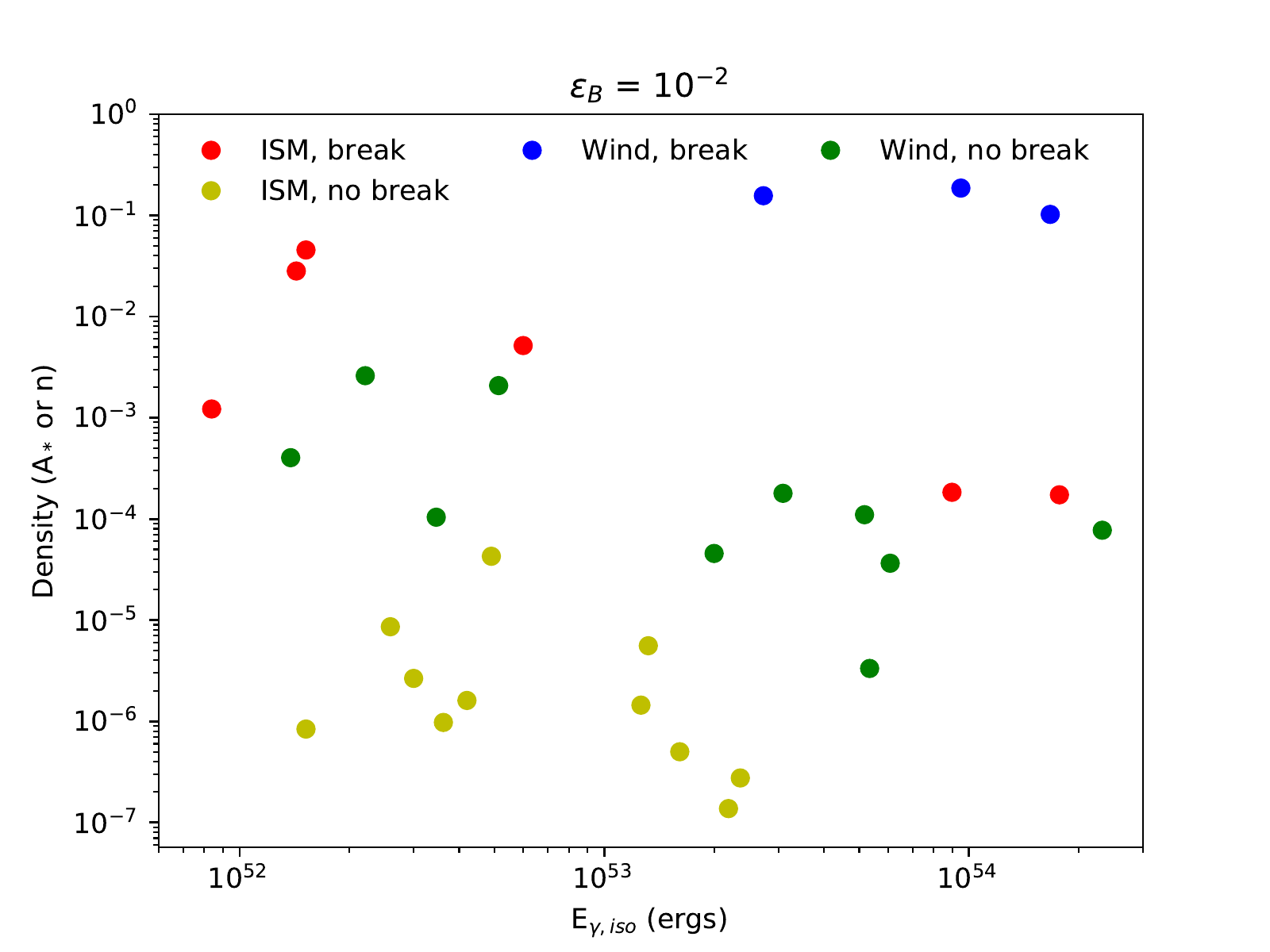}
\includegraphics[width=8.9cm]{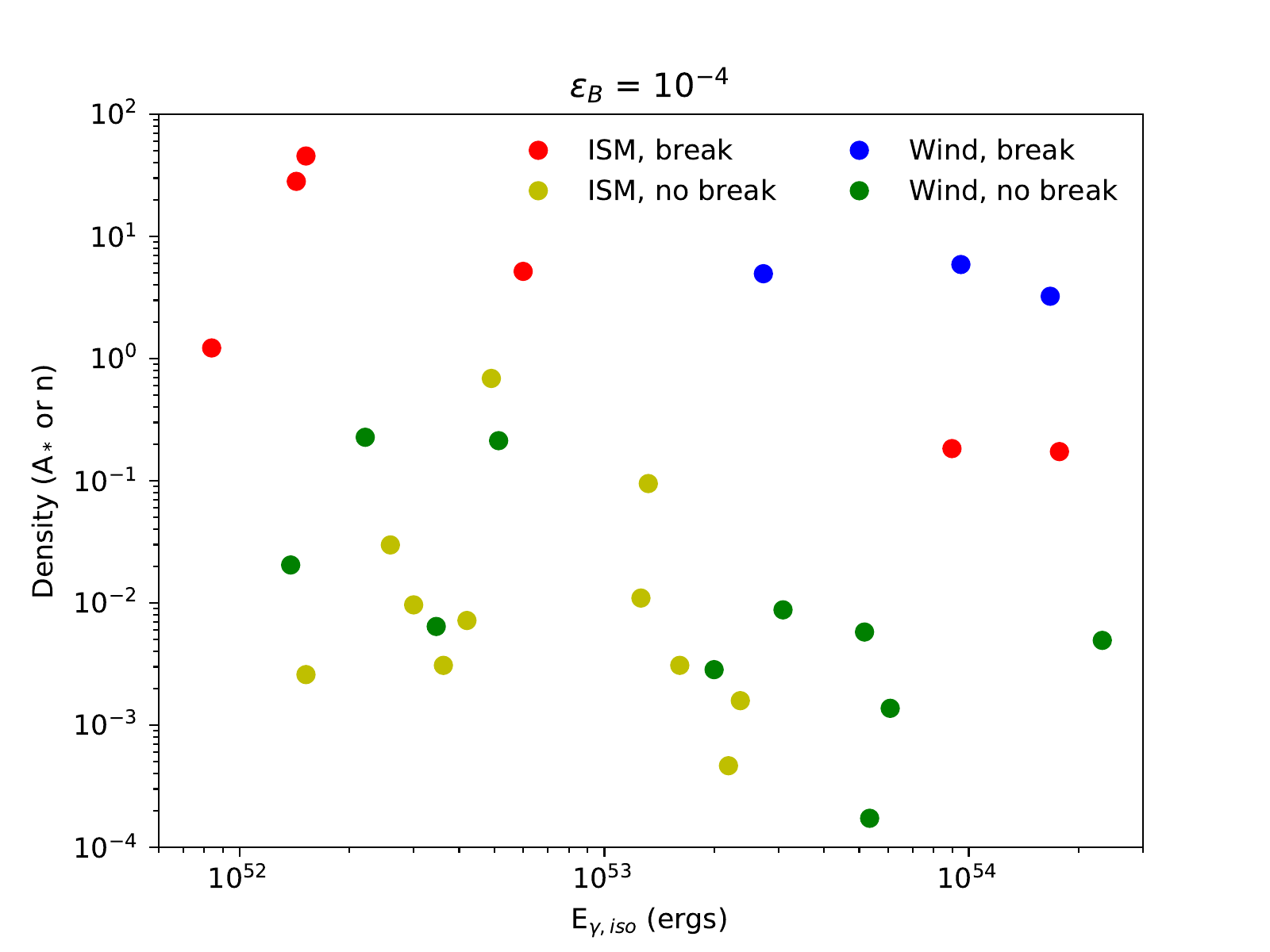}
\caption{Estimates of environment density vs $E_{\gamma,iso}$ for an assumed value of $\epsilon_B$ and $\epsilon_e = 0.1$. Errors are found to be of a similar order of magnitude to the values. The units of the y-axis are mixed, and are cm$^{-3}$ for the ISM-like bursts, or $5\times10^{11}$A for wind-like bursts, where A is in g~cm$^{-1}$. The left plot shows the inferred densities for $\epsilon_B = 10^{-2}$, and the right plot for $\epsilon_B = 10^{-4}$. Bursts in which $\nu_c < \nu_x$ are found to be in the densest environments, as would be expected. It is noticeable that $\epsilon_B = 10^{-2}$ leads to many ISM bursts having densities of just $10^{-6}$~cm$^{-3}$ or lower, which is the intergalactic medium value. Lower values of $\epsilon_B$ are therefore favored. \label{fig:density}}
\end{center}
\end{figure*}

\subsection{Implications for the Blast-Wave}
As discussed in Section~\ref{sec:params}, once the environment type is known and we have a measurement for $p$, we can solve the appropriate equation from Section~\ref{sec:params} to find a value for the degenerate combination of $\epsilon_B$ and $n$ (or $A_*$). In cases IV and V we must also assume a value for $\epsilon_e$. Again, we take this value to be $\epsilon_e = 0.1$, following \citet{Beniamini17}.

\begin{figure}
\begin{center}
\includegraphics[width=8.9cm]{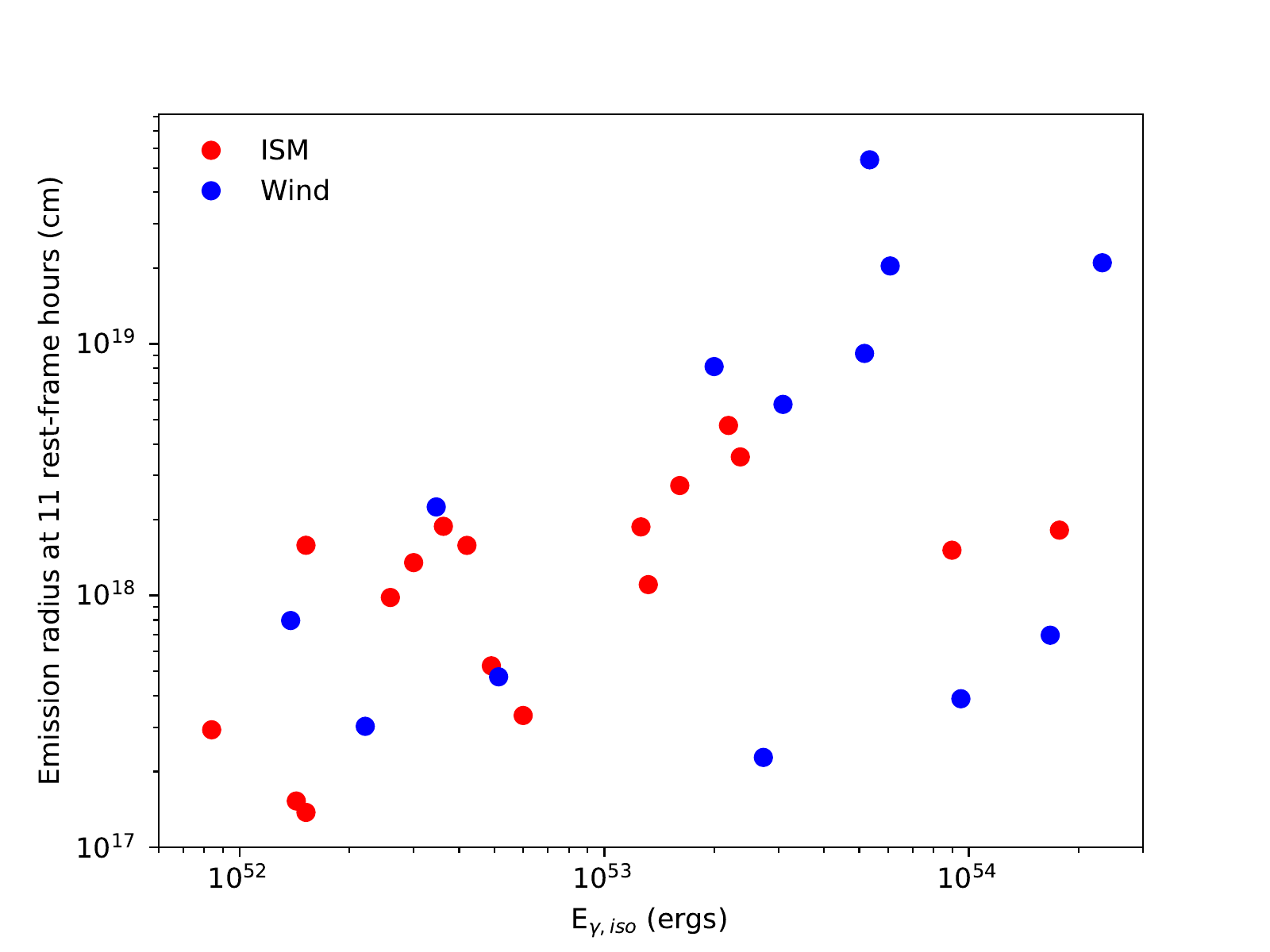}
\includegraphics[width=8.9cm]{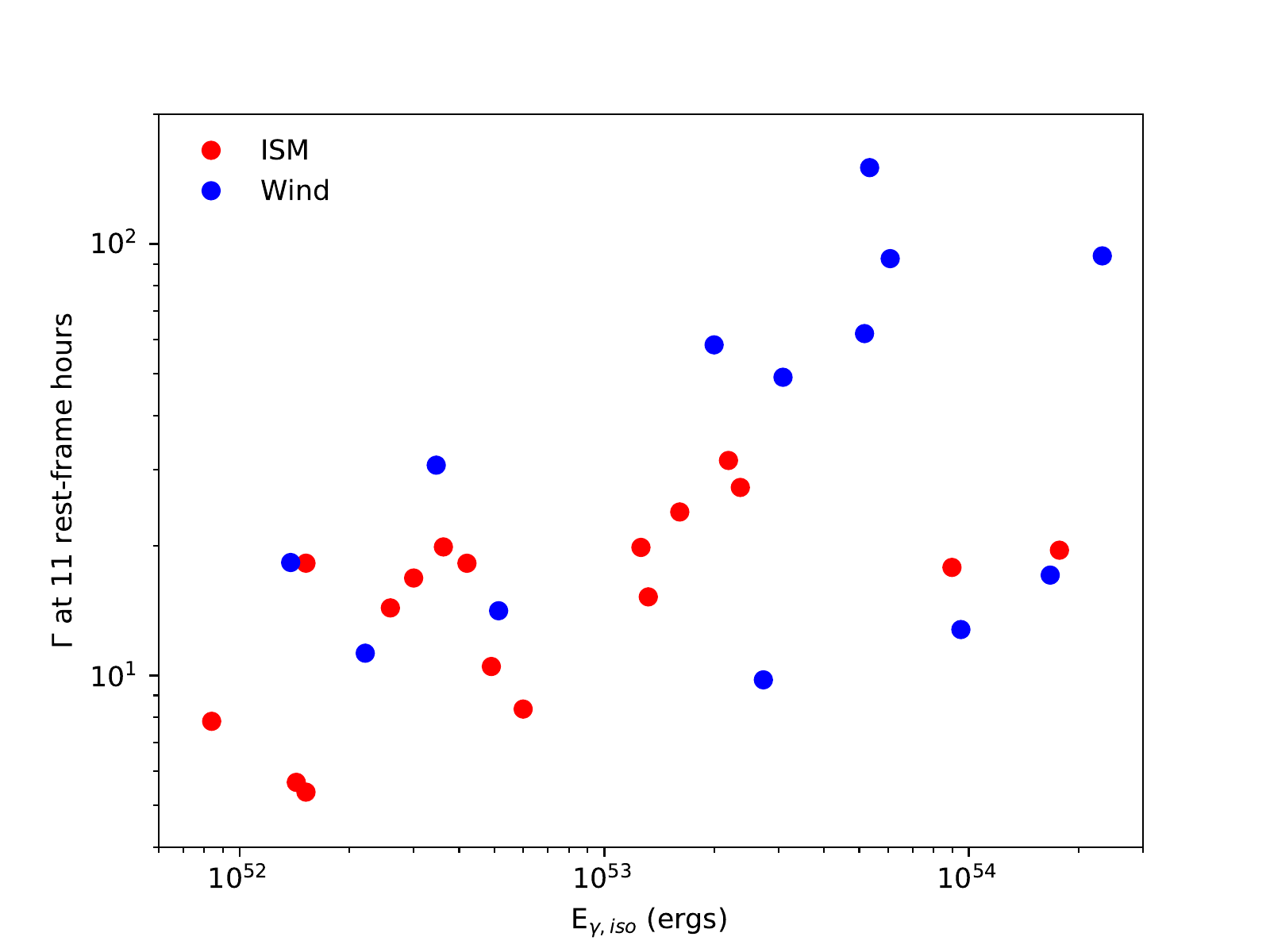}
\caption{The estimated emission radius and Lorentz factor at 11 hours rest-frame after trigger. We assume $\epsilon_B = 10^{-4}$ and $\epsilon_e = 0.1$. The cluster of bursts in the upper right of the lower panel have unusually high Lorentz factors for this time in the burst evolution. They reinforce the need for $\epsilon_B \ll 10^{-2}$, since lower assumed $\epsilon_B$ leads to lower implied $\Gamma$ \label{fig:shock}}
\end{center}
\end{figure}

Figure~\ref{fig:density} shows estimates of the density from two different assumed values of $\epsilon_B$. We find that densities associated with $\epsilon_B = 10^{-2}$ are extremely low; in many cases $10^{-6}$~cm$^{-3}$ or lower. Densities as low as this are associated with the intergalactic medium (IGM), and are therefore too low to be plausible around massive stars and in star forming regions. For this reason, lower values of $\epsilon_B$ are favored by our analysis, and we find $\epsilon_B = 10^{-4}$ gives a range of $10^{-4} < n < 10^2$~cm$^{-3}$ and $10^{-3} < A_* < 10^1$. This finding suggests that $\epsilon_B = 10^{-2}$, which is commonly assumed in GRB literature, over-estimates the magnetic field energy.

It is important to note that in our analysis, we use $E_{\gamma,iso}$ as a proxy for the total GRB energy. In fact, $E_{\gamma,iso}$ is the energy radiated by the electrons, and so ought to be $\sim \epsilon_e \times E_{tot}$. However, a large fraction of the prompt emission is likely to be thermal \citep[at least in very high energy GRBs;][]{Peer15}, so our assumption is an acceptable estimate for the GRB energy.

Taking our density analysis a step further, we can use our results to estimate the emission radius and Lorentz factor of the shock front at a given time after trigger. Emission radius is given by
\begin{equation}
r = 5.85\times10^{17} \bigg(\frac{1+z}{2}\bigg)^{-1/4} E_{53}^{1/4} n_0^{-1/4} t_{day,obs}^{1/4} \mbox{cm}
\end{equation}
for the ISM and
\begin{equation}
r = 3.2\times10^{17} \bigg(\frac{1+z}{2}\bigg)^{-1/2} E_{53}^{1/2} A_*^{-1/2} t_{day,obs}^{1/2} \mbox{cm}
\end{equation}
for the wind. The Lorentz factor is give by
\begin{equation}
\Gamma = 10.6 \bigg(\frac{1+z}{2}\bigg)^{3/8} E_{53}^{1/8} n_0^{-1/8} t_{day,obs}^{-3/8}
\end{equation}
in the ISM, and by
\begin{equation}
\Gamma = 11.1 \bigg(\frac{1+z}{2}\bigg)^{1/4} E_{53}^{1/4} A_*^{-1/4} t_{day,obs}^{-1/4}
\end{equation}
for the wind \citep[cf.][]{Blandford76,Sari98,Peer05}. Figure~\ref{fig:shock} shows the values for both at 11 hours after trigger in the rest frame. We use the densities implied for $\epsilon_B = 10^{-4}$. No clear separation is seen between the wind-like and ISM-like bursts in our sample, though the largest radii and highest Lorentz factors are all found in wind environments.

\subsection{Intrinsic Error} 
\begin{figure}
\begin{center}
\includegraphics[width=9.cm]{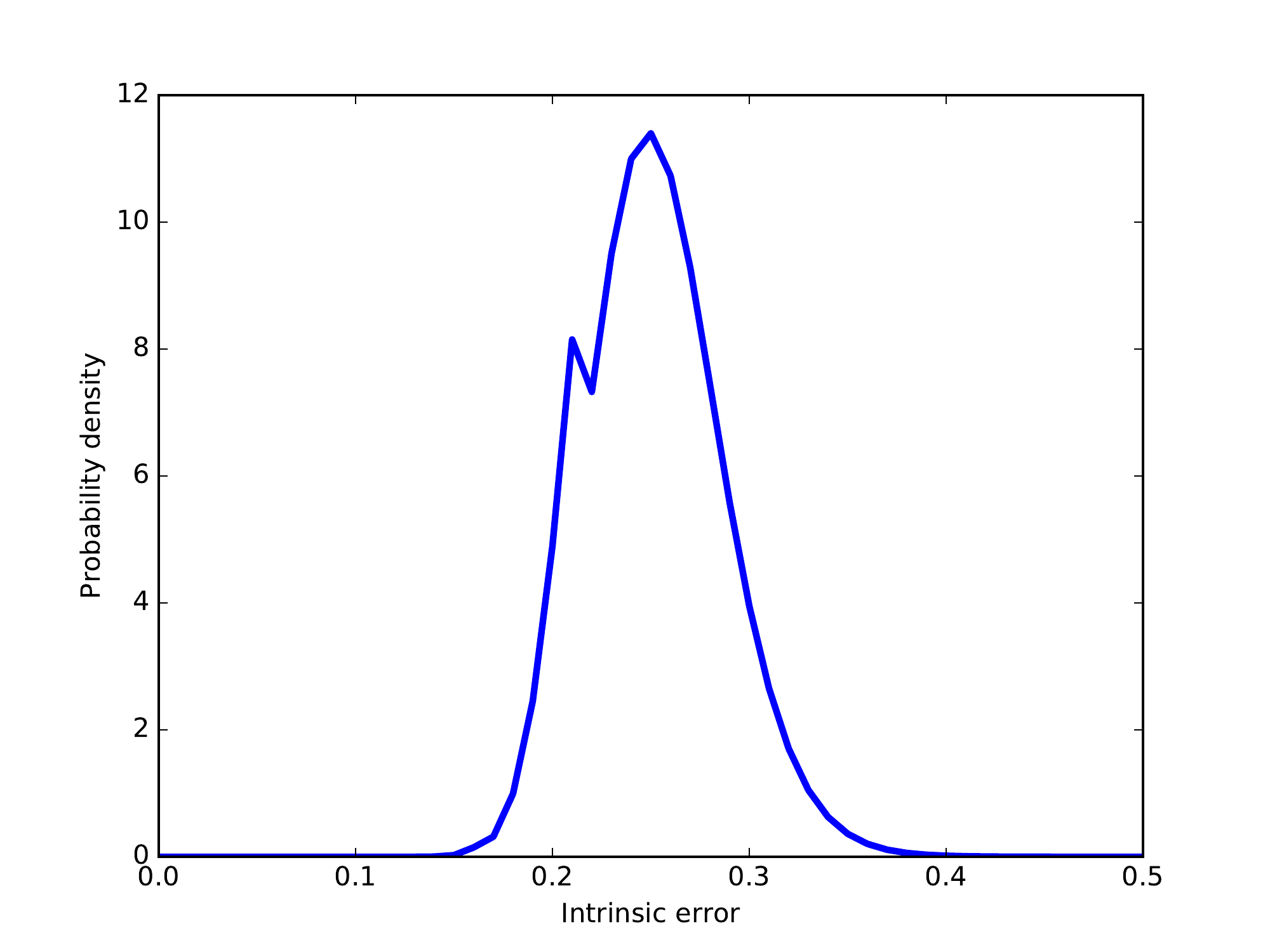}
\caption{The normalized product of the best fit probabilities from the $\chi^2$ probability density function for the $56$ GRBs as a function of the standard error added in quadrature to the measured errors. The peak, indicating the maximum likelihood value, is at a standard error of $0.25 \pm 0.04$. \label{fig:chidist}}
\end{center}
\end{figure}

As discussed in Section~\ref{sec:interr}, we assume that the intrinsic error can be represented by an error in $p$, caused by the derived $p$ not behaving as predicted by the standard power-law closure relations. The likelihood function for our estimate of the intrinsic error are plotted in Figure~\ref{fig:chidist}. We find a best estimate of the standard error of $0.25 \pm 0.04$, or roughly a ten percent error given the typical value of $p$ of 2.5. Table~\ref{tab:bestfits+se} shows the updated values of $p$ and $\chi^2_{\nu}$ for our best fit sample from Table~\ref{tab:bestfits} when this standard error is included. Individual fits are shown in Appendix~\ref{app:fits_se}.

\LTcapwidth=8.5cm
\begin{longtable}{lccccc}
\hline\hline
GRB & $p$ & $\sigma_p$ & SE$_p$ & $\chi^2_{\nu}$ & Prob \\
\hline
\endhead
\bf{LAT} \\
090323 & 3.28 & 0.37 & 0.06 & 2.62 & 0.62 \\
090328A & 3.10 & 0.34 & 0.06 & 1.79 & 0.77 \\
090902B & 2.66 & 0.22 & 0.05 & 1.18 & 0.88 \\
090926A & 2.37 & 0.30 & 0.07 & 1.95 & 0.74 \\
091003 & 2.69 & 0.20 & 0.07 & 0.89 & 0.83 \\
091208B & 1.81 & 0.15 & 0.05 & 0.60 & 0.90 \\
100728A* & 2.35 & 0.19 & 0.13 & 0.70 & 0.71 \\
110731A & 2.78 & 0.31 & 0.08 & 2.16 & 0.71 \\
120711A & 2.52 & 0.31 & 0.07 & 2.31 & 0.68 \\
130427A & 2.56 & 0.19 & 0.05 & 1.05 & 0.90 \\
130907A & 2.94 & 0.19 & 0.07 & 0.75 & 0.86 \\
131108A & 2.44 & 0.29 & 0.14 & 1.09 & 0.78 \\
131231A & 2.66 & 0.14 & 0.06 & 0.47 & 0.92 \\
141028A & 2.00 & 0.16 & 0.06 & 0.38 & 0.98 \\
150403A & 2.56 & 0.16 & 0.12 & 0.49 & 0.78 \\
160623A & 2.83 & 0.19 & 0.06 & 0.88 & 0.83 \\
160625B & 2.06 & 0.27 & 0.13 & 1.04 & 0.79 \\
\hline
\bf{GBM} \\
080916A* & 2.23 & 0.13 & 0.17 & 0.19 & 0.91 \\
081121 & 2.74 & 0.35 & 0.07 & 2.22 & 0.70 \\
090424 & 2.62 & 0.10 & 0.05 & 0.24 & 0.99 \\
090618 & 2.61 & 0.20 & 0.05 & 0.99 & 0.91 \\
091127 & 2.61 & 0.39 & 0.04 & 4.10 & 0.39 \\
100906A* & 2.49 & 0.3 & 0.08 & 2.14 & 0.34 \\
101219B* & 1.97 & 0.21 & 0.05 & 1.04 & 0.90 \\
110213A* & 3.21 & 0.26 & 0.06 & 1.32 & 0.86 \\
111228A & 2.66 & 0.15 & 0.06 & 0.55 & 0.97 \\
120119A & 3.04 & 0.32 & 0.07 & 2.45 & 0.65 \\
120729A* & 2.32 & 0.22 & 0.09 & 0.63 & 0.73 \\
130420A & 2.21 & 0.06 & 0.16 & 0.05 & 0.98 \\
140213A & 1.98 & 0.19 & 0.05 & 0.86 & 0.83 \\
140506A & 1.84 & 0.12 & 0.15 & 0.15 & 0.99 \\
140512A & 2.44 & 0.25 & 0.09 & 1.31 & 0.73 \\
140703A & 3.30 & 0.34 & 0.07 & 2.25 & 0.52 \\
140801A & 1.97 & 0.25 & 0.09 & 1.08 & 0.90 \\
150301B & 2.65 & 0.19 & 0.15 & 0.51 & 0.77 \\
151027A & 3.11 & 0.38 & 0.06 & 3.24 & 0.52 \\
161017A* & 2.45 & 0.40 & 0.08 & 3.31 & 0.35 \\
170113A & 2.60 & 0.17 & 0.08 & 0.56 & 0.97 \\
\hline\hline
\caption{Same as Table~\ref{tab:bestfits}, but with our derived standard error included. Bursts marked with an asterisk have best fits that remain a factor of three times better than the next best fit. Note the addition of a solution for GRB 161017A, which occurred because a competing solution drifted outside of our allowed range of $1.8 \leq p \leq 3.0$. \label{tab:bestfits+se}}
\end{longtable}

\subsection{Parameter Correlations}
A number of measured parameters are available for correlative analysis: $E_{\rm \gamma,iso}$, $z$, $E_{\rm peak}$, $F_R/F_X$ and $T_{\rm 90}$. In addition, we have the derived parameter $p$. We tested them against each other using the ranked Spearman correlation test (we use the Spearman test because it does not assume that the data are normally distributed). Firstly, we clearly recover the well-established correlation between $E_{\rm \gamma,iso}$ and $E_{\rm peak, rest}$ \citep{Amati06} at a significance much greater than $3 \sigma$. We also see a correlation between $E_{\rm \gamma,iso}$ and $z$ at greater than $3 \sigma$ significance. This is likely due to a selection effect in which fainter bursts at greater distances do not trigger GRB satellites, so that the faint-distant population goes unobserved.

No other compelling correlations are seen between the data. There is a $> 3 \sigma$ correlation between $F_R/F_X$ and $p$, but only in the sample of wind-like bursts for which $\nu_c > \nu_x$. While this correlation is expected in this spectral regime because a higher $F_R/F_X$ implies a steeper spectral index and therefore a greater value of $p$, the trend is not seen in the equivalent ISM sample, nor in the sample overall. In addition, the wind-like $\nu_c > \nu_x$ sub-sample contains only $11$ GRBs. We also see a $> 3 \sigma$ correlation between $E_{\rm \gamma,iso}$ and rest-frame $T_{\rm 90}$ for the $17$ bursts that were designated ISM-like, but not in the wind-like bursts or the sample overall.

\section{Discussion}\label{sec:discussion}
\subsection{Environments}\label{sec:environments}
\begin{figure}
\begin{center}
\includegraphics[width=9.cm]{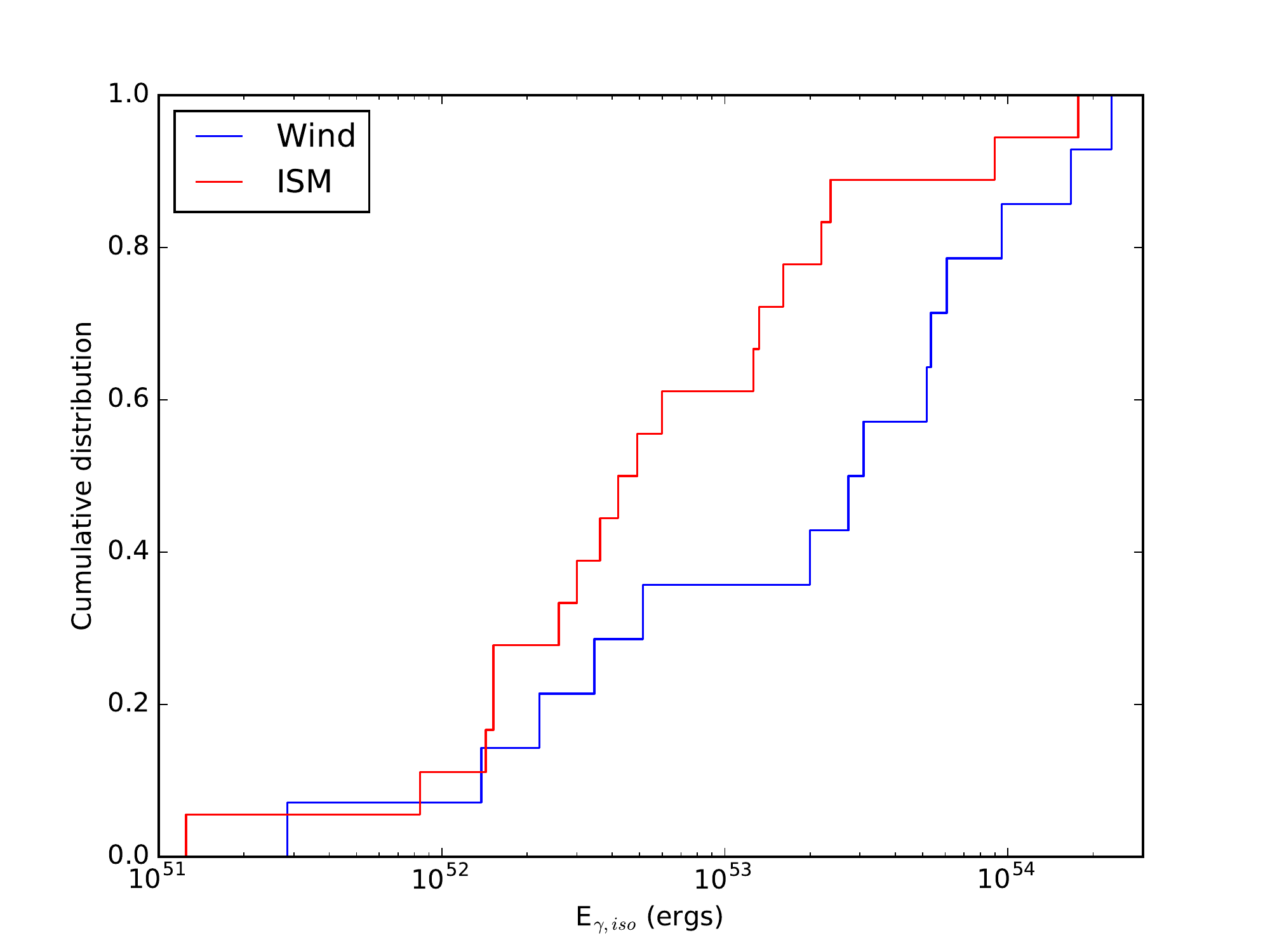}
\caption{The cumulative distributions of the energies of bursts identified as being in wind-like (blue) or ISM-like (red) environments. The distributions do exhibit some separation, but a KS test finds a $p = 0.048$ probability that they are drawn from the same overall population, so any separation is not statistically significant above $2\sigma$. \label{fig:cumdist}}
\end{center}
\end{figure}

Through interpreting the synchrotron spectrum using the closure relations, as is standard in the field, we find a roughly equal split between wind-like and ISM-like environments within our population of GRBs. This raises the question of how these two distinct environments arise. There are two likely possibilities for this:
\begin{enumerate}
\item{The two environments are due to two distinct progenitors.}
\item{Both originate from the same progenitor type, but are seen at different stages of evolution.}
\end{enumerate}
In practice these scenarios are not so different; a star with a weak stellar wind in the latter stages of life will influence its local environment to a lesser radial extent than a star with a strong stellar wind, so that an expanding jet from a GRB will cross into the ISM much faster as it will need to travel less distance. Since a more massive star is likely to release more energy on collapse, and also more likely to radiate a strong stellar wind, finding wind-like environments associated with more energetic GRBs perhaps makes intuitive sense. We tested this hypothesis by examining the cumulative distribution functions of $E_{\rm \gamma,iso}$ for the bursts we identified as wind or ISM (Figure~\ref{fig:cumdist}). Interestingly, we do indeed find a separation, with wind-like bursts typically exhibiting higher energies. A KS test reveals that this dichotomy is significant to $2\sigma$, with a $p = 0.048$ probability that the two are drawn from the same overall population. However, since the test statistic falls short of $3\sigma$ significance, our finding is not fully conclusive. An increased sample size may help to confirm the result. The existence of two populations in LGRBs has previously been suggested via analysis of their radio emission. First through an apparent dichotomy in radio flux \citep{Hancock13}, and furthermore between the $\Gamma$-ray durations of radio-bright and radio-quiet bursts \citep{Lloyd-Ronning17}. Both of these works focused on the GRB radio observation catalog of \citet{Chandra12}. The overlap with our sample is small, but an updated comparison remains a promising avenue for testing our findings in future works.

If the two types are from the same progenitor and we just observe them before/after they cross from a wind-like profile close to the parent star to an ISM-like profile beyond the termination shock, then it is reasonable to expect to see a number of cases where this transition is observed. \citet{Dai02} performed hydrodynamic simulations of GRB afterglows crossing a termination shock for ISM-ISM and wind-ISM transitions. Their results show that the expected signature is a sharp drop in the light curve, followed by a flare, then by the afterglow settling back into a power-law. For the wind-ISM transition, this second power-law is shallower than the pre-transition index. Such a signature has been claimed in a number of GRBs: 030226 \citep{Dai03}, 081109A \citep{Jin09}, 080916C, 090902B, 090926A \citep{Feng11} and 130907A \citep{Veres15}. While suggestive, none of these are conclusive; in some cases the supposed transition occurs in gaps between the data, and in others the light curve behaviour could be explained by the transition between the prompt emission tail and the rise of the underlying afterglow, or the passage of $\nu_c$ through the X-ray bandpass.

The lack of identified termination shock crossings is concerning for a single LGRB progenitor scenario; previous studies of much smaller samples than ours \citep[e.g.][]{Panaitescu02,Starling08,Curran09} find that 25 - 50 per cent of GRBs exhibit ISM-like environments (consistent with our own findings), and so termination shocks should be common, yet none are seen. We note that since we find a roughly even wind/ISM split when searching close to 11 rest-frame hours after trigger, this must be close to the mean termination shock crossing time if this is the cause for the two observed environment types. However, while we may be close to the mean value, the spread of termination shock radii is likely to be quite high, perhaps deviating by as much as an order of magnitude \citep{Peer06}.

The wind and ISM environments are approximations of the true environment, in which a density with a perfect power-law dependence is assumed. Indeed, some previous works have attempted to fit $k$, the power-law index of this dependence, as a free parameter \citep[e.g.][]{Starling08}. Our method does not enable us to treat $k$ as a free parameter because the extra degree of freedom would prevent fits from converging. However, to test whether our assumptions of $k = 0$ (ISM) or $k = 2$ (wind) are robust, we include an intermediate case in which $k = 1$ and re-fit our sample. We find that our results are largely unchanged; $5$ GRBs do find best fits with $k = 1$, but none of these are high probability, and all are GRBs for which the wind/ISM dichotomy provided poor quality fits originally, thus enabling the $k = 1$ model to `fill in the gaps'. Additionally, when repeating our measurement of intrinsic error we find it almost unchanged, with a new value of $0.21 \pm 0.04$. This is completely consistent with our original value of $0.25 \pm 0.04$, and indicates that our assumption of wind/ISM environments is most likely not a significant contributor to the intrinsic error.

\subsection{Constraints on the Plasma Parameters}
Our estimates of the local density in Figure~\ref{fig:density} show that bursts with $\nu_c < \nu_x$ are found in the highest density environments. This is an expected result, since the frequency of $\nu_c$ is inversely proportional to density in the ISM (Equation~\ref{eqn:case2}), or the square of the density in a wind (Equation~\ref{eqn:case3}). Higher densities therefore more readily lead to $\nu_c < \nu_x$. 

For similar progenitors, we may expect the emission radius to be higher for the ISM-like bursts, since these are supposed to be post-termination shock. However, Figure~\ref{fig:shock} shows that this is not the case for our sample; the highest inferred emission radii are all for wind-like media. This finding is consistent with the idea that the energy release and/or local density varies several orders of magnitudes in LGRBs. If the progenitors are the same for wind and ISM (i.e. emit a wind of a comparable strength), then the local environment density will determine the radial extent to which this stellar wind exerts influence. Following this, variations in the explosion energy determine how long it takes the forward shock to cross this region. If the progenitors are not the same, then this is further complicated by the varying wind strengths. Finding wind-like environments with larger emission radii and Lorentz factors than are found for the ISM-like environments tells us that the wind does not look the same in all LGRBs \citep{Peer06,vanMarle06}.

We note that the cluster of bursts in the upper-right of the right-hand panel in Figure~\ref{fig:shock} have very high Lorentz factors for 11 hours after trigger. They are GRB 090323 ($\Gamma = 150$), GRB 090926A ($\Gamma = 94$) and GRB 090902B ($\Gamma = 93$). The same bursts are also seen in the lower-right of the right-hand panel of Figure~\ref{fig:density} in green. They have high measured $E_{\gamma,iso}$ and were found to have low densities. The implication is that their jets were launched with a high Lorentz factor, and the tenuous surrounding environment did little to slow them down. However, converting their inferred $A_*$ to $n$ at 11 hours after trigger reveals particle densities in the range $10^{-8} < n < 10^{-5}$~cm$^{-3}$, which is extremely low. This can potentially be alleviated by a lower $\epsilon_B$, and further reinforces our earlier finding that $\epsilon_B \ll 10^{-2}$. This finding is in agreement with a number of previous works that find similarly low values of $\epsilon_B$ \citep[e.g.][]{Santana14,Beniamini15}.

\subsection{The Effects of Intrinsic Error}\label{sec:erreffects}
While our findings hold under a standard analysis of the GRB afterglows, we note a lack of success in fitting those GRBs with the most precise spectral and temporal index measurements. In these cases, the indicators of the underlying electron energy distribution, $p$, do not agree on a single value within their own errors. As discussed in Section~\ref{sec:interr}, this may indicate complexity in nature that goes beyond the simple equivalence drawn between the measured indices and $p$ through the closure relations. This intrinsic error may go unnoticed when the measured errors are large, but cause a falloff in goodness-of-fit with increasing precision. It is already known that standard synchrotron theory is a simplified model of a complex system, and it has been previously noted in the literature that the theory does not always do well against observations of GRB afterglows \citep[e.g.][]{Wang15}.

\begin{figure*}
\begin{center}
\includegraphics[width=8.9cm]{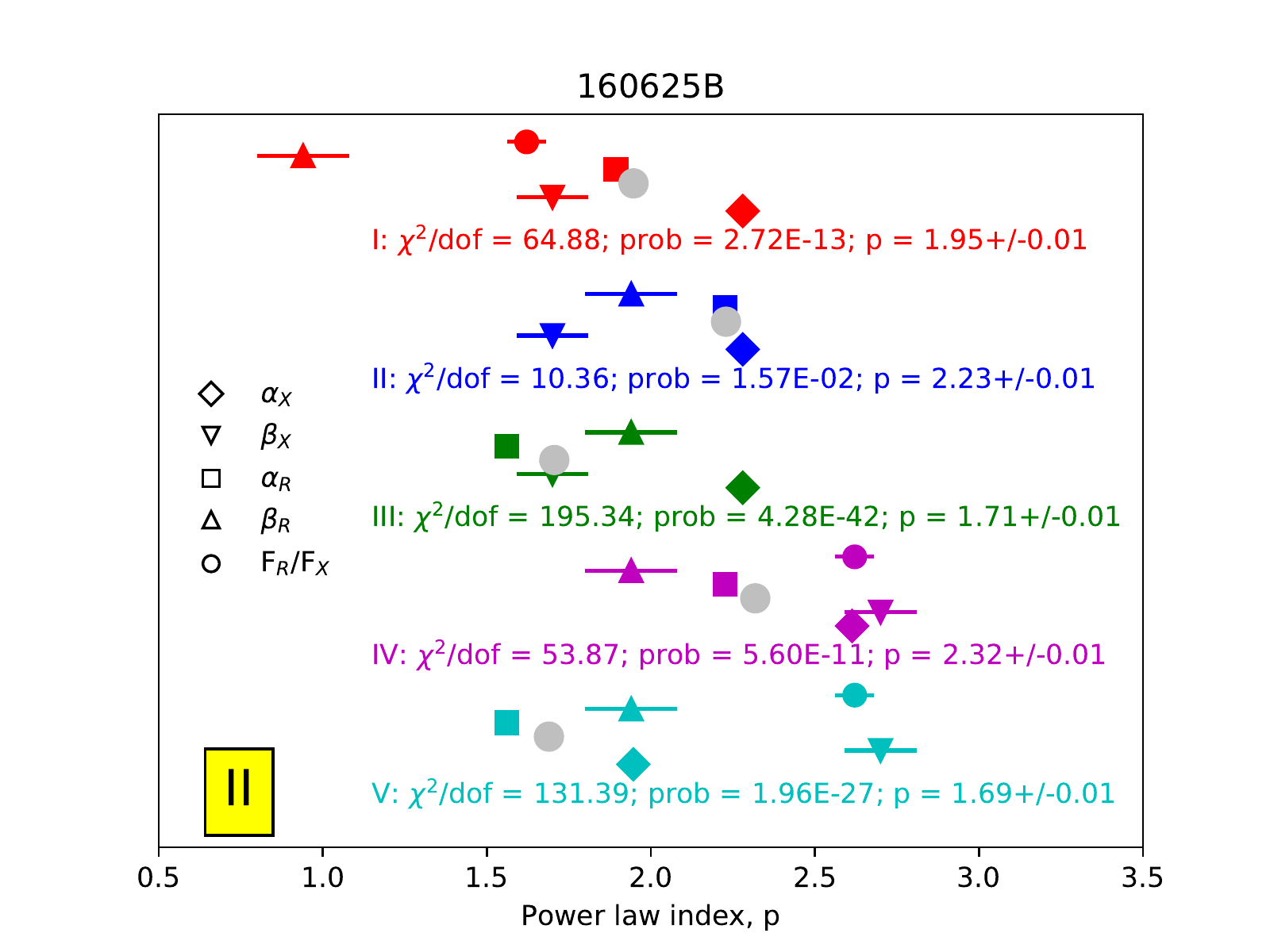}
\includegraphics[width=8.9cm]{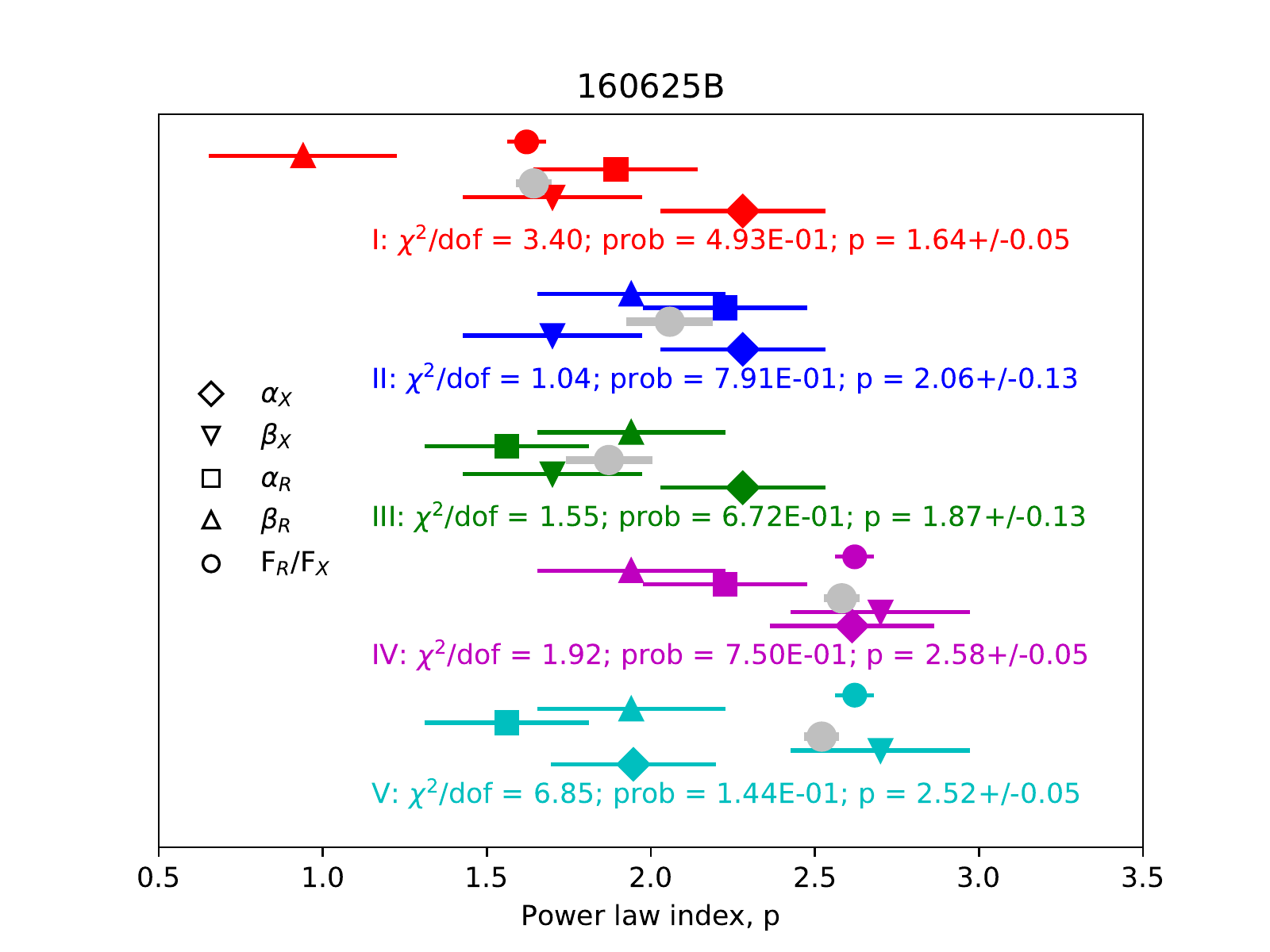}
\caption{Fits for each of the 5 spectral regimes listed in Section~\ref{sec:environment}. The measured values of $\alpha_x$, $\beta_x$, $\alpha_o$, $\beta_o$ and $F_R/F_X$ have been converted to $p$ using the relevant closure relations, and the best fitting value of $p$ is shown in grey for each spectral regime. The error on $p$ is the uncertainty of the weighted mean, not the standard deviation of the data.\\ {\it Left:} Best fits with no added intrinsic error. Model II ($\nu_R < \nu_c < \nu_X$ in an ISM environment) is the best fit. However, with $\chi^2_{\nu} = 10.36$, it is not a good fit.\\ {\it Right:} Best fits with an intrinsic error of $0.25$ added in quadrature to the measured errors for each index (except $F_R/F_X$). Model II is still the best fit, now with $\chi^2_{\nu} = 1.04$, but is no longer a factor of three times higher in probability than the best fit of an alternative environment type (model III; $\nu_R < \nu_c < \nu_X$ in a wind environment). Therefore, although an ISM is favored, we cannot say with certainty which environment surrounds GRB 160625B. \label{fig:secompare}}
\end{center}
\end{figure*}

\begin{figure}
\begin{center}
\includegraphics[width=9.cm]{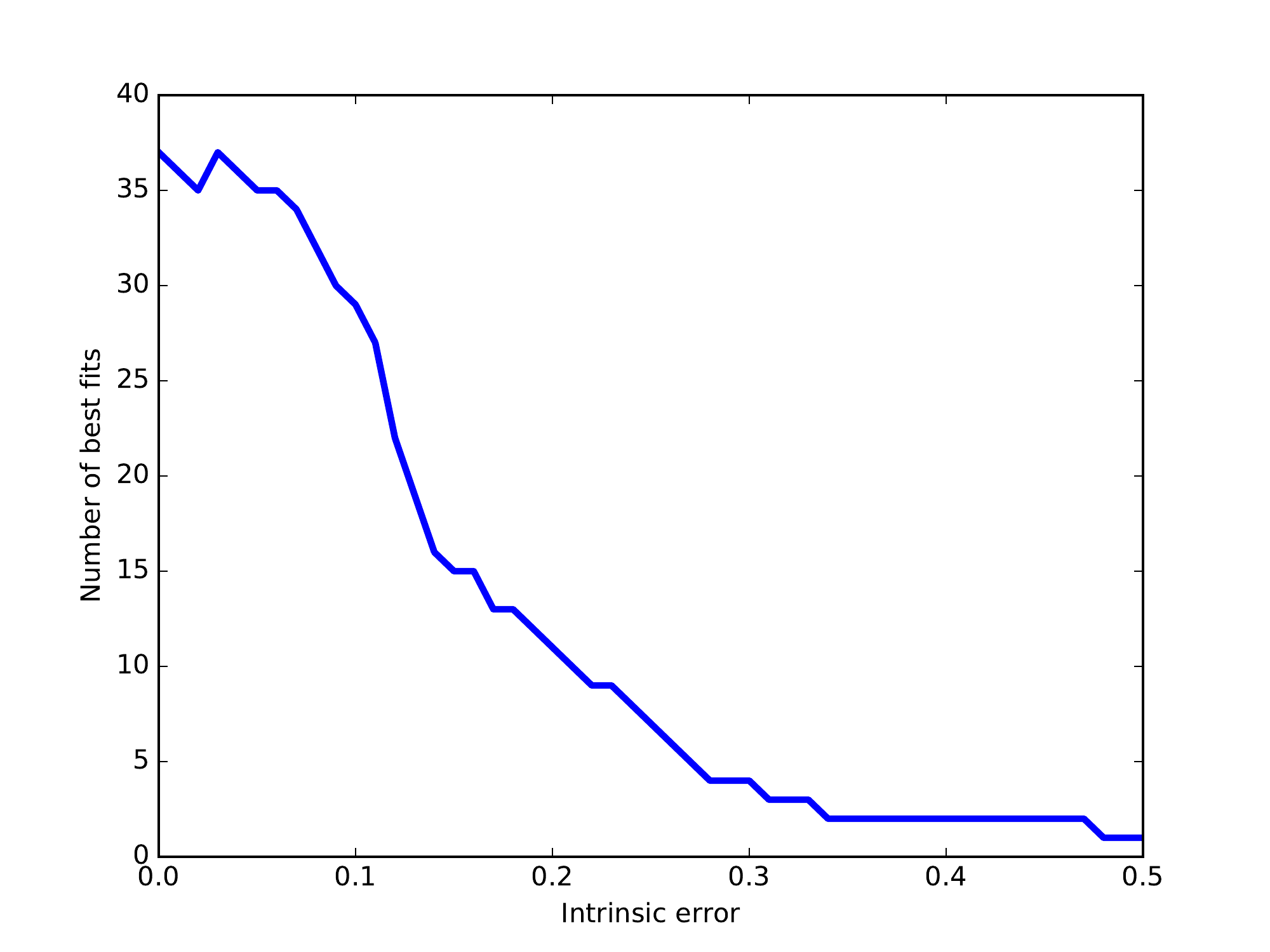}
\caption{The number of surviving best fits (where the fit probability is at least three times better than the best fit of an opposing environment type) as a function of the added intrinsic error. \label{fig:fitloss}}
\end{center}
\end{figure}

Figure~\ref{fig:secompare} illustrates the difference adding a standard error makes using GRB 160625B. This burst had very high precision measurements of both the X-ray and r-band temporal indices, and as a result the best fit must stick very close to them. However, their slight disagreement with one another, and more pronounced disagreement with other indicators of $p$, means that this ``best" fit had a $\chi^2_{\nu} = 10.36$; too high to be considered a good fit. With the standard error included, the best fit has a $\chi^2_{\nu} = 1.04$, but the most likely model is now not more than three times better than its nearest competitor of a different environment type. Losing definite best fits is an inevitable side effect of increasing the intrinsic error. An intrinsic error of $0.25$ reduces the number of GRBs with identifiable best fit environments to $7$ (down from $37$). $2$ wind environments, $1$ ISM environment and $4$ bursts where $\nu_c < \nu_R$ remain. The persistence of this last type is due to the fact that they occupy a niche by requiring steep optical spectral indices, whereas the other spectral regimes are more uniform in terms of their manifestations in the indices and so are easily confused with one another. They therefore more readily fall short of the requirement of being three times more probable than the next best fit. The number of identifiable best fits as a function of intrinsic error is shown in Figure~\ref{fig:fitloss}.

Ours is the first attempt to measure this uncertainty through the data, and we find that an intrinsic error of $0.25 \pm 0.04$ (about 10 per cent of p) added in quadrature is required to maximise the likelihood of the $\chi^2$ probability density function of our results. Worryingly, this intrinsic error negates our ability to identify the environment type in almost all cases - a result that could have sweeping consequences across the field if true. However, Table~\ref{tab:bestfits} shows that even without this intrinsic error, the majority of bursts do find acceptable best fits. Additionally, when a minimum value for an acceptable best-fit probability is imposed on the sample (not including intrinsic error), the split between wind and ISM-like bursts remains constant. This was true for all minimum best-fit probabilities tested, from $10^{-12}$ up to $0.01$ in order of magnitude increments. We re-emphasize that even with the intrinsic error included, the majority of bursts find the same environment types to be the best fit as without. The only difference is that they no longer exceed our (semi-arbitrary) threshold of a factor of 3 in probability better than the next-best fit of a different environment type.

The leading candidates for the source of any intrinsic error are ongoing energy injection, Compton scattering, varying microphysical parameters, deviations of the electron energy distribution from a power-law, non-power-law density profiles around the progenitor star, and intrinsic scatter in the data. Intrinsic scatter in the data is the only non-physical element that potentially contributes to our measured intrinsic error, and we discuss its influence in the next section. These effects are likely bundled together in a non-trivial way, and almost certainly differ in their individual degrees of influence from burst to burst.

\subsection{Assessing Intrinsic Scatter in the Data}\label{sec:intscat}
We have investigated our fitting results for bursts for which $\chi_{\nu}^2 > 10$ and all five indicators of $p$ were available. This sub-sample consists of GRBs 090902B, 091127, 101219B, 110731A, 120119A, 120711A, 130427A, 151027A, and 160625B. These bursts represent the population that drives the inclusion of a standard error, as their very precisely measured spectral and temporal indices do not agree well on a single value of $p$. Possible reasons for this disagreement include the misidentification of absorption in the spectral indices, or flares and other variability in the temporal indices. Both of these effects would result in an incorrect power-law measurement, and incidents should be identifiable as outliers in their derived $p$ when compared to the other measurements.

After inspecting the light curves and SEDs in the literature, we find no consistent cause for the disagreement on $p$. In cases where an underestimated absorption may help reduce the tension (090902B, 091127, 120119A, 151027A, 160625B), the amount required is physically implausible. Other cases show spectral indices that would need to be shallowed somehow, but already contain negligible absorption. In one case (120711A), the temporal and spectral indices from individual bands agree with one another, but the values of $p$ indicated by each band do not agree. In addition, our sub-sample contains some of the brightest bursts with the best sampled data, and inspection confirms that no flaring or variability that might affect the temporal index is apparent.

While on a case-to-case basis, misidentified absorption and variability may contribute to our measured standard error, inspection of the data shows that it cannot be playing more than a marginal role. Quantitatively, to replace the need for a standard error of the magnitude we measure, a measurement error in the temporal index of $0.02$ or less (which is true for almost half the sub-sample) would have to be an under-estimate of the true error by a factor of $10$; when the tension in $p$ is removed by adding a standard error in quadrature with the measured errors, the effective error becomes $10$ times larger for this subset bursts with extremely well-measured afterglows.

\section{Conclusions}\label{sec:conclusions}
We have assessed the environments of a large sample of \emph{Fermi} detected LGRBs by fitting their spectral and temporal indices with the synchrotron closure relations. We find a roughly even split between wind-like and ISM-like environments in cases where an environment could be assigned a fit probability at least three times higher than the next-best fit probability for an environment of a different type. This division persists when our factor three constraint is lessened, or when a minimum best-fit probability is imposed. The identification of two environment types indicates that either we see a single population of LGRBs before and after their forward shock emission sites have transitioned across a termination shock at the edge of a stellar wind-dominated bubble into the ISM, or that LGRBs are in fact the product of two distinct progenitors that occur in different environments.

If the two environment types are due to the crossing of the termination shock, then our chosen observation time of 11 hours in the rest-frame must be quite close to the mean shock crossing time, based on the roughly even division between wind (pre-crossing) and ISM (post-crossing) environments measured. However, we find a $2\sigma$ separation in the distributions of $\gamma$-ray isotropic equivalent energies between bursts identified as being in wind environments and those identified as being in ISM environments, with the wind bursts being systematically more energetic. This could be evidence for two distinct populations, though a larger sample will be required to make this claim with confidence.

Having identified the environment types and a best-fit value for $p$, we are able to analyze the physical parameters of the shock. Our results provide a value for the degenerate values of $\epsilon_B \times n$ (or $A_*$), though in some cases we must assume $\epsilon_e = 0.1$. The densities inferred by different values of assumed $\epsilon_B$ indicate that $\epsilon_B$ must be very low; $\epsilon_B \sim 10^{-4}$ or lower is required to avoid densities as sparse as the IGM, or unnaturally high Lorentz factors.

Finally, we find that those GRBs with the most precise measurements of their spectral and temporal indices (i.e. those with the smallest error bars) do not result in more precise convergences of these indices to a single value of an underlying power-law distribution of electron energies, as is typically assumed in synchrotron theory. This strongly indicates an intrinsic error in the GRB population that must be accounted for when assigning environment types using the closure relations. Our best fit value for this error is $0.25 \pm 0.04$, obtained by maximising the likelihood of all the fits in our sample across a range of consistent intrinsic errors added in quadrature to the measured errors. This is the first measurement of the discrepancy between theory and nature made from the data. With the intrinsic error included, our ability to determine the GRB environment type to a threshold of a factor of three in probability is diminished, but we nonetheless retain a roughly even split in wind/ISM best fits with/without this condition applied.

\section{Acknowledgements}
We thank David Rubin for an independent check of the intrinsic error presented in this paper. We thank the anonymous referee for useful comments and suggestions that helped to improve this manuscript. BPG has received funding from the European Research Council (ERC) under the European Union's Horizon 2020 research and innovation programme (grant agreement no 725246, TEDE, PI A. Levan). AP acknowledges support by the European Union Seventh Framework Program (FP7/2007-2013) under grant agreement no. 618499. This work made use of data supplied by the UK Swift Science Data Centre at the University of Leicester.

\bibliographystyle{aasjournal}
\bibliography{ref}

\appendix
\section{Best Fit Plots}\label{app:fits}
\subsection{LAT}
\noindent \includegraphics[width=8.9cm]{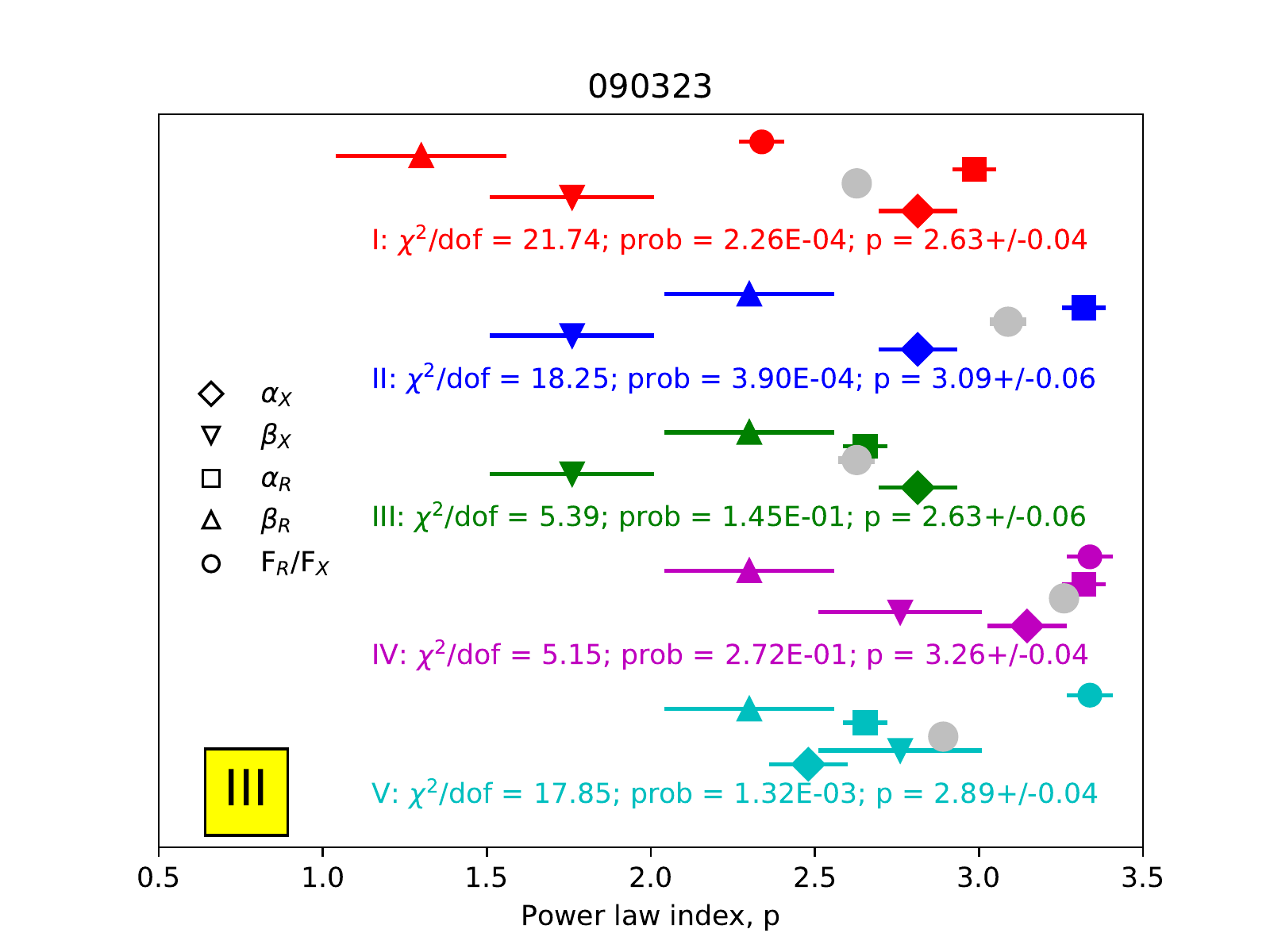}
\includegraphics[width=8.9cm]{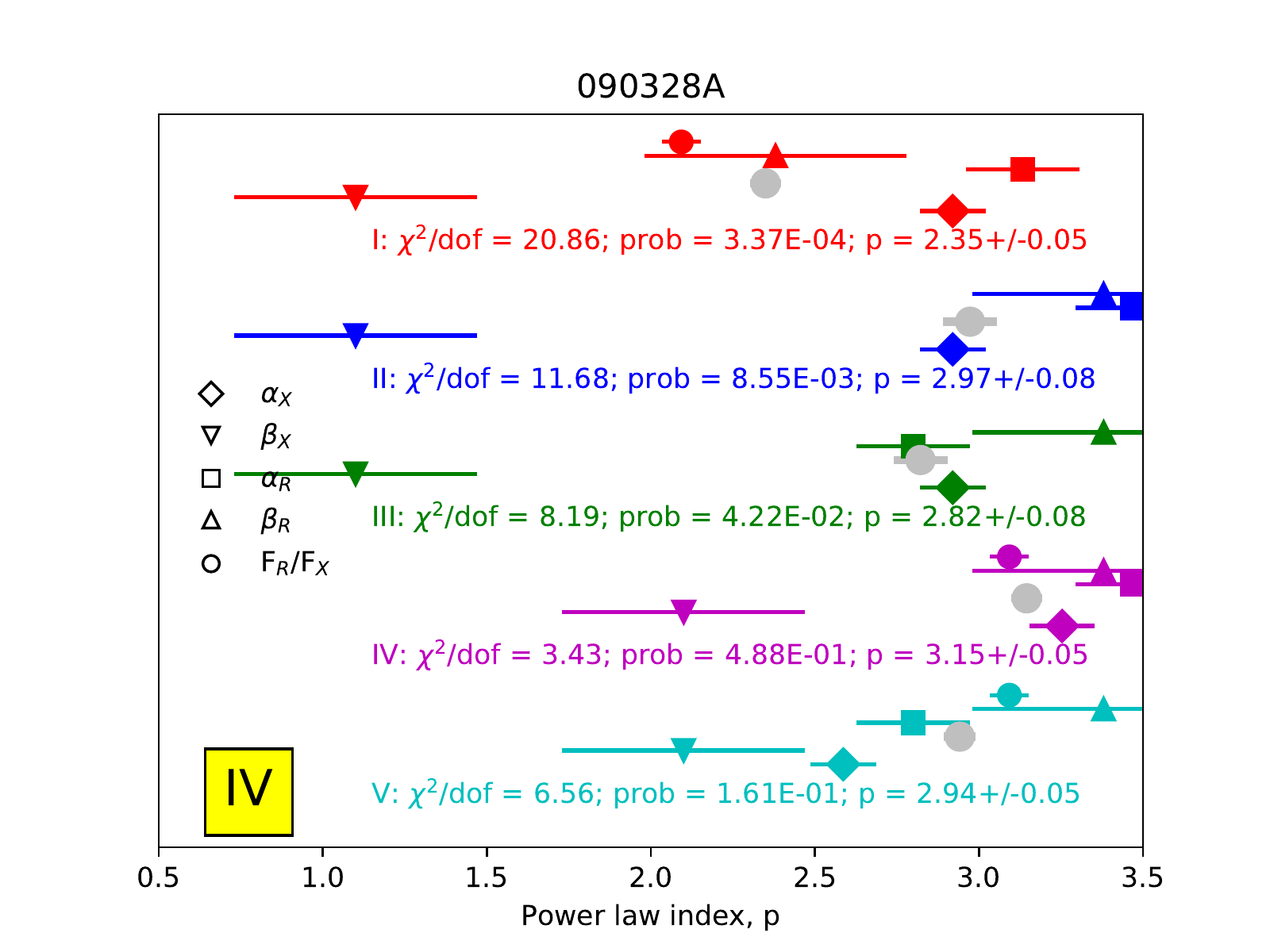}
\includegraphics[width=8.9cm]{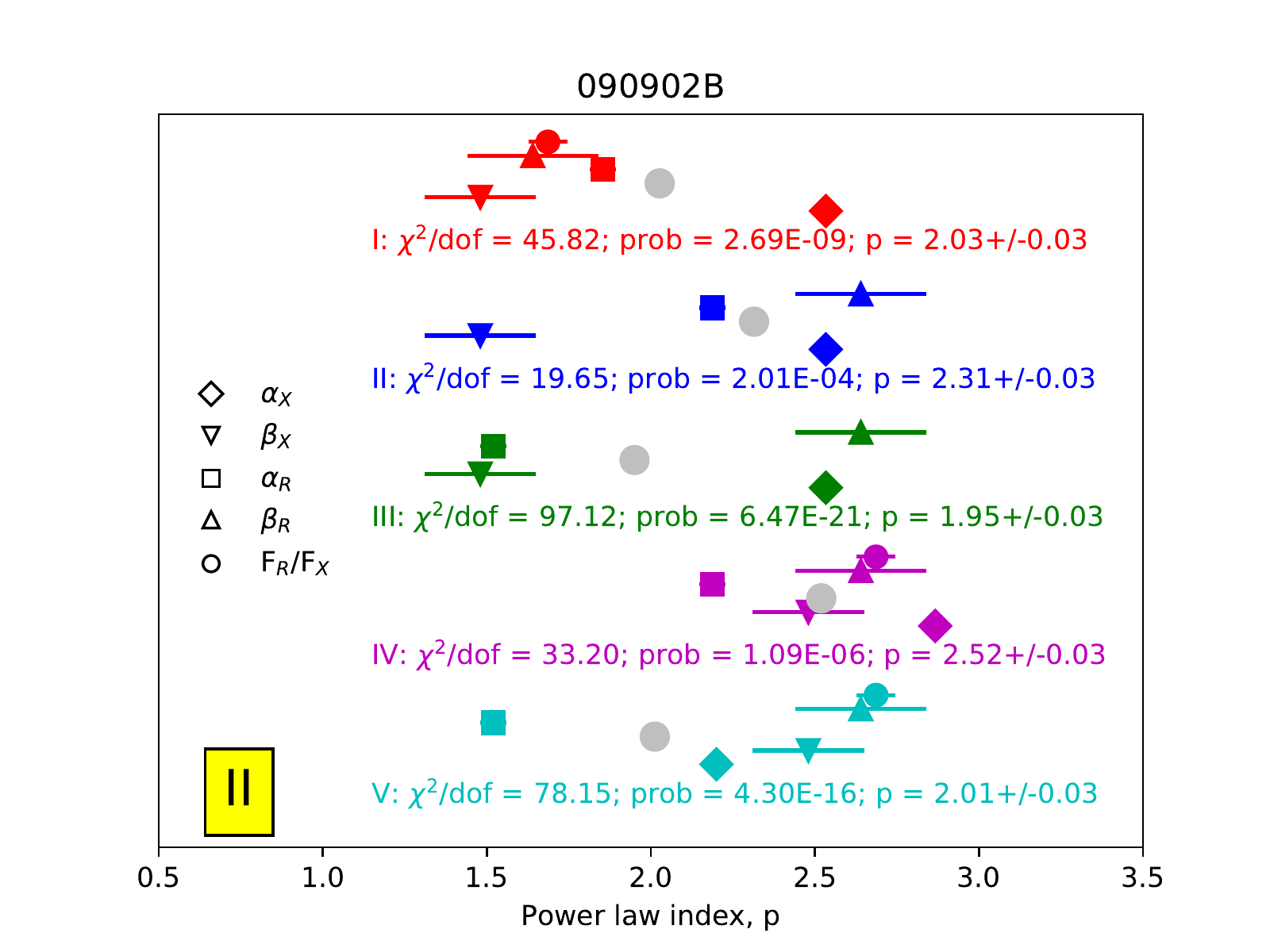}
\includegraphics[width=8.9cm]{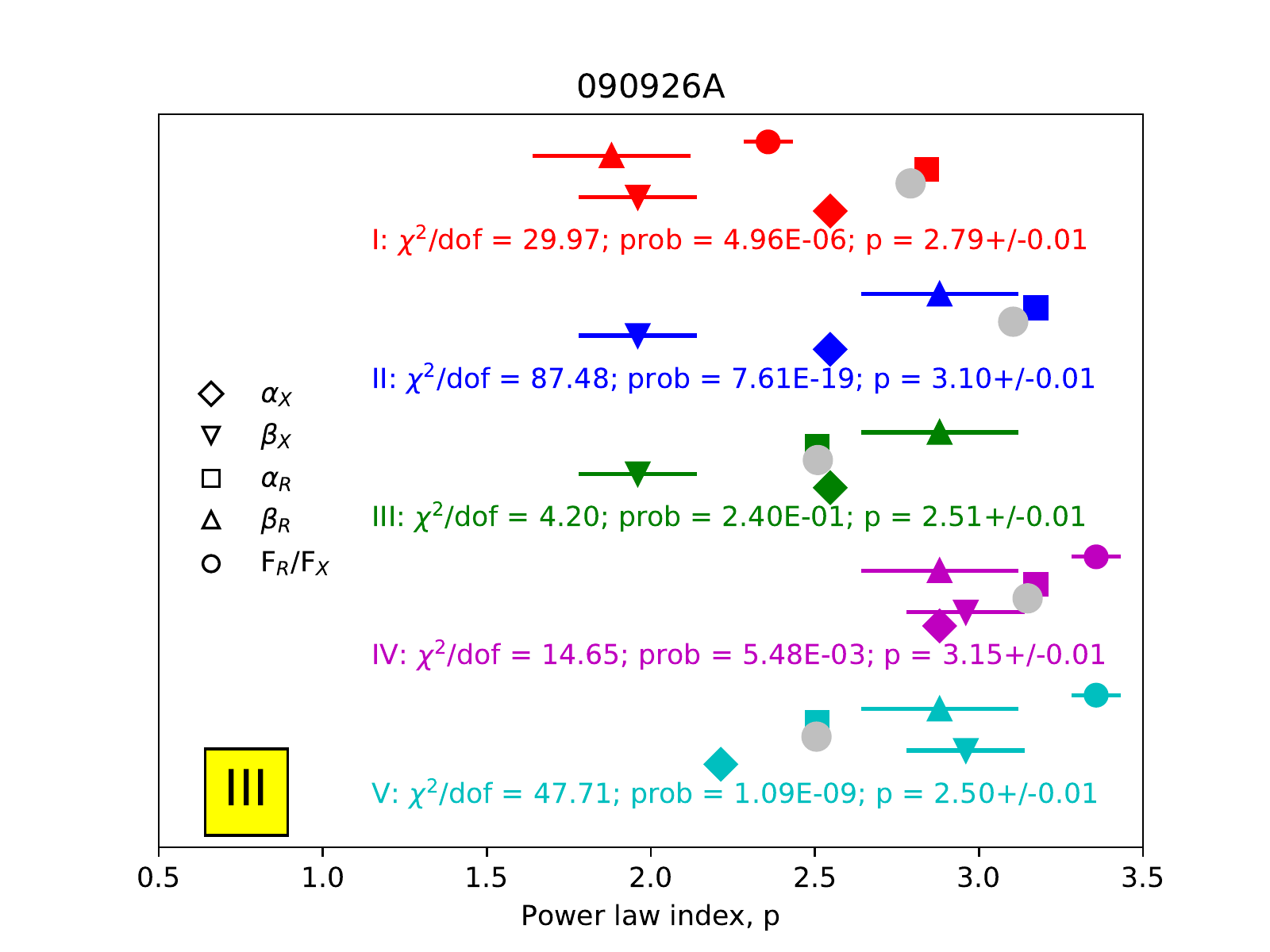}
\includegraphics[width=8.9cm]{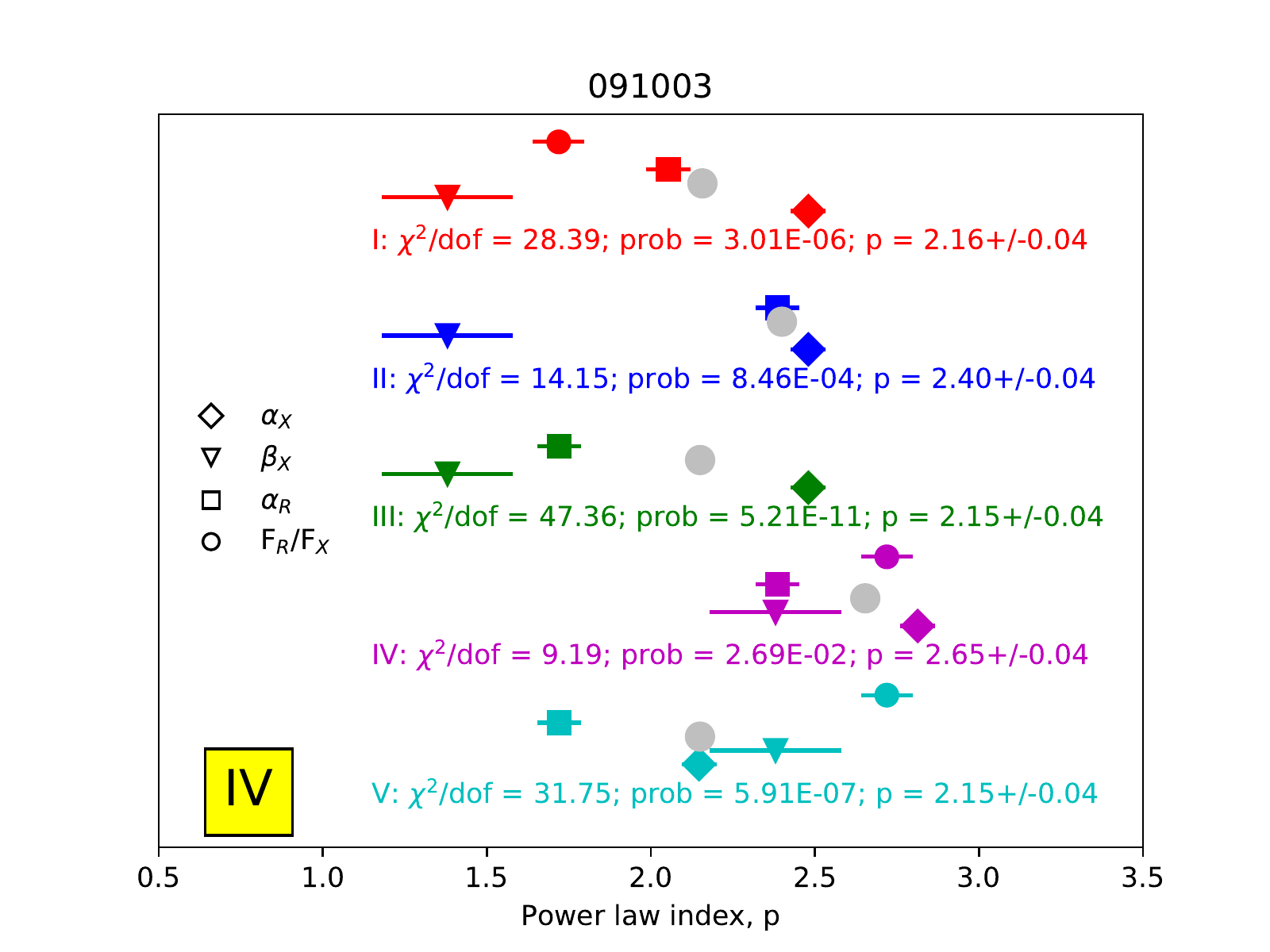}
\includegraphics[width=8.9cm]{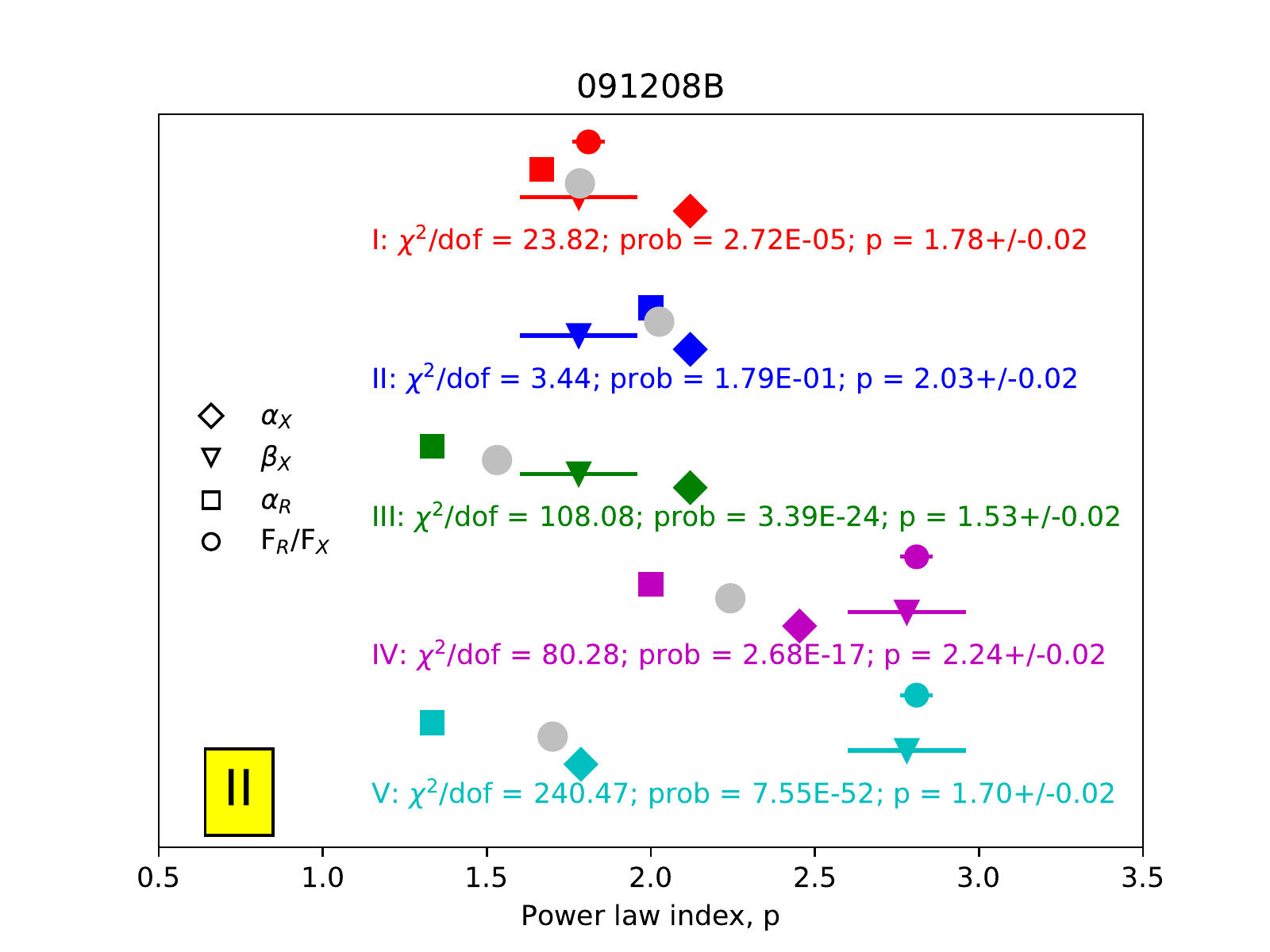}
\includegraphics[width=8.9cm]{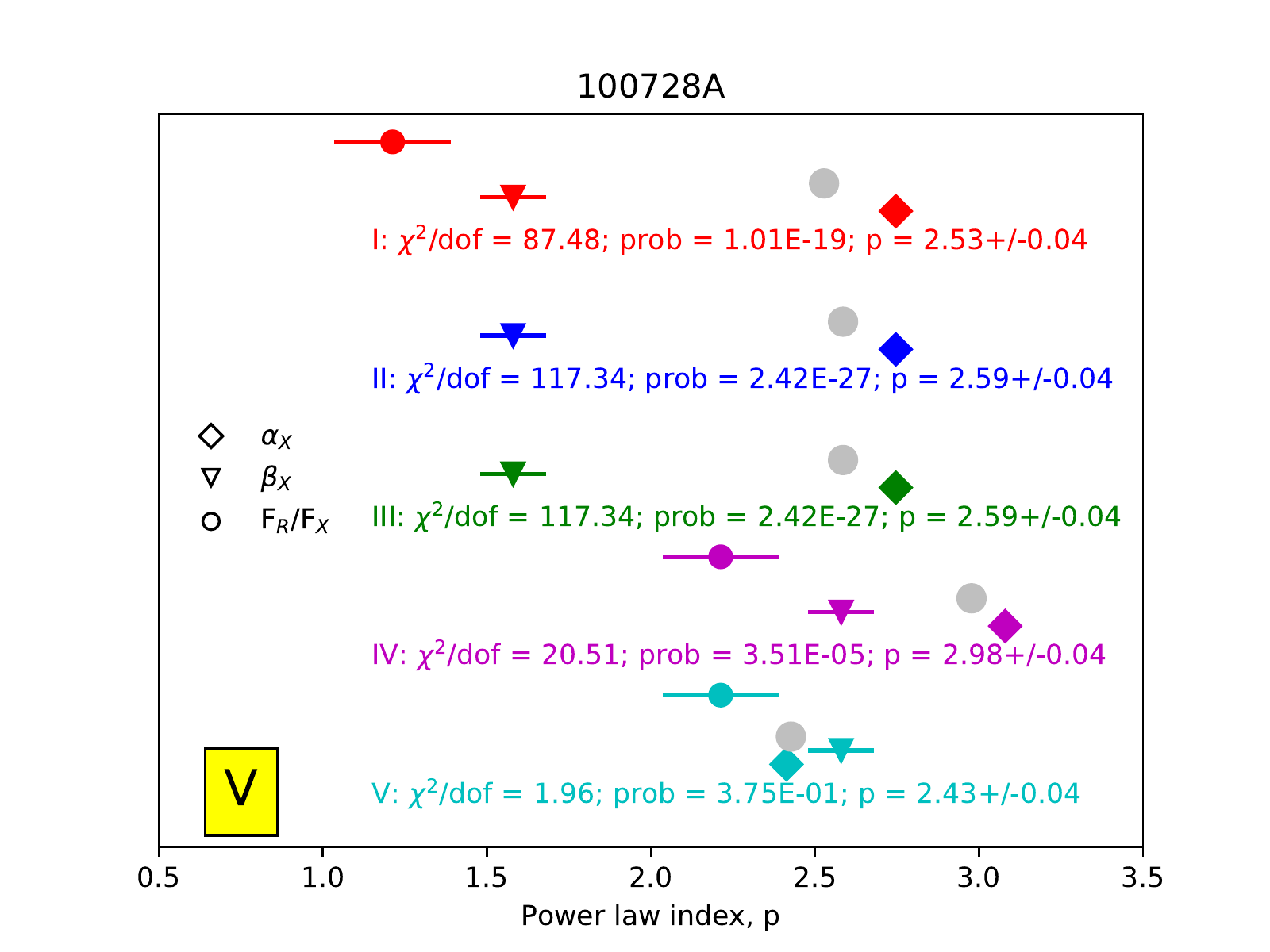}
\includegraphics[width=8.9cm]{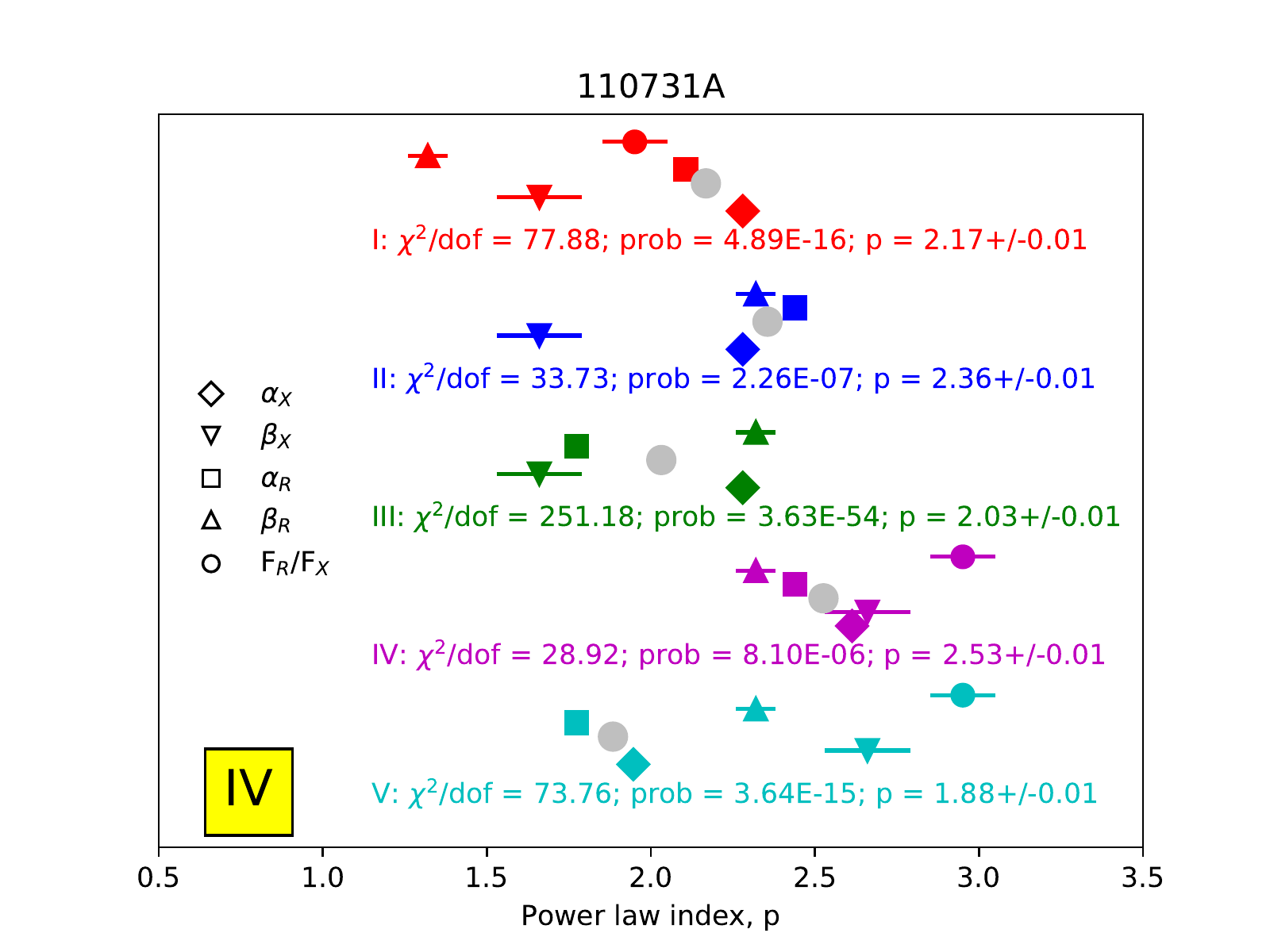}
\includegraphics[width=8.9cm]{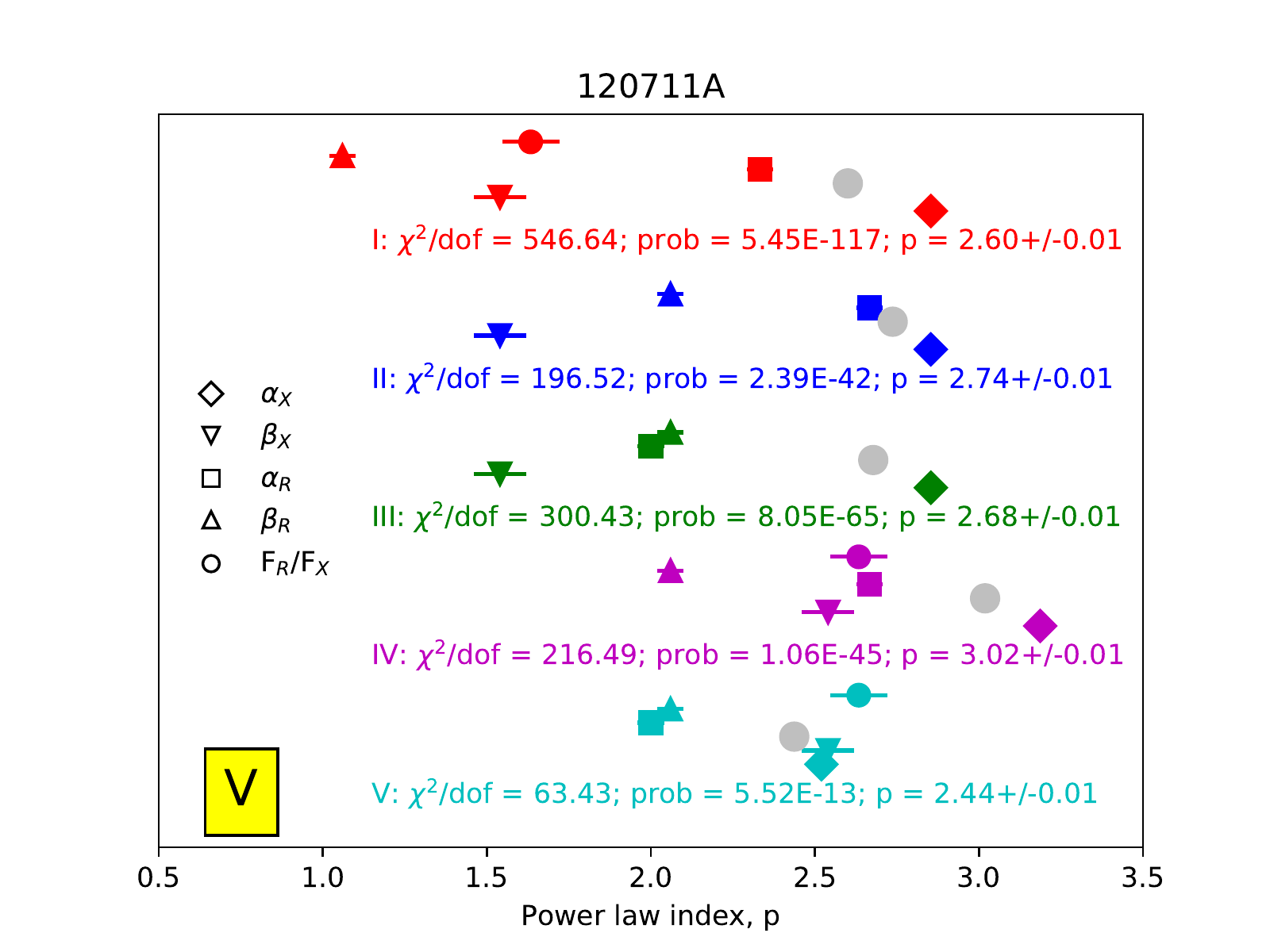}
\includegraphics[width=8.9cm]{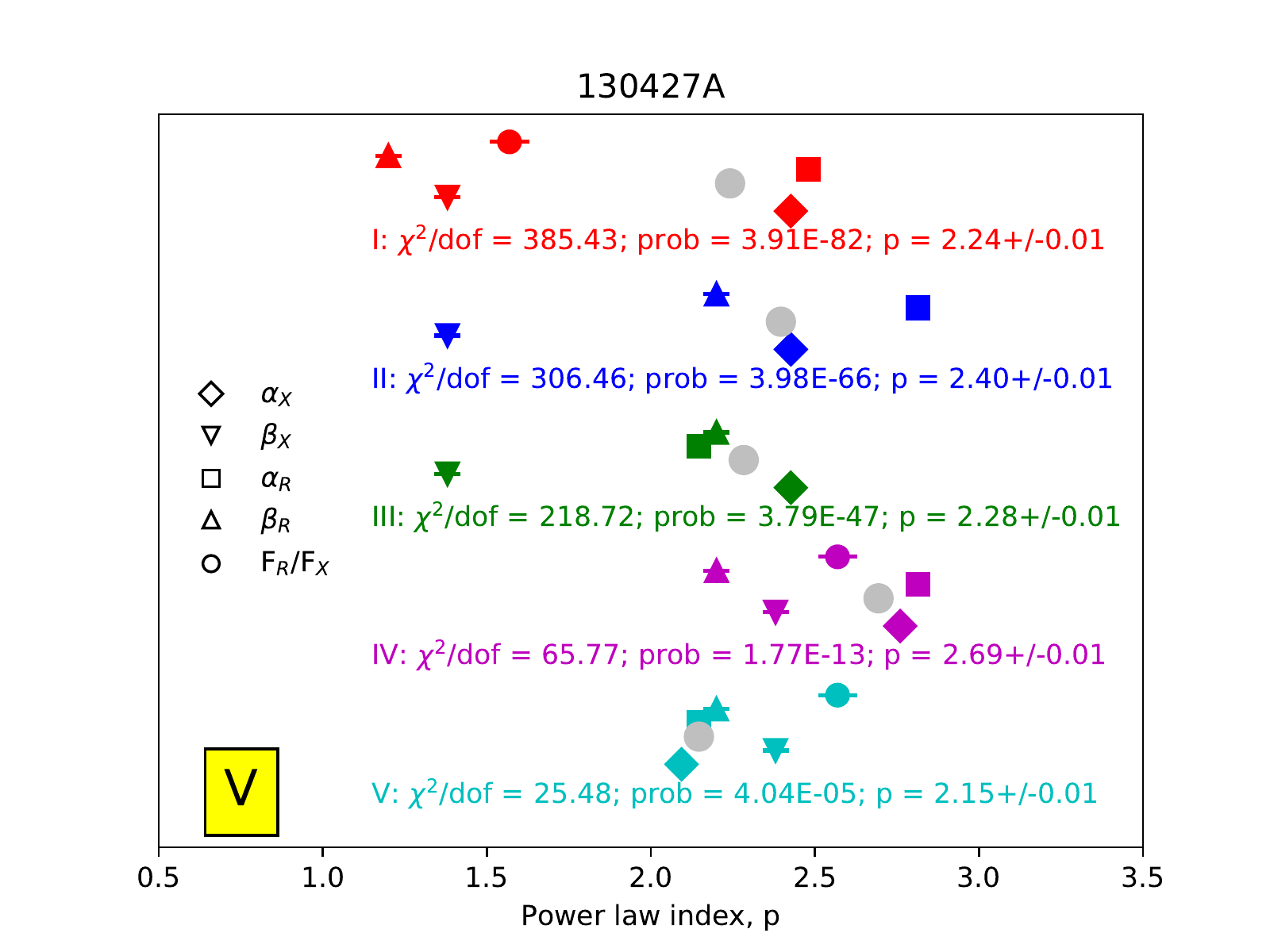}
\includegraphics[width=8.9cm]{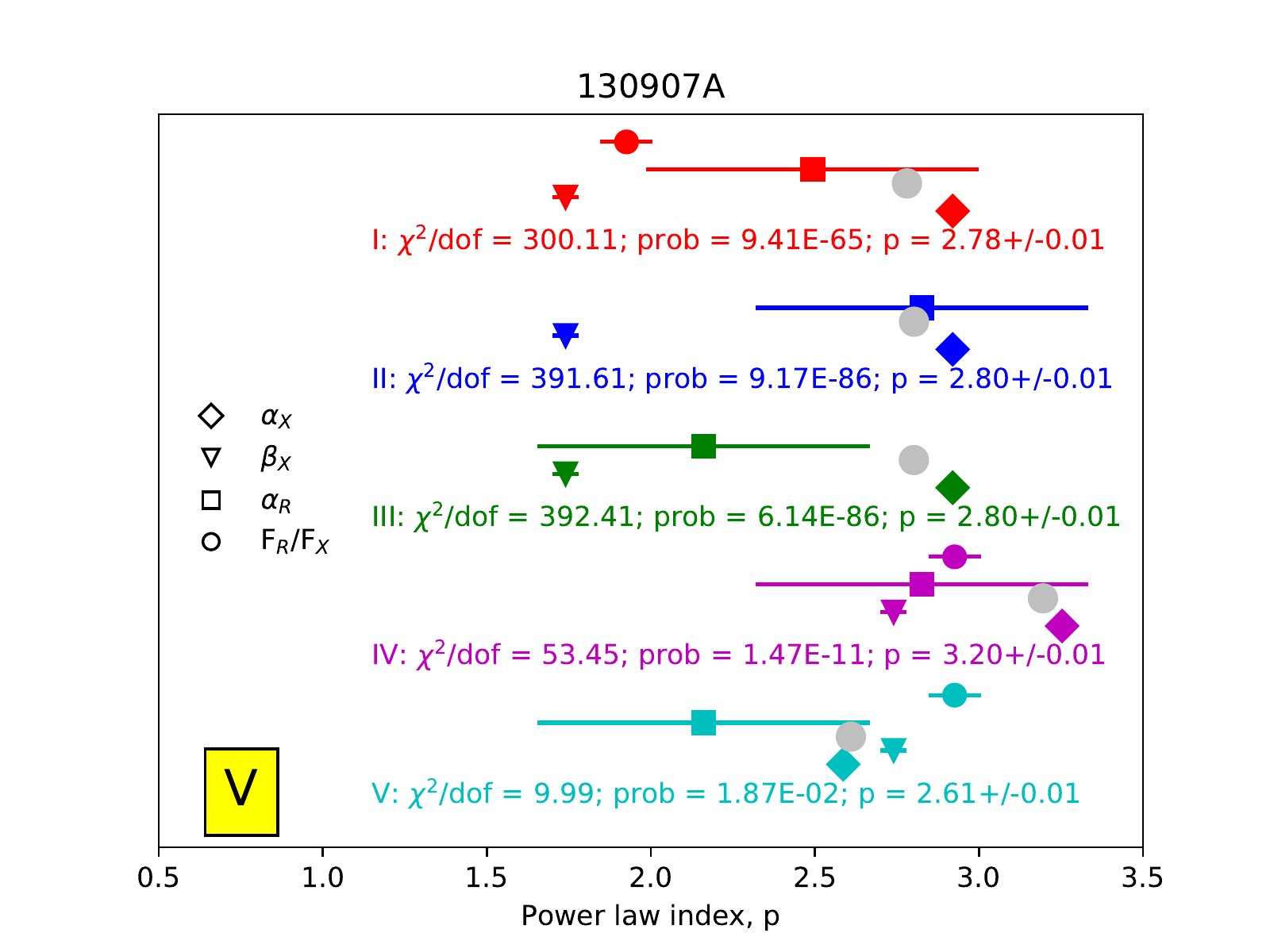}
\includegraphics[width=8.9cm]{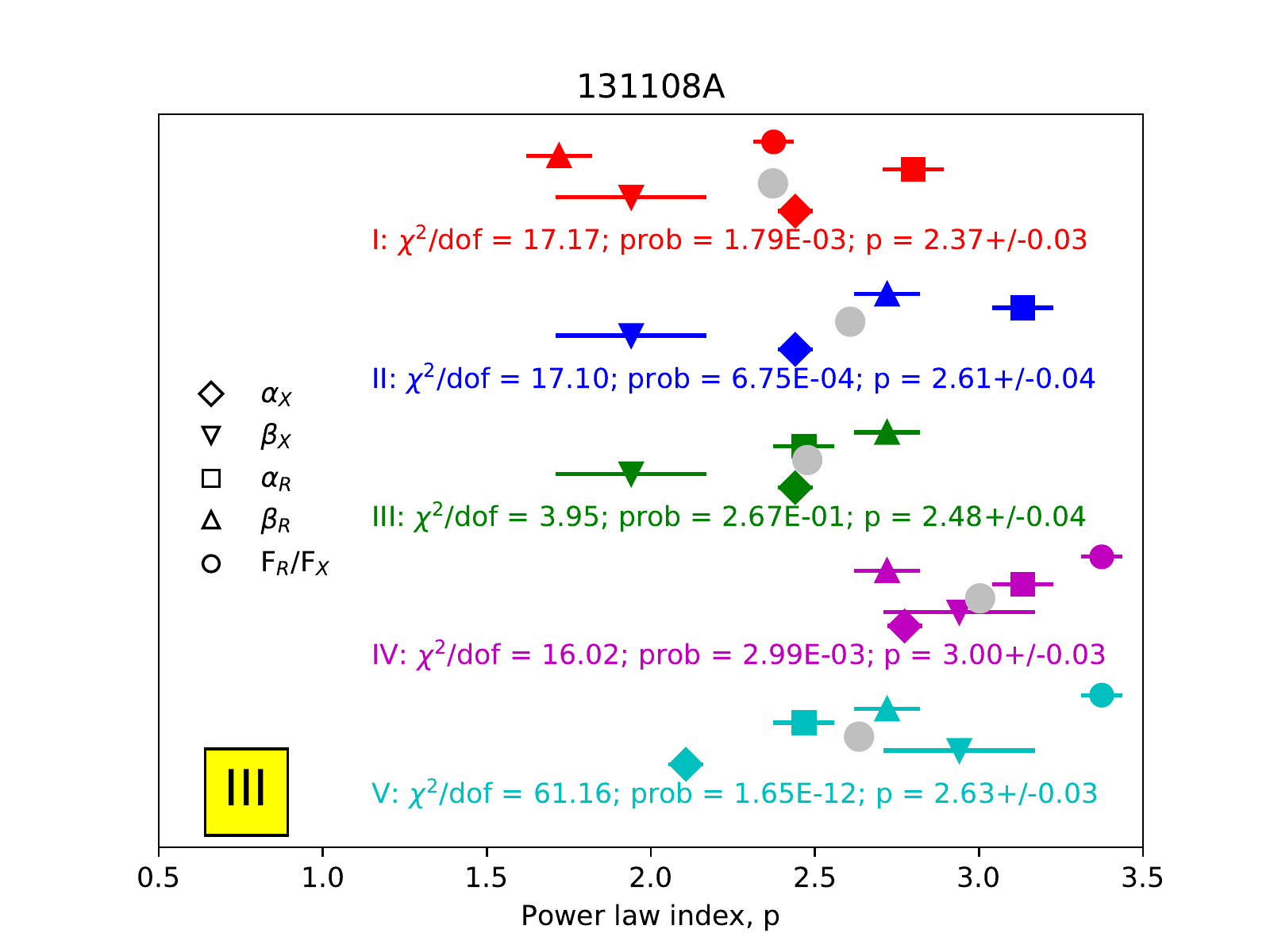}
\includegraphics[width=8.9cm]{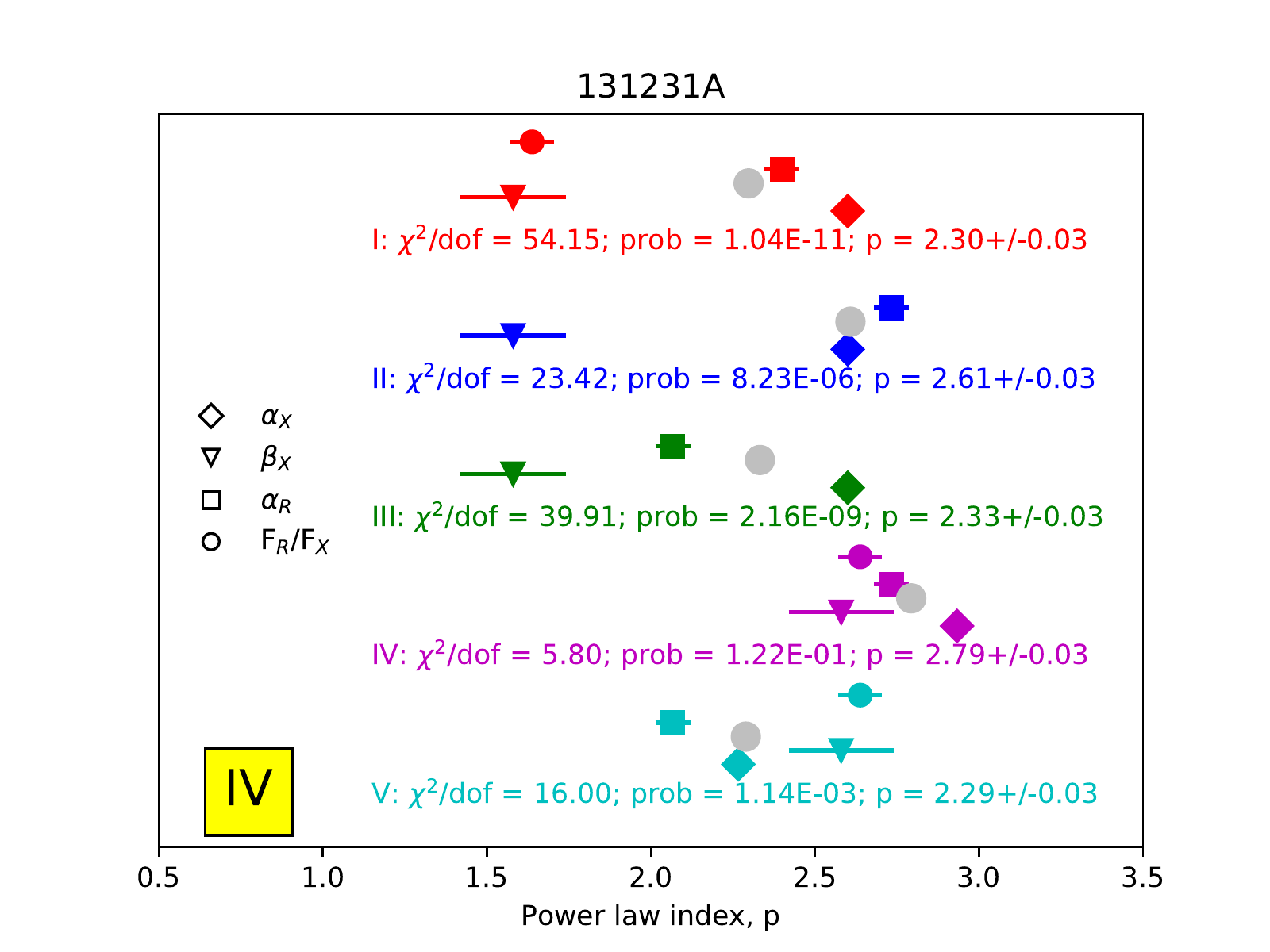}
\includegraphics[width=8.9cm]{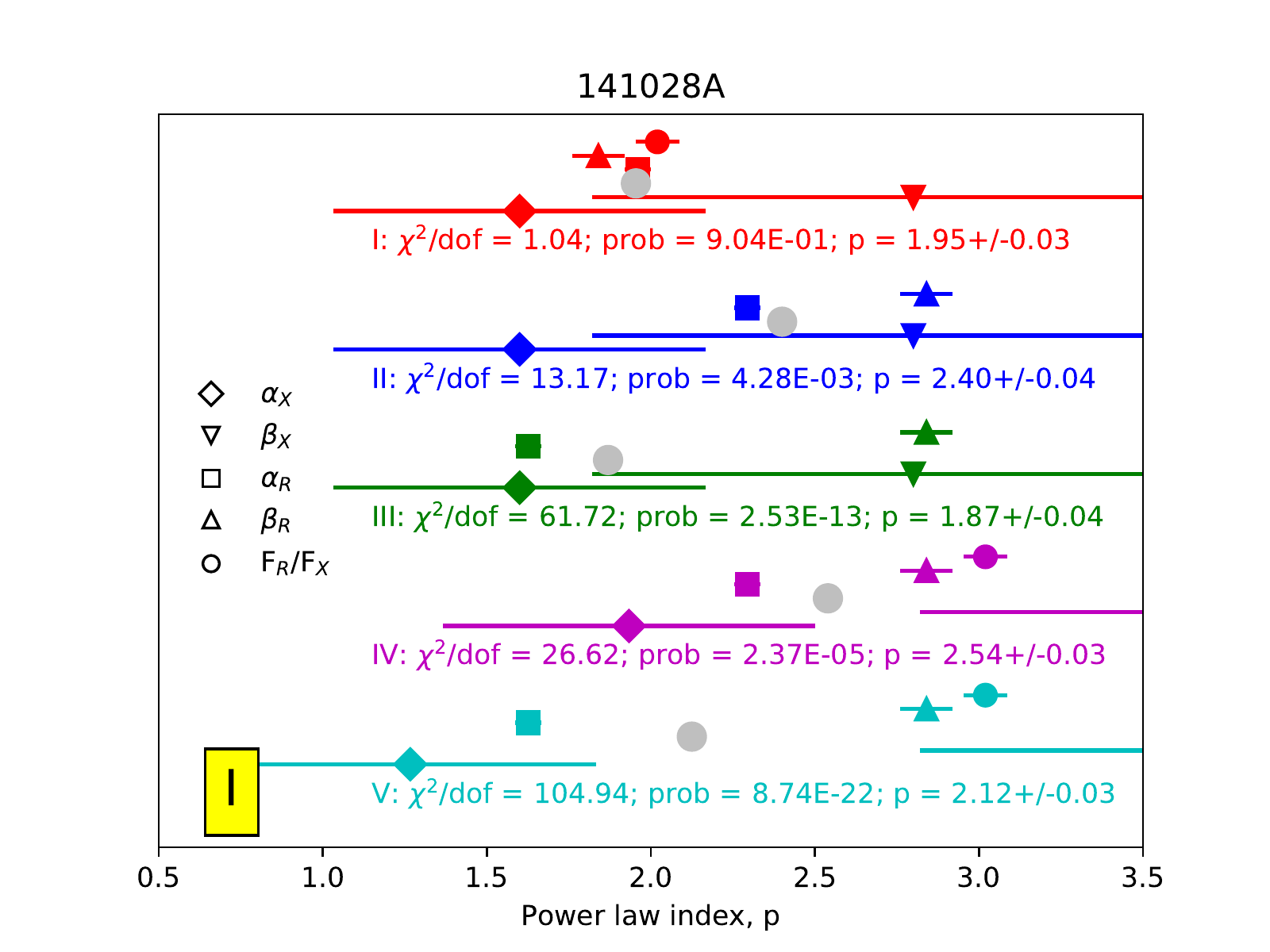}
\includegraphics[width=8.9cm]{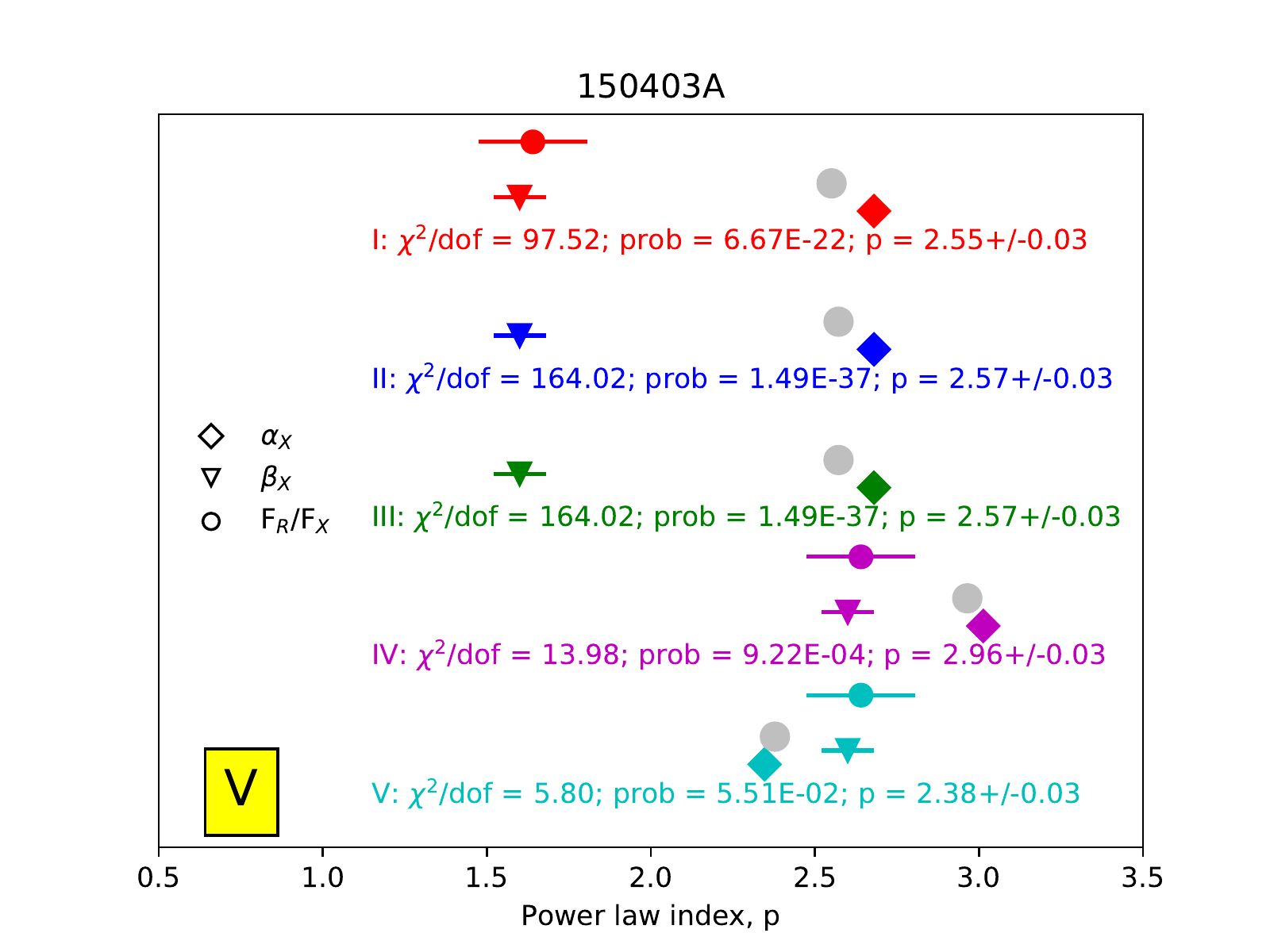}
\includegraphics[width=8.9cm]{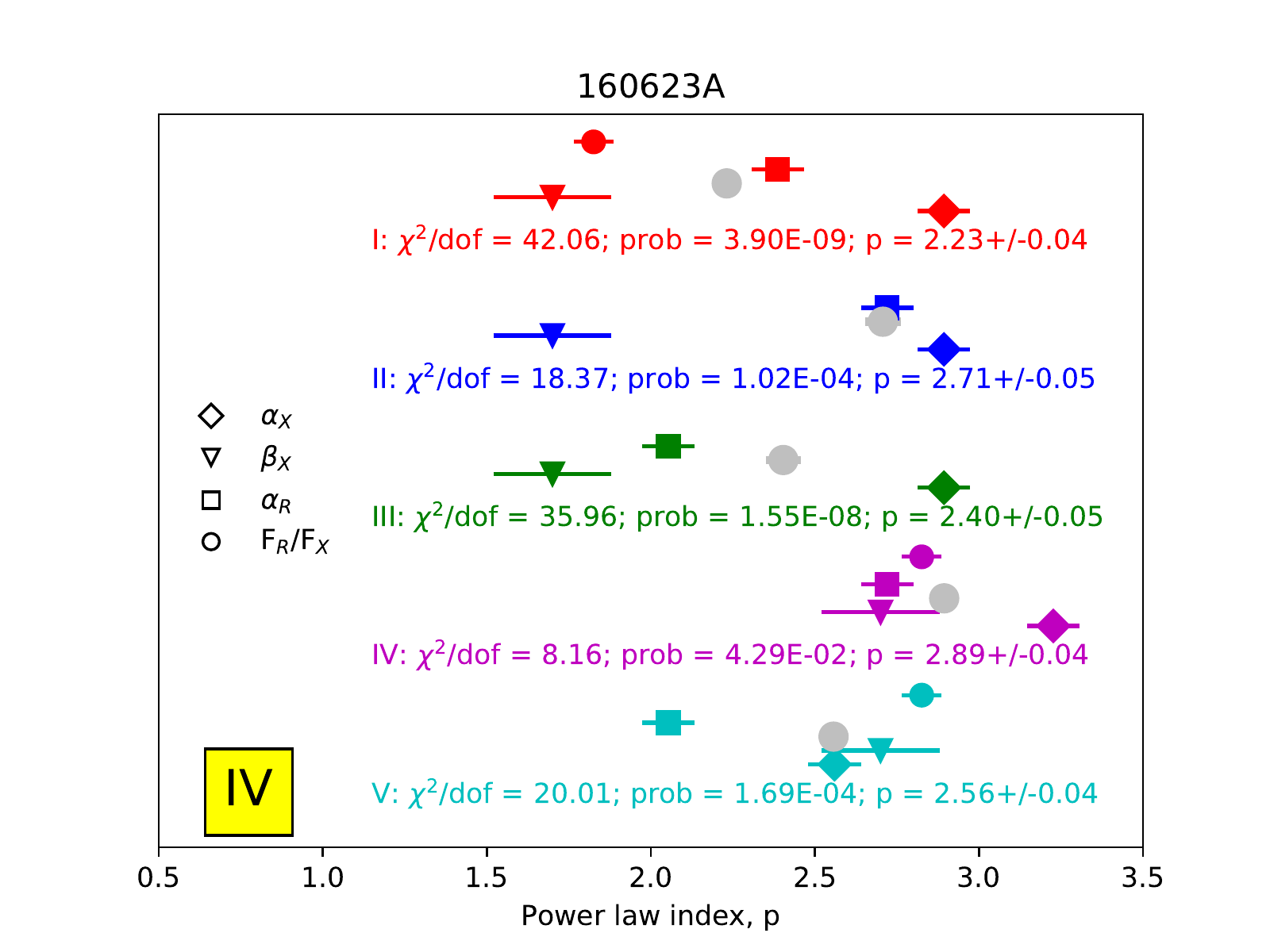}
\includegraphics[width=8.9cm]{160625B.pdf}
\subsection{GBM}
\noindent \includegraphics[width=8.9cm]{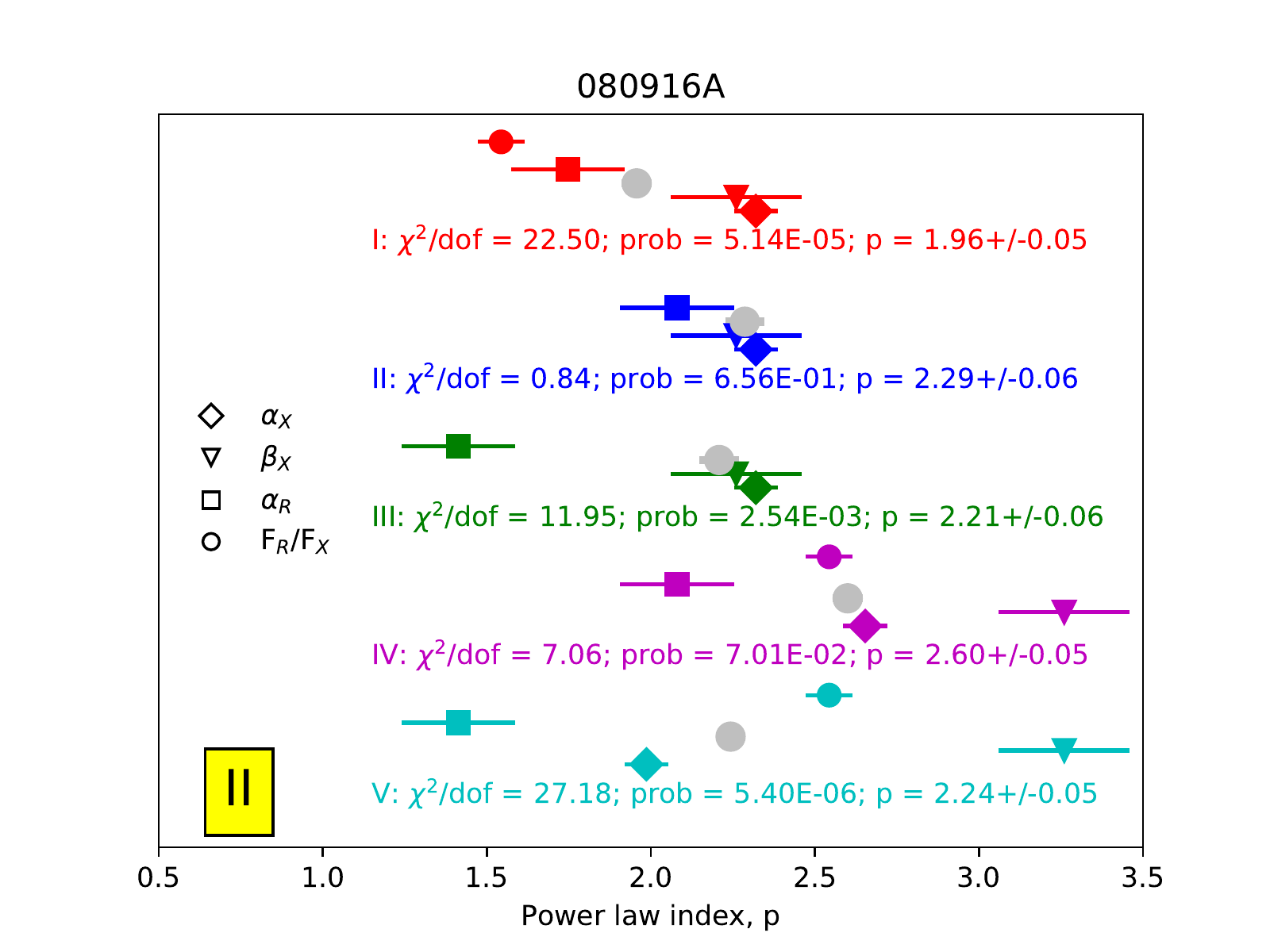}
\includegraphics[width=8.9cm]{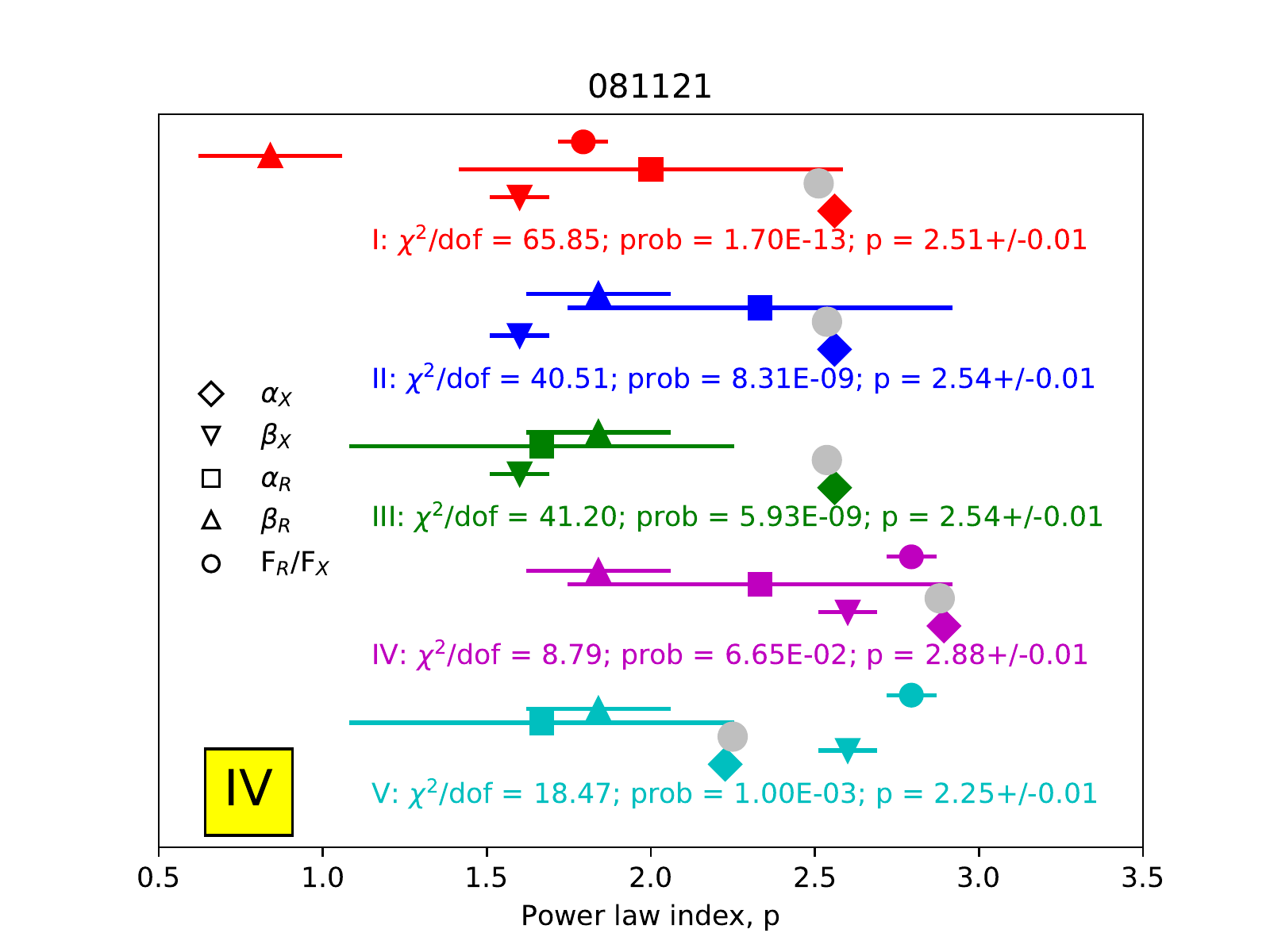}
\includegraphics[width=8.9cm]{090424.pdf}
\includegraphics[width=8.9cm]{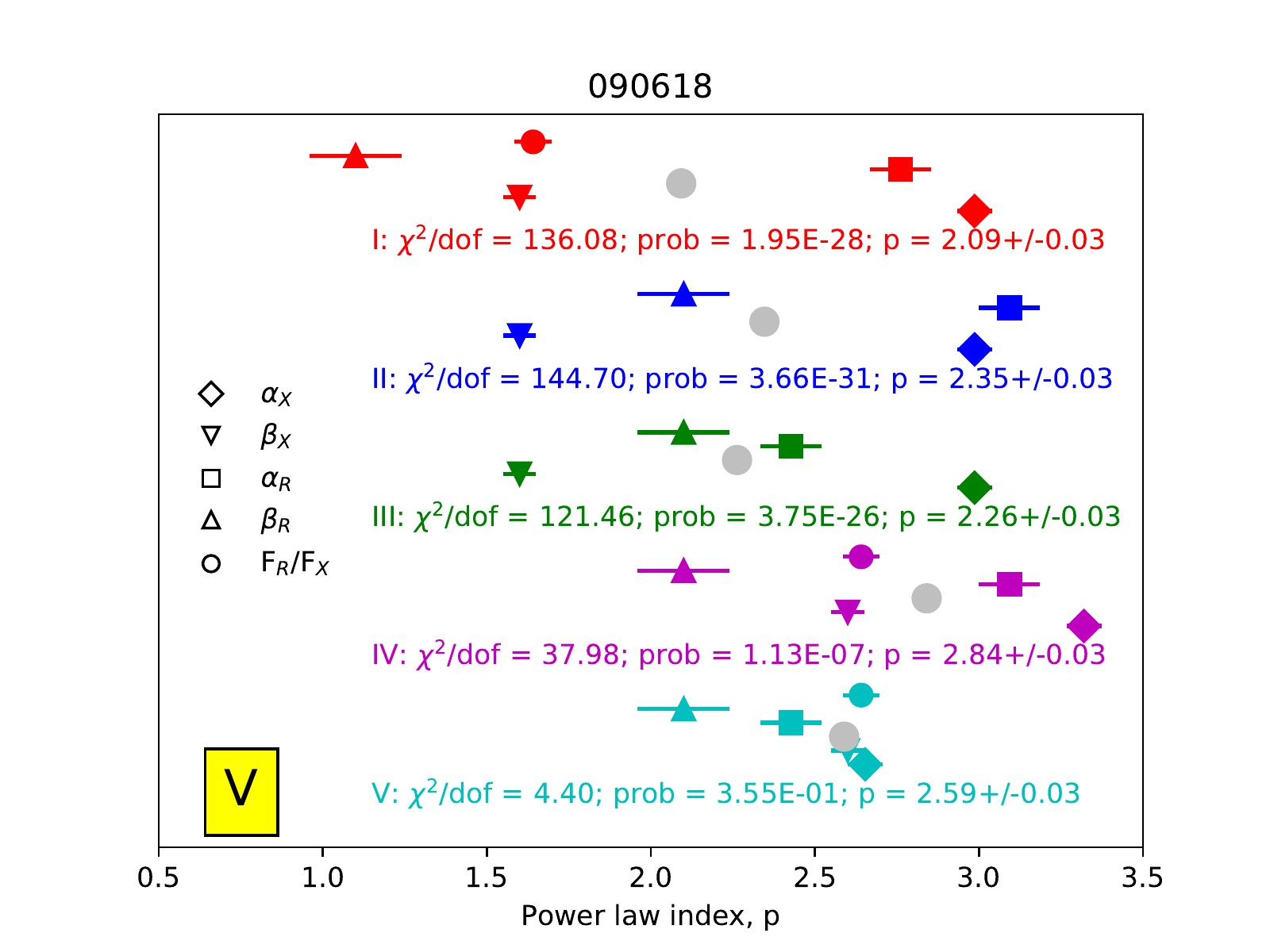}
\includegraphics[width=8.9cm]{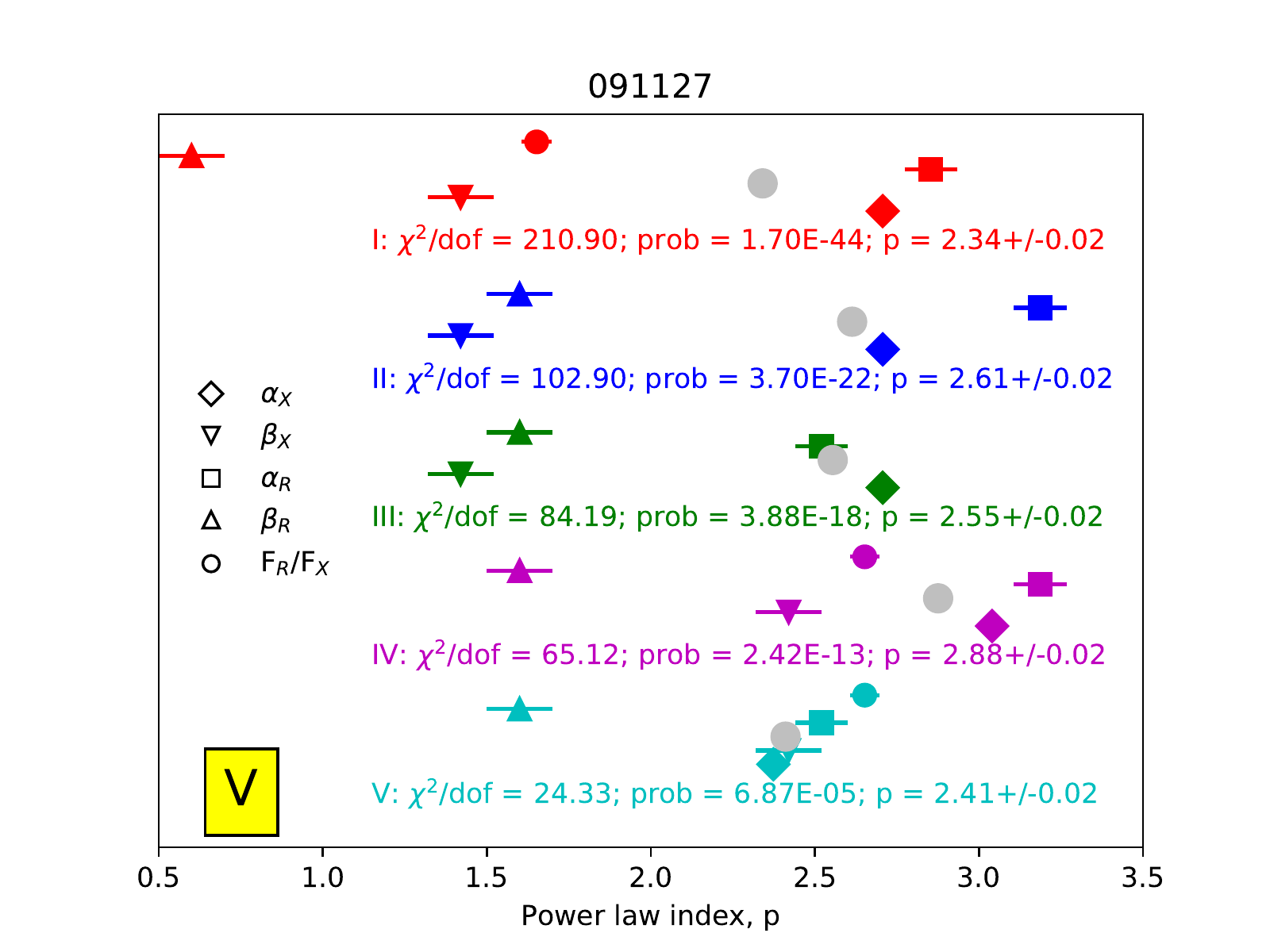}
\includegraphics[width=8.9cm]{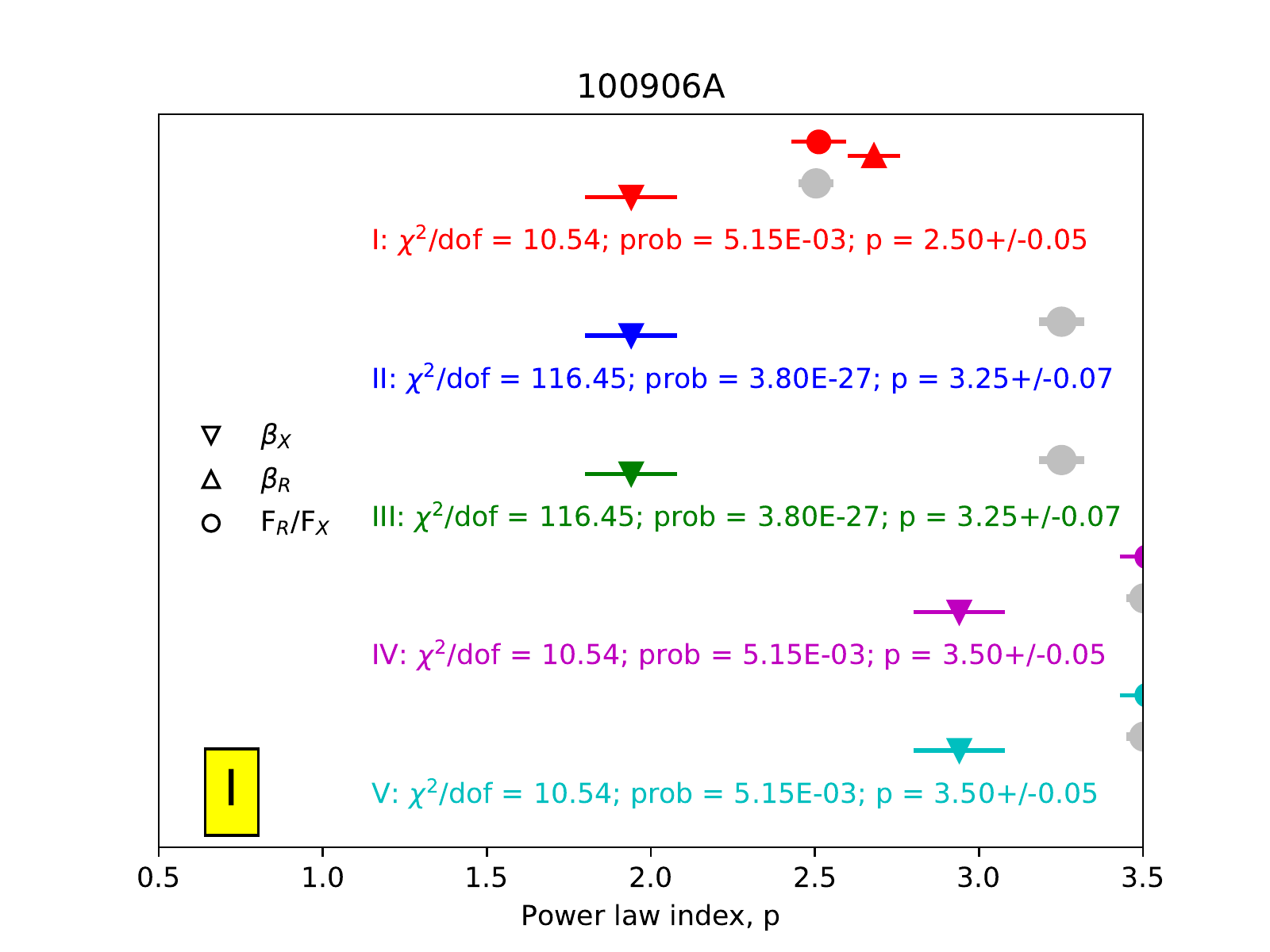}
\includegraphics[width=8.9cm]{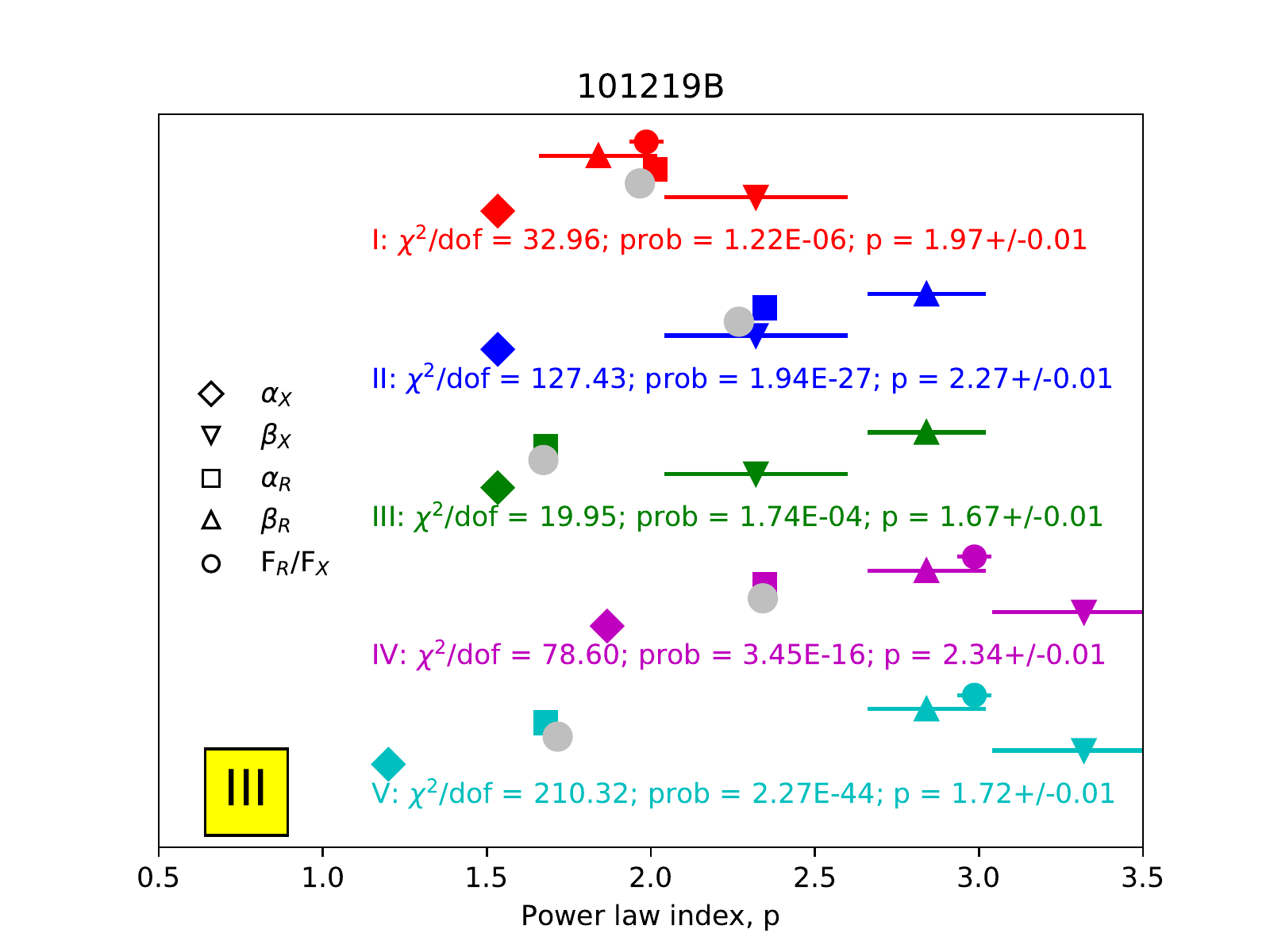}
\includegraphics[width=8.9cm]{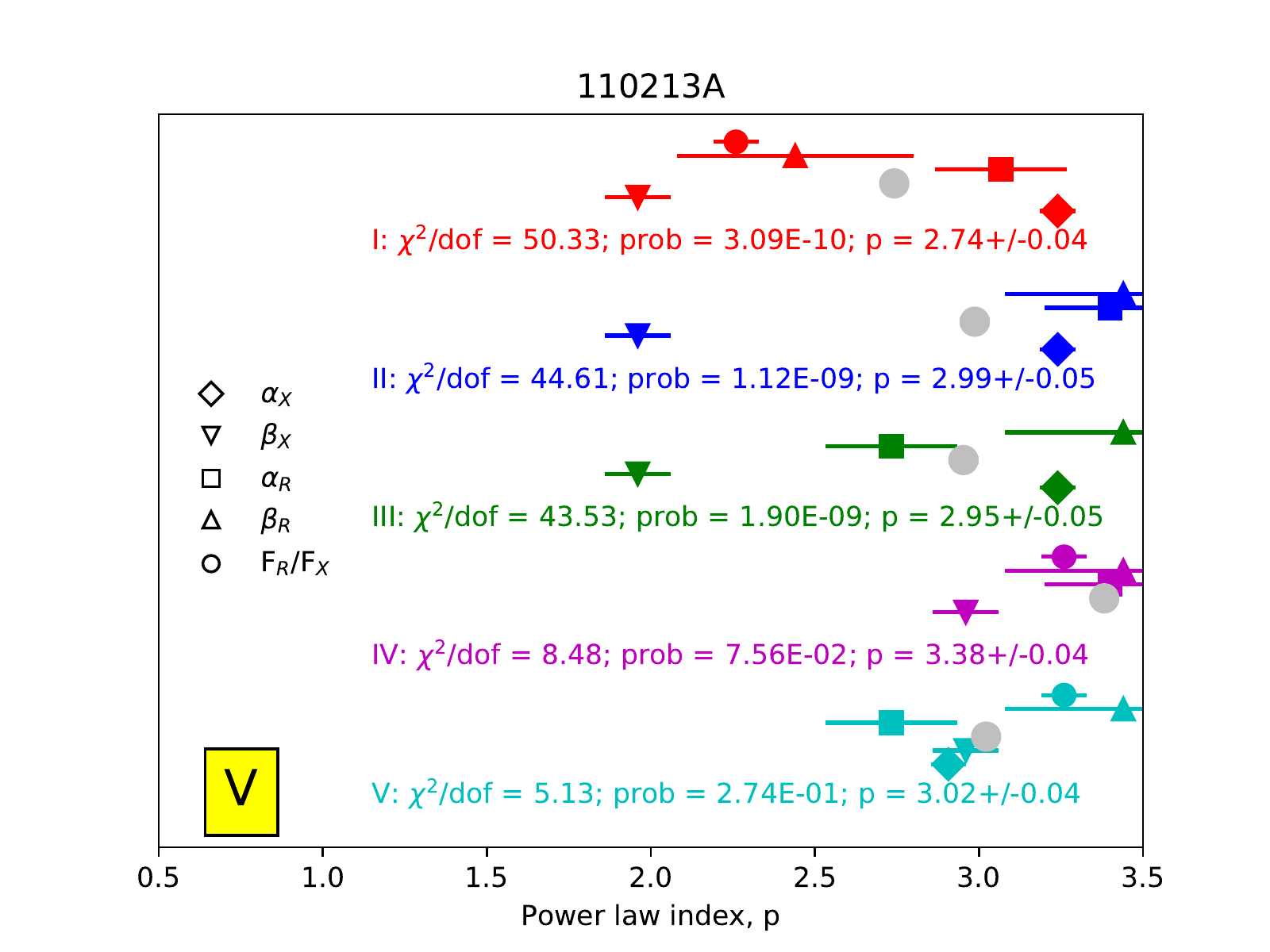}
\includegraphics[width=8.9cm]{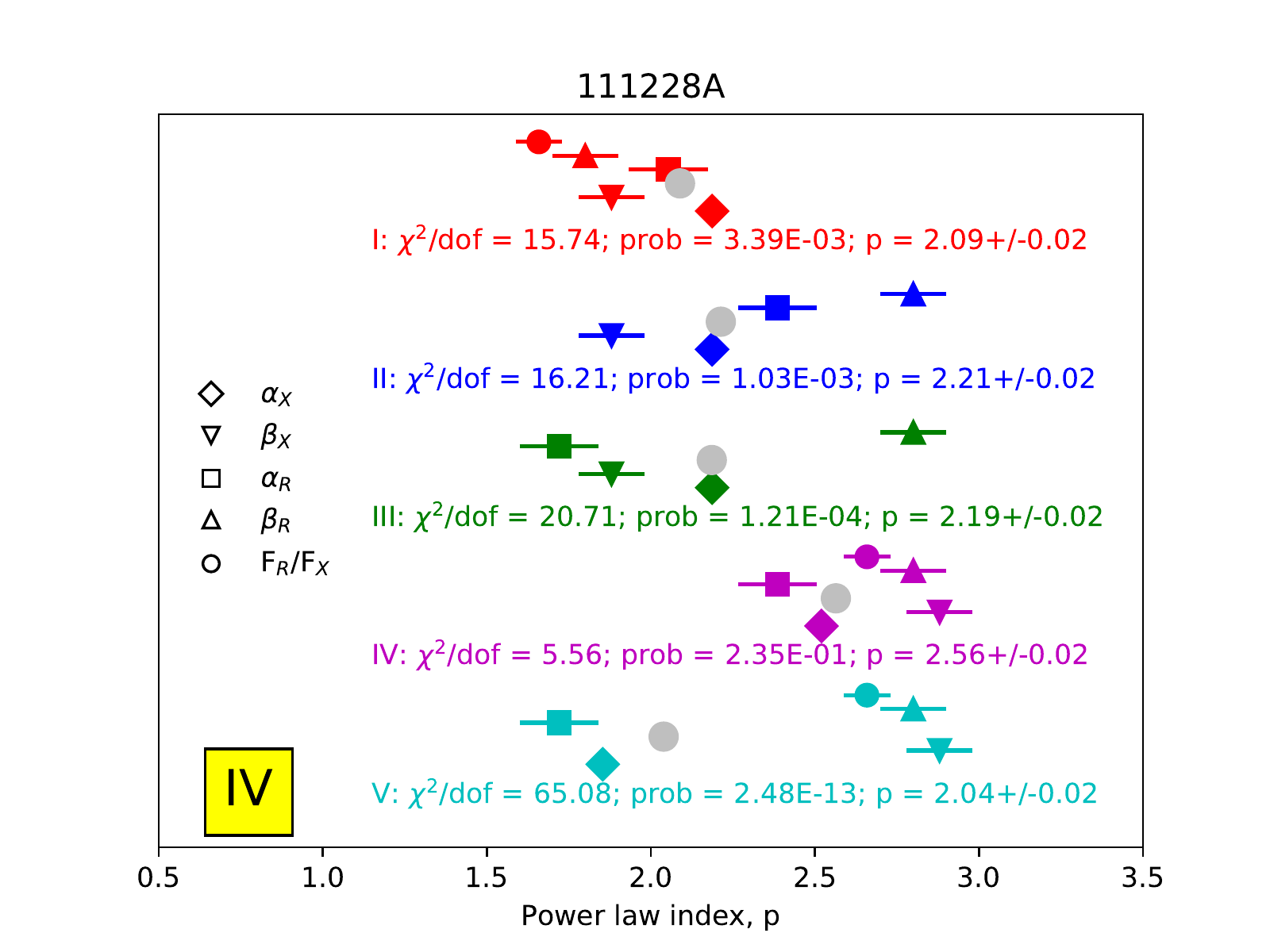}
\includegraphics[width=8.9cm]{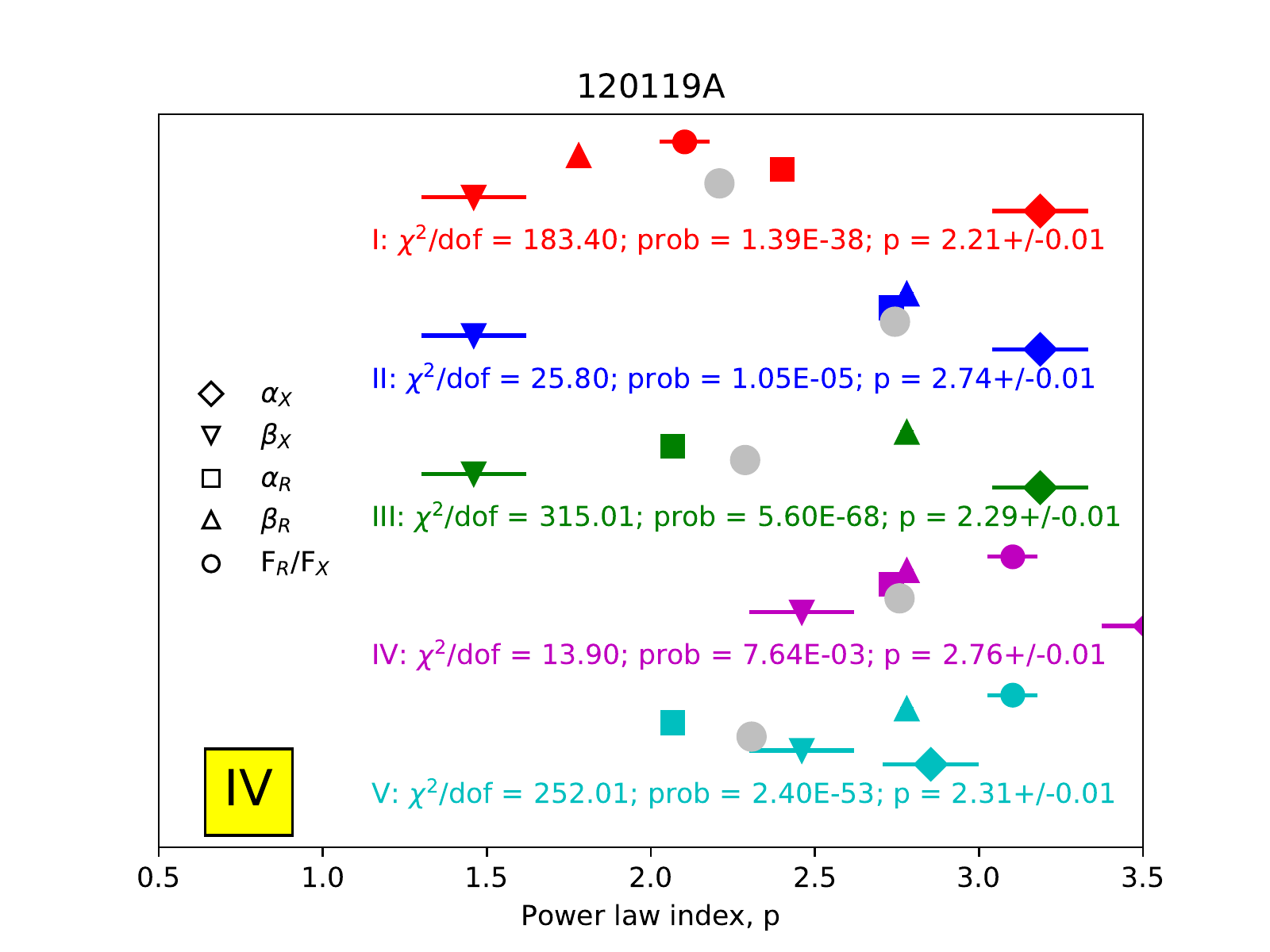}
\includegraphics[width=8.9cm]{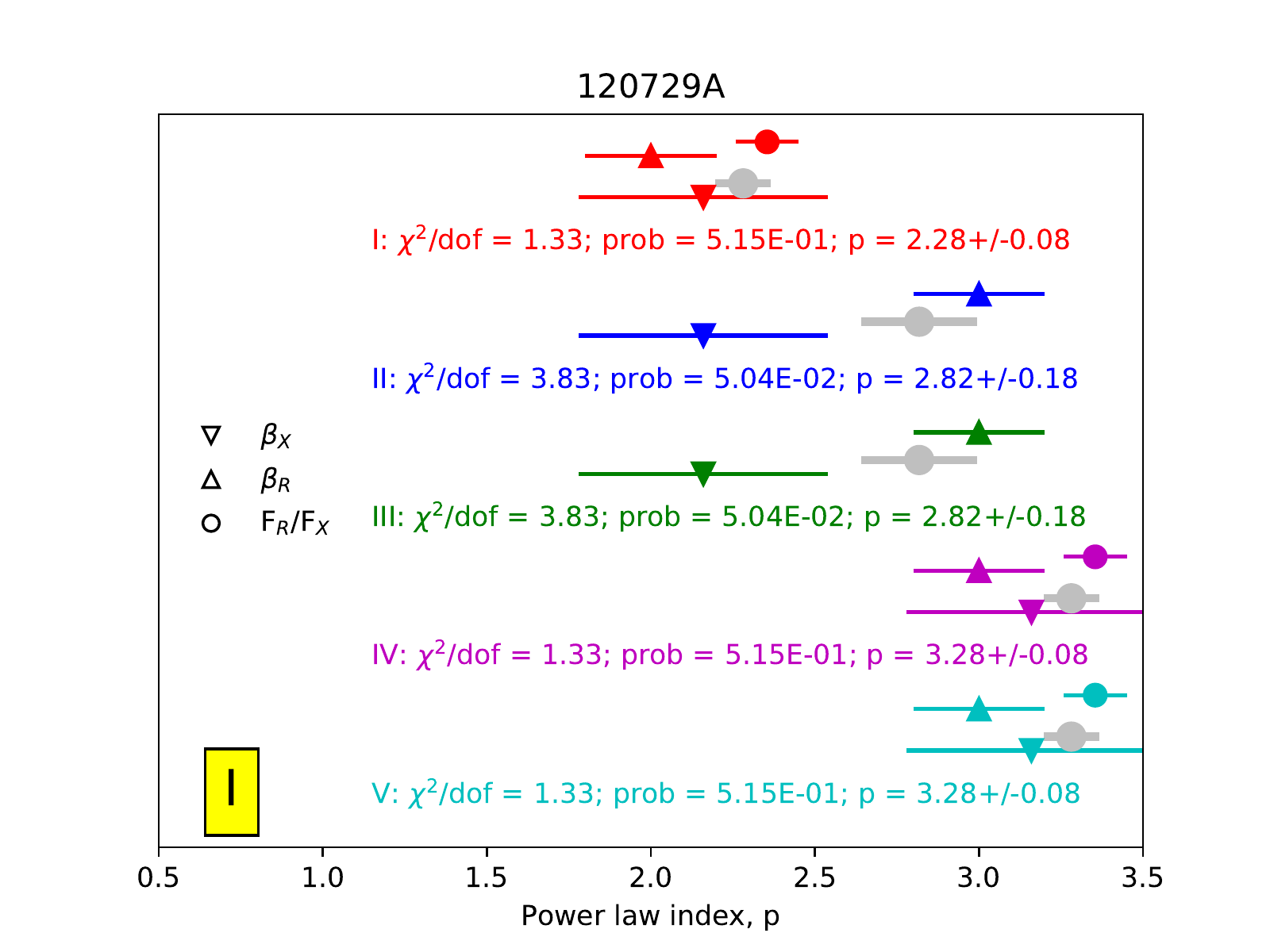}
\includegraphics[width=8.9cm]{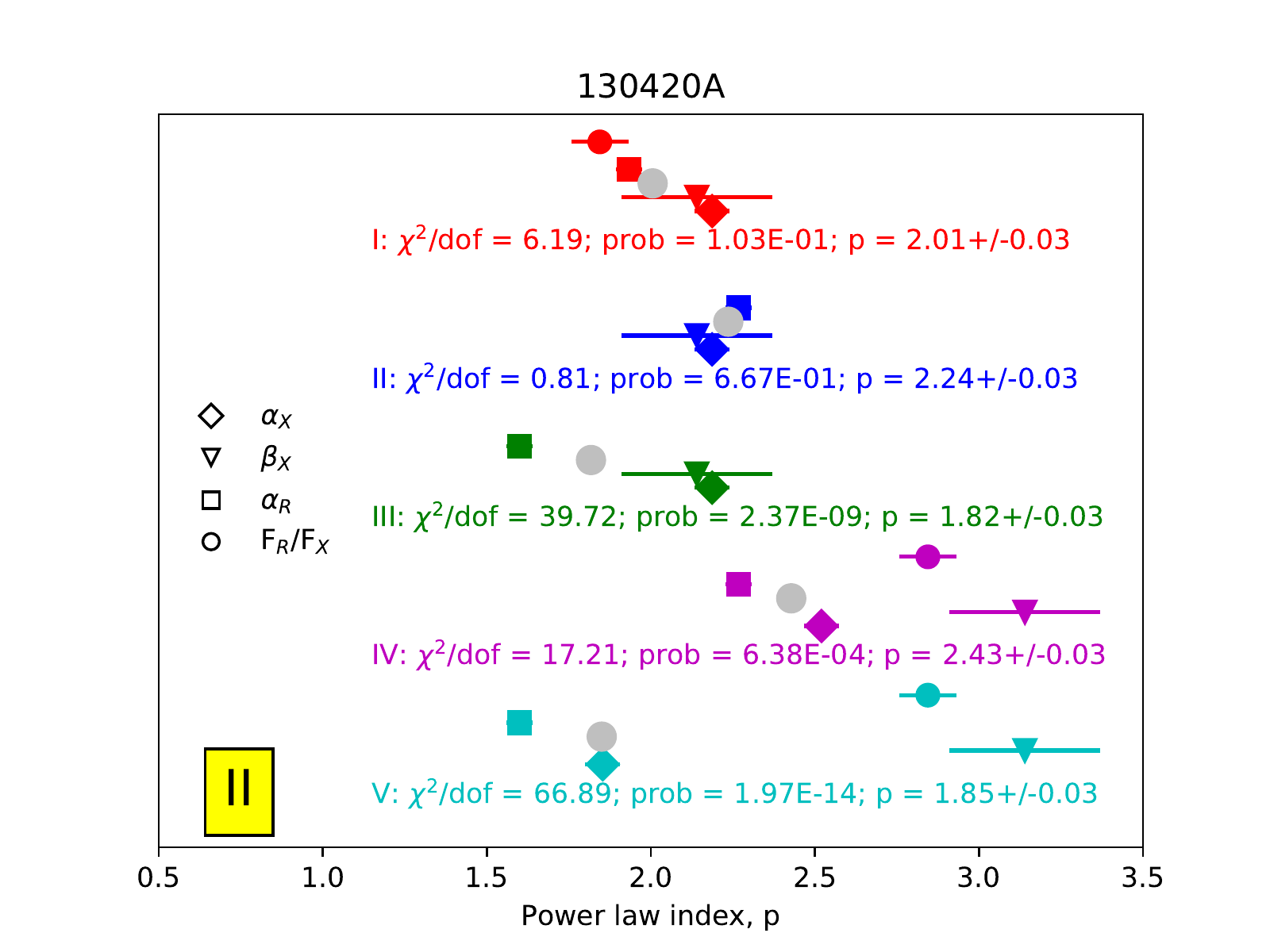}
\includegraphics[width=8.9cm]{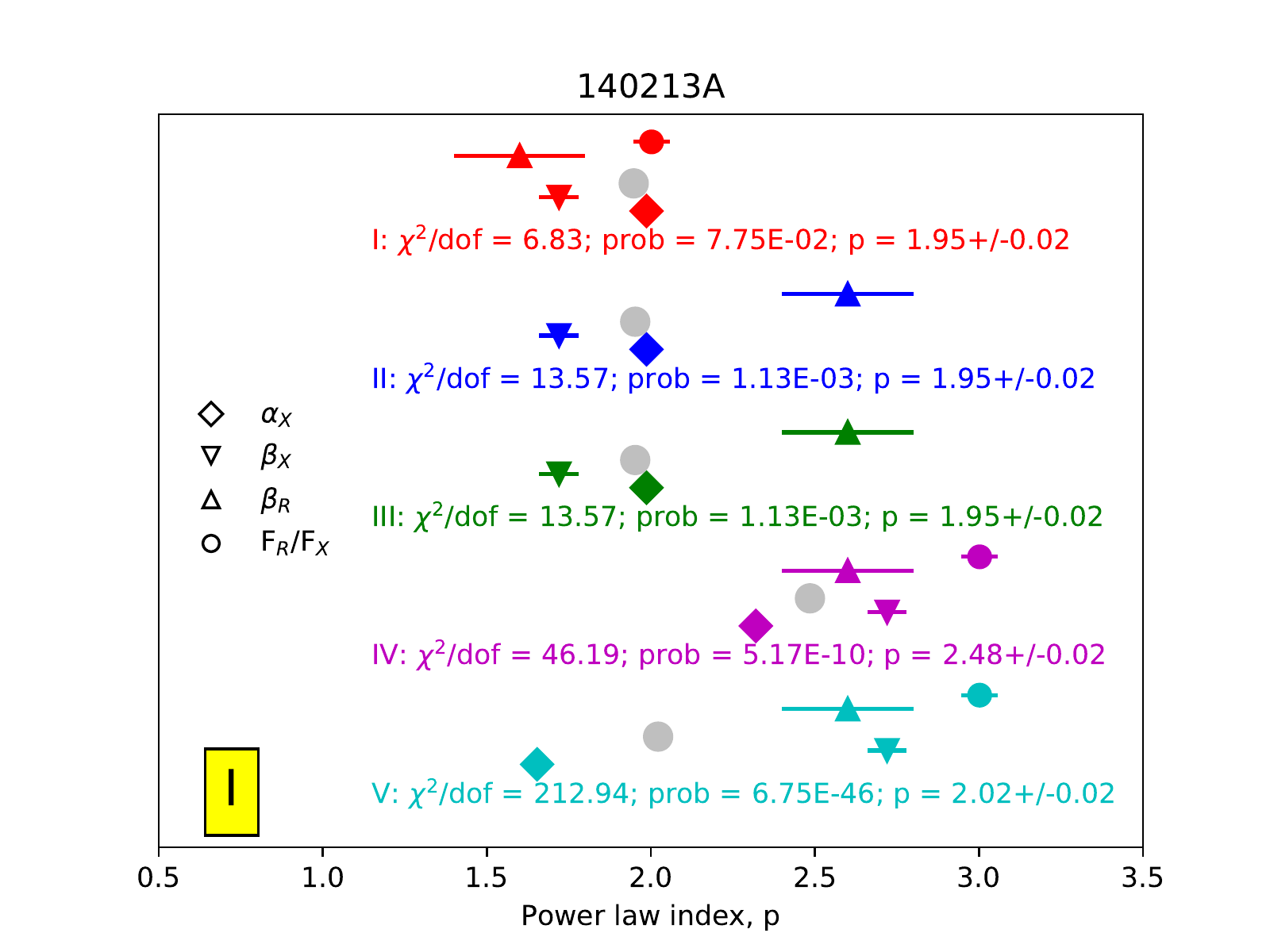}
\includegraphics[width=8.9cm]{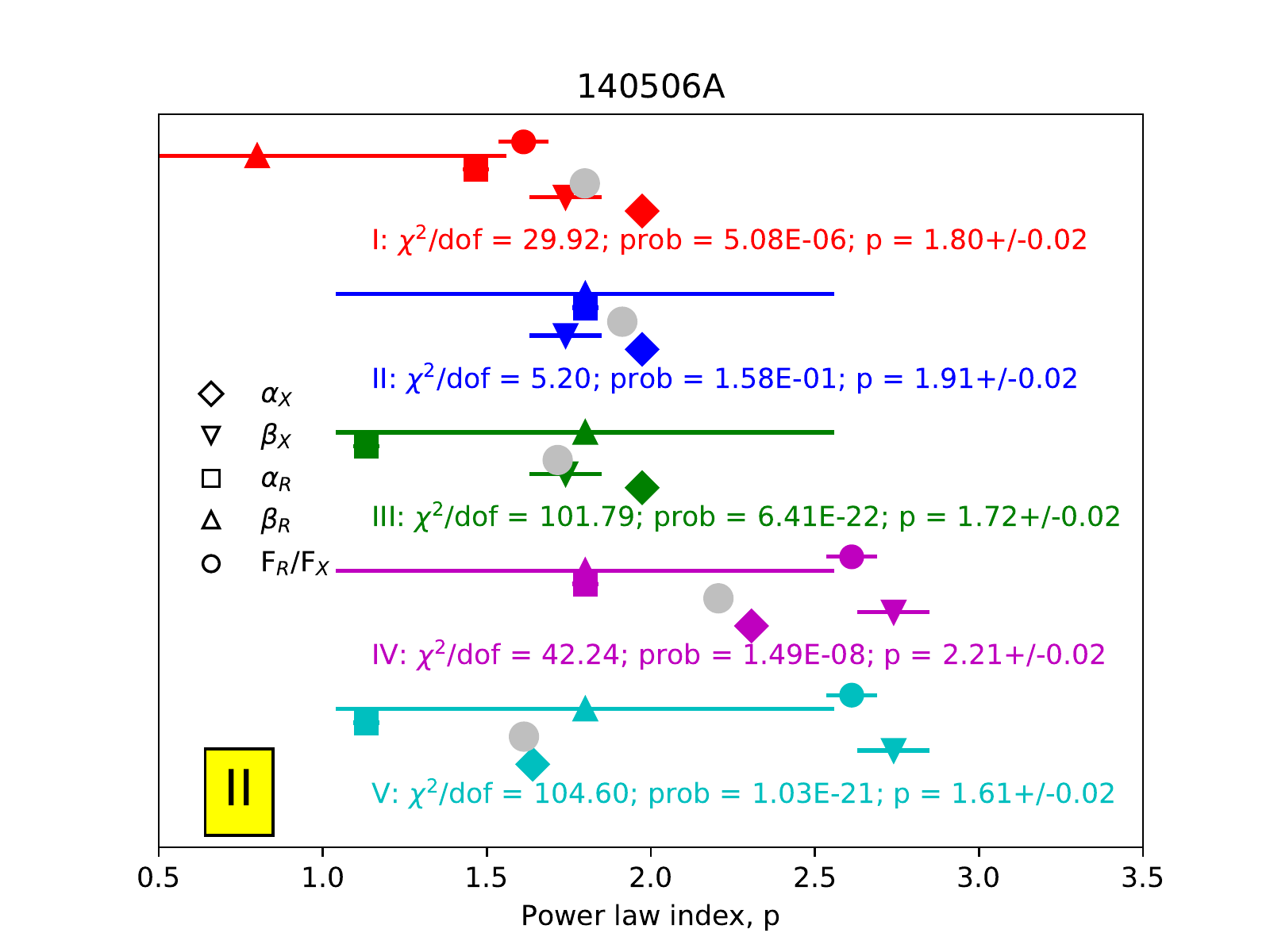}
\includegraphics[width=8.9cm]{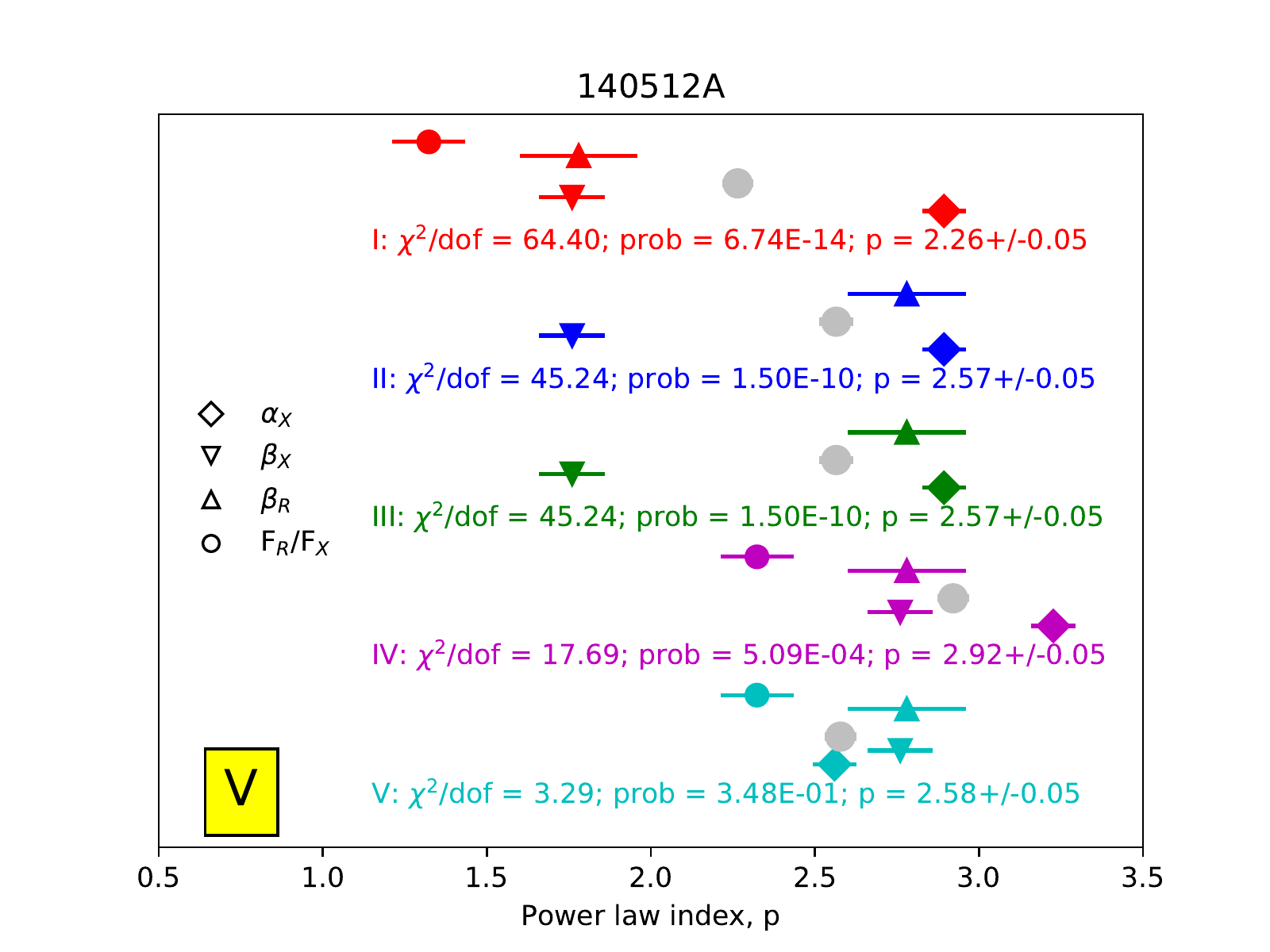}
\includegraphics[width=8.9cm]{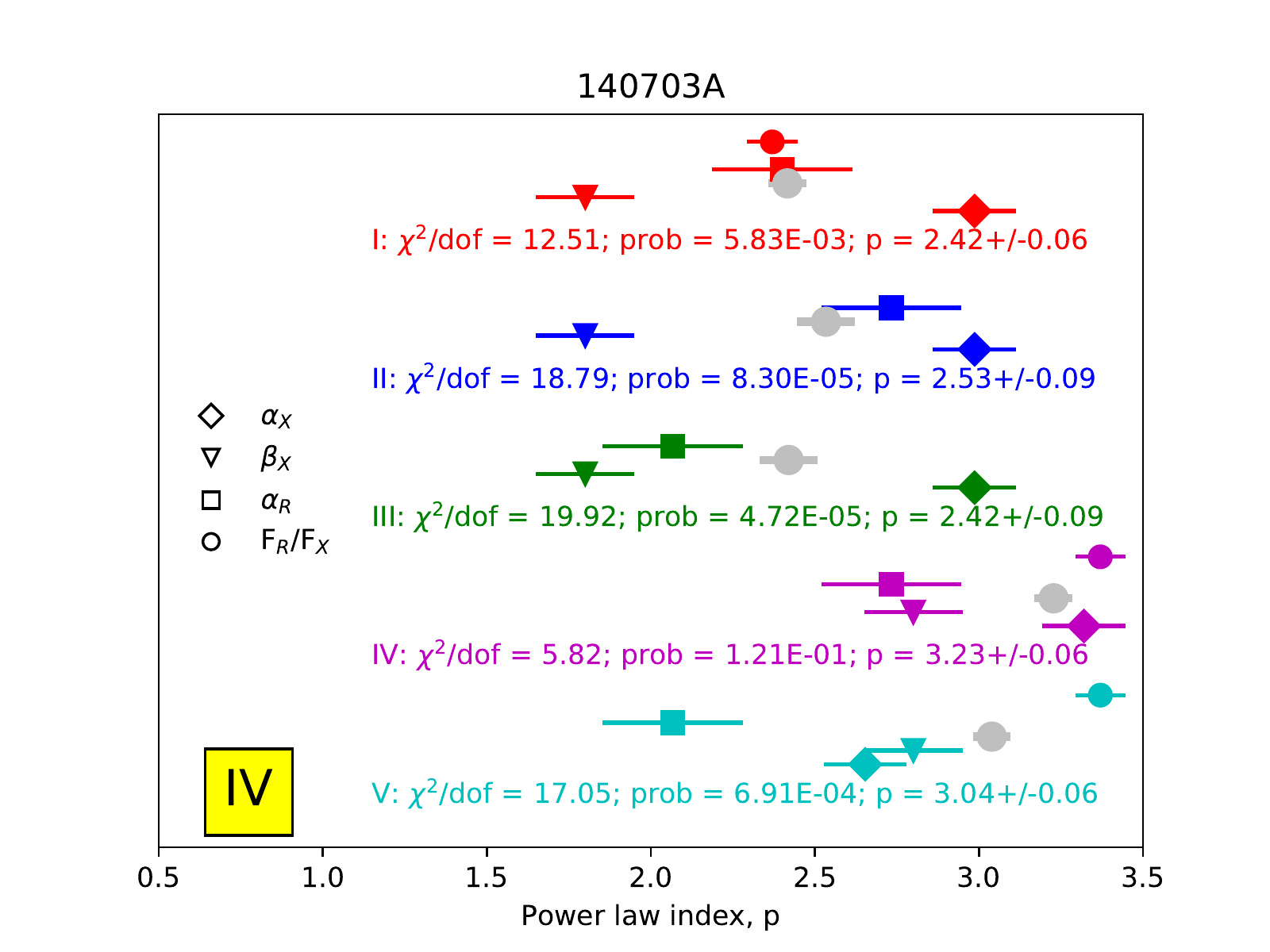}
\includegraphics[width=8.9cm]{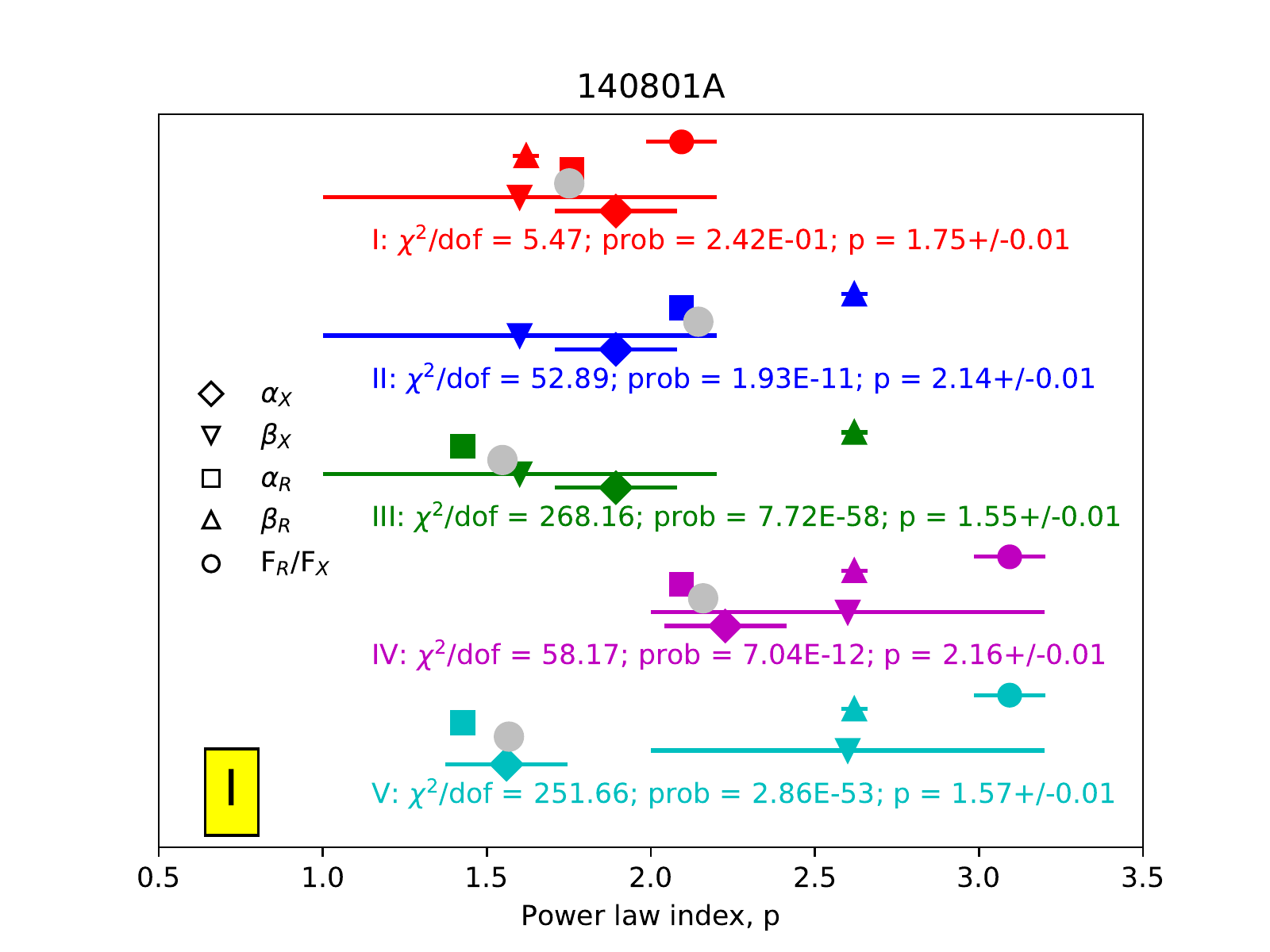}
\includegraphics[width=8.9cm]{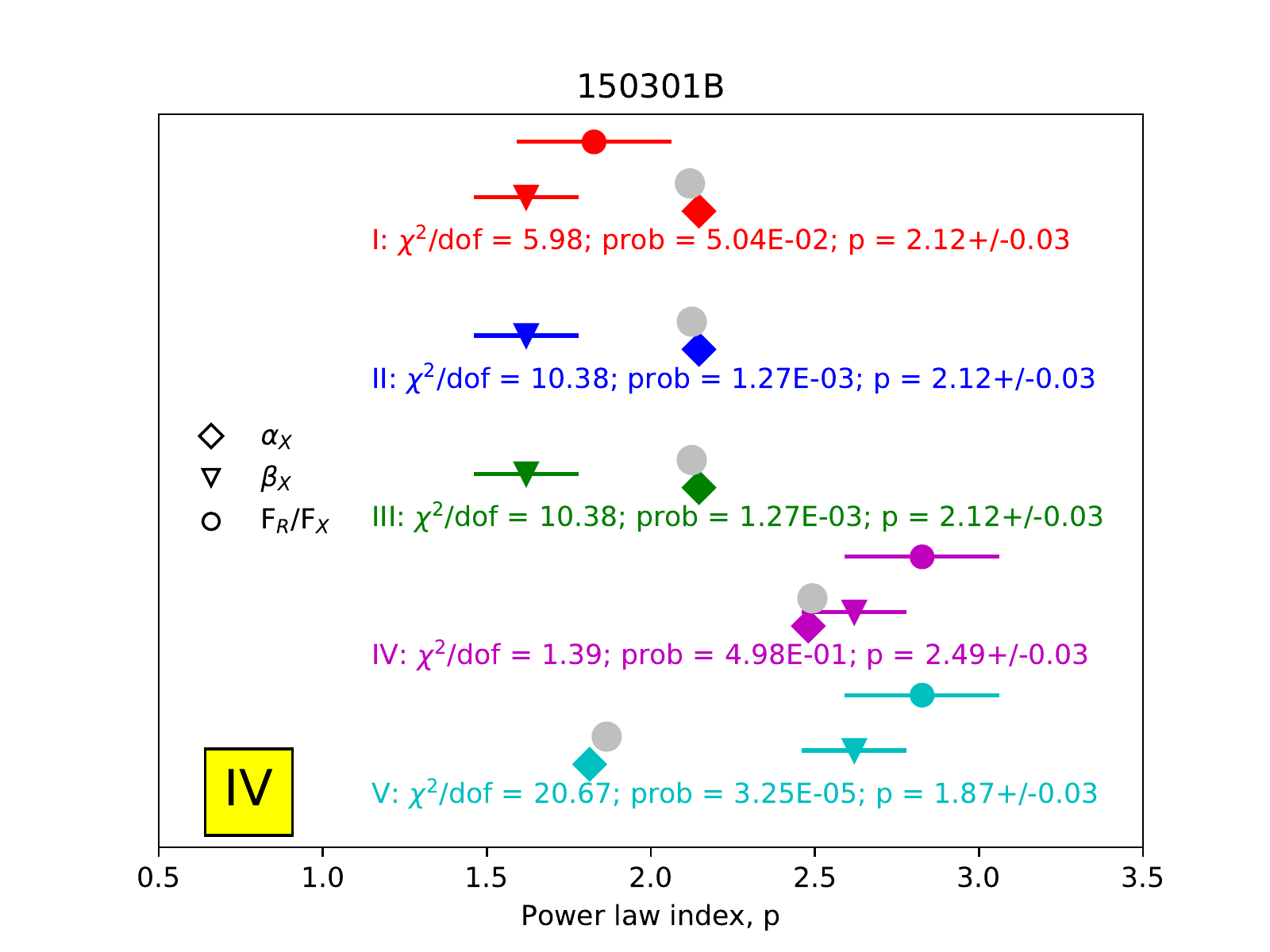}
\includegraphics[width=8.9cm]{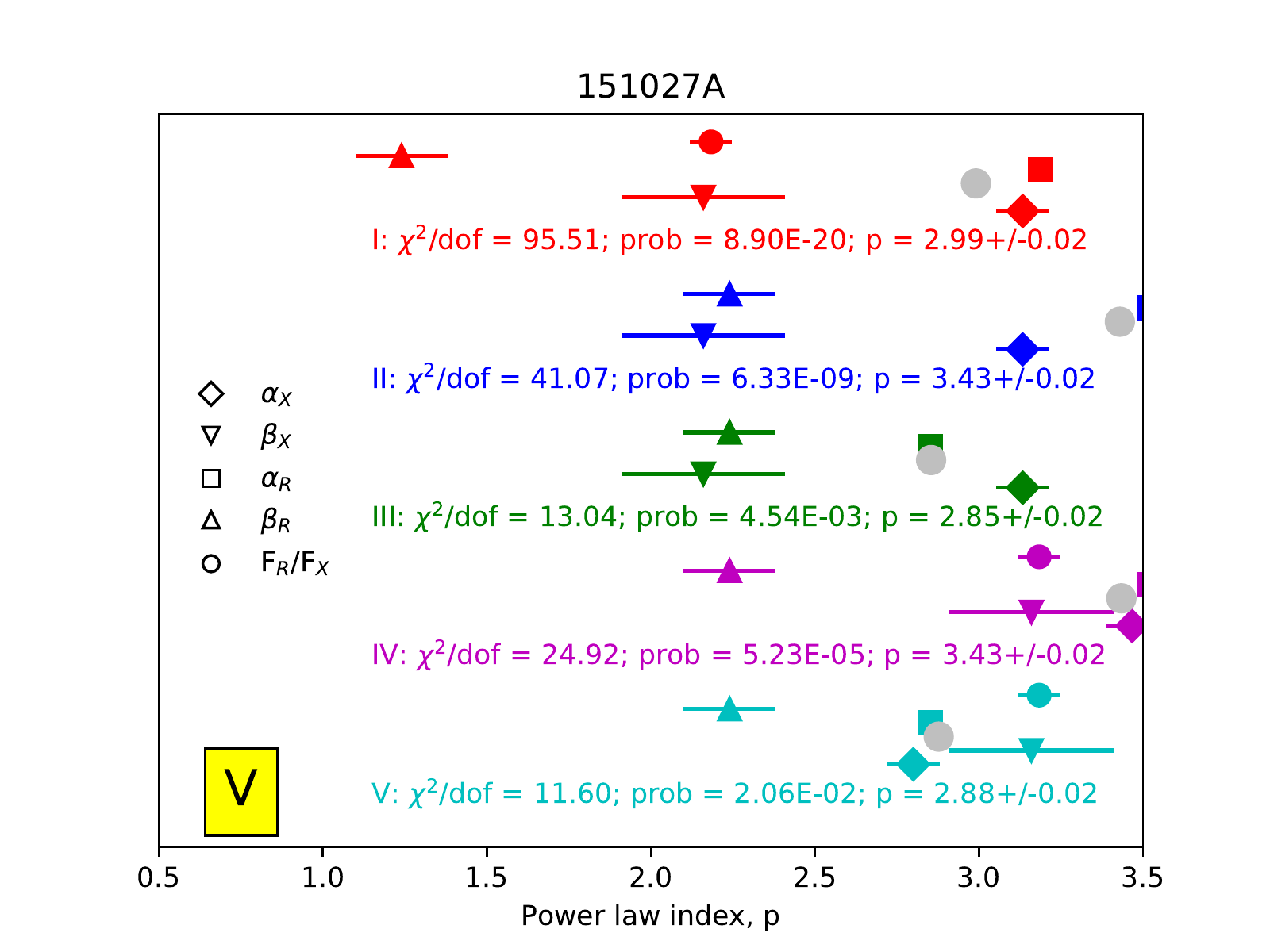}
\includegraphics[width=8.9cm]{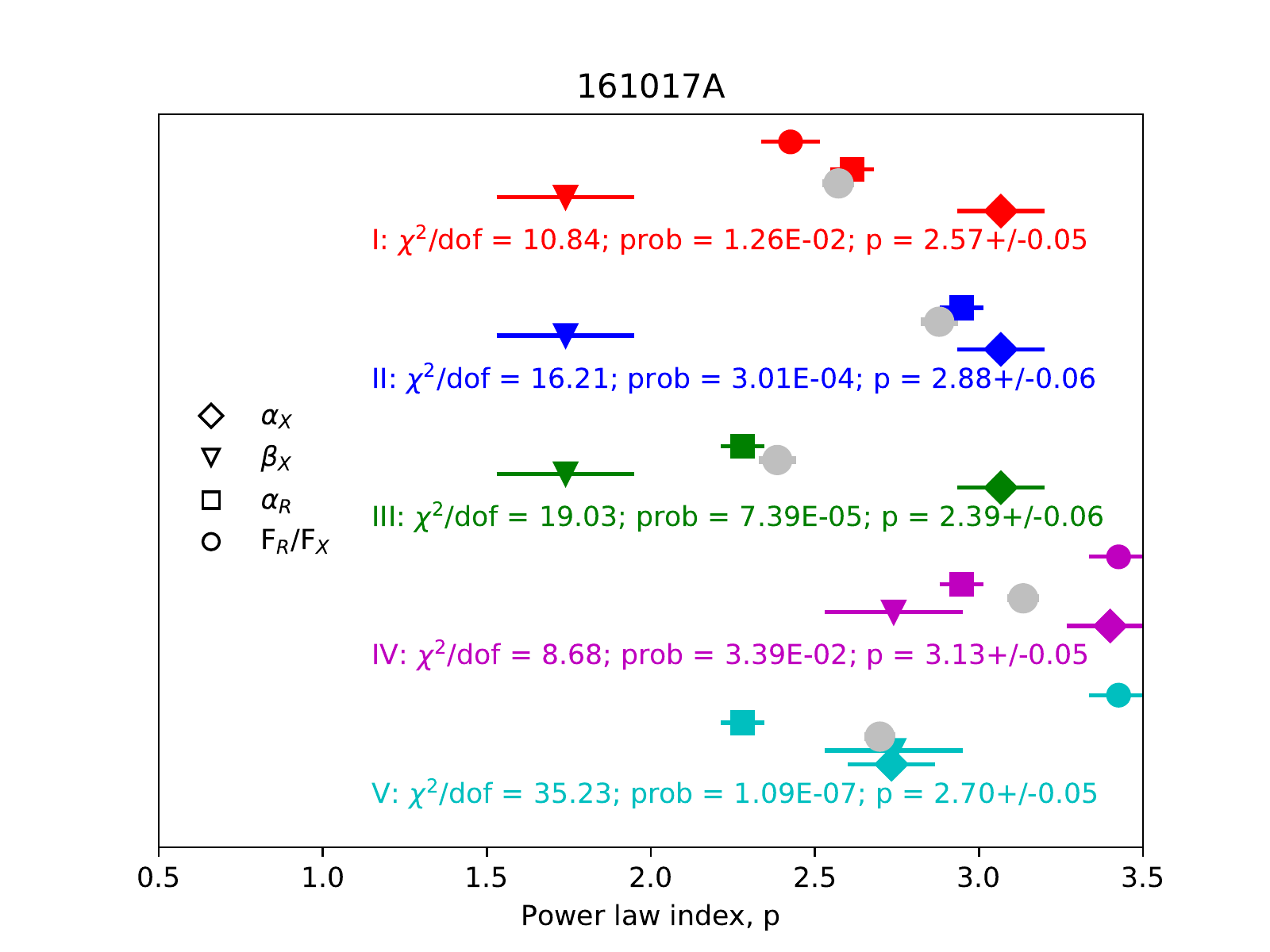}
\includegraphics[width=8.9cm]{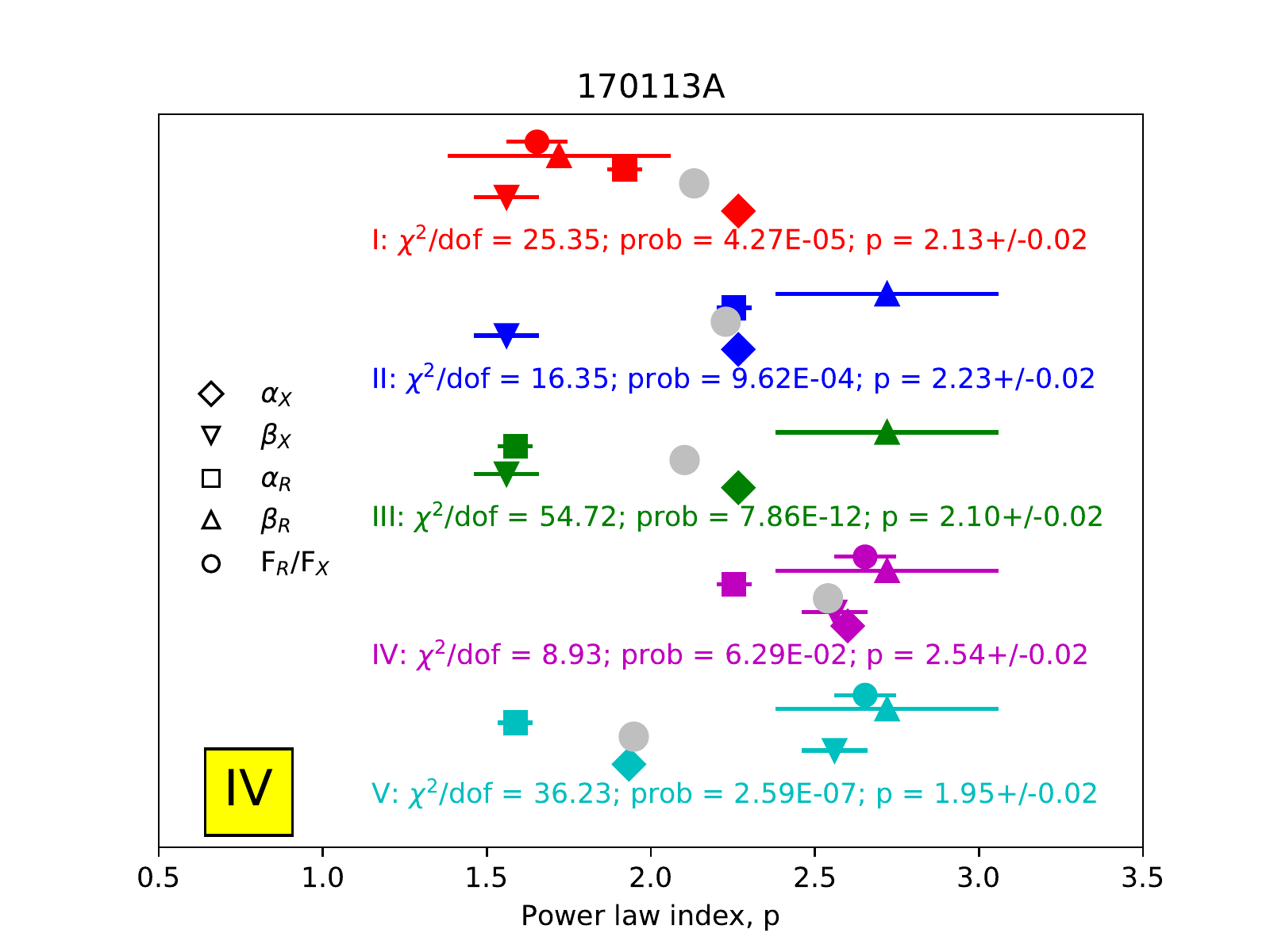}

\section{Best Fits Including the Standard Error}\label{app:fits_se}
\subsection{LAT}
\noindent \includegraphics[width=8.9cm]{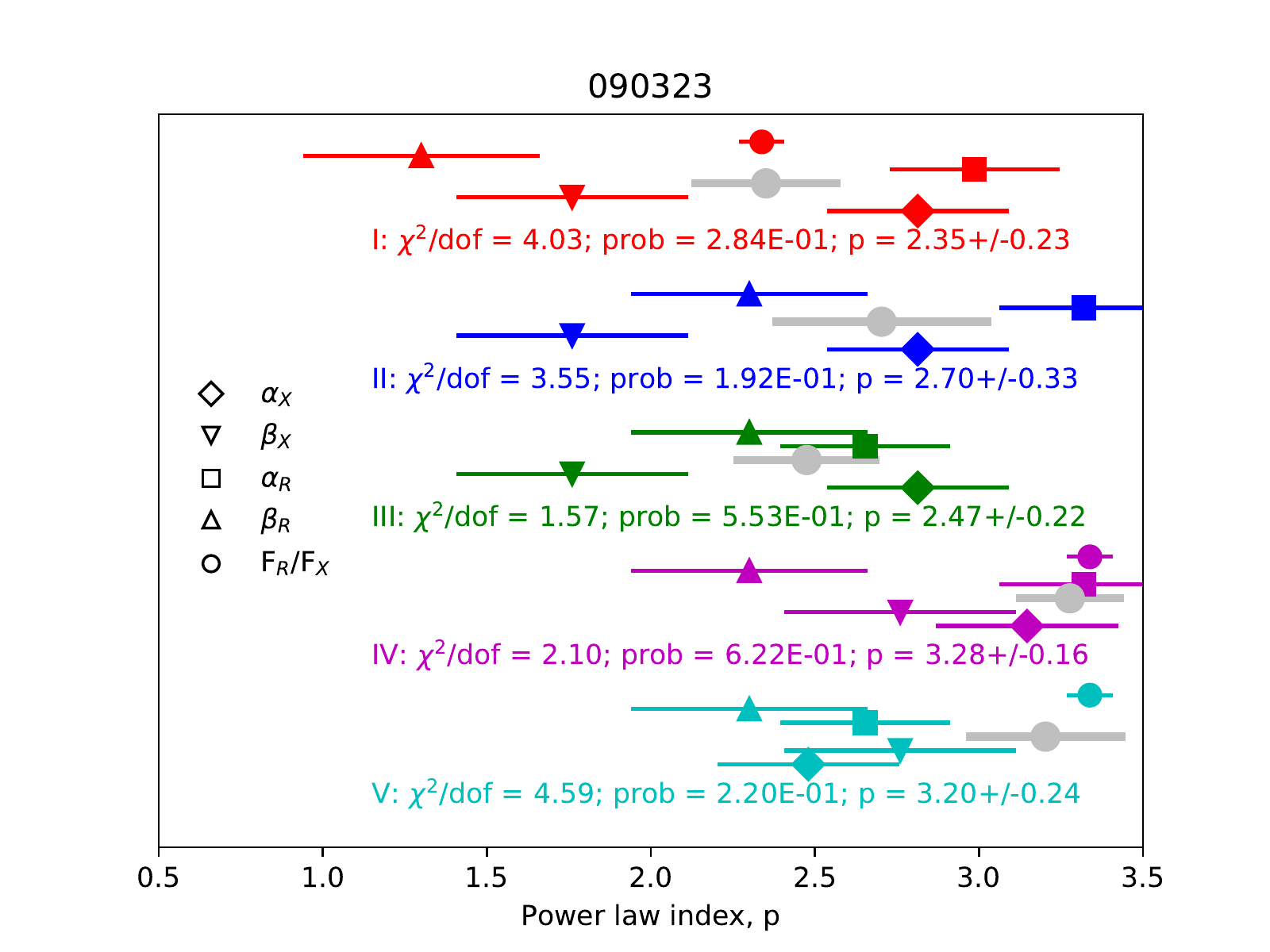}
\includegraphics[width=8.9cm]{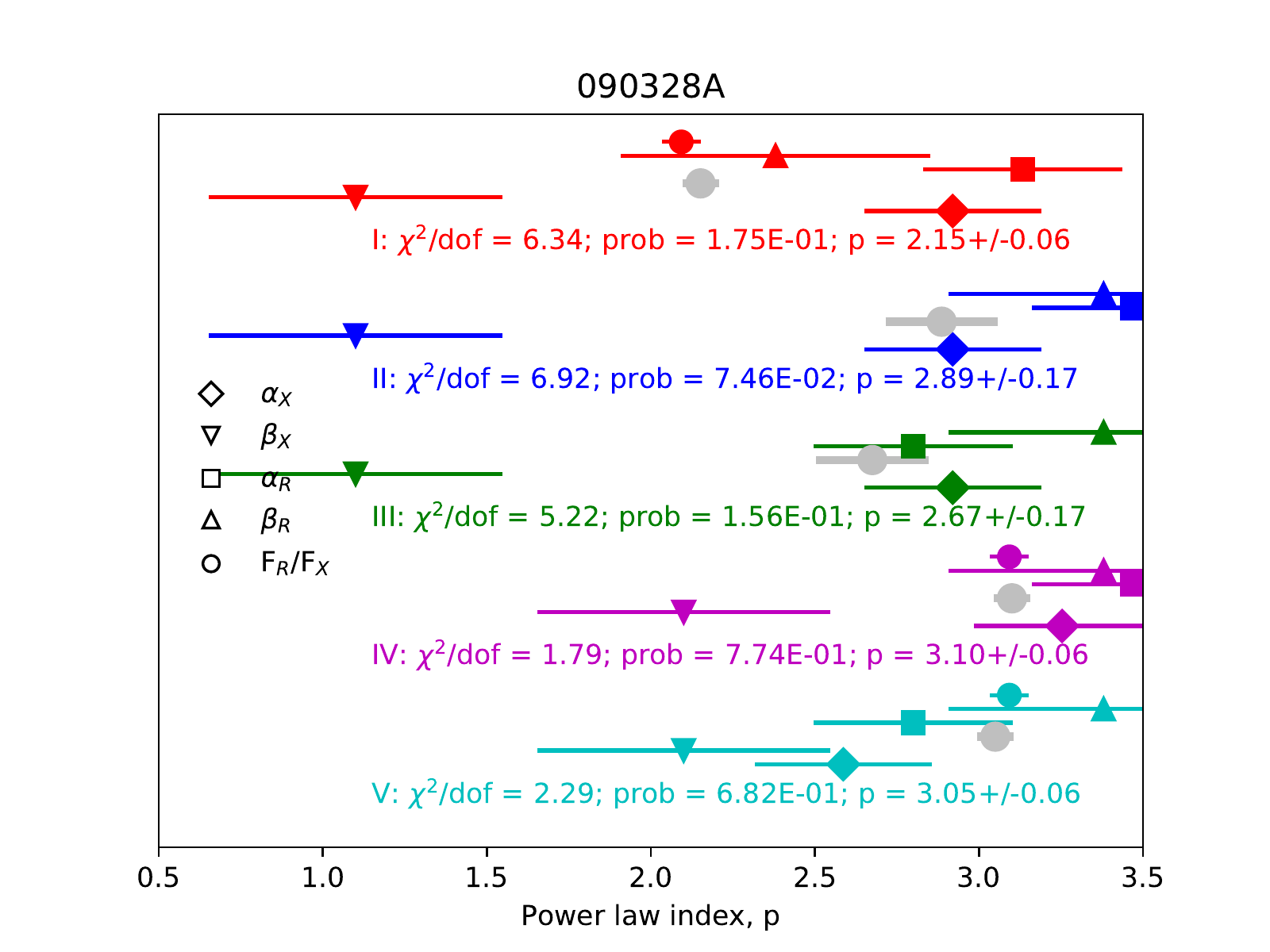}
\includegraphics[width=8.9cm]{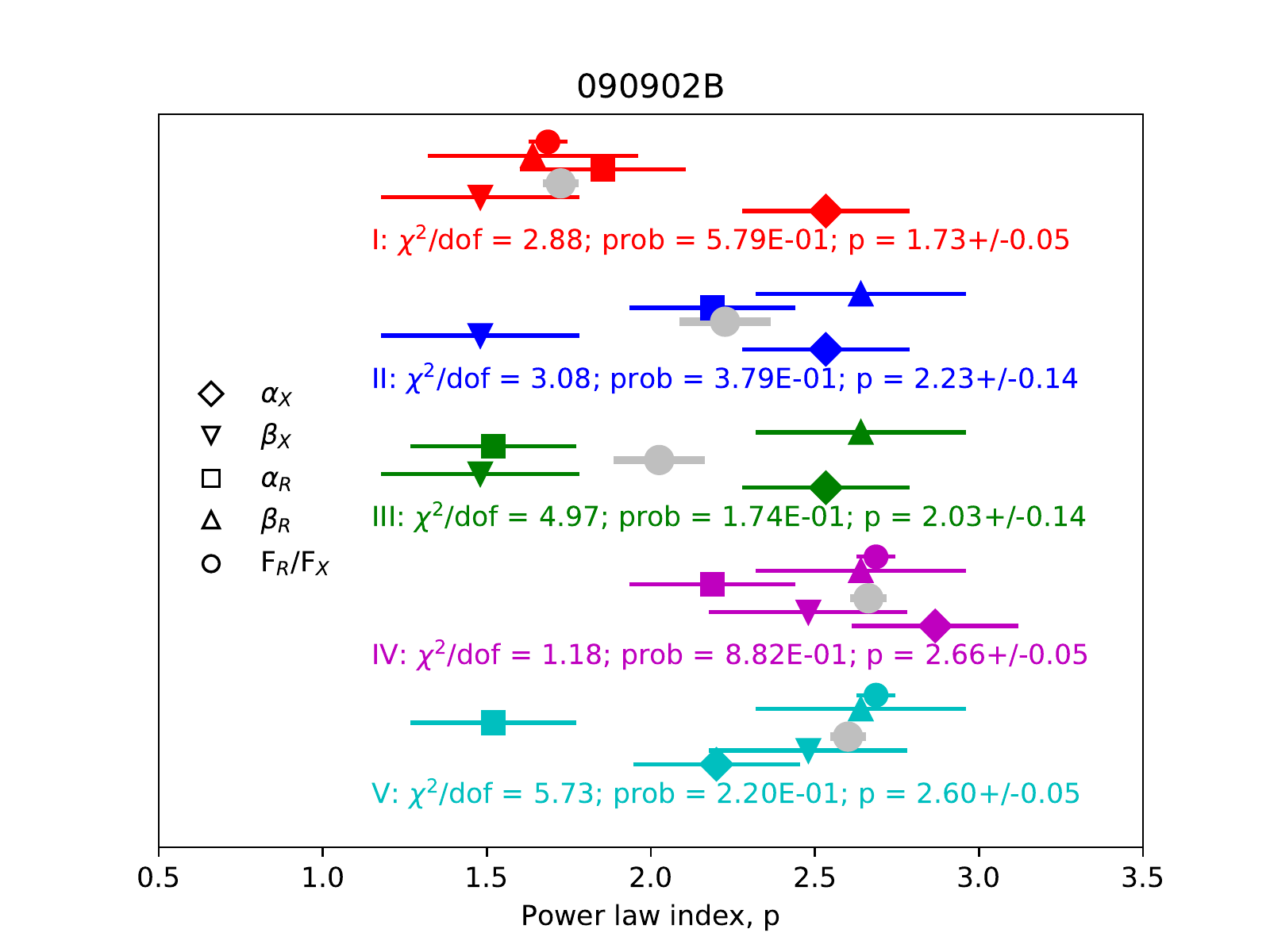}
\includegraphics[width=8.9cm]{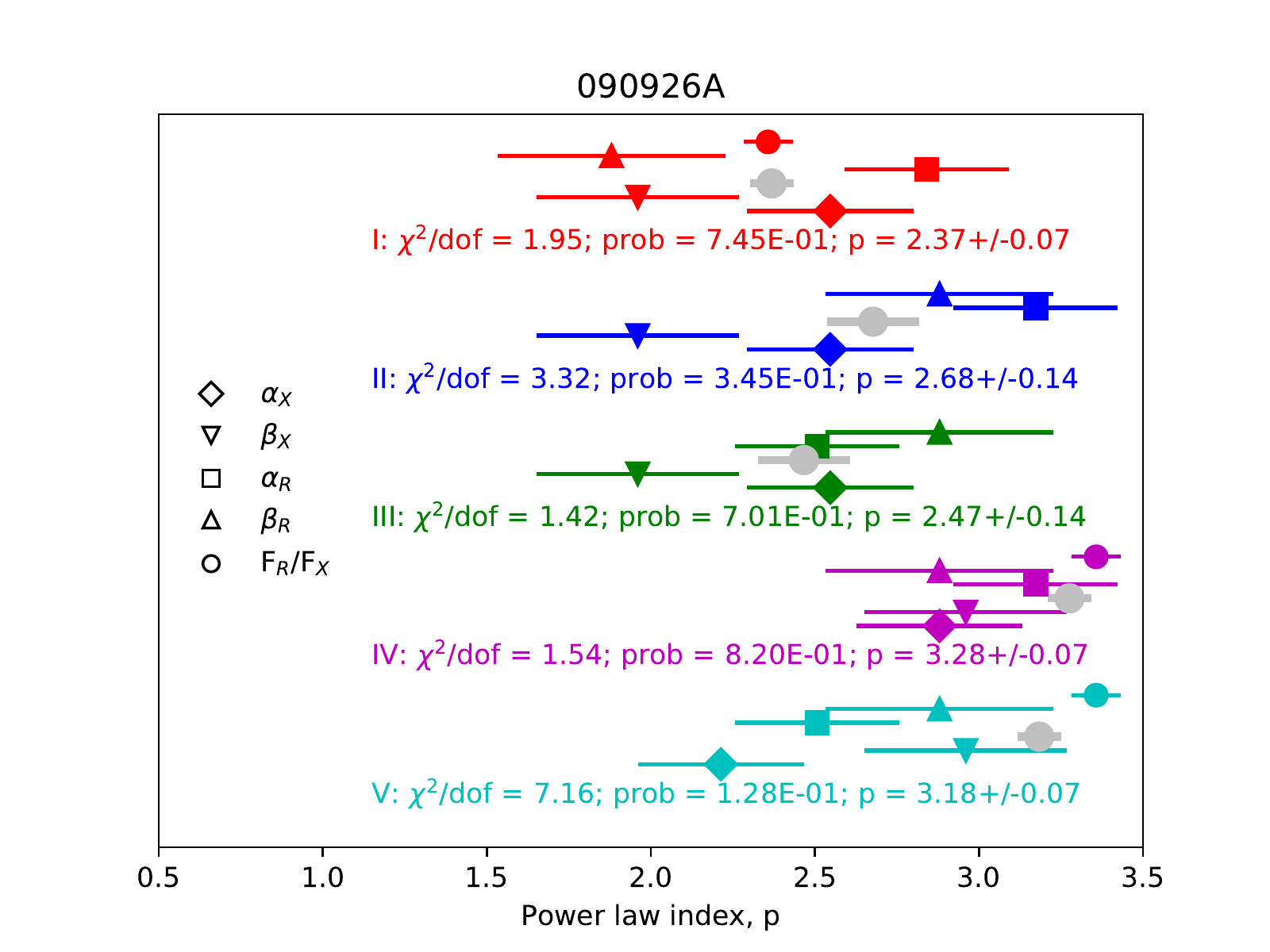}
\includegraphics[width=8.9cm]{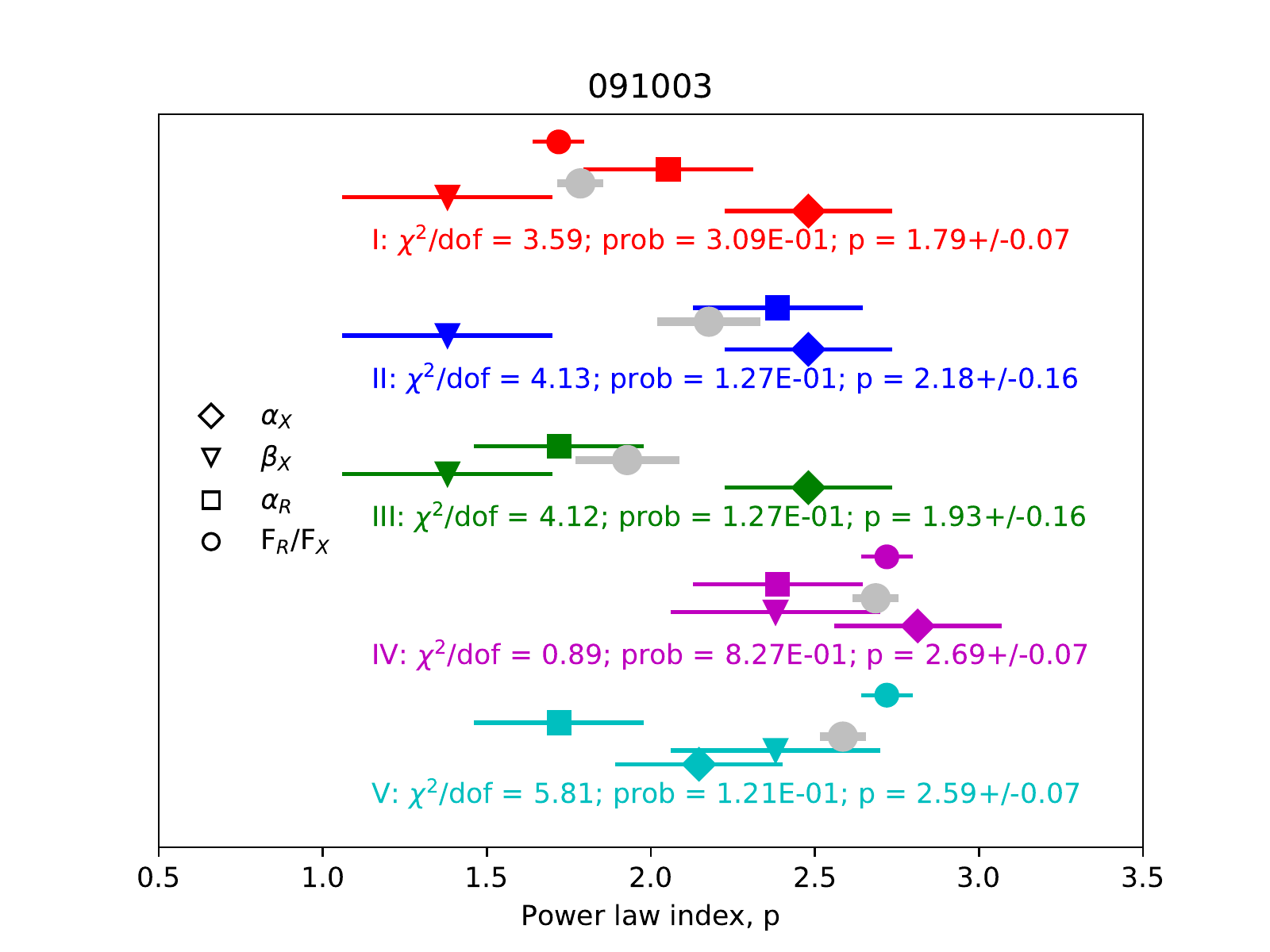}
\includegraphics[width=8.9cm]{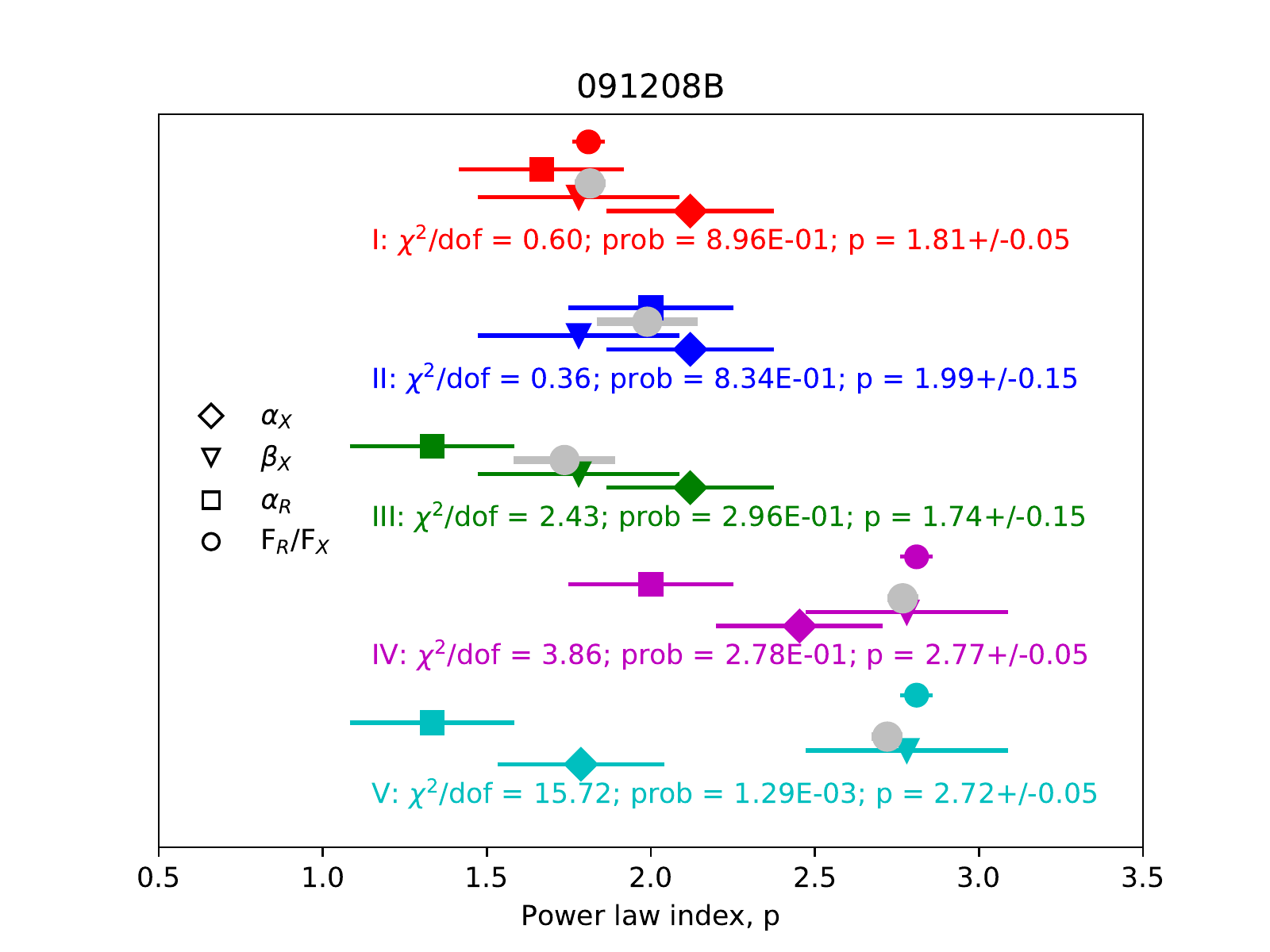}
\includegraphics[width=8.9cm]{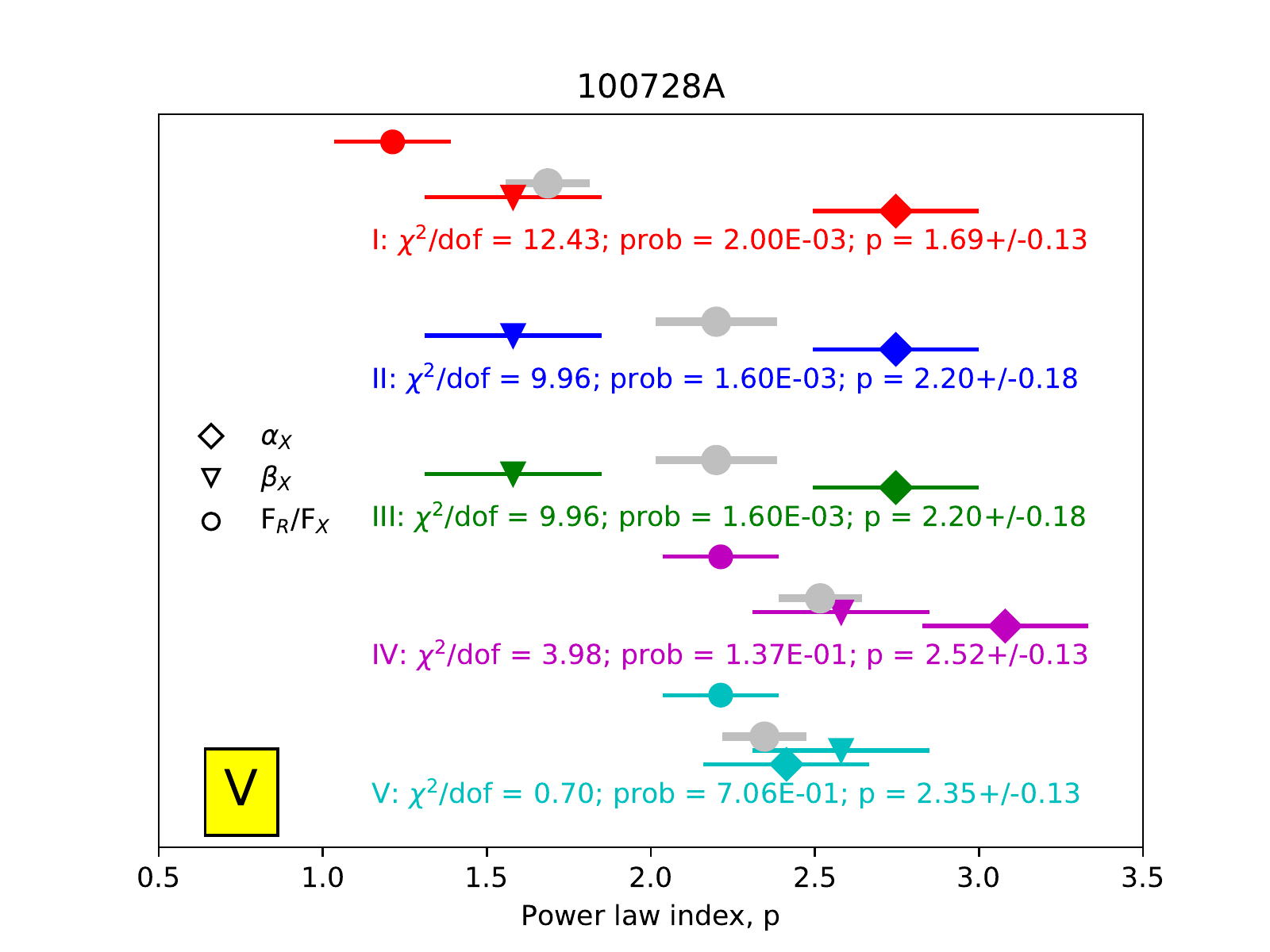}
\includegraphics[width=8.9cm]{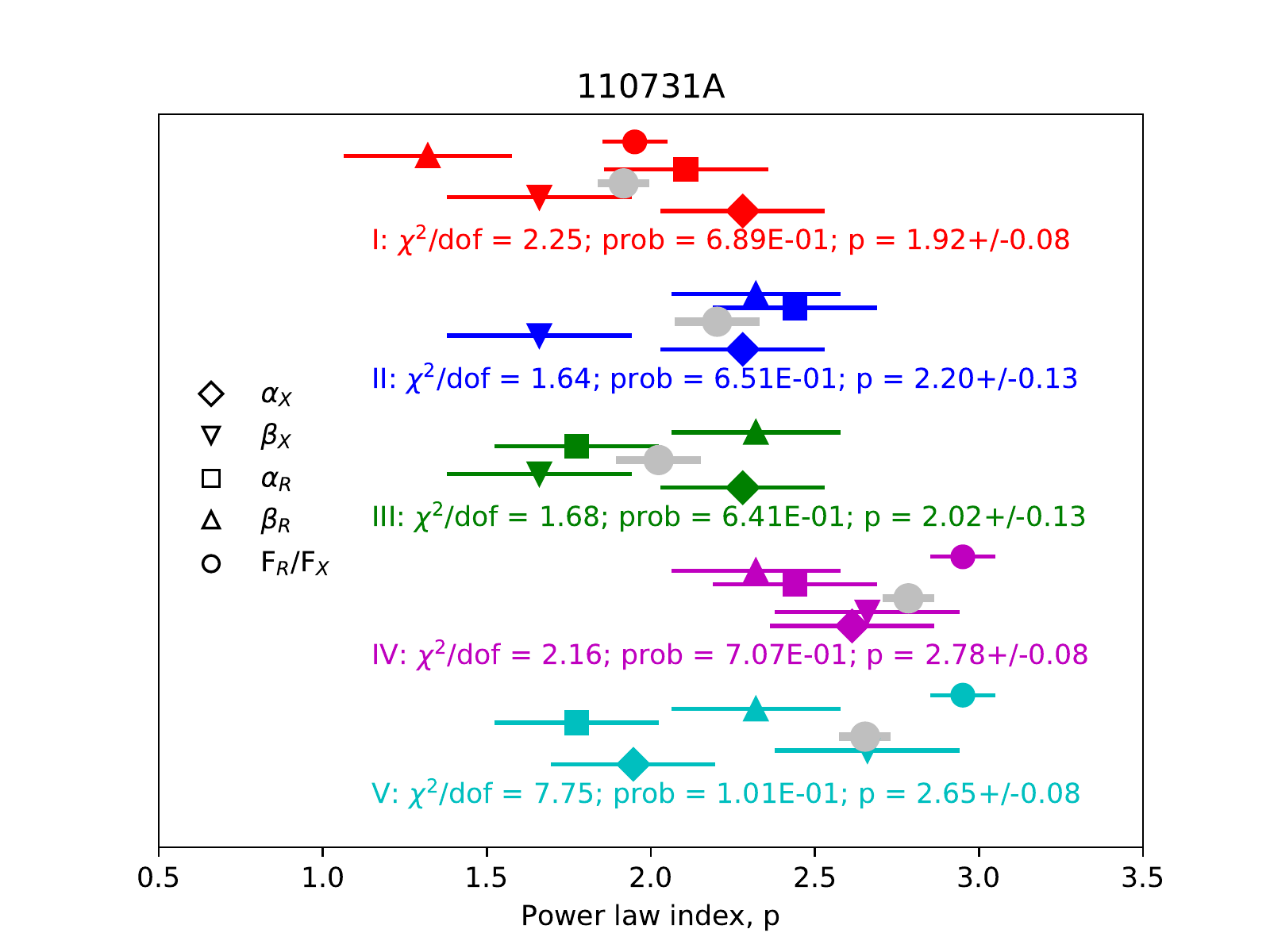}
\includegraphics[width=8.9cm]{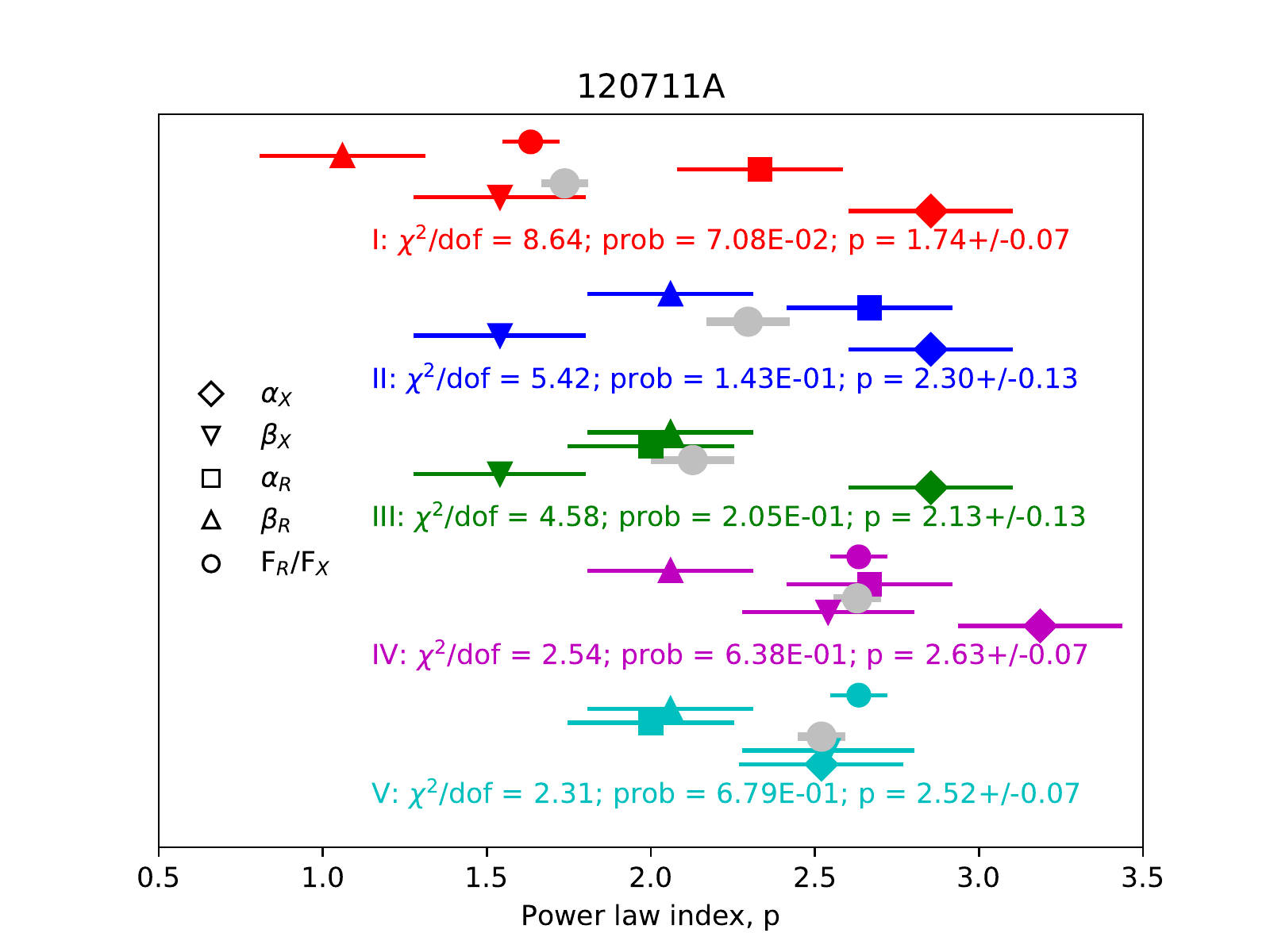}
\includegraphics[width=8.9cm]{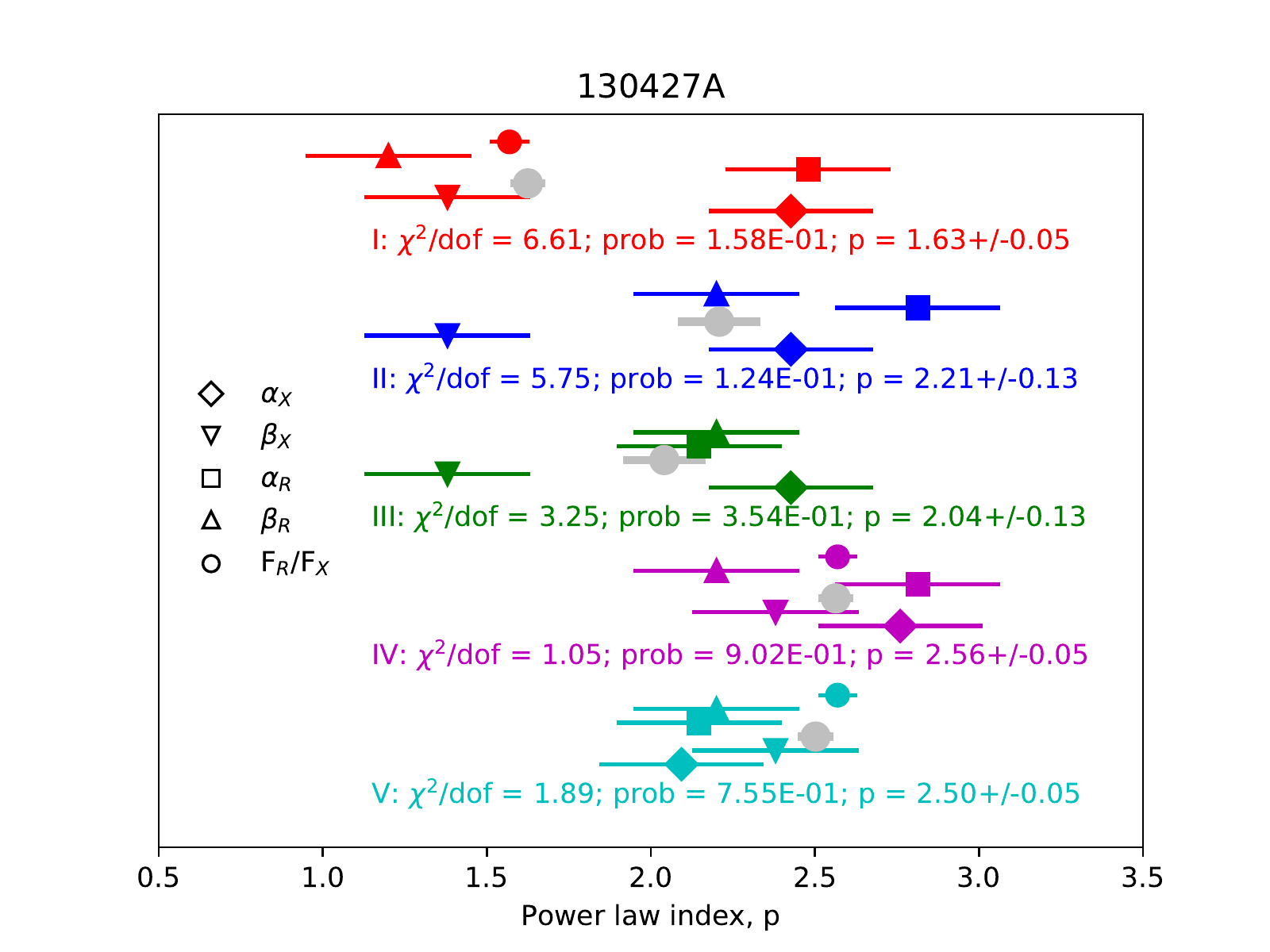}
\includegraphics[width=8.9cm]{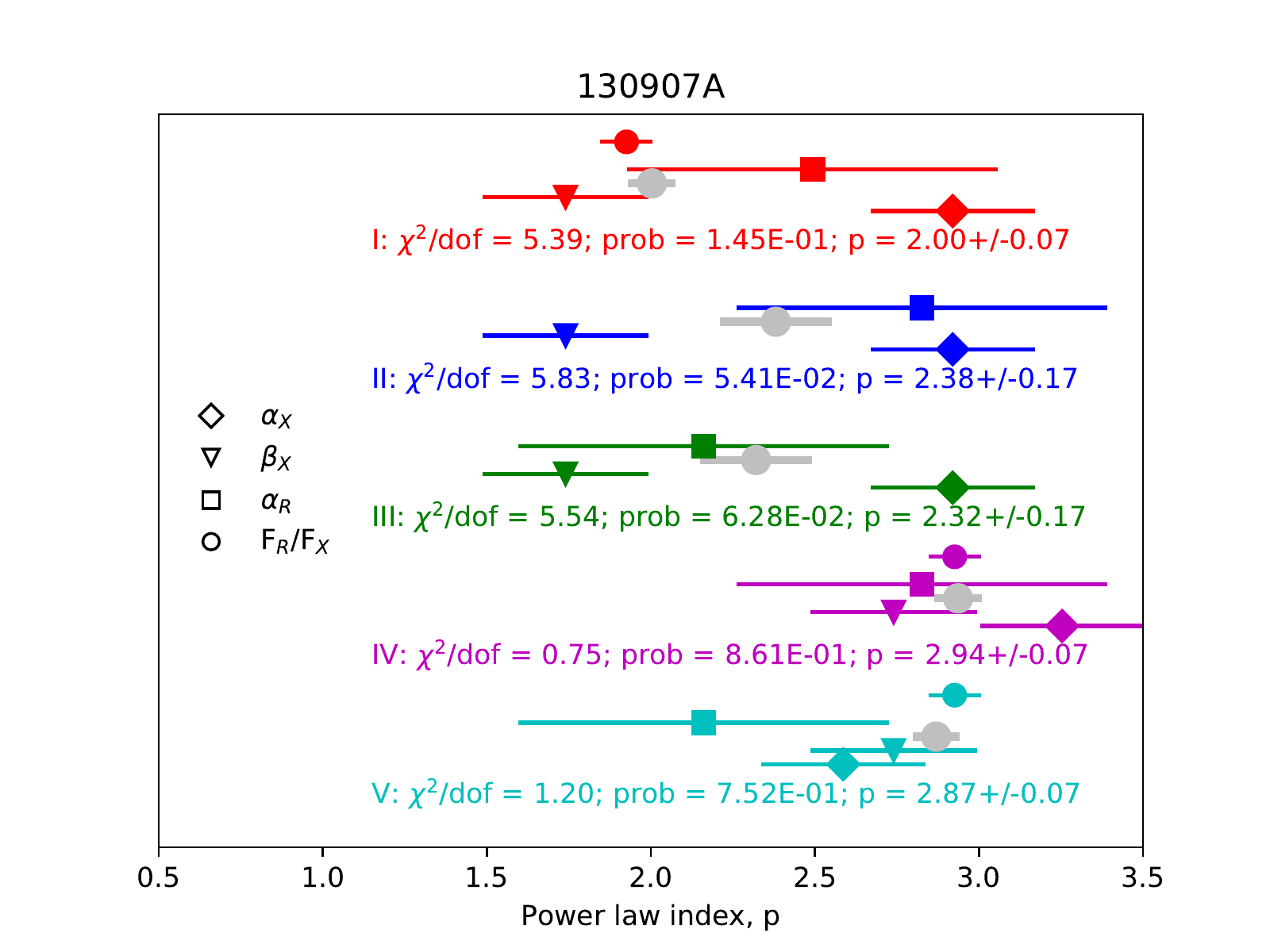}
\includegraphics[width=8.9cm]{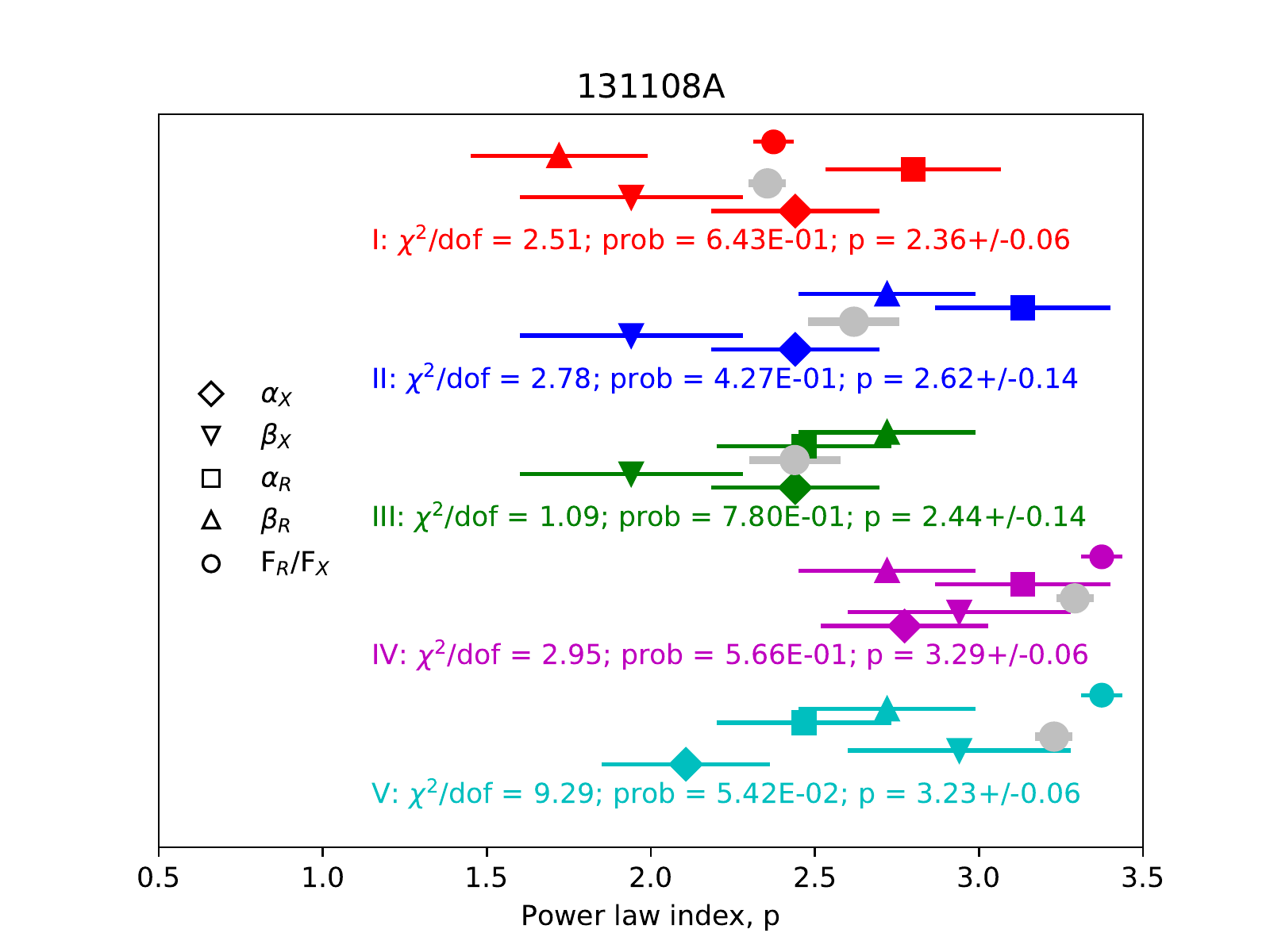}
\includegraphics[width=8.9cm]{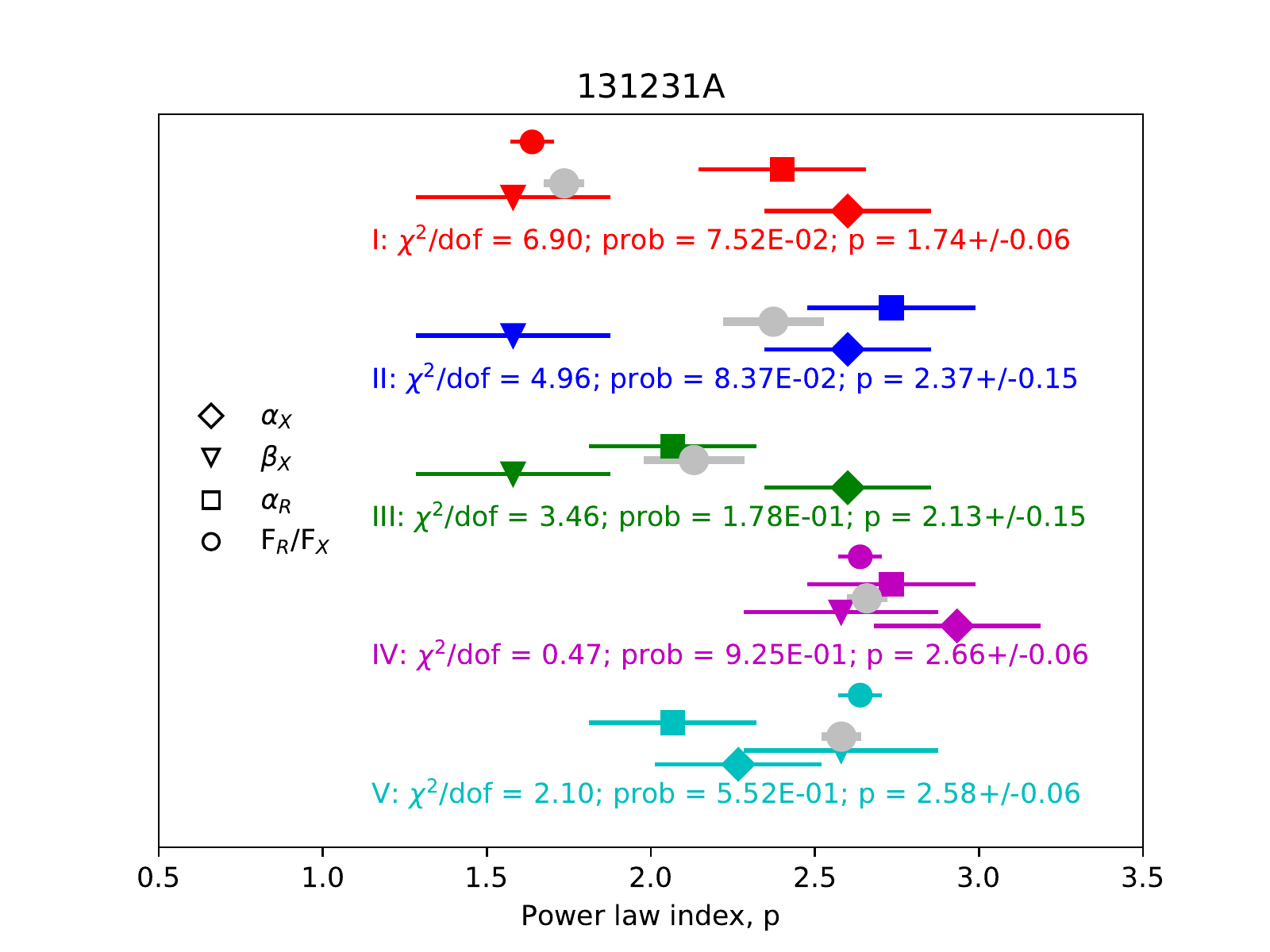}
\includegraphics[width=8.9cm]{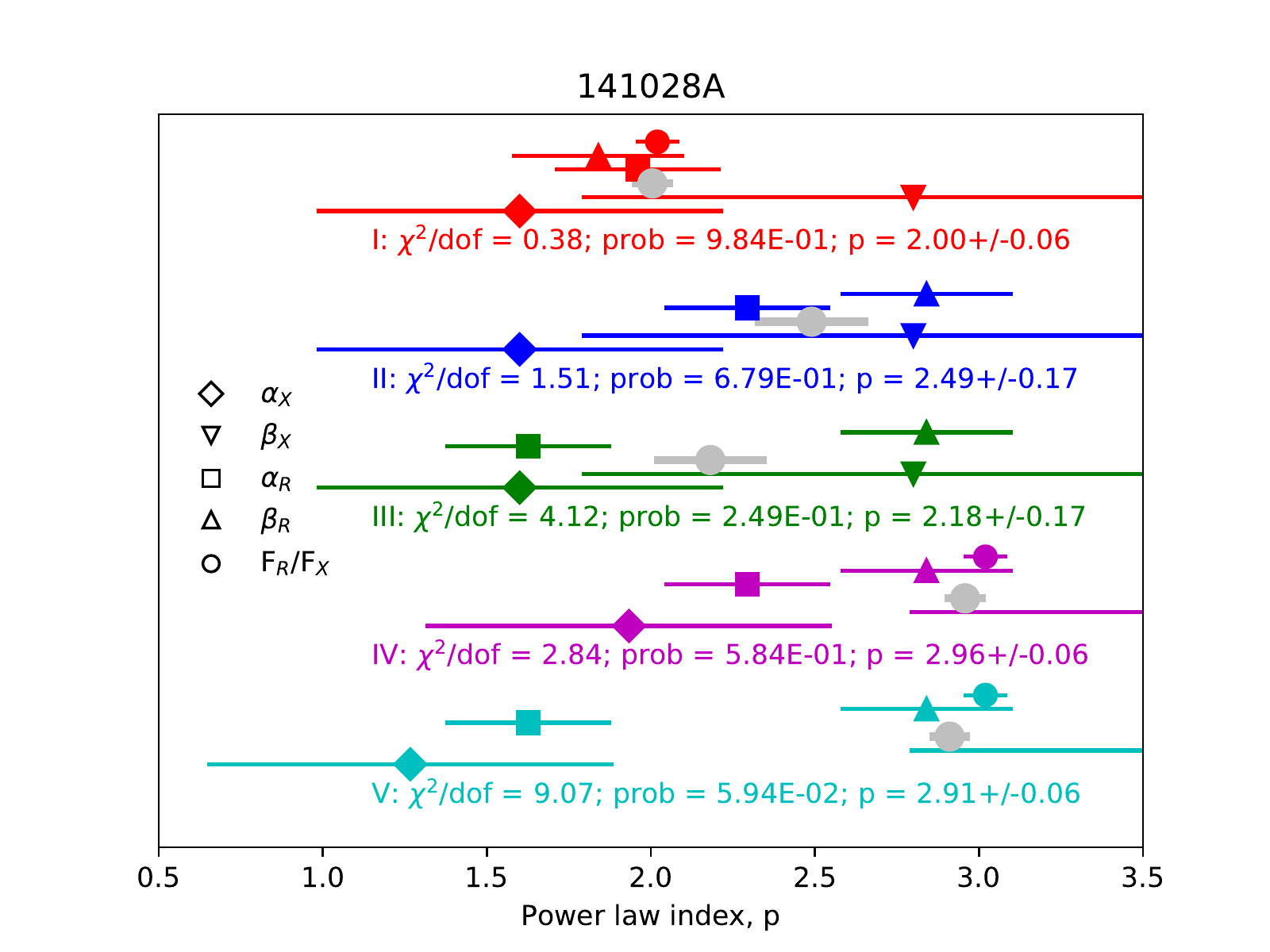}
\includegraphics[width=8.9cm]{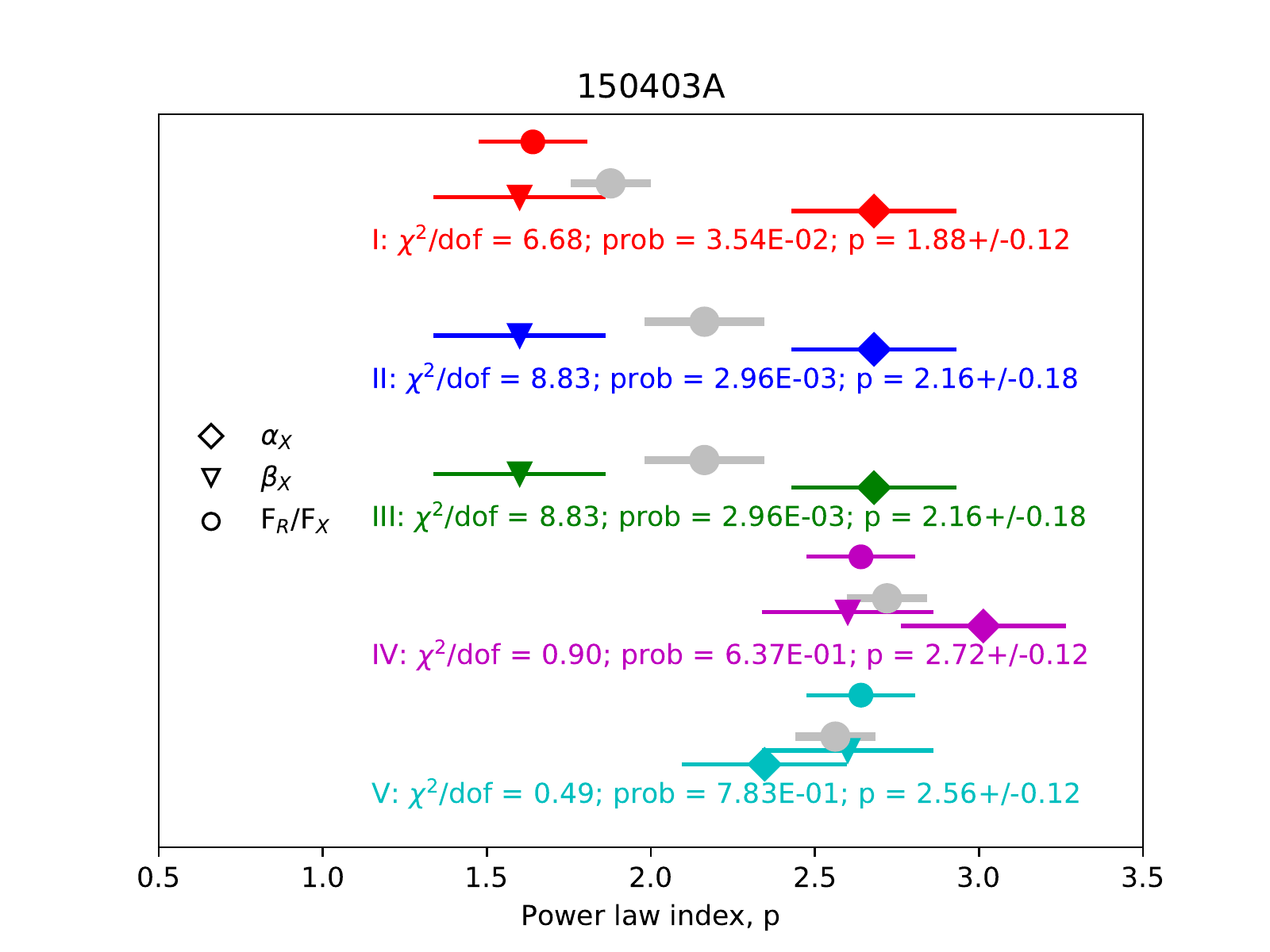}
\includegraphics[width=8.9cm]{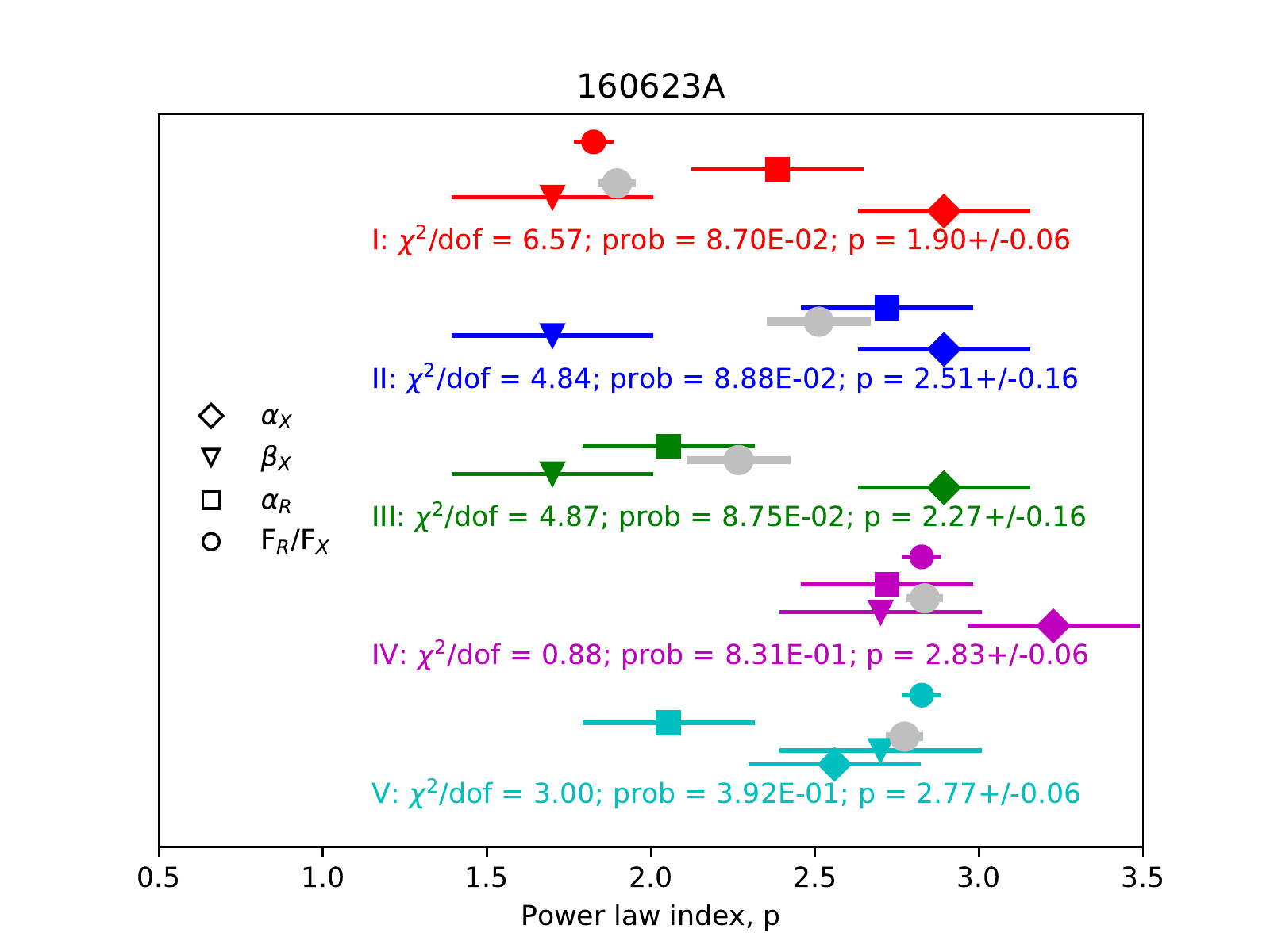}
\includegraphics[width=8.9cm]{160625B_se.pdf}
\subsection{GBM}
\noindent \includegraphics[width=8.9cm]{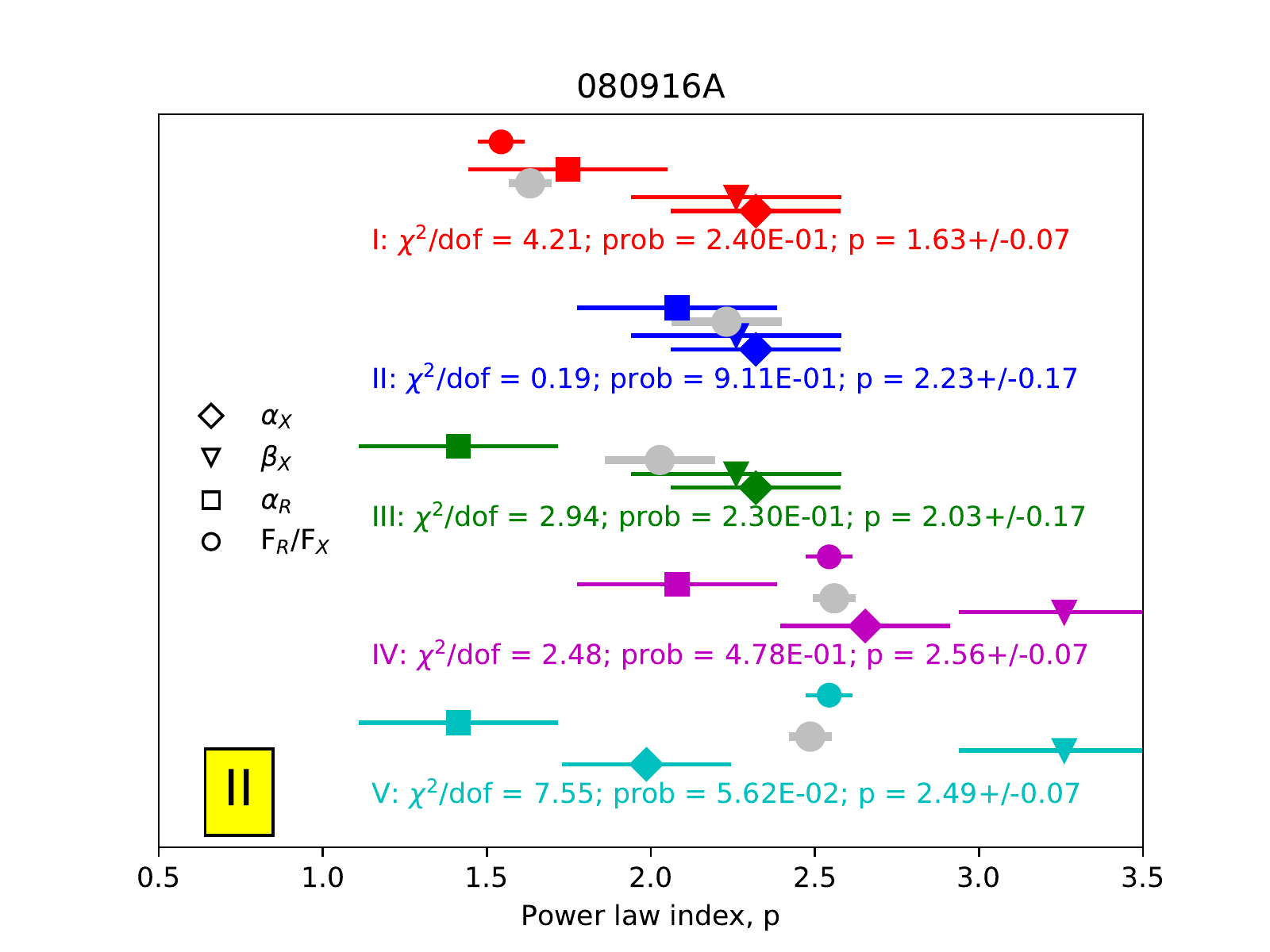}
\includegraphics[width=8.9cm]{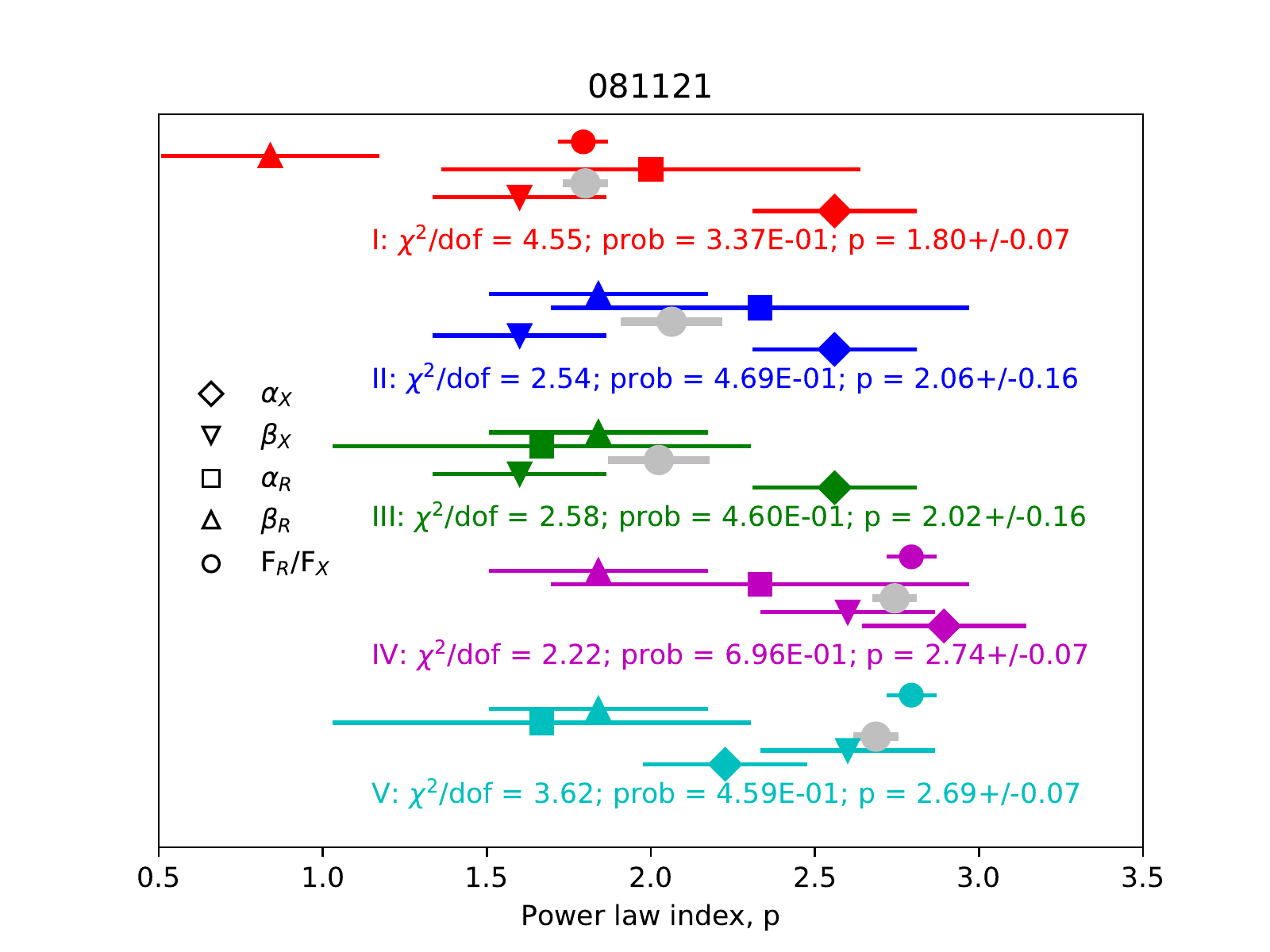}
\includegraphics[width=8.9cm]{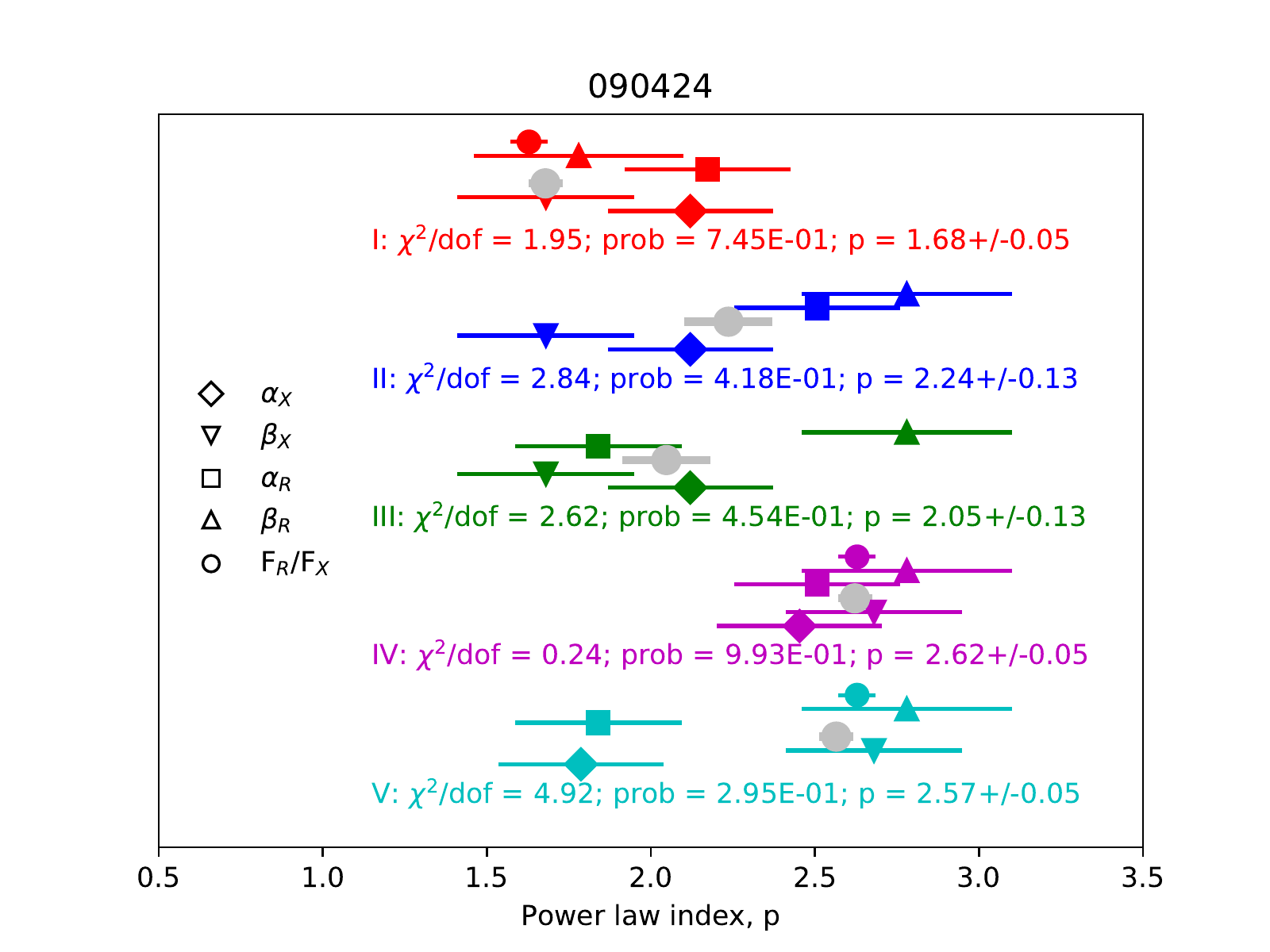}
\includegraphics[width=8.9cm]{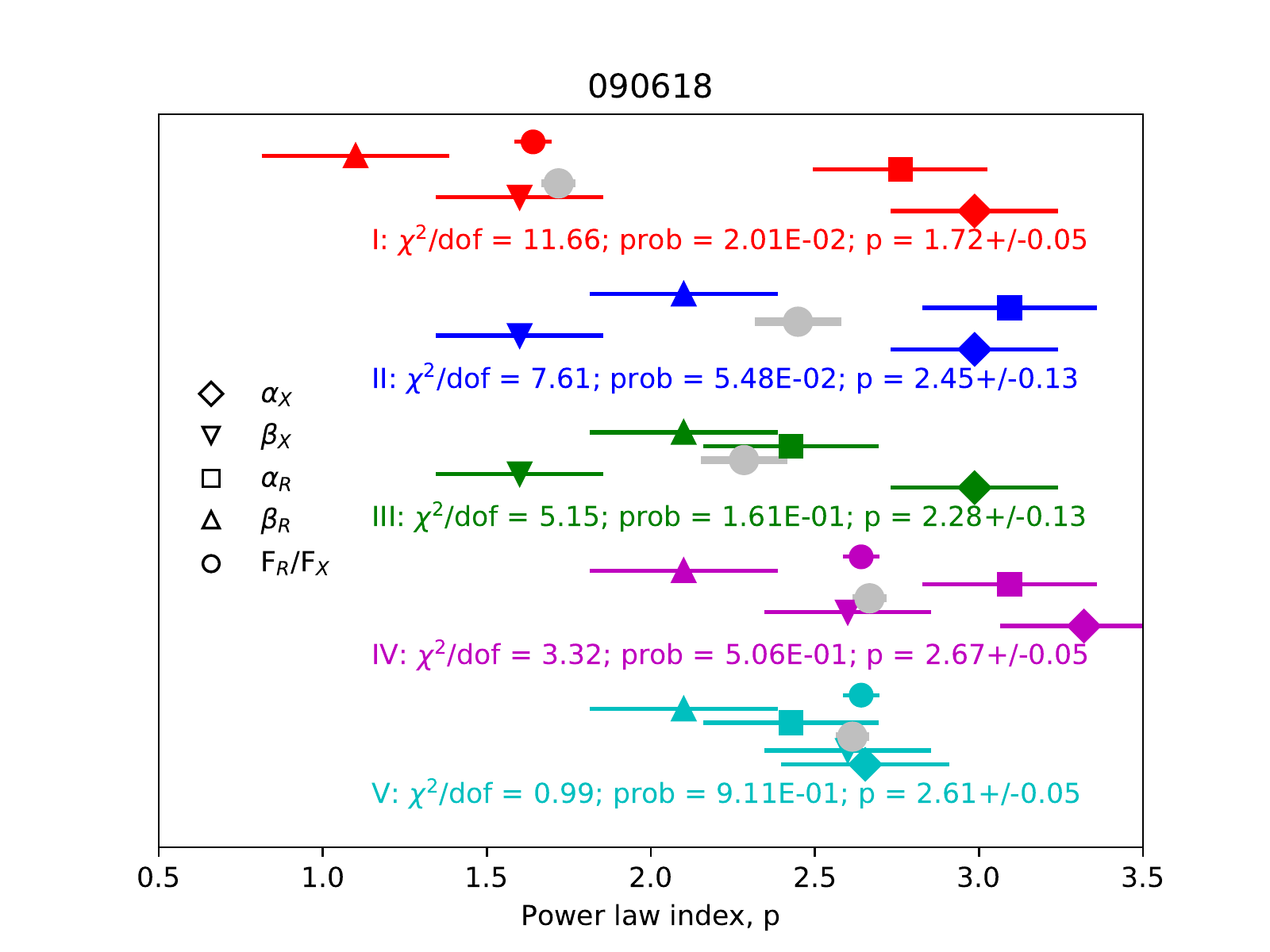}
\includegraphics[width=8.9cm]{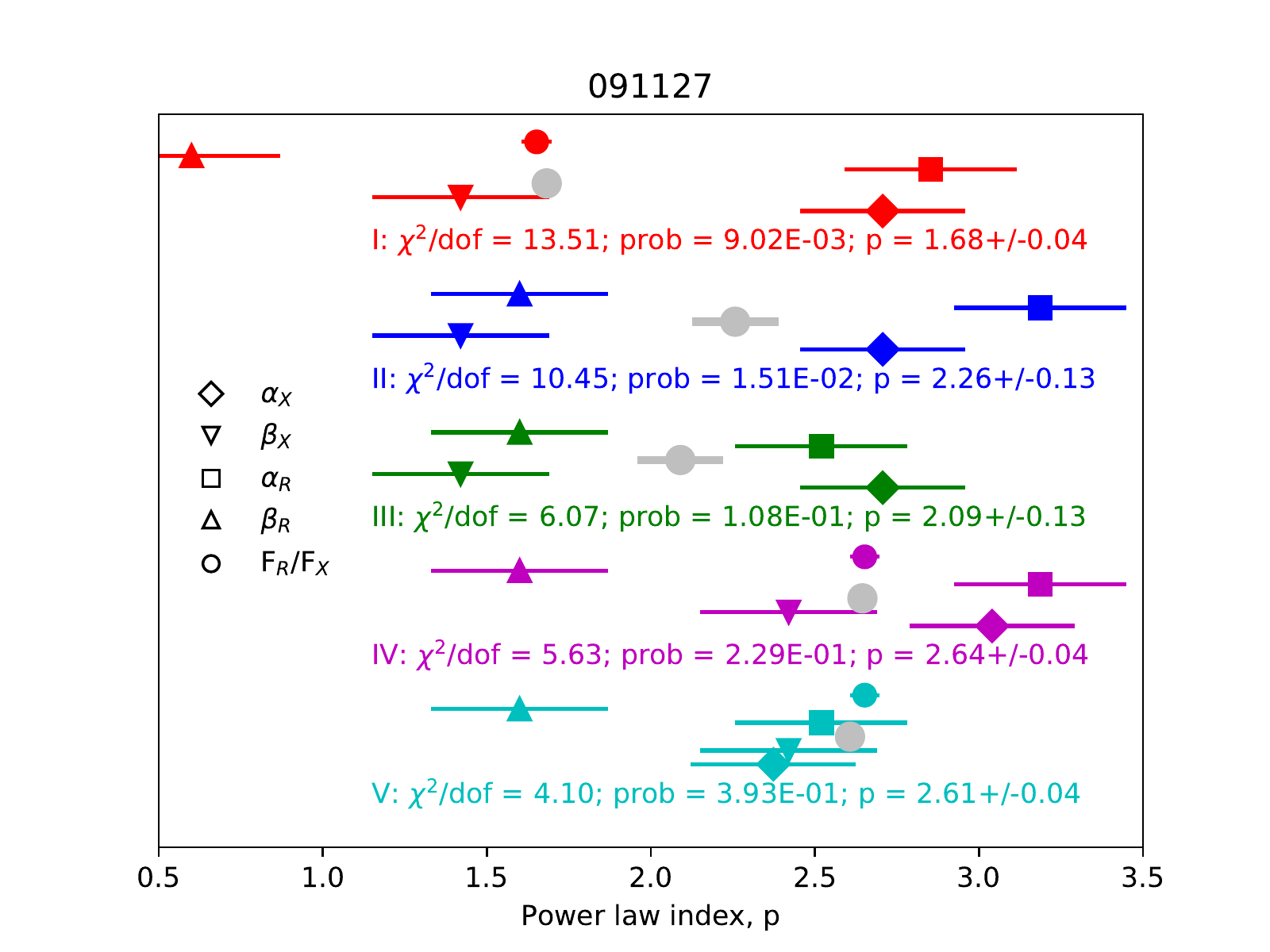}
\includegraphics[width=8.9cm]{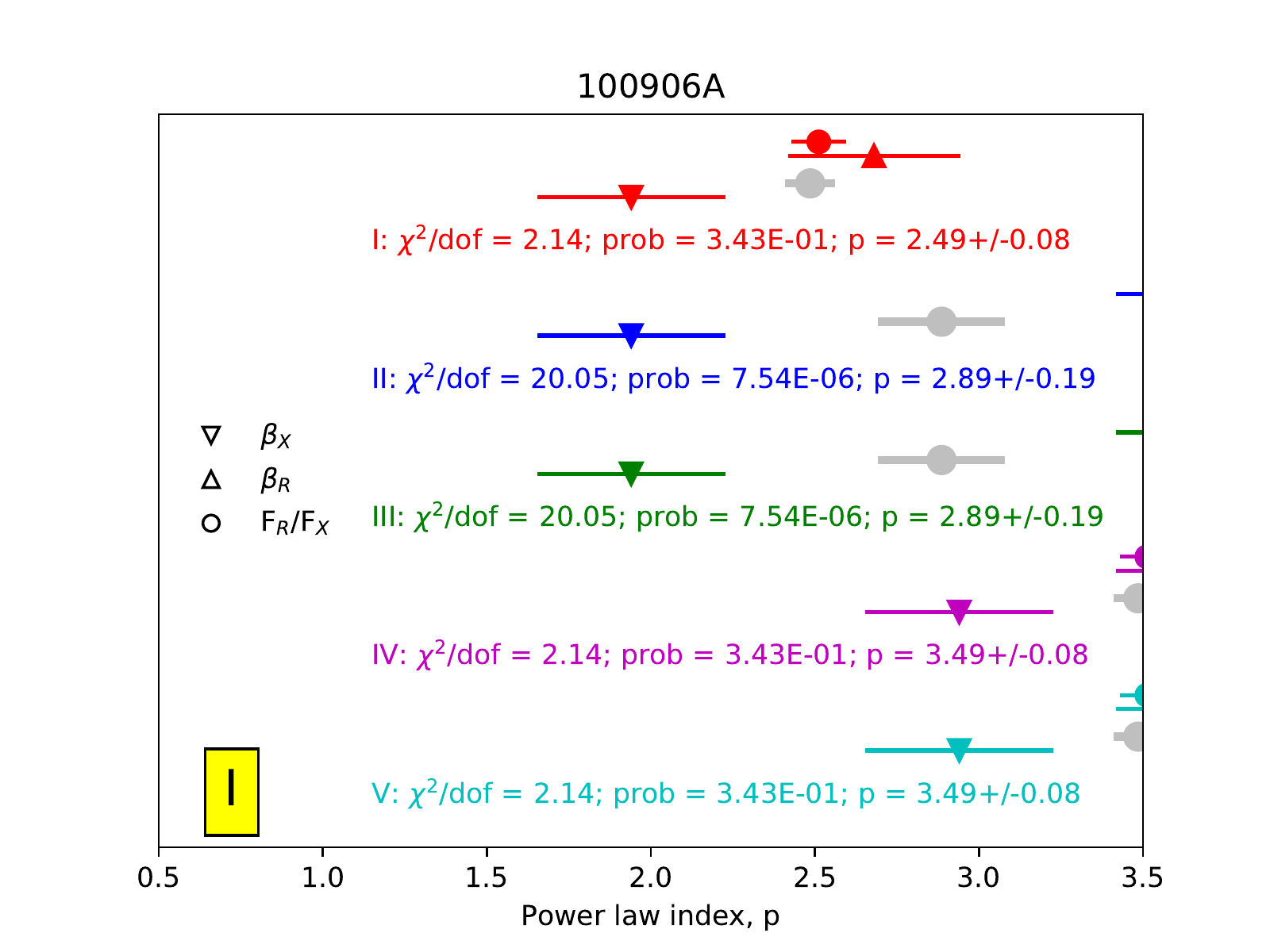}
\includegraphics[width=8.9cm]{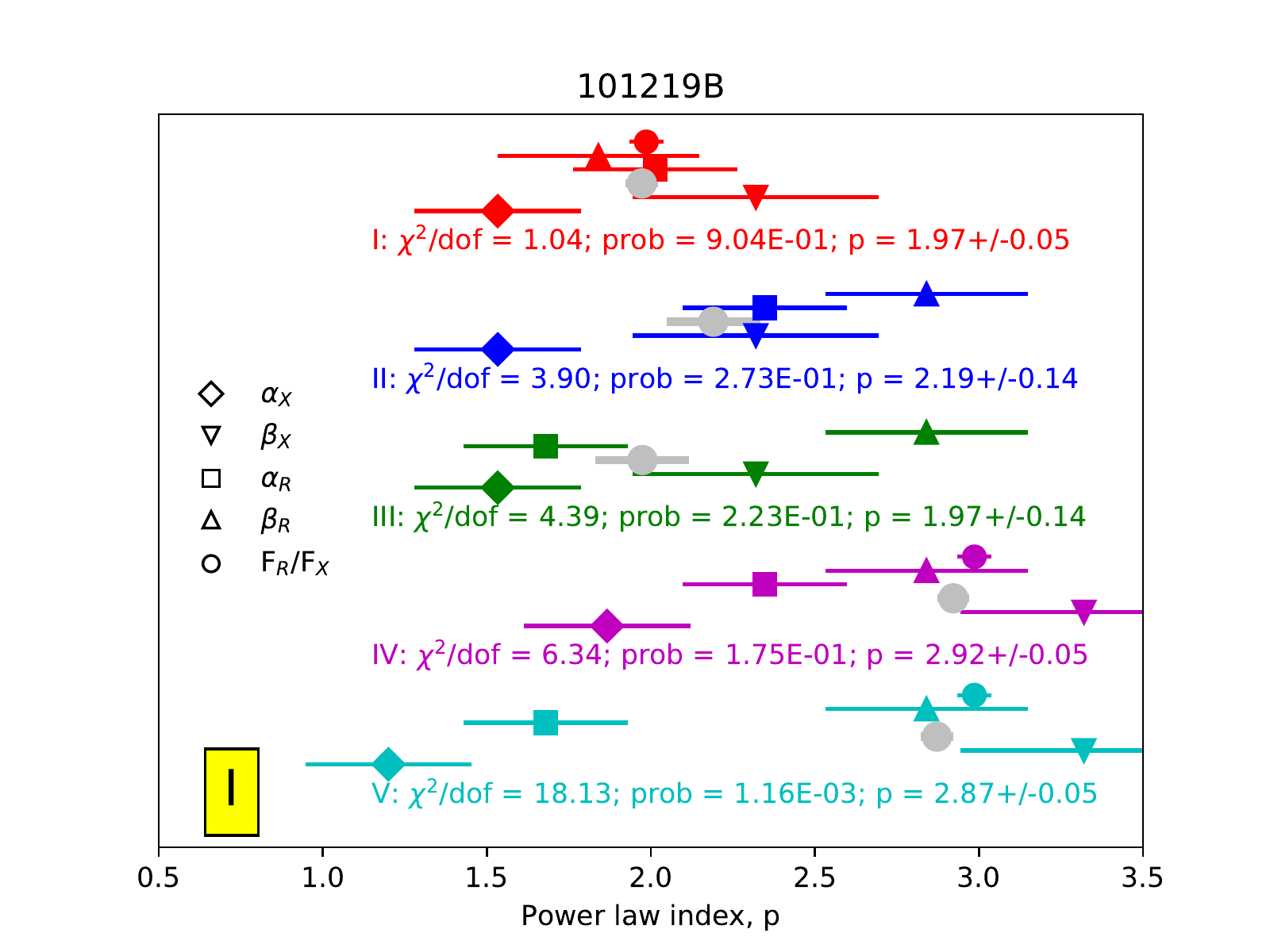}
\includegraphics[width=8.9cm]{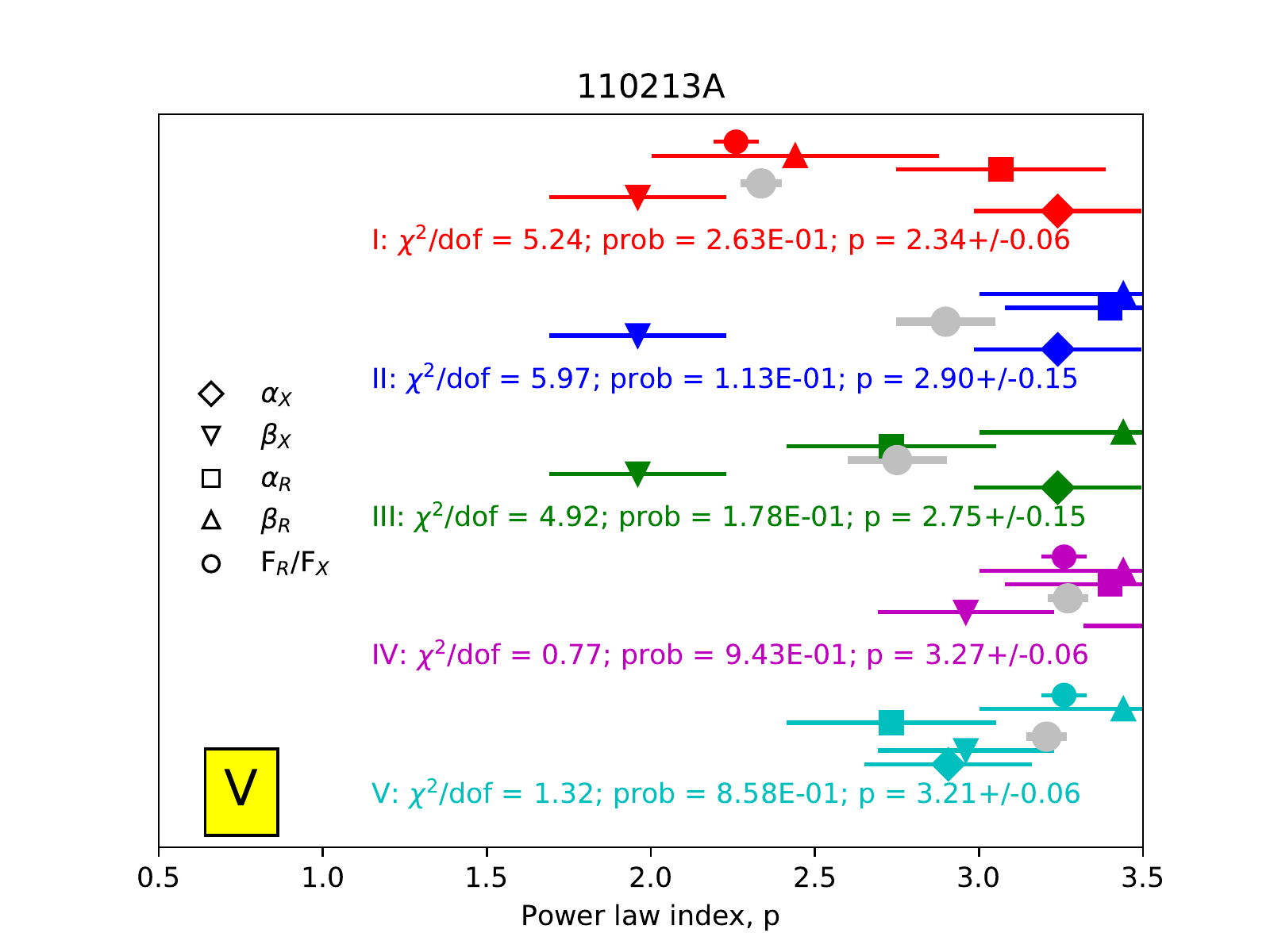}
\includegraphics[width=8.9cm]{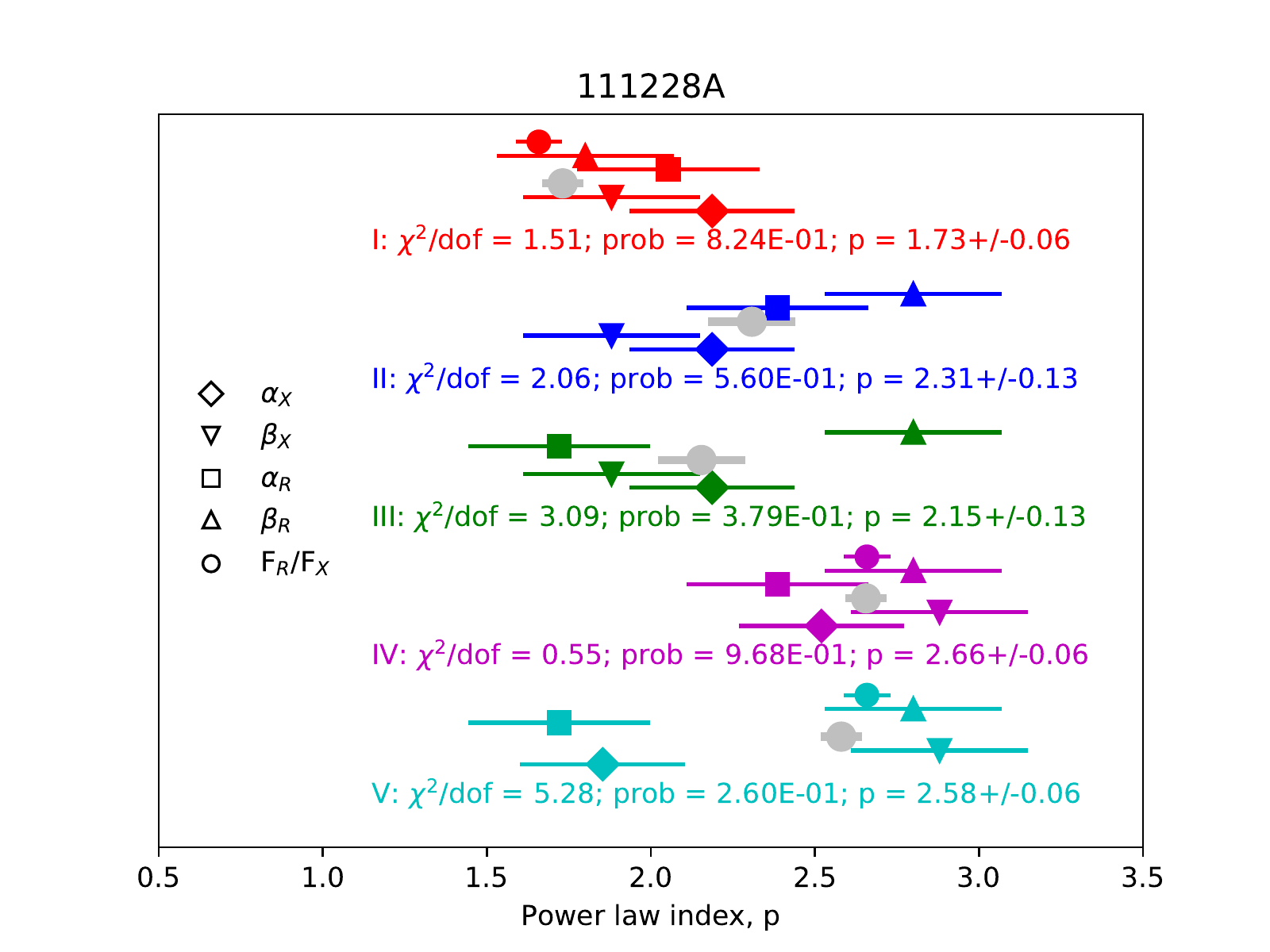}
\includegraphics[width=8.9cm]{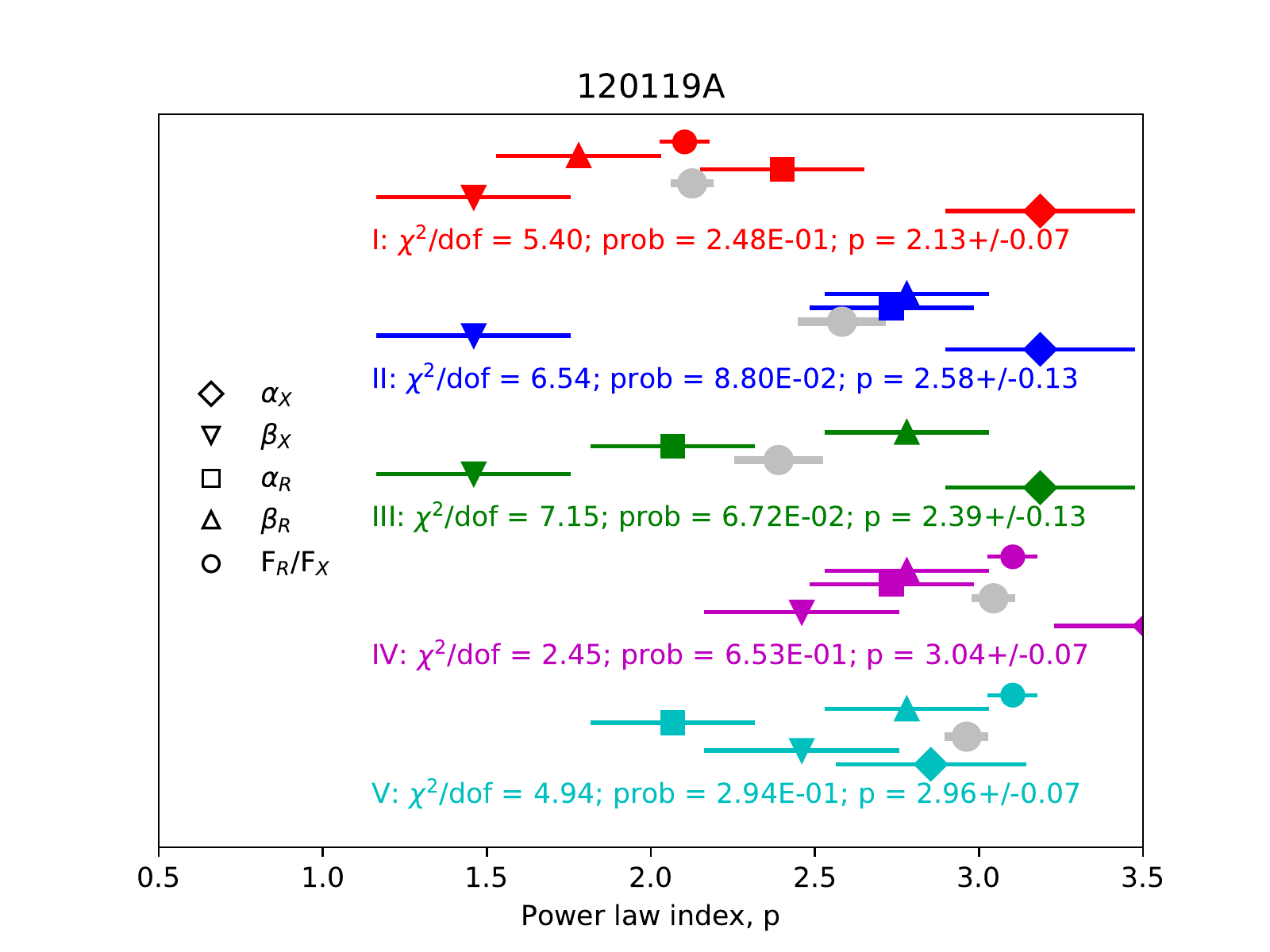}
\includegraphics[width=8.9cm]{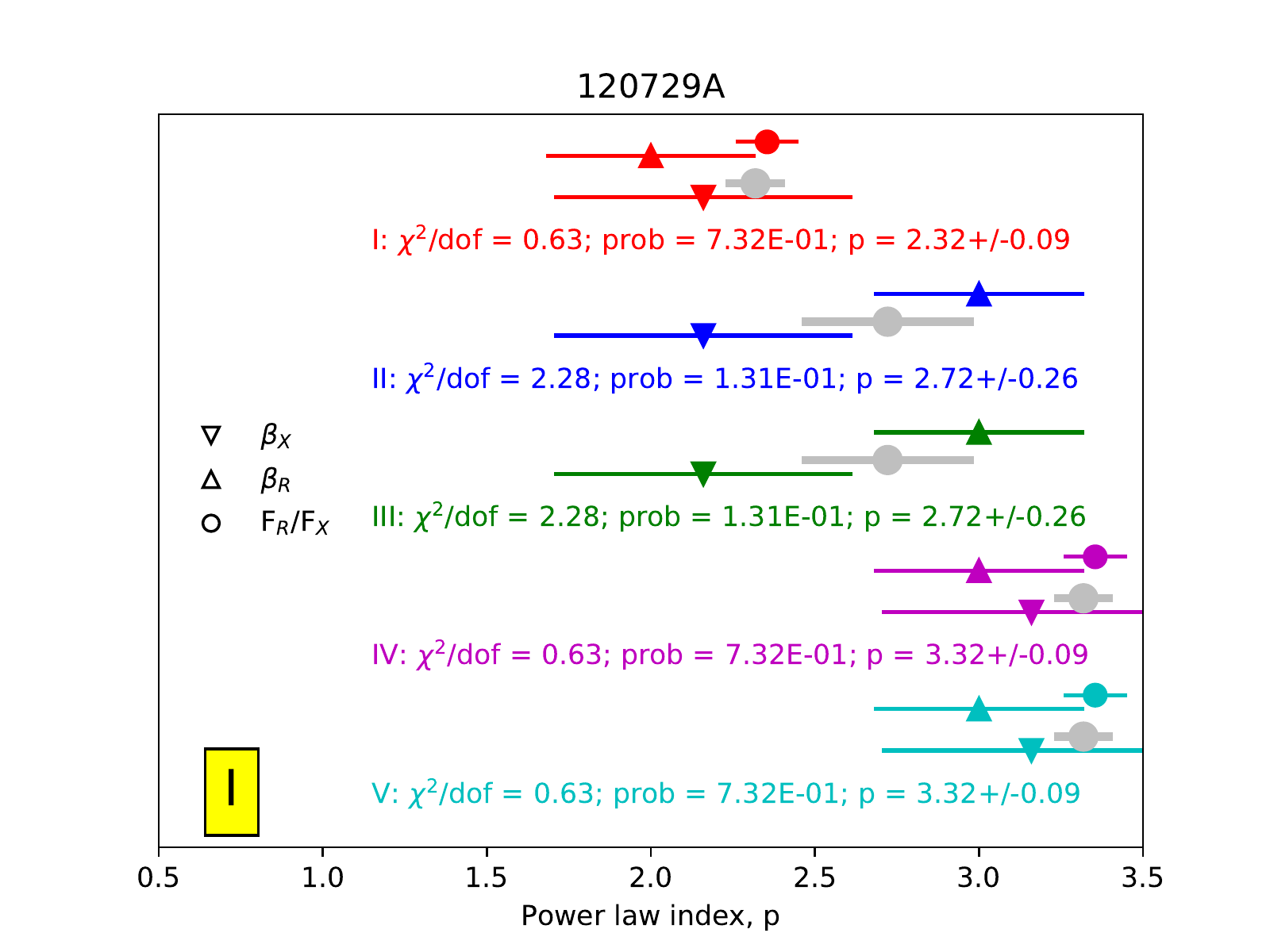}
\includegraphics[width=8.9cm]{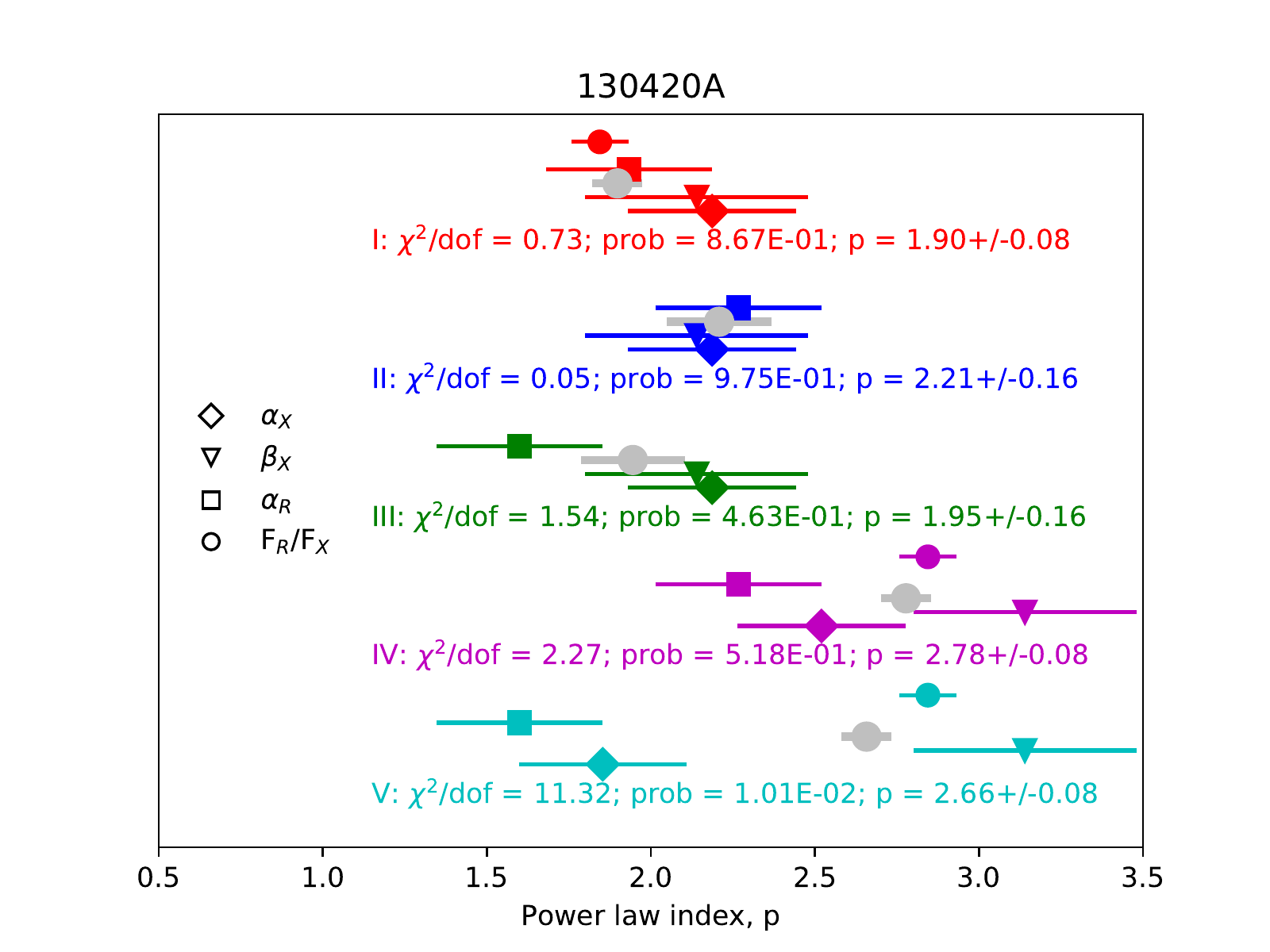}
\includegraphics[width=8.9cm]{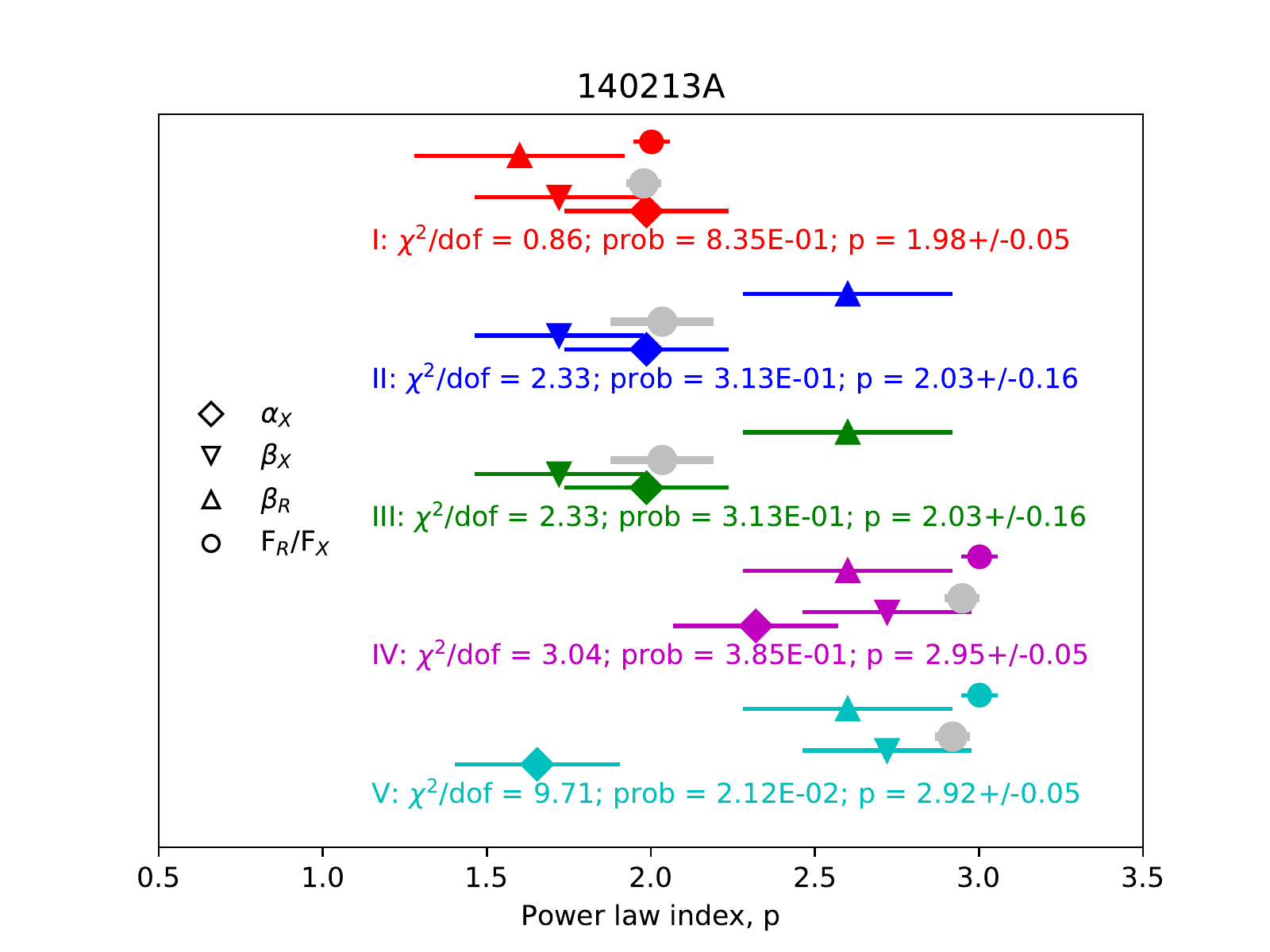}
\includegraphics[width=8.9cm]{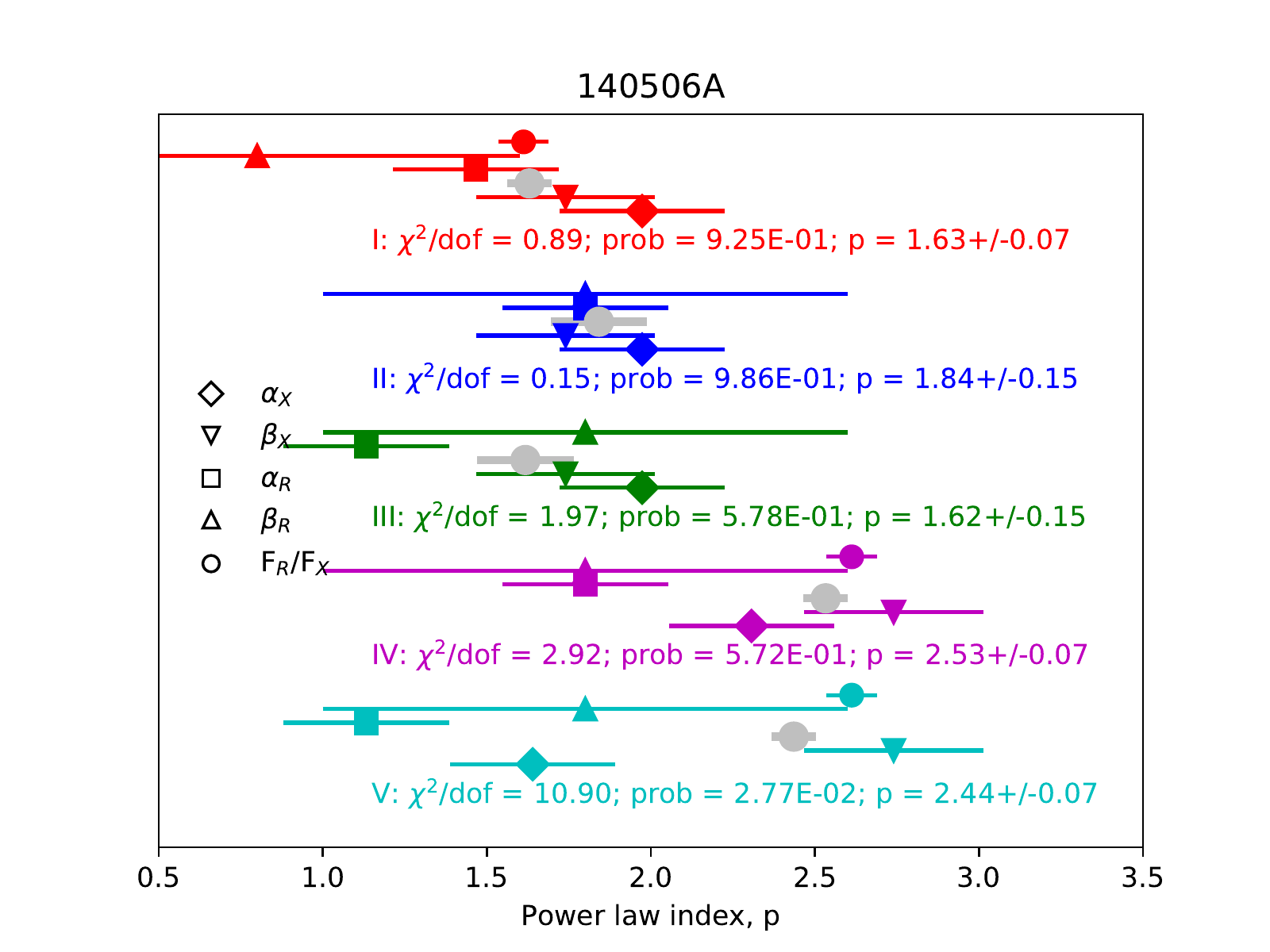}
\includegraphics[width=8.9cm]{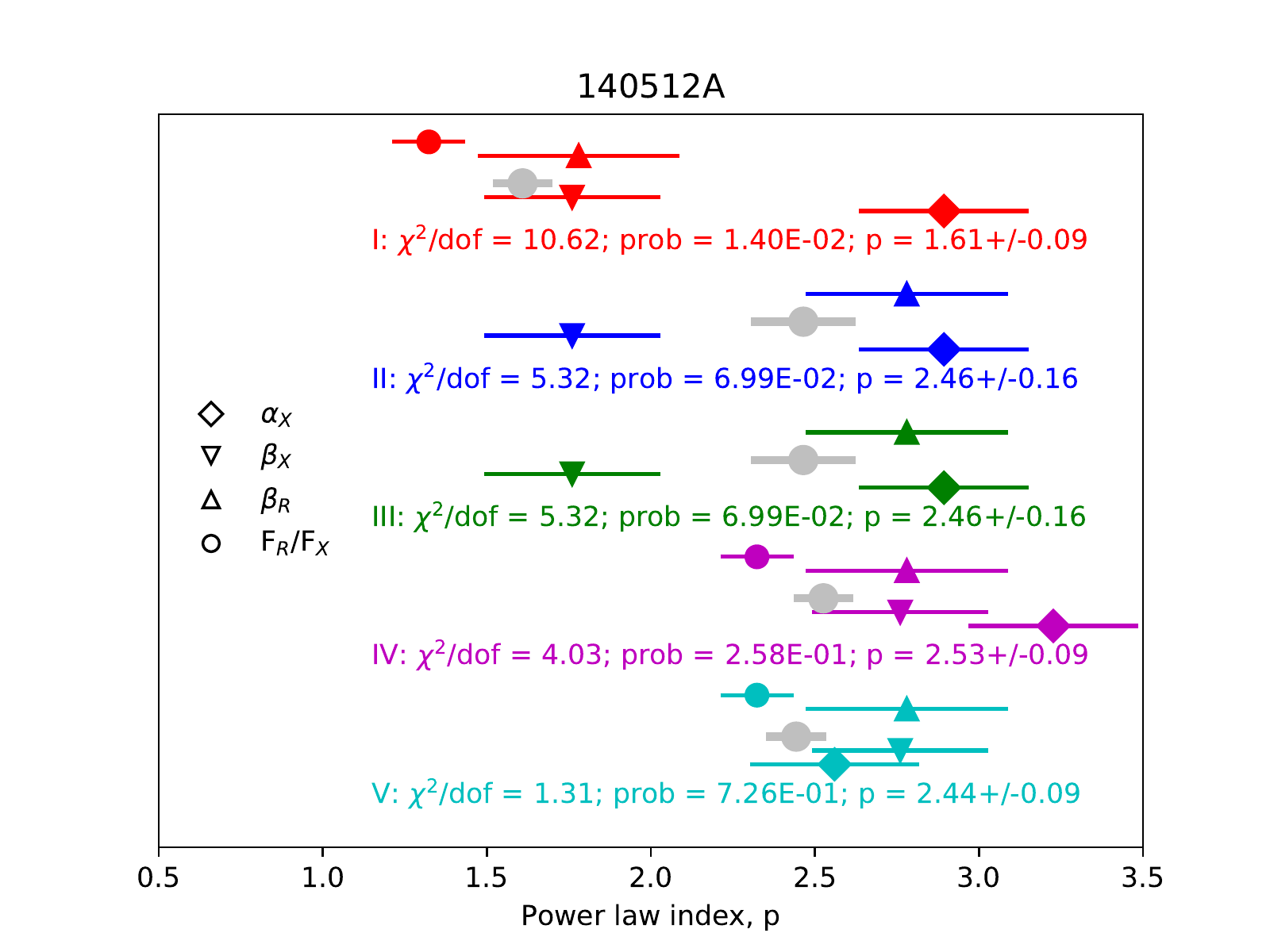}
\includegraphics[width=8.9cm]{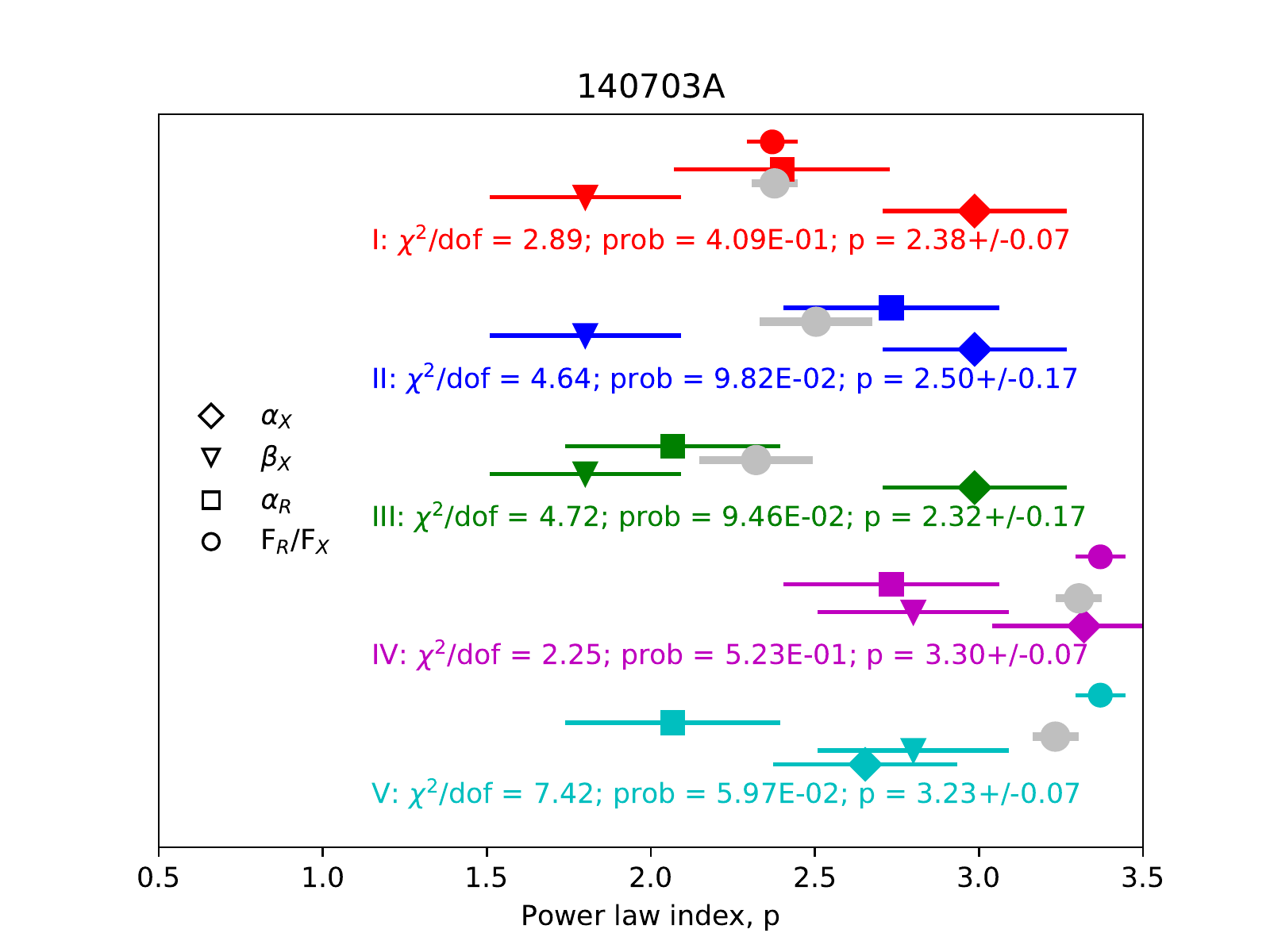}
\includegraphics[width=8.9cm]{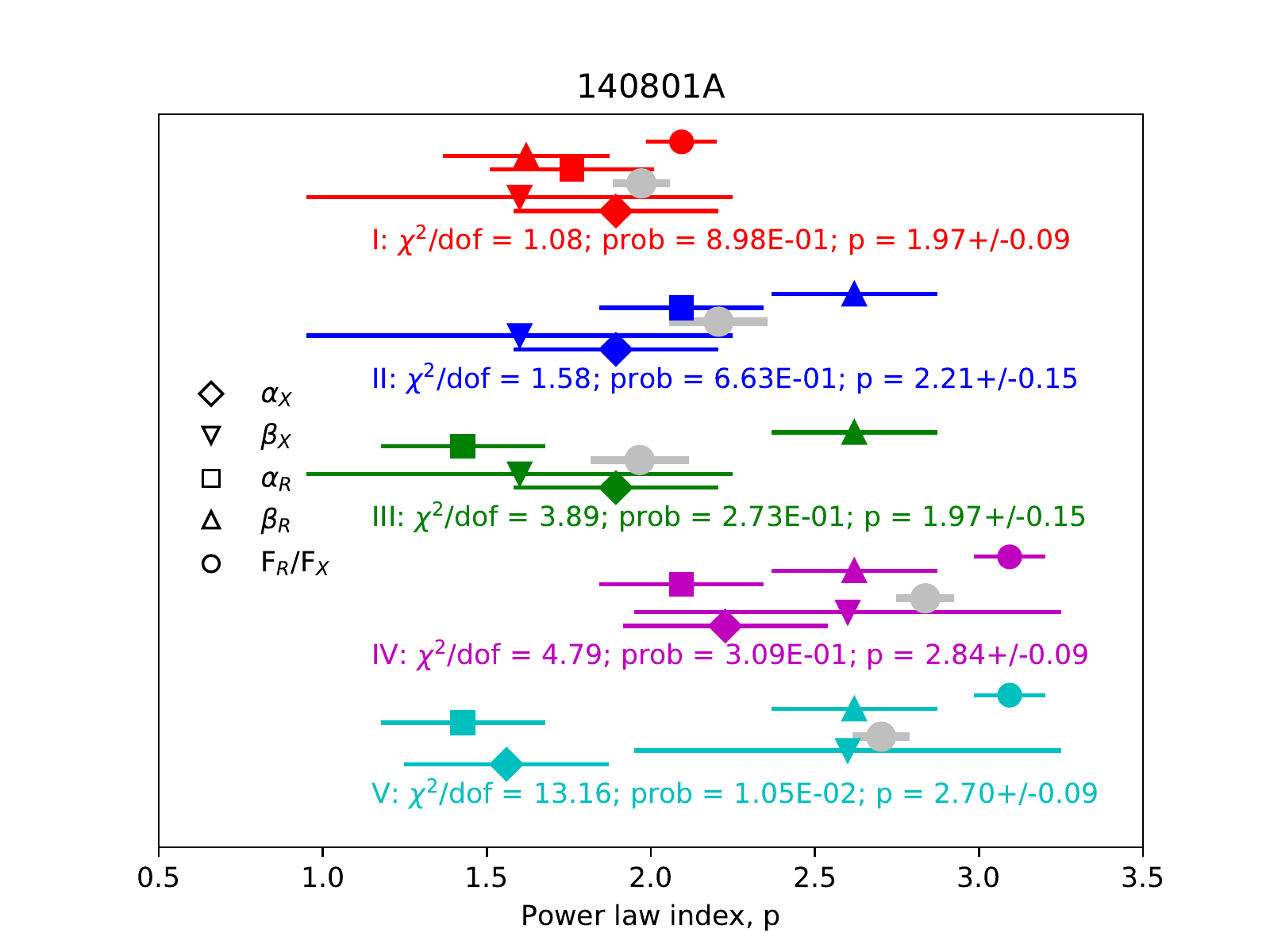}
\includegraphics[width=8.9cm]{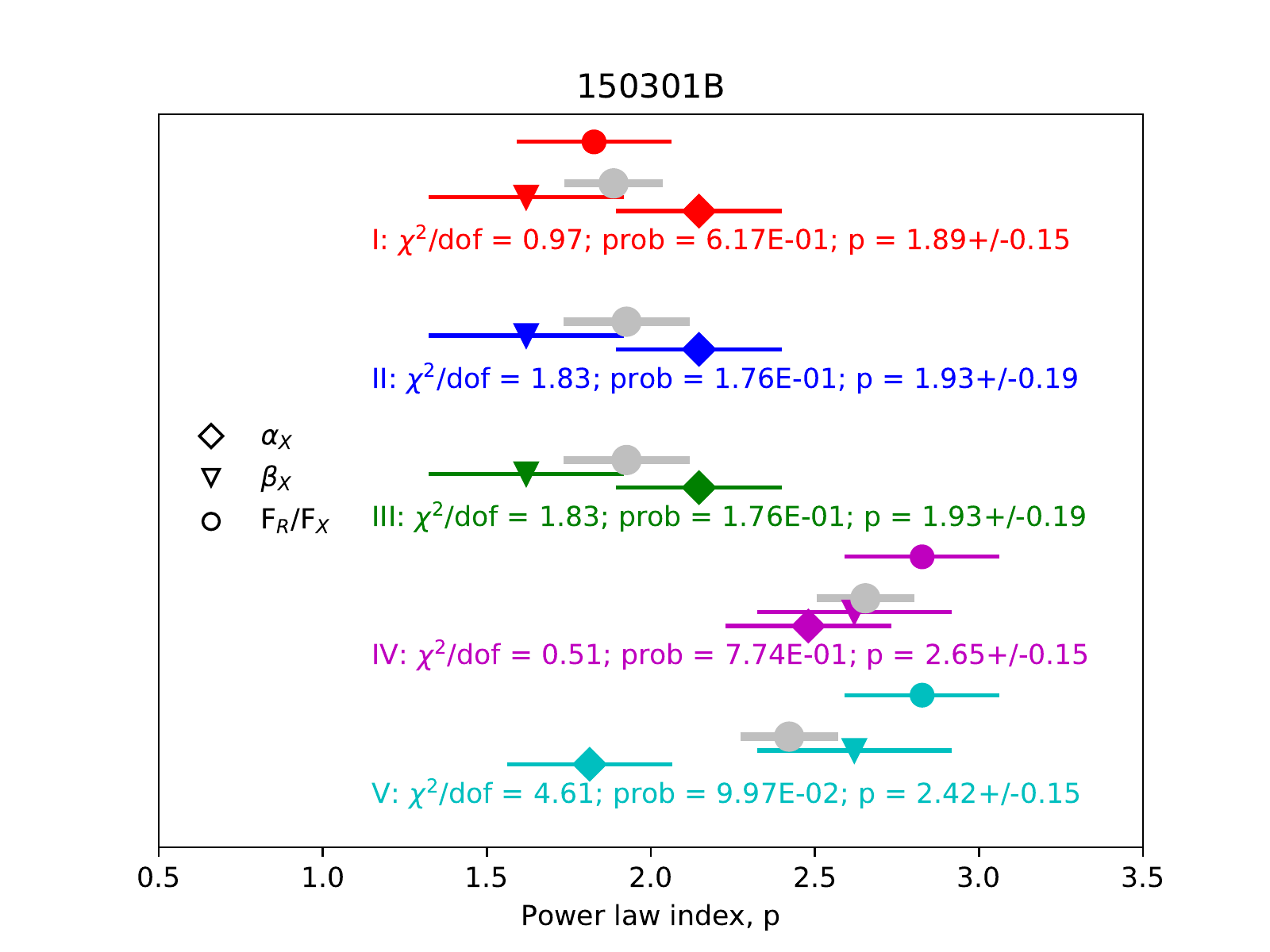}
\includegraphics[width=8.9cm]{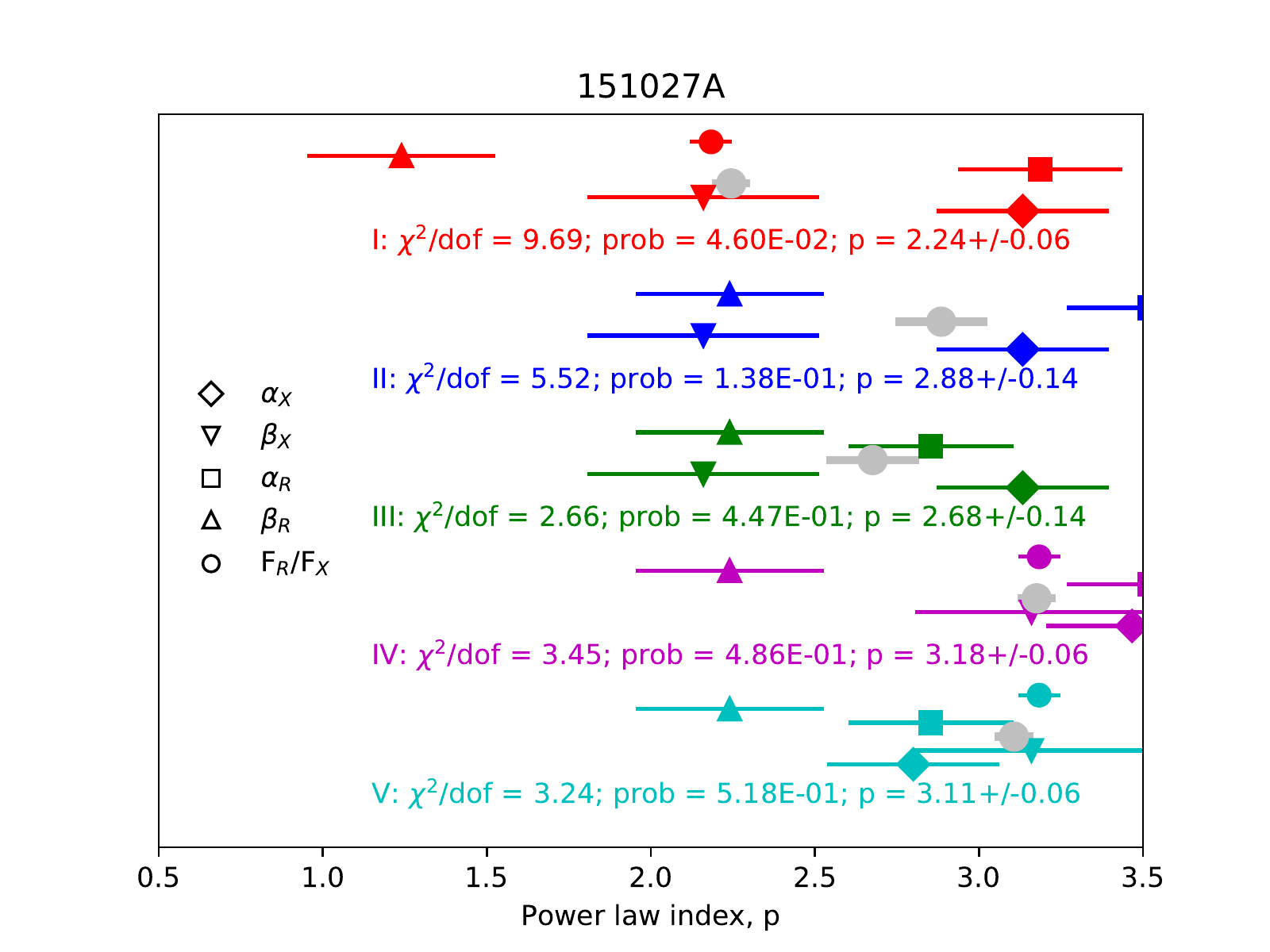}
\includegraphics[width=8.9cm]{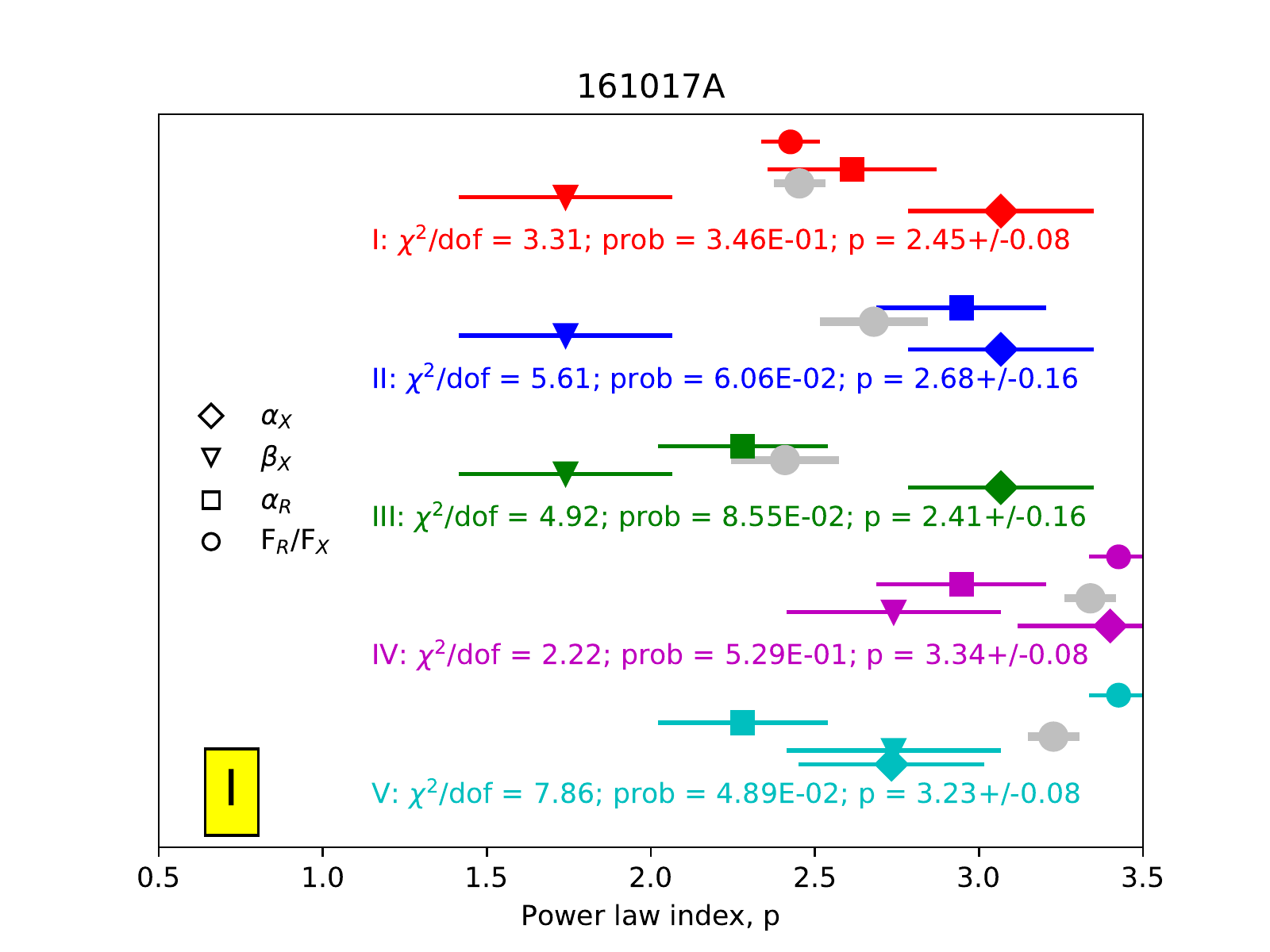}
\includegraphics[width=8.9cm]{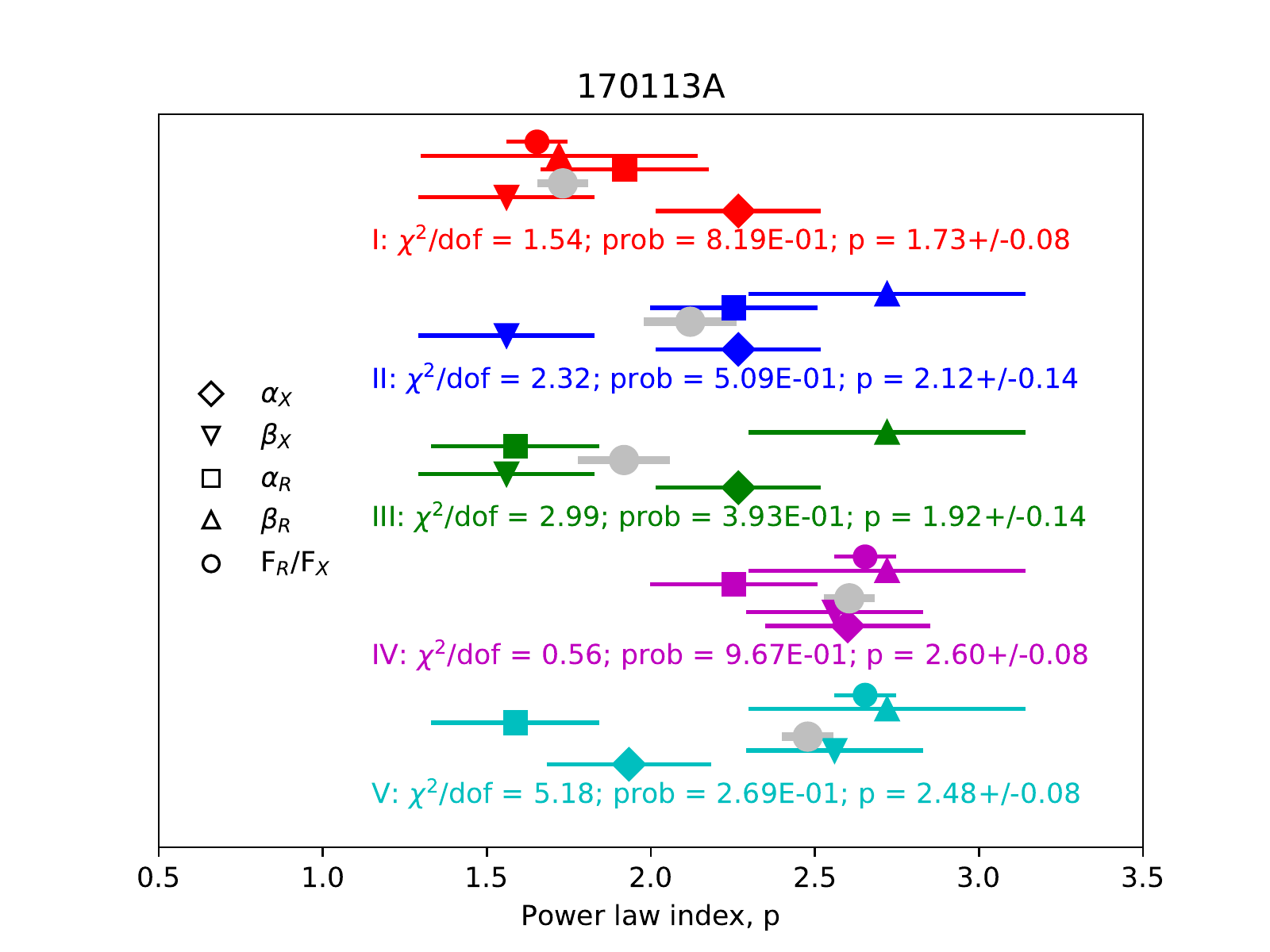}

\end{document}